%% file: P2CDR.tex
\documentclass[epj]{svjour}

\pdfoutput=1

\usepackage{graphicx}
\usepackage{xspace}
\usepackage{amsmath}
\usepackage{amssymb}
\usepackage{booktabs}
\usepackage{siunitx}
\usepackage{color}
\usepackage{url}
\usepackage[section]{placeins}

\newcommand{\sw}{$\sin^2\theta_W$\xspace}
\newcommand{\ICDs}{integrating Cherenkov detectors\xspace}

\begin{document}
\title{The P2 Experiment}
\subtitle{A future high-precision measurement of the electroweak
mixing angle at low momentum transfer}
\author{        Dominik Becker\inst{1,2} \and
				Razvan Bucoveanu\inst{1,3} \and
				Carsten Grzesik\inst{1,2} \and
				Kathrin Imai\inst{1,2} \and
				Ruth Kempf\inst{1,2} \and
				Matthias Molitor\inst{1,2} \and
				Alexey Tyukin\inst{1,2} \and
				Marco Zimmermann\inst{1,2}  \and
                David Armstrong\inst {4} \and
                Kurt Aulenbacher\inst{1,2,5} \and
				Sebastian Baunack\inst{1,2} \and
                Rakitha Beminiwattha\inst{6} \and
				Niklaus Berger\inst{1,2} \and
				Peter Bernhard\inst{1,7} \and
				Andrea Brogna\inst{1,7} \and
                Luigi Capozza\inst{1,2,5} \and
				Silviu Covrig Dusa\inst{8} \and
		        Wouter Deconinck\inst{4} \and
                J\"urgen Diefenbach\inst{1,2} \and
                                 James Dunne\inst{9} \and
				Jens Erler\inst{10} \and
                Ciprian Gal \inst{11} \and
				Michael Gericke\inst{12} \and
                Boris Gl\"aser \inst{1,2} \and
				Mikhail Gorchtein\inst{1,2} \and
                Boxing Gou \inst{1,2,5} \and
				Wolfgang Gradl\inst{1,2} \and
                Yoshio Imai \inst{1,2} \and
				Krishna S. Kumar\inst{13} \and
				Frank Maas\inst{1,2,5,}\thanks{Corresponding author, \texttt{maas@uni-mainz.de}} \and
                Juliette Mammei\inst{12} \and
                Jie Pan\inst{12} \and
                Preeti Pandey\inst{12} \and
                Kent Paschke\inst{11} \and
				Ivan Peri\'c\inst{14} \and
                Mark Pitt\inst{15} \and
                Sakib Rahman\inst{12} \and
				Seamus Riordan\inst{16} \and
 				David Rodr\'iguez Pi\~neiro\inst{1,2,5} \and
				Concettina Sfienti\inst{1,2,5,7} \and
				Iurii Sorokin\inst{1,2} \and
				Paul Souder\inst{17} \and
				Hubert Spiesberger\inst{1,3} \and
				Michaela Thiel\inst{1,2} \and
				Valery Tyukin\inst{1,2} \and
				Quirin Weitzel\inst{1,7}
}                     
%
%
\institute{                          
						PRISMA Cluster of Excellence, Johannes Gutenberg-Universit\"at Mainz, Germany    
          \and Institute of Nuclear Physics, Johannes Gutenberg-Universit\"at Mainz, Germany 
					\and Institute of Physics, Johannes Gutenberg-Universit\"at, Mainz, Germany 
					\and College of William and Mary, Williamsburg,  Virginia, USA 
					\and Helmholtz Institute Mainz, Johannes Gutenberg-Universit\"at Mainz, Germany 
					\and Louisiana Tech University, Ruston, Louisiana, USA 
					\and Detector Laboratory, PRISMA Cluster of Excellence, Johannes Gutenberg-Universit\"at Mainz, Germany 
					\and Thomas Jefferson National Accelerator Facility, Newport News, Virginia, USA 
					\and Mississippi State University, Mississippi State, MS, USA
					\and Departamento de F\'isica Te\'orica, Instituto de F\'isica, Universidad Nacional Aut\'onoma de M\'exico, CDMX, M\'exico 
					\and University of Virginia, Charlottesville, Virginia, USA 
					\and Department of Physics and Astronomy, University of Manitoba, Winnipeg, Canada 
                                         \and Department of Physics and Astronomy, Stony Brook University, Stony Brook, USA 
					\and Institute for Data Processing and Electronics, Karlsruhe Institute of Technology, Karlsruhe, Germany 
					\and Virginia Tech University, Blacksburg, Virginia, USA 
					\and Physics Division, Argonne National Laboratory, Argonne, USA 
					\and Physics Department, Syracuse University, Syracuse, USA
}                                        
\date{Received: date / Revised version: date}
%
\abstract{
This article describes the research and development work for 
the future P2 experimental facility at the upcoming MESA 
accelerator in Mainz. The facility is optimized for the detection 
of order $10^{-8}$ parity-violating cross section asymmetries in 
electron scattering. The physics program of the facility comprises 
indirect, high precision search for physics beyond the Standard 
Model, measurement of the neutron distribution in nuclear physics, 
single-spin asymmetries stemming from two-photon exchange and a 
possible future extension to the measurement of hadronic parity 
violation. 
\\
The first measurement of the P2 experiment for which the research 
and development work is most advanced is described here in detail. 
It aims for a high precision determination of the weak mixing 
angle \sw to a precision of \SI{0.14}{\percent} at a four-momentum 
transfer of $Q^2 = 4.5\times10^{-3}$~GeV$^2$. The accuracy is 
comparable to existing measurements at the $Z$ pole. 
It comprises a sensitive test of the standard model up to a mass 
scale of \SI{50}{TeV}, extendable to \SI{60}{TeV}. 
This requires a measurement of the parity violating cross section 
asymmetry $\langle A^\text{exp}\rangle = -39.94 \times 10^{-9}$ 
in the elastic electron-proton scattering with a total accuracy 
of $\Delta \langle A^\text{exp}\rangle_\text{Total} = 0.56 \times 
10^{-9} $ (1.4\%) in \SI{10000}{h} of \SI{150}{\micro A} polarized 
electron beam impinging on a \SI{60}{cm} $\ell H_2$ target 
allowing for an extraction of the weak charge of the proton 
which is directly connected to the weak mixing angle \sw. 
Contributions from $\gamma Z$-box graphs become small at the 
small beam energy of $E_\text{beam} = \SI{155}{MeV}$. 
\\
The P2 asymmetry is smaller than any asymmetry measured so far 
in electron scattering with an unprecedented goal for the accuracy. 
The use of a solenoid-spectrometer with \SI{100}{\percent} 
$\phi$-acceptance as well as an atomic H trap polarimeter are 
some new features, which have never before been used in 
parity-violation experiments, and which we describe among 
others, here. In order to collect the enormous statistics 
required for this measurement, the new Mainz Energy Recovery 
Superconducting Accelerator (MESA) is under construction. The 
plans on the associated beam control system and the polarimetry 
is described in this article as well. A new $\ell H_2$ high-power 
target design with an enormously low noise level of \SI{10}{ppm} 
needs to be constructed. We report here in addition on the 
conceptual design of the P2 spectrometer, its Cherenkov detectors, 
the integrating read-out electronics as well as the ultra-thin, fast tracking detectors. 
There has been substantial theory work done 
in preparation of the determination of \sw. The further physics 
program in particle and nuclear physics is described here as well.
}

\maketitle
\tableofcontents
\section{Introduction and physics motivation}
\label{sec:Intro}
\input{introduction}


\section{Determining the Weak Mixing Angle from Parity Violating Electron Scattering}
\label{sec:SinThetaW}
\input{SinThetaW}

\section{The MESA Accelerator}
\label{sec:Accelerator}
\input{accelerator}

\subsection{Polarized source}
\label{sec:Source}
\input{source}

\subsection{Polarimetry}
\label{sec:Polarimetry}
\input{polarimetry}

\subsection{Beam control}
\label{sec:Beam control}
\input{beamcontrol}

\section{High Power Liquid Hydrogen Target}
\label{sec:Target}
\input{target}

\section{The P2 Spectrometer}
\label{sec:Spectrometer}
\input{spectrometer}

\subsection{Monte Carlo simulations}
\label{sec:Simulation}
\input{simulation}

\subsection{Integrating Cherenkov detectors}
\label{sec:IntegratingDetectors}
\input{cherenkovdetector}

\subsection{High resolution ADCs}
\label{sec:HighResolutionADCs}
\input{highresolutionadc}

\subsection{Tracking detectors}
\label{sec:TrackingDetectors}
\input{tracker}

\section{Theory input}
\label{sec:Theory}
\input{theory}

\subsection{Box graph and hadronic uncertainties}
\label{sec:Boxgraph}
\input{boxgraph}

\subsection{QED corrections}
\label{sec:QED}
\input{qedcorrections}

\subsection{Theory summary}
\label{sec:Theorysummary}
\input{theorysummary.tex}

\section{Further physics programme}
\label{sec:FurtherPhysics}

\subsection{Measurements with Carbon-12}
\label{sec:Carbon}
\input{carbon}

\subsection{Neutron skin measurement}
\label{sec:NeutronSkin}
\input{neutronskin}

\subsection{Backward angle measurement}
\label{sec:BackwardAngle}
\input{backwardangle}

\section{Conclusions and Outlook}
\label{sec:Summary}
\input{summary}

\section{Acknowledgements}
\label{sec:Acknowledgements}
\input{Acknowledgements}

\appendix

\FloatBarrier
\section{Nucleon form factor fit parameters}
\label{sec:NucleonFormFactorFitParameters}
\input{SinThetaW_appendix}
\FloatBarrier

%
 \bibliographystyle{unsrt_collab_comma}
 \bibliography{P2CDR}
%
%
%

\end{document}

%% file: introduction.tex

In the Standard Model of Elementary Particle Physics (SM) the 
weak interaction is the only force that violates parity. 
Over the past 30 years, the measurement of parity violation 
in weak interactions has been a well established experimental 
technique in atomic as well as particle and nuclear physics. 
The violation of parity had been postulated by the theoreticians 
Lee \& Yang in 1956 \cite{Lee:1956qn}. It was proven to be an 
experimental fact in nuclear physics in $1957$ in the course of the
Wu experiment \cite{Wu:1957my} by a careful analysis of the beta-decay 
of $^{60}\mathrm{Co}$. In addition Garwin, Lederman and Weinrich had 
shown that the $\mu$-decay violates parity \cite{Garwin:1957hc}. 
As first pointed out by Zeldovich in 1959~\cite{bib:Zeldovich:59}, 
the existence of a neutral partner of the charged weak interaction 
responsible for $\beta$-decay, should lead to observable parity 
violation in atomic physics and in electron scattering. These 
ideas preceded the development of the electroweak theory, and 
were confirmed experimentally by Prescott in electron scattering 
at SLAC~\cite{Prescott:1978tm} and in cesium atoms by 
Bouchiat~\cite{Bouchiat:1982um}. In the rest of this article we 
concentrate on parity violation in electron scattering. 

\begin{figure}[t]
  \centering
  \resizebox{0.45\textwidth}{!}{\includegraphics{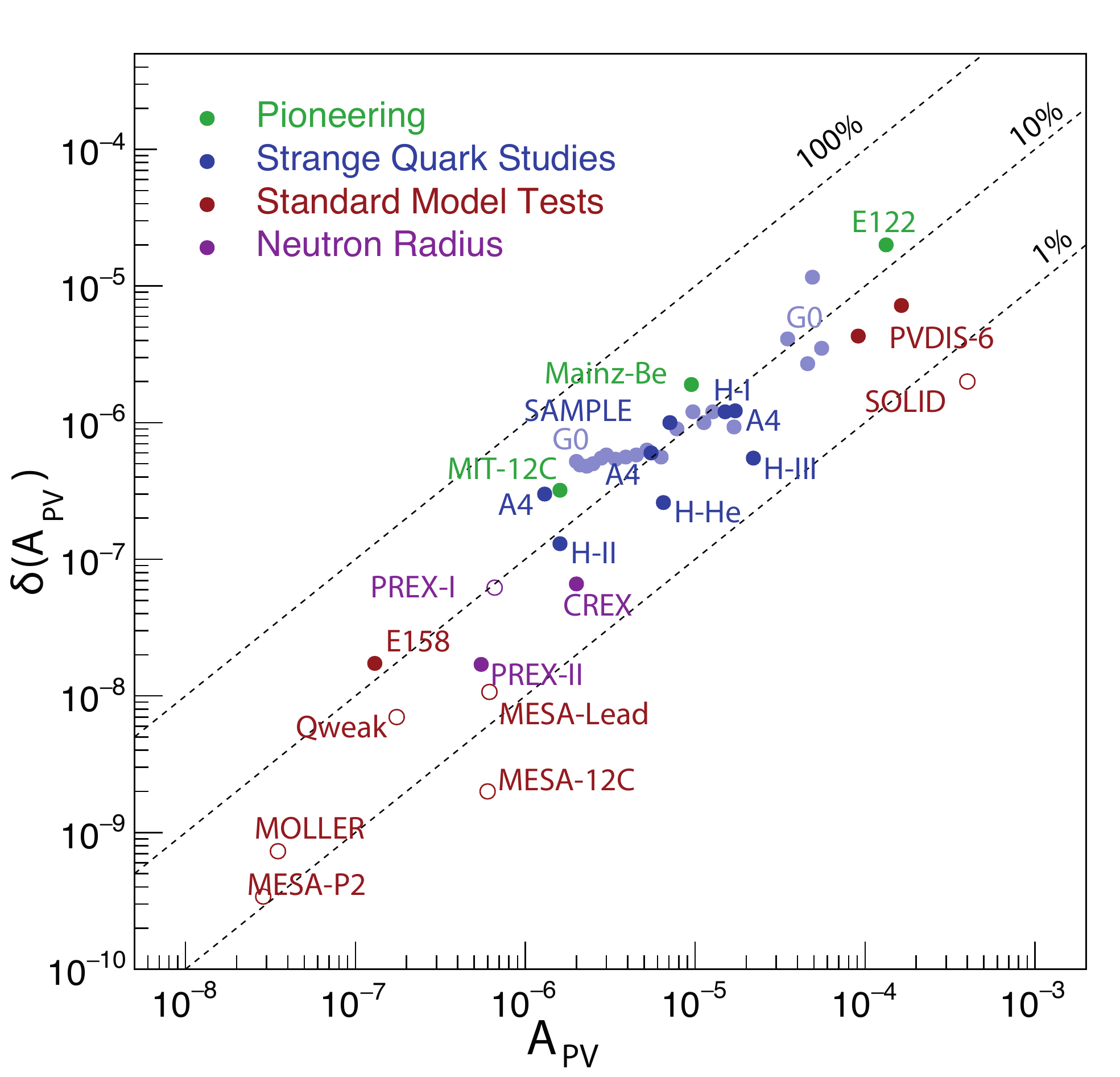}}
  \caption{
  Overview over past (full points) and future (open circles) 
  electron scattering experiments. From the very 
  early measurements at SLAC, at Bates and in Mainz up to today, 
  parity-violating electron scattering has become a well established 
  technique to explore hadron physics, nuclear physics and particle 
  physics, depending on kinematics and target. The point labelled 
  MESA-P2 is the P2 experiment at the MESA accelerator employing 
  a $\ell H_2$-Target. The point labeled MESA-12C denotes the P2 
  experimental facility with a $^{12}$C target. 
  }
  \label{fig:parity_experiments}
\end{figure}

Since then, many parity-violating electron scattering experiments 
have been performed, all summarized in Fig.~\ref{fig:parity_experiments}. 
Prescott's experiment was followed by an experiment of the Mainz group 
of Otten and Heil~\cite{Heil:1989dz} and another one at MIT-Bates on 
a $^{12}$C target~\cite{Souder:1990ia}. Their experimental 
techniques were pioneering and are used still today. They were also 
ground-breaking in establishing parity-violation and making the first 
measurements of SM parameters from electron scattering (see the green 
points in Fig.~\ref{fig:parity_experiments} labeled ``Pioneering''). 

It was first pointed out by Kaplan and Manohar in 
1988~\cite{Kaplan:1988ku} that one can get 
access to a possible contribution of strange quarks to the 
electromagnetic form factors of the nucleon by measuring its 
weak electric and magnetic form factors in parity-violating 
electron scattering. This triggered a whole series of parity-violation 
electron scattering experiments at the MIT-Bates accelerator 
\cite{Mueller:1997mt,Spayde:1999qg,Wells:2000rx,Hasty:2001ep,Spayde2004,Ito:2003mr}, at 
the MAMI accelerator in Mainz \cite{Maas:2004ta,Baunack:2011xva,Achenbach:1998yv,Achenbach:2001kf,Maas:2004pd,Maas:2004dh,Hammel:2005sr,Altarev:2005gp,Baunack:2009gy,BalaguerRios:2016ftd,Rios:2017vsw}
as well as at JLab's CEBAF in Newport 
News \cite{Aniol:1998pn,Falletto:2000mu,Aniol:2000at,Aniol:2004hp,Aniol:2005zf,Aniol:2005zg,Acha:2006my,Ahmed:2011vp,Armstrong:2005hs,Armstrong:2007vm,Androic:2009aa,Androic:2011rha,Androic:2011rh,G0:2011aa} (see in addition  \cite{Musolf:1993tb,Kumar:2000eq,Lyubovitskij:2002ng,Kumar:2013yoa,Maas:2017snj} for 
review articles, blue points in Fig.~\ref{fig:parity_experiments} 
labeled ``Strange Quark Studies''). An accurate measurement of the 
neutron distribution in heavier nuclei and especially the so called 
``neutron skin'' can be obtained from parity-violating electron 
scattering on heavy nuclei like lead \cite{Abrahamyan:2012gp,Abrahamyan:2012cg}. The associated parity-violation 
experiments are labeled ``Neutron Radius'' in Fig.~\ref{fig:parity_experiments}.
In recent years, experiments have been performed and new proposals have been worked out to measure the weak charge 
of the proton or of the electron, or the ratio of quark charges. Those are 
labeled ``Standard Model Tests'' in Fig.~\ref{fig:parity_experiments} \cite{Anthony:2003ub,Anthony:2005pm,Wang:2014bba,Benesch:2014bas,Chen:2014psa}.
The parity-violating electron scattering experiments at the new Mainz 
MESA accelerator \cite{Milner:2013yua} are the subject of this manuscript.

\begin{figure}[t]
  \centering
  \resizebox{0.45\textwidth}{!}{\includegraphics{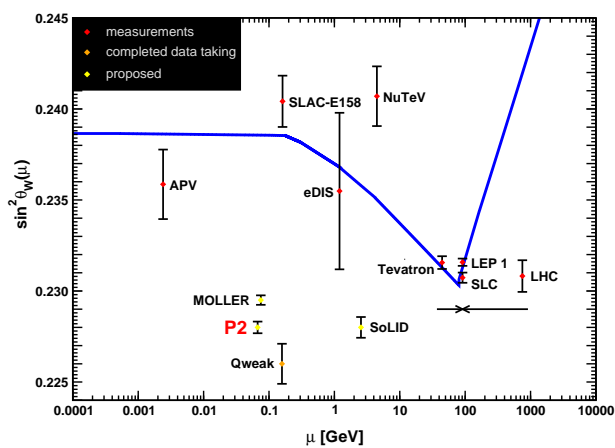}}
  \caption{
  Scale dependence of the weak mixing angle, 
  $\sin^2\hat\theta_\text{W}(\mu)$ compared with existing and 
  forthcoming measurements. 
  The anticipated QWeak point is shown with error bars 
  obtained with the full data set, but moved to an arbitrary 
  position next to the anticipated MOLLER and P2 points 
  \cite{Carlini-PANIC}.
  }
  \label{fig:s2wplot}
\end{figure}

In the P2 experiment, parity-violation in elastic elec\-tron-proton 
scattering at low momentum transfer, $Q^2$, will provide 
experimental access to the proton's weak charge $Q_\text{W}(\text{p})$, 
the analog of the electric charge which determines the 
strength of the neutral-current weak interaction. In the 
SM, $Q_\text{W}(\text{p})$ is related to the electroweak mixing angle, 
$\sin^2 \theta_\text{W}$. The weak charge of the proton is 
particularly interesting, compared to that of other 
nuclei, since it is suppressed in the SM and therefore 
sensitive to hypothetical new physics effects. The SM also 
provides a firm prediction for the energy-scale dependence 
of the running of $\sin^2 \theta_\text{W}$. This scale dependence, 
defined in the $\overline{\text{MS}}$ scheme, is shown in 
Fig.~\ref{fig:s2wplot} together with the anticipated 
sensitivity of the measurement of the weak mixing angle 
at P2 compared to other forthcoming determinations (blue 
points) and existing measurements (red points).

A precise measurement of the weak charge provides, therefore, 
a precision test of the SM and its predictions. The envisaged 
measurement of the P2 experiment at low momentum transfer will 
complement other high-precision determinations, like those of 
the LEP and SLC experiments at the $Z$ pole. The P2 experiment 
may thus help to resolve differences between previous 
measurements, or find interesting new effects. 

Extensions of the SM lead to modified predictions for 
various observables, visible only in high-precision 
measurements. For example, models with dark photons 
predict a small shift of the running weak mixing angle 
at low mass scales, i.e., a change of $\sin^2 \theta_W$ 
visible at P2, but not at the $Z$-pole measurements. 
Other models, like supersymmetry, lead to characteristic 
deviations for different observables and only a combined 
analysis can reveal the type of new physics. A convenient 
way to compare the reach of different observables to the 
mass scale, $\Lambda$, of new physics is based on a 
description with effective 4-fermion operators. Following 
the convention that the relevant coupling constant is 
normalized by $g^2 = 4\pi$, one can estimate the reach 
by $\Lambda^2 = 8\pi\sqrt{2} / (G_\text{F} \Delta Q_\text{W})$ 
where $G_\text{F}$ is the Fermi constant and $\Delta Q_\text{W}$ 
the precision for the measurement of the weak charge. At P2, the 
measurement of the weak charge of the proton is expected 
to reach mass scales close to \SI{50}{TeV}~\cite{Erler:2014fqa}. 
Other targets, like ${}^{12}$C may increase this limit even further. 

Our understanding of the electroweak interactions will improve 
considerably by the forthcoming high-precision experiments. 
The measurement of the weak charge is expected to improve by 
a factor of 3 compared to the anticipated precision achieved 
by the QWeak collaboration at JLab \cite{Carlini-PANIC}. 
Similarly, the MOLLER experiment at JLab 
will provide us with a factor of 5 improvement of the 
determination of the weak charge of the electron, compared 
with the result of the E158 experiment at SLAC. Also the 
ratio of $u$- and $d$-quark weak charges will be measured 
more precisely than before at the SOLID experiment at JLab. 
If completed, the combination of these experiments will 
offer us a handle to distinguish extensions of the SM, which 
contribute in different ways to these experiments. 
\\

The Mainz MESA electron accelerator with the P2 experimental 
facility for parity-violation experiments opens a door to a 
rich parity-violation measurement program including different 
targets and kinematics. This is partly described in the 
Sect.~\ref{sec:FurtherPhysics}. This research program has its 
roots in discussions at a workshop at MIT, organised 
by MIT, JLab, and Mainz in 2013. For more information, 
see~\cite{Milner:2013yua}.

The second experimental facility at MESA is MAGIX. It will be 
equipped with two magnetic spectrometers and a hydrogen cluster 
jet target. MAGIX has a rich program in nucleon and nuclear 
physics, including measurements of the proton radius, the 
electromagnetic form factors of the nucleon, measurements of 
nuclear cross sections relevant for open questions in astrophysics, 
and dark photon searches in scattering and in a beam dump experiment. 
Also the MAGIX research program was discussed at the 
aforementioned MIT workshop.
\\

This manuscript is organized as follows: In 
Section~\ref{sec:SinThetaW} we describe how -- and how accurate -- 
we can extract the weak mixing angle from the measurement of 
the parity-violating cross section asymmetry. In 
Section~\ref{sec:Accelerator} we describe the new MESA 
accelerator which will be installed in a new accelerator 
hall from 2020 on. This Section also describes the polarimetry 
at MESA as well as the control of helicity-correlated and 
uncorrelated beam fluctuations. For the measurement proposed 
here, a liquid hydrogen target with the lowest possible level 
of density fluctuations (\SI{10}{ppm} in \SI{4}{ms}) from 
boiling in the volume of the $\ell$H$_2$ or from other sources 
will represent one of the centerpieces. The design 
approach, the method to calculate the density fluctuations beforehand 
and the experience from the design of the $\ell$H$_2$ target at 
the former QWeak experiment are described in Sect.~\ref{sec:Target}. 
The P2 spectrometer, consisting of the $\ell$H$_2$ target, the large 
solenoid magnet, the results of the full Geant4 simulations of the 
spectrometer, the integrating Cherenkov detectors, the 
high-resolution ADC system for the read-out as well as the 
tracking detector is presented in 
Sect.~\ref{sec:Spectrometer}. The new level of experimental 
accuracy required in the past, and still requires, corresponding 
theory efforts in order to get effects from QED corrections, box 
graph and hadronic uncertainties and other electroweak radiative 
corrections under control. The relevant recent theory work is 
described in Sect.~\ref{sec:Theory}. The P2 experimental facility for 
parity-violating electron scattering allows for a rich measurement 
program, like an additional backward-angle measurement to 
further reduce the uncertainty from the axial form factor and 
the contribution from strangeness to the magnetic form factor. 
Both are not sizeable quantities, but still have large error bars. 
Another example is an additional measurement with a ${}^{12}$C 
target. This allows for an even more sensitive search for beyond 
Standard Model physics. A very sensitive measurement of the neutron 
skin thickness in lead is possible with the P2 spectrometer as well. 
This exciting further physics program at the P2 experimental facility 
is described in Sect.~\ref{sec:FurtherPhysics}. The manuscript 
presented here closes with a summary.

%% file: SinThetaW.tex

In this chapter, the experimental method for measuring the proton's weak charge
$Q_\text{W}(\text{p})$ is presented and the achievable precision in the
determination of the electroweak mixing angle $\sin^2 \theta_\text{W}$ is
discussed.

\subsection{Experimental method}
\label{sec:ExperimentalMethod}

\begin{figure}[htb]
  \centering
  \resizebox{0.48\textwidth}{!}{\includegraphics{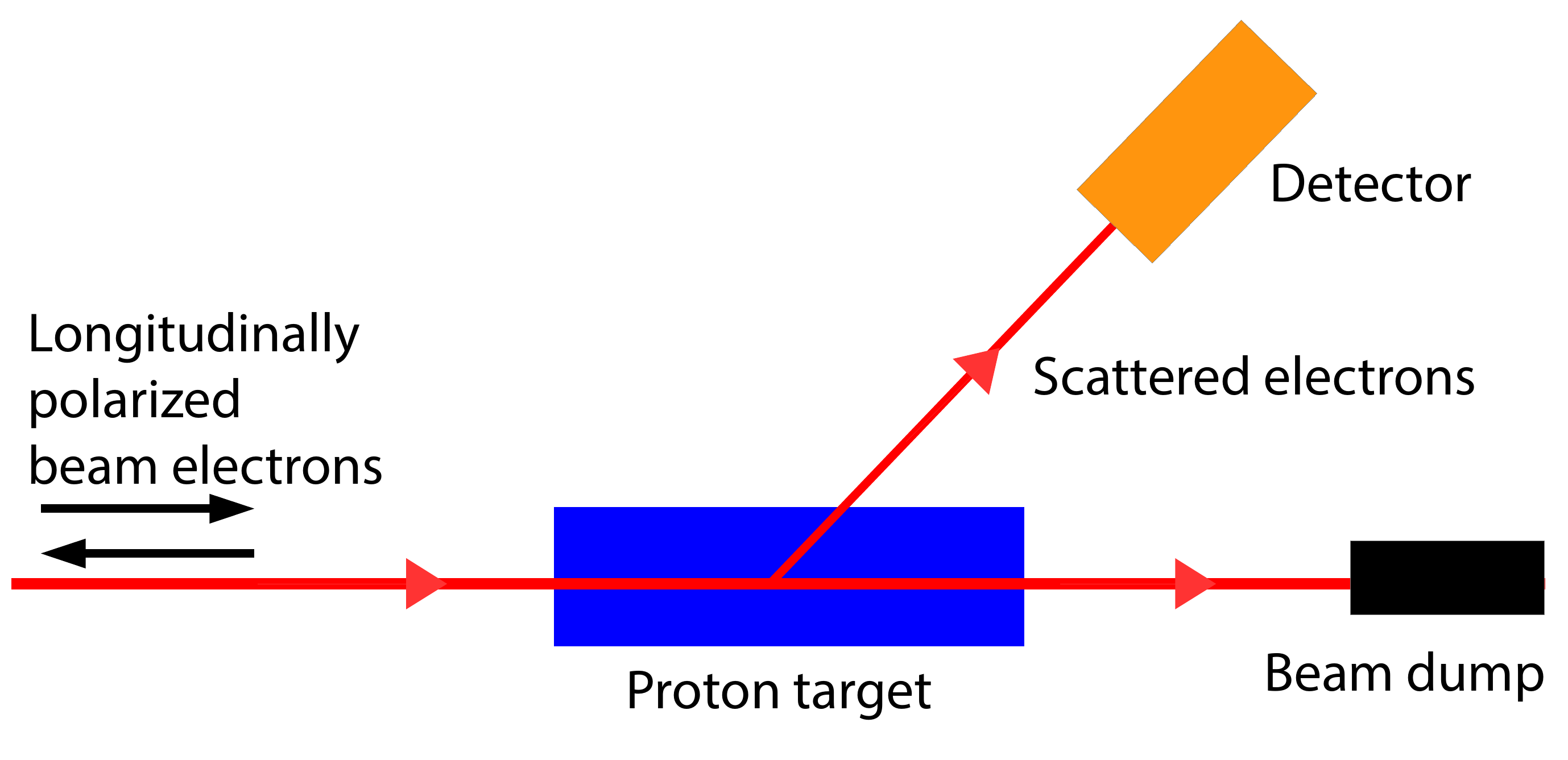}}
  \caption{Experimental method to be used in the P2 experiment: A longitudinally
  polarized beam of electrons is impinged on a long proton target. For each
  helicity state of the beam electrons the elastically scattered electrons are
  detected.}
  \label{fig:experimental_method}
\end{figure}

For the P2 experiment, MESA will provide a beam of longitudinally polarized
electrons. The beam energy will be
\begin{equation}
  E_\text{beam} = \SI{155}{MeV}
\end{equation}
and the beam current is scheduled to be
\begin{equation}
  I_\text{beam} = \SI{150}{\micro A}.
\end{equation}
The helicity of the beam electrons will be switched with a frequency $f \sim
\SI{1}{kHz}$. The beam electrons impinge on an unpolarized
$\ell\mathrm{H}_2$-target with a length of $L = \SI{600}{mm}$ oriented along the
beam direction. The electrons, which are scattered elastically off the protons,
are detected in an azimuthally symmetric Cherenkov detector.
Figure~\ref{fig:experimental_method} illustrates the measurement principle.
Since the luminosity $\mathcal{L}$ of the P2 experiment is projected to be
\begin{equation}
  \mathcal{L} = I_\text{beam}/e \cdot \rho_\text{part} \cdot L =
  \SI{2.38E39}{cm^{-2} s^{-1}},
\end{equation}
where $e$ is the elementary charge and $\rho_\text{part}$ is the proton density
in $\ell\mathrm{H}_2$, the total rate of the electrons scattered elastically off
protons which needs to be detected is in the order of \SI{0.1}{THz}. This makes
an integrating measurement of the event rates necessary.

\subsubsection{Parity-violating asymmetry in elastic electron-proton scattering}
\label{sec:Parityviolatingasymmetry}

The main observable in the P2 experiment is the parity-violating asymmetry
$A^\text{PV}$ in elastic electron-proton scattering. It is an asymmetry in the
cross section which may be defined by
\begin{equation}
  A^\text{PV} \equiv \frac{\mathrm{d}\sigma_\text{ep}^+ -
  \mathrm{d}\sigma_\text{ep}^-}{\mathrm{d}\sigma_\text{ep}^+ +
  \mathrm{d}\sigma_\text{ep}^-}.
  \label{eq:Apv_differential_form}
\end{equation}
In this equation, $\mathrm{d}\sigma_\text{ep}^\pm$ is the differential cross
section for the elastic scattering of electrons with helicity 
$\pm 1/2$ off unpolarized protons. 

\begin{figure}[htb]
  \centering
  \resizebox{0.3\textwidth}{!}{\includegraphics{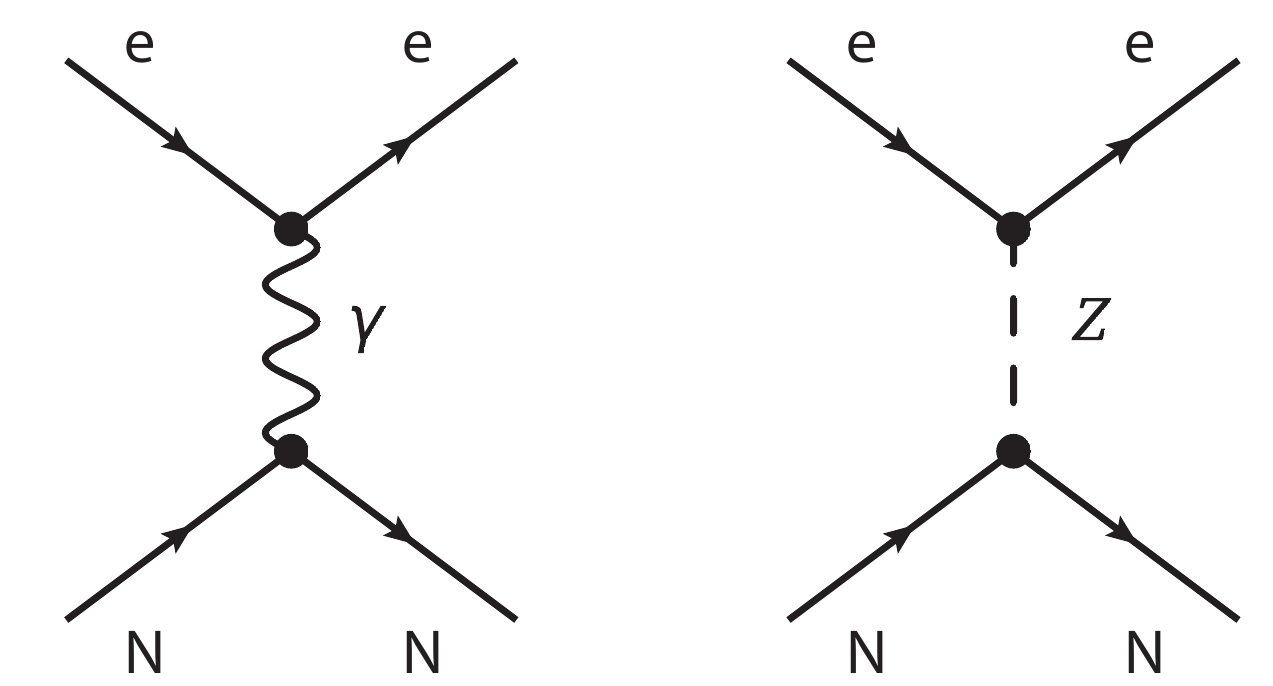}}
  \caption{Feynman diagrams showing the exchange of a virtual photon and
  $Z$-boson in the process of electron-nucleon scattering.}
  \label{fig:graphs_eN_scattering_tree_level}
\end{figure}
$A^\text{PV}$ is due to the interference between the exchange of a virtual
photon and a $Z$-boson in the scattering process, both of which are illustrated
in Fig.~\ref{fig:graphs_eN_scattering_tree_level}. The differential cross
section of the scattering process can be written
\begin{equation}
  {\left( \frac{\mathrm{d}\sigma^\pm_\text{ep}} {\mathrm{d}\Omega} \right)} =
  \left( \frac{\alpha_\text{em}}{4 m_\text{p} Q^2} \frac{E_\text{f}}{E_\text{i}} \right)^2
  {\left\vert \mathcal{M}_\text{ep}^\pm \right\vert}^2,
  \label{eq:cross_section_general}
\end{equation}
where $\alpha_\text{em}$ is the electromagnetic coupling, $m_\text{p}$ is the
proton mass, and
\begin{equation}
  Q^2 \approx 4 E_\text{i} E_\text{f} \sin^2\left(\theta_\text{f} / 2\right)
  \label{eq:Q2_electron}
\end{equation}
is the negative square of the 4-momentum transfer between electron and proton.
Here, the electron mass can be neglected. $E_\text{i}$ is the electron's initial
state energy, $E_\text{f}$ the energy of the scattered electron and
$\theta_\text{f}$ the scattering angle with respect to the beam direction.
$\mathcal{M}^{\pm}_\text{ep}$ is the transition matrix element, at leading order
given by the Feynman diagrams shown in 
Fig.~\ref{fig:graphs_eN_scattering_tree_level}. 

The resulting parity-violating helicity asymmetry is written as
\begin{equation}
  A^\text{PV} = \frac{-G_\text{F} Q^2}{4\pi\alpha_\text{em}\sqrt{2}} \left[
  Q_\text{W}(\text{p}) - F(E_\text{i}, Q^2) \right],
  \label{eq:Apv}
\end{equation}
where $G_\text{F}$ is the Fermi coupling constant. Here, the weak charge of the
proton, $Q_\text{W}(\text{p})$, is defined as the limit of the asymmetry at
zero-momentum transfer, normalized such that Eq.\ (\ref{eq:Apv}) holds, i.e.,
$F(E_\text{i}, Q^2 = 0) = 0$. At non-zero momentum transfer, the hadronic
structure of the proton has to be taken into account, parametrized by the $Q^2$-
and energy-dependent function $F(E_\text{i}, Q^2)$. The function $F$ is
often written as $F(E_\text{i}, Q^2) = Q^2 B(Q^2)$ and the energy-dependence 
not shown explicitly.

Based on a flavour decomposition of the matrix elements of the electromagnetic
and weak neutral currents, the form factor contribution $F(Q^2)$ is usually
written as a sum of three terms
\begin{equation}
  F(E_\text{i}, Q^2) \equiv F^\text{EM}(E_\text{i}, Q^2) +
  F^\text{A}(E_\text{i}, Q^2) + F^\text{S}(E_\text{i}, Q^2),
  \label{eq:F(Q2)}
\end{equation}
where
\begin{equation}
  F^\text{EM}(E_\text{i}, Q^2) \equiv \frac{\epsilon
  G^{\text{p},\gamma}_\text{E} G^{\text{n},\gamma}_\text{E} + \tau
  G^{\text{p},\gamma}_\text{M} G^{\text{n},\gamma}_\text{M}}{\epsilon
  {\left(G^{\text{p},\gamma}_\text{E} \right)}^2 + \tau {\left(
  G^{\text{p},\gamma}_\text{M} \right)}^2}
  \label{eq:Fem}
\end{equation}
is given by the proton's electric and magnetic form factors
$G^{\text{p},\gamma}_\text{E}$ and $G^{\text{p},\gamma}_\text{M}$ as well as the
neutron's electric and magnetic form factors $G^{\text{n},\gamma}_\text{E}$ and
$G^{\text{n},\gamma}_\text{M}$. $F^\text{A}(E_\text{i}, Q^2)$ depends on the
proton's axial form factor $G^\text{p,Z}_\text{A}$ and is denoted as
\begin{equation}
  F^\text{A}(Q^2) \equiv \frac{\left(1 - 4 \sin^2 \theta_\text{W} \right)
  \sqrt{1 - \epsilon^2} \sqrt{\tau(1 - \tau)} G^{\text{p},\gamma}_\text{M}
  G^\text{p,Z}_\text{A}}{\epsilon {\left( G^{\text{p},\gamma}_\text{E}
  \right)}^2 + \tau{\left( G^{\text{p},\gamma}_\text{M} \right)}^2}.
  \label{eq:Faxial}
\end{equation}
$F^\text{S}(Q^2)$ depends on the nucleon's strange electric and magnetic form
factors $G^\text{s}_\text{E}$ and $G^\text{s}_\text{M}$ as well as the
isospin-breaking form factors $G^\text{u,d}_\text{E}$ and
$G^\text{u,d}_\text{M}$:
\begin{equation}
  \begin{split}
    F^\text{S}(E_\text{i}, Q^2) \equiv& \frac{\epsilon
    G^{\text{p},\gamma}_\text{E} G^\text{s}_\text{E} + \tau
    G^{\text{p},\gamma}_\text{M} G^\text{s}_\text{M}}{\epsilon {\left(
    G^{\text{p},\gamma}_\text{E} \right)}^2 + \tau {\left(
    G^{\text{p},\gamma}_\text{M} \right)}^2} \\
    +& \frac{\epsilon G^{\text{p},\gamma}_\text{E} G^\text{u,d}_\text{E} + \tau
    G^{\text{p},\gamma}_\text{M} G^\text{u,d}_\text{M}}{\epsilon
    {\left(G^{\text{p},\gamma}_\text{E}\right)}^2 + \tau
    {\left(G^{\text{p},\gamma}_\text{M}\right)}^2}.
  \end{split}
  \label{eq:Fstrange}
\end{equation}
In these expressions we have used the abbreviations
\begin{equation}
  \epsilon \equiv {\left[ 1 + 2(1 + \tau) \tan^2\left( \frac{\theta_\text{f}}{2}
  \right)\right]}^{-1}
\end{equation}
and
\begin{equation}
  \tau \equiv \frac{Q^2}{4 m_\text{p}^2}.
  \label{eq:tau}
\end{equation} 

\begin{figure}[htb]
  \centering
  \resizebox{0.5\textwidth}{!}{\includegraphics{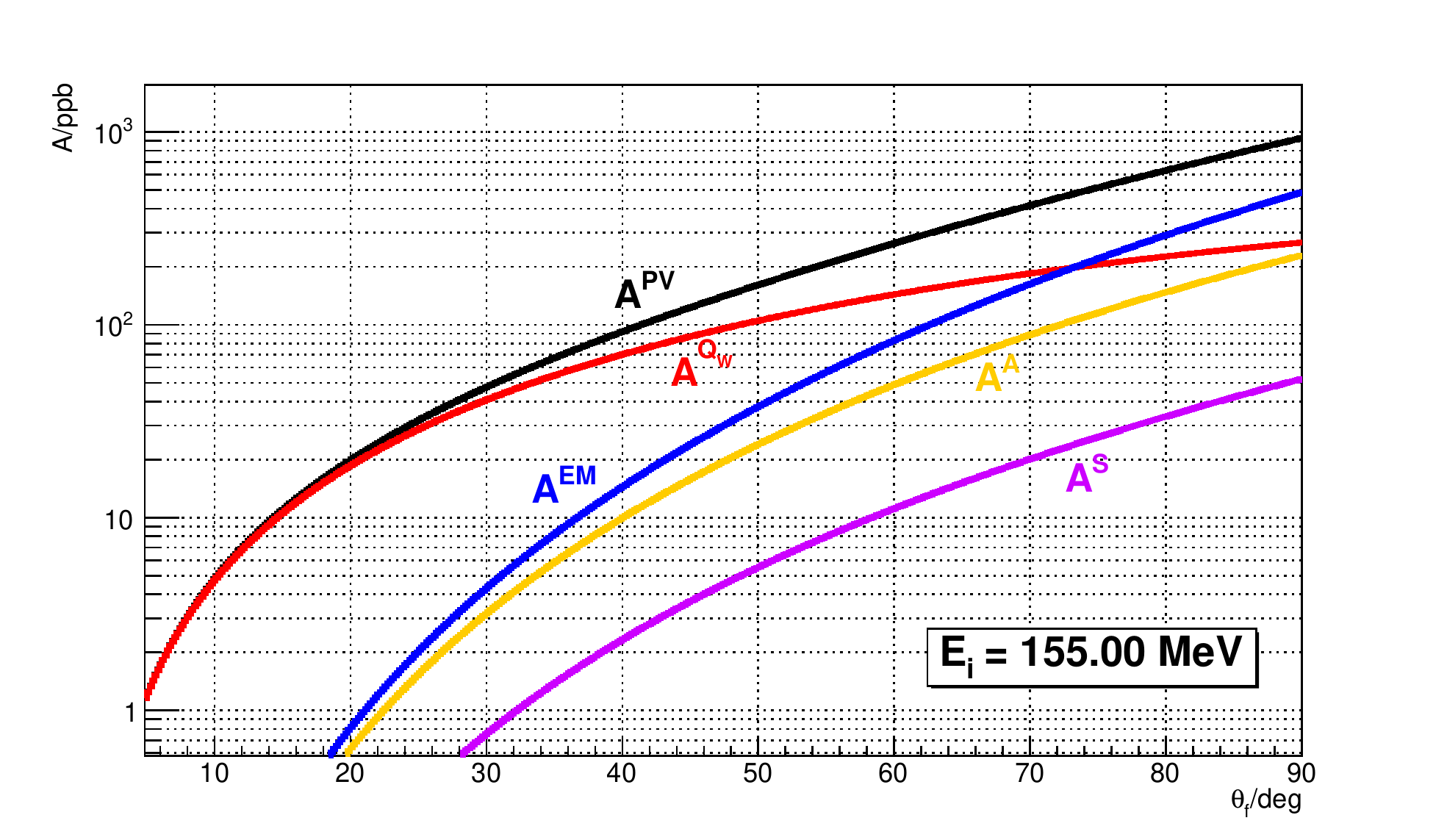}}
  \caption{Dependence of $A^\text{PV}$ on $\theta_\text{f}$ for $E_\text{i} =
  \SI{155}{MeV}$. Plotted are the absolute values of the asymmetry and the
  contributions by the proton's weak charge and the nucleon form factors to
  $A^\text{PV}$. For low values of $Q^2$, $A^\text{PV}$ is dominated by the weak
  charge contribution $A^{\text{Q}_\text{W}}$. At the central scattering angle
  of the P2 experiment, $A^\text{PV}(\theta_\text{f} = \ang{35}) =
  \SI{-67.34}{ppb}$.}
  \label{fig:Apv_vs_theta}
\end{figure}

According to Eq.~(\ref{eq:Apv}), $A^\text{PV}$ is proportional to $Q^2$. In
Fig.~\ref{fig:Apv_vs_theta} we show the dependence of $A^\text{PV}$ on
$\theta_\text{f}$ for $E_\text{i} = \SI{155}{MeV}$, which equals the beam energy
to be used in the P2 experi\-ment. The picture also shows the separate
contributions
\begin{equation}
  \begin{split}
    A^{\text{Q}_\text{W}} \equiv& \frac{-G_\text{F}
    Q^2}{4\pi\alpha_\text{em}\sqrt{2}} \cdot Q_\text{W}(\text{p}), \\
    A^\text{EM} \equiv& \frac{G_\text{F} Q^2}{4\pi\alpha_\text{em}\sqrt{2}}
    \cdot F^\text{EM}, \\
    A^\text{A} \equiv& \frac{G_\text{F} Q^2}{4\pi\alpha_\text{em}\sqrt{2}} \cdot
    F^\text{A}, \\
    A^\text{S} \equiv& \frac{G_\text{F} Q^2}{4\pi\alpha_\text{em}\sqrt{2}} \cdot
    F^\text{S}
  \end{split}
\end{equation}
to $A^\text{PV}$. One can see that at low $Q^2$, $A^\text{PV}$ is dominated by
$A^{\text{Q}_\text{W}}$, while the hadronic contributions are small. A
measurement of $A^\text{PV}$ at low $Q^2$ is therefore sensitive to the weak
charge of the proton, $Q_\text{W}(\text{p})$.

\subsubsection{The proton's weak charge and the electroweak mixing angle}
\label{sec:The protonWeakCharge}

Neglecting radiative corrections, the tree-level expression for the proton's
weak charge in (\ref{eq:Apv}) is
\begin{equation}
  Q_\text{W}(\text{p}) = 1 - 4 \sin^2 \theta_\text{W},
  \label{eq:QWp_tree_level}
\end{equation}
where $\theta_\text{W}$ is the electroweak mixing angle or Weinberg-angle. In
the following we will often use the abbreviation $s_\text{W} = \sin
\theta_\text{W}$. Since $\sin^2 \theta_\text{W} \approx 0.23$,
$Q_\text{W}(\text{p})$ is small in the SM. From (\ref{eq:QWp_tree_level}), using
Gaussian error propagation, it follows that
\begin{equation}
  \frac{\Delta\sin^2 \theta_\text{W}}{\sin^2 \theta_\text{W}} = \frac{1 - 4
  \sin^2 \theta_\text{W}}{4 \sin^2 \theta_\text{W}} \cdot \frac{\Delta
  Q_\text{W}(\text{p})}{Q_\text{W}(\text{p})} \approx 0.09 \cdot \frac{\Delta
  Q_\text{W}(\text{p})}{Q_\text{W}(\text{p})}.
\end{equation}
Therefore, a precise measurement of $Q_\text{W}(\text{p})$ will result in an
approximately $10$ times more precise determination of the electroweak mixing
angle. The weak charge of the proton is therefore highly sensitive to the 
electroweak mixing angle. 

Even small corrections to $\sin^2 \theta_\text{W}$ may modify 
$Q_\text{W}(\text{p})$ significantly. A wide range of beyond-SM 
effects can lead to such corrections. They need to be disentangled 
from SM radiative corrections. Higher-order corrections to 
$A^\text{PV}$ will be discussed later in Sect.~\ref{sec:Theory}.

\subsection{Achievable precision in the determination of the weak mixing angle}
\label{sec:Achievableprecisioninthedeterminationoftheweakmixingangle}

In order to predict the achievable precision in the determination of
$\sin^2\theta_\text{W}$, error propagation calculations based on the Monte Carlo
method have been carried out. The goal of these calculations was to determine
the achievable uncertainty $\Delta s_\text{W}^2$ as a function of the beam
energy $E_\text{beam}$, the central electron scattering angle
$\bar{\theta}_\text{f}$ and the acceptance of the azimuthally symmetric detector
in $\theta_\text{f}$, denoted $\delta\theta_\text{f} \equiv
[\bar{\theta}_\text{f} - \delta\theta_\text{f}/2, ~\bar{\theta}_\text{f} +
\delta\theta_\text{f}/2]$. In the following, the method used to calculate
$\Delta s_\text{W}^2$ will be discussed and results of the calculations will be
presented. Based on these results, the beam energy and detector acceptance to be
used in the P2 experiment are chosen.

\subsubsection{Method to determine the achievable uncertainty of $\sin^2 \theta_\text{W}$}
\label{sec:MethodUncertainty}

The analysis presented in this section is based on the leading-order prediction
of $A^\text{PV}$. Therefore we use the generic symbol $s_\text{W}$ for the sine
of the weak mixing angle. Only if higher-order corrections are included, one
will have to distinguish between the on-shell or the $\overline{\text{MS}}$
definitions of $s_\text{W}$ and, in the latter case, specify its
scale-dependence. 

In a parity-violation electron scattering experiment, one measures the asymmetry
\begin{equation}
  A^\text{exp} \equiv \frac{N^+ - N^-}{N^+ + N^-},
\end{equation}
which is determined by the total numbers $N^\pm$ of detected scattering events
for the two helicity states of the beam electrons. One can write
\begin{equation}
  A^\text{exp} = P \cdot \langle A^\text{PV} \rangle_{L, ~\delta\theta_\text{f}} +
  A^\text{app},
  \label{eq:Aexp}
\end{equation}
where $P$ is the average polarization of the electron beam,
\begin{equation}
  \langle A^\text{PV}\rangle_{L, ~\delta\theta_\text{f}} = \frac{\int\limits_0^L
  \mathrm{d}z
  \int\limits_{\theta_\text{f}^\text{min}}^{\theta_\text{f}^\text{max}}
  \mathrm{d}\cos\theta_\text{f} \left(
  \frac{\mathrm{d}\sigma^0_\text{ep}}{\mathrm{d}\Omega} \right) \cdot
  A^\text{PV} } { \int\limits_0^L \mathrm{d}z
  \int\limits_{\theta_\text{f}^\text{min}}^{\theta_\text{f}^\text{max}}
  \mathrm{d}\cos\theta_\text{f} \left( \frac{ \mathrm{d} \sigma^0_\text{ep} } {
  \mathrm{d}\Omega} \right) }
  \label{eq:Apv_averaged}
\end{equation}
is the expected value of $A^\text{PV}$ after averaging over the target length
$L$ and the detector acceptance in $\theta_\text{f}$. In addition, the
polarization independent part of the differential cross section
\begin{equation}
  \begin{split}
    \frac{\mathrm{d}\sigma^0_\text{ep}}{\mathrm{d}\Omega} =&
    \frac{\mathrm{d}\sigma^\text{Mott}}{\mathrm{d}\Omega} \cdot \left[
    \frac{{(G^{\text{n},\gamma}_\text{E})}^2 + \tau
    {(G^{\text{n},\gamma}_\text{M})}^2 } { 1 + \tau } \right. \\
    + & \left. 2 \tau {\left( G^{\text{n},\gamma}_\text{M}
    \tan\left(\theta_\text{f}/2\right) \right)}^2 \right]
  \end{split}
  \label{eq:Rosenbluth_formula}
\end{equation}
given by the Rosenbluth formula \cite{Rosenbluth:1950yq} is taken into account.
Furthermore, $A^\text{app}$ is an apparative asymmetry induced by
helicity-correlated fluctuations of the electron beam's properties. In
Eq.~(\ref{eq:Apv_averaged}), the averaging over the target's length is done to
include the mean energy loss of the beam electrons due to collision and
radiation processes as they travel through the target volume. At leading-order,
both $\mathrm{d}\sigma_\text{ep}^0$ and $A^\text{PV}$ may be regarded as
functions of the electron's mean initial state energy $\langle E_\text{i}
\rangle(z)$ and scattering angle $\theta_\text{f}$. $\langle E_\text{i}
\rangle(z)$ depends on the $z$-coordinate of the interaction vertex's position
in the target and decreases with increasing penetration depth of the beam
electrons.

To predict $\Delta s_\text{W}^2$, Eq.~(\ref{eq:Aexp}) is solved for
$s_\text{W}^2$, and an error propagation calculation based on the resulting
expression is performed. This expression has the general structure
\begin{eqnarray}
  s_\text{W}^2 & = & s_\text{W}^2 \left( A^\text{exp}, ~P, ~E_\text{beam},
  ~\bar{\theta}_\text{f}, ~\delta\theta_\text{f}, \right. \nonumber \\
  && \left. ~~~~~ F(\{\kappa_k\}, Q^2), \Box_{\gamma\text{Z}}\right),
  \label{eq:sw2_functional}
\end{eqnarray}
where $F(\{\kappa_k\}, Q^2)$ represents parametrizations of the nucleon's form
factors which depend on a set of independent, real parameters $\{\kappa_k\}$,
and $\Box_{\gamma\text{Z}}$ are the contributions from $\gamma Z$-box graphs.
Therefore, one may consider $s_\text{W}^2$ a function of a set of independent
parameters $\{\lambda_l\}$:
\begin{equation}
  s_\text{W}^2 = s_\text{W}^2(\{\lambda_l\}),
  \label{eq:sZ2(lambda_l)}
\end{equation}
with
\begin{equation}
  \lambda_l \in \{A^\text{exp}, ~P, ~E_\text{beam}, ~\bar{\theta}_\text{f},
  ~\delta\theta_\text{f}, ~\{\kappa_k\}, ~\Box_{\gamma\text{Z}}\}.
  \label{eq:lambda_l}
\end{equation}
The Gaussian error propagation based on Eq.~(\ref{eq:sZ2(lambda_l)}) is 
not straightforward, since the integrations in Eq.~(\ref{eq:Apv_averaged})
cannot be done analytically. 

Therefore, Monte Carlo based error propagation calculations have been carried
out numerically. The idea is to treat the parameters $\lambda_l$ as independent
random variables and to calculate a distribution of $s_\text{W}^2$ values by
inserting the randomized parameters into Eq.~(\ref{eq:sZ2(lambda_l)}). The
first step to achieve this is to assign a Gaussian probability distribution to
each parameter $\lambda_l$ by defining the distribution's mean value $\langle
\lambda_l \rangle$ and standard deviation $\Delta\lambda_l$. One then samples a
set of random values $\{\lambda^\prime_l\}$ according to the assigned probability
distributions using a random number generator. Substituting the randomized
values into Eq.~(\ref{eq:sZ2(lambda_l)}) leads to
\begin{equation}
  \left(s_\text{W}^2\right)^\prime = s_\text{W}^2 \left( \{\lambda^\prime_{l}\}
  \right).
\end{equation}
Iterating this procedure leads to a distribution of
$\left(s_\text{W}^2\right)^\prime$ values, from which the expected value
$\langle s_\text{W}^2 \rangle$ and standard deviation $\Delta s_\text{W}^2$ can
be extracted. While the mean value is expected to match the input value of
$s_\text{W}^2$ used to calculate $A^\text{PV}$, $\Delta s_\text{W}^2$ is in the
following regarded as the achievable precision in the determination of the
electroweak mixing angle. 

The algorithm allows to calculate the contributions of individual parameters
$\lambda_l$ to $\Delta s_\text{W}^2$ by sampling only the parameter whose
contribution is of interest while all other parameters are kept constant at
their expected values. Furthermore, it is possible to compute the expected value
$\langle \Delta s_\text{W}^2 \rangle$ and the uncertainty $\Delta(\Delta
s_\text{W}^2)$ of $\Delta s_\text{W}^2$ by sampling a distribution of $\Delta
s_\text{W}^2$-values. This distribution is generated by repeatedly calculating
values of $\Delta s_\text{W}^2$ for the same choice of
$\{\langle\lambda_l\rangle\}$ and $\{\Delta\lambda_l\}$. From the resulting
distribution of $\Delta s_\text{W}^2$ values, both the mean value $\langle
\Delta s_\text{W}^2 \rangle$ and the standard deviation $\Delta(\Delta
s_\text{W}^2)$ are extracted.

\subsubsection{Input parameters to the calculation of $\Delta s_\text{W}^2$}
\label{sec:InputparameterstothecalculationsofDeltaSz2}

The mean values and standard deviations of the parameters $\lambda_l$ are chosen
according to the projected experimental conditions of the experiment.
Table~\ref{tab:input_parameters_ep} lists all parameter values related to the
experimental conditions at the MESA facility. A systematic scan of the mean
values of $(E_\text{beam}, \theta_\text{f}, \delta\theta_\text{f})$ has been
performed in order to determine appropriate values for these parameters for the
design of the experimental apparatus.
\begin{table}
  \centering
  \begin{tabular}{ccc}
    \toprule[1.5pt]
    $\lambda_l$ & $\langle \lambda_l \rangle$ & $\Delta \lambda_l$ \\
    \midrule[1.5pt]
    $E_\text{beam}$ & variable & \SI{0.13}{MeV} \\
    \midrule
    $\bar{\theta}_\text{f}$ & variable & \ang{0} \\
    \midrule
    $\delta\theta_\text{f}$ & variable & \ang{0.1} \\
    \midrule
    $\delta\phi_\text{f}$ & \ang{360} & \ang{0} \\
    \midrule
    $I_{beam}$ & \SI{150}{\micro A} & \SI{0.001}{\micro A}\\
    \midrule
    $P$ & $0.85$ & $0.00425$ \\
    \midrule
    $L$ & \SI{600}{mm} & \SI{0}{mm} \\
    \midrule
    $T$ & \SI{1E4}{h} & \SI{0}{h}\\
    \midrule
    $A^\text{app}$ & $0$ & \SI{0.1}{ppb} \\
    \bottomrule[1.5pt]
  \end{tabular}
  \caption{Mean values and standard deviations chosen for performing the error
  propagation calcuations. $\delta\phi_\text{f}$ denotes the azimuthal
  acceptance of the detector and $T$ is the measuring time. Parameters for
  which $\Delta\lambda_l = 0$ is shown have been kept constant during the
  calculations.}
  \label{tab:input_parameters_ep}
\end{table} 

The expected value of the asymmetry $\langle A^\text{exp} \rangle$ is calculated
by inserting the mean values of the relevant parameters into
Eq.~(\ref{eq:Aexp}). As standard deviation $\Delta A^\text{exp}$ the statistical
uncertainty of $A^\text{exp}$ is chosen. Assuming Poisson-statistics and
starting from Eq.~(\ref{eq:Aexp}), one finds that
\begin{equation}
  \Delta A^\text{exp} \equiv \sqrt{\frac{1}{N}},
\end{equation}
where it has been assumed that
\begin{equation}
  N \equiv N^+ + N^- \approx 2 N^+ \approx 2 N^-.
\end{equation}

In addition to the uncertainty contributions to $\Delta s_\text{W}^2$ 
originating from the experiment-related parameters listed in 
Tab.~\ref{tab:input_parameters_ep},
one expects a significant contribution from the $\gamma Z$-box graph. At
$E_\text{i} = \SI{155}{MeV}$, one expects
\begin{eqnarray}
  \Box_{\gamma \text{Z}} &=& \SI{1.07e-3}{}, 
  \\
  \Delta \Box_{\gamma \text{Z}}  &=& \SI{0.18e-3}{}
\end{eqnarray}
for the central value and the $1\sigma$ uncertainty of $\Box_{\gamma\text{Z}}$
\cite{Gorchtein:2015naa}. Details of the box-graph corrections and the
definition of $\Box_{\gamma\text{Z}}$ are discussed in Sect.~\ref{sec:Boxgraph}.

It has been assumed in Eq.~(\ref{eq:sw2_functional}) that the nucleon form
factors can be parametrized as functions $F(\{\kappa_k\}, ~Q^2)$, where
$\{\kappa_k\} \subset \{\lambda_l\}$ is a set of independent, real parameters.
In the following, we present the parametrizations of the nucleon form factors
that have been used to carry out the error propagation calculations.

\paragraph{Electromagnetic form factors of the proton.}
\label{sec:Electromagneticformfactorsoftheproton}

The form factors $G^{\text{p},\gamma}_\text{E}$ and
$G^{\text{p},\gamma}_\text{M}$ have been parametrized using the ``Dipole 
$\times$ Polynomial-Model'', which has been developed by Bernauer et al.
\cite{BernauerPhD2010}. In this model, the standard dipole term
\begin{equation}
  G^\text{std}_\text{dipole}(Q^2) = \left( 1 +
  \frac{Q^2}{\SI{0.71}{GeV^2}}\right)^{-2}
\end{equation}
is multiplied by a polynomial
\begin{equation}
  G^\text{poly}_\text{E,M}(Q^2) = 1 + \sum\limits_{i=1}^8 \left(
  \kappa_i^\text{E,M} \cdot Q^{2i} \right)
\end{equation}
such that
\begin{eqnarray}
  G^{\text{p},\gamma}_\text{E}(Q^2) =& G^\text{std}_\text{dipole}(Q^2) \cdot
  G^\text{poly}_\text{E}(Q^2), \nonumber \\ & \nonumber \\
  G^{\text{p},\gamma}_\text{M}(Q^2) =& \mu_\text{P}/\mu_\text{N} \cdot
  G^\text{std}_\text{dipole}(Q^2) \cdot G^\text{poly}_\text{M}(Q^2),
  \label{eq:GpE_and_GpM}
\end{eqnarray}
where $\mu_\text{P} = \SI{2.792847356}{} \cdot \mu_\text{N}$ is the proton's
magnetic moment and $\mu_\text{N} = (e\hbar)/(2m_\text{p})$ is the nuclear
magneton. In order to retrieve the parameters $\kappa_l$ of the parametrization,
Eq.~(\ref{eq:GpE_and_GpM}) has been fitted to the data sets given in
section K 2.2.3 of \cite{BernauerPhD2010}, leading to the parameter values
listed in the Appendix, Tabs.~\ref{tab:kappa_GpE} and \ref{tab:kappa_GpM}.

\paragraph{Electromagnetic form factors of the neutron.}
\label{sec:Electromagneticformfactorsoftheneutron}

The functions used to parametrize $G^{\text{n},\gamma}_\text{E}$ and
$G^{\text{n},\gamma}_\text{M}$ have been chosen as in 
Ref.~\cite{YakoubiPhD2007}. For $G^{\text{n},\gamma}_\text{E}$, a fit 
function according to Galster \cite{Galster:1971kv} has been used:
\begin{equation}
  G^{\text{n},\gamma}_\text{E}(Q^2) = \frac{\kappa_1\tau}{1+\kappa_2\tau}
  \cdot G^\text{std}_\text{dipole}(Q^2),
  \label{eq:GnE}
\end{equation}
where $G^\text{std}_\text{dipole}(Q^2)$ is defined in
Eq.~(\ref{eq:GpE_and_GpM}) and $\tau$ in Eq.~(\ref{eq:tau}). The fit of
Eq.~(\ref{eq:GnE}) to the data set given in \cite{YakoubiPhD2007} leads to the
parameter values listed in Tab.~\ref{tab:kappa_GnE}. In order to parametrize the
neutron's magnetic form factor $G^{\text{n},\gamma}_\text{M}$ a polynomial of
degree $9$ is used:
\begin{equation}
  G^{\text{n},\gamma}_\text{M}(Q^2) = \sum\limits_{i = 0}^9 \kappa_i Q^{2i}.
  \label{eq:GnM}
\end{equation}
Fitting Eq.~(\ref{eq:GnM}) to the data given in \cite{YakoubiPhD2007} results in
the parameters listed in Tab.~\ref{tab:kappa_GnM}.

\paragraph{Strangeness form factors.}
\label{sec:Strangenessformfactors}

The experimental determination of the strangeness form factors
$G^\text{s}_\text{E}$ and $G^\text{s}_\text{M}$ has been the subject of a
comprehensive measurement program for $15$ years in three major research
facilities. Measurements with $Q^2 = \SI{0.1}{GeV^2}$, $Q^2 =
\SI{0.23}{GeV^2}$ and $Q^2 = \SI{0.63}{GeV^2}$ have been carried out by
the SAMPLE, HAPPEX, G0 and the A4 Collaborations
\cite{Spayde2004,Armstrong:2005hs,Acha:2006my,Androic:2009aa,Maas:2004ta,Baunack:2009gy}. 

Like $G^{\text{n},\gamma}_\text{E}$, $G^\text{s}_\text{E}$ has been parametrized
using Eq.~(\ref{eq:GnE}). Fitting this expression to the available world data
leads to the parameter values presented in Tab.~\ref{tab:kappa_GsE}. For
$G^\text{s}_\text{M}$, the parametrization
\begin{equation}
  G^\text{s}_\text{M} = \kappa_0 + \kappa_1 \cdot Q^2,
\end{equation}
which was taken from Ref.~\cite{Wang:1900ta}, has been used along with 
the parameter values given in Tab.~\ref{tab:kappa_GsM}. 

In the following it is assumed that the uncertainties of $G^\text{s}_\text{E}$
and $G^\text{s}_\text{M}$ can be reduced by factors of $4$ and $12$, respectively, 
in the $Q^2$ region of relevance to the P2 experiment. This reduction can be
achieved by an additional backward-angle measurement, see 
Sect.~\ref{sec:BackwardAngle}.

\paragraph{Axial form factor of the proton.}
\label{sec:Axial form factor of the proton}

The axial form factor of the proton $G^\text{p,Z}_\text{A}$ can be determined
from results of parity-violation electron scattering experiments with
$\ell\mathrm{H}_2$- and $\ell\mathrm{D}_2$-targets, which have been carried out
at backward scattering angles at the same values of $Q^2$. Appropriate
measurements have been done by the SAMPLE, G0 and A4 Collaborations
\cite{Ito:2003mr,Armstrong:2005hs,Androic:2009aa,BalaguerRios:2016ftd}. 

For the purpose of the error propagation calculations presented in this section,
$G^\text{p,Z}_\text{A}$ has been parametrized as suggested by Musolf et
al.\ in Ref.~\cite{Musolf:1993tb}:
\begin{equation}
  G^\text{p,Z}_\text{A}(Q^2) = \kappa_0 \cdot {\left( 1 + \frac{Q^2}{\kappa_1^2}
  \right)}^{-2}
  \label{eq:GpA}
\end{equation}
This parametrization is used together with the parameter values given in
\cite{Musolf:1993tb} and listed in Tab.~\ref{tab:kappa_GpA}. 

For the error propagation calculations presented in this section it has been
assumed that the global uncertainty of the parametrization given by
Eq.~(\ref{eq:GpA}) can be reduced by a factor of $10$. This reduction can be
achieved by a backward-angle measurement of $G^\text{p,Z}_\text{A}$ (see 
Sect.~\ref{sec:BackwardAngle}). 

The requirement of reducing the uncertainties of $G^\text{s}_\text{E}$,
$G^\text{s}_\text{M}$ and $G^\text{p,Z}_\text{A}$ in order to achieve the
envisaged precision in the determination of $\sin^2 \theta_\text{W}$ 
renders the form factor measurement within the scope of the P2 experiment 
mandatory.

\paragraph{Isospin breaking electromagnetic form factors.}
\label{sec:Isospin breaking electromagnetic form factors}

The para\-metrizations of the isospin-breaking form factors
$G^\text{ud}_\text{E}$ and $G^\text{ud}_\text{M}$ have been done using the
dataset quoted in the bachelor thesis of P. Larin \cite{LarinBT2011}. Larin has
extracted data from the predictions for the $Q^2$-dependence of the form factors
given in \cite{Kubis:2006cy}. In order to parametrize $G^\text{ud}_\text{E}$ and
$G^\text{ud}_\text{M}$, polynomials of degree $4$ have been used such that
\begin{equation}
  G^\text{ud}_\text{E,M} = \sum\limits_{i=0}^4 \kappa_i^\text{E,M} \cdot
  Q^{2i}.
  \label{eq:GudE_and_GudM}
\end{equation}
The fits of these functions to the data given in Ref.~\cite{LarinBT2011} 
result in the parameter values collected in Tabs.~\ref{tab:kappa_GudE} 
and \ref{tab:kappa_GudM}.

\subsubsection{Results of the error propagation calculations}
\label{sec:Results of the error propagation calculations}

An extensive scan in the mean values of $E_\text{beam}$, $\bar{\theta}_\text{f}$
and $\delta\theta_\text{f}$ has been performed using the input parameters
discussed in the preceding section in order to determine suitable values of
these variables to carry out the P2 experiment. In this section, selected
results are presented and discussed. 

\begin{figure}[tb]
  \centering
  \resizebox{0.5\textwidth}{!}{\includegraphics{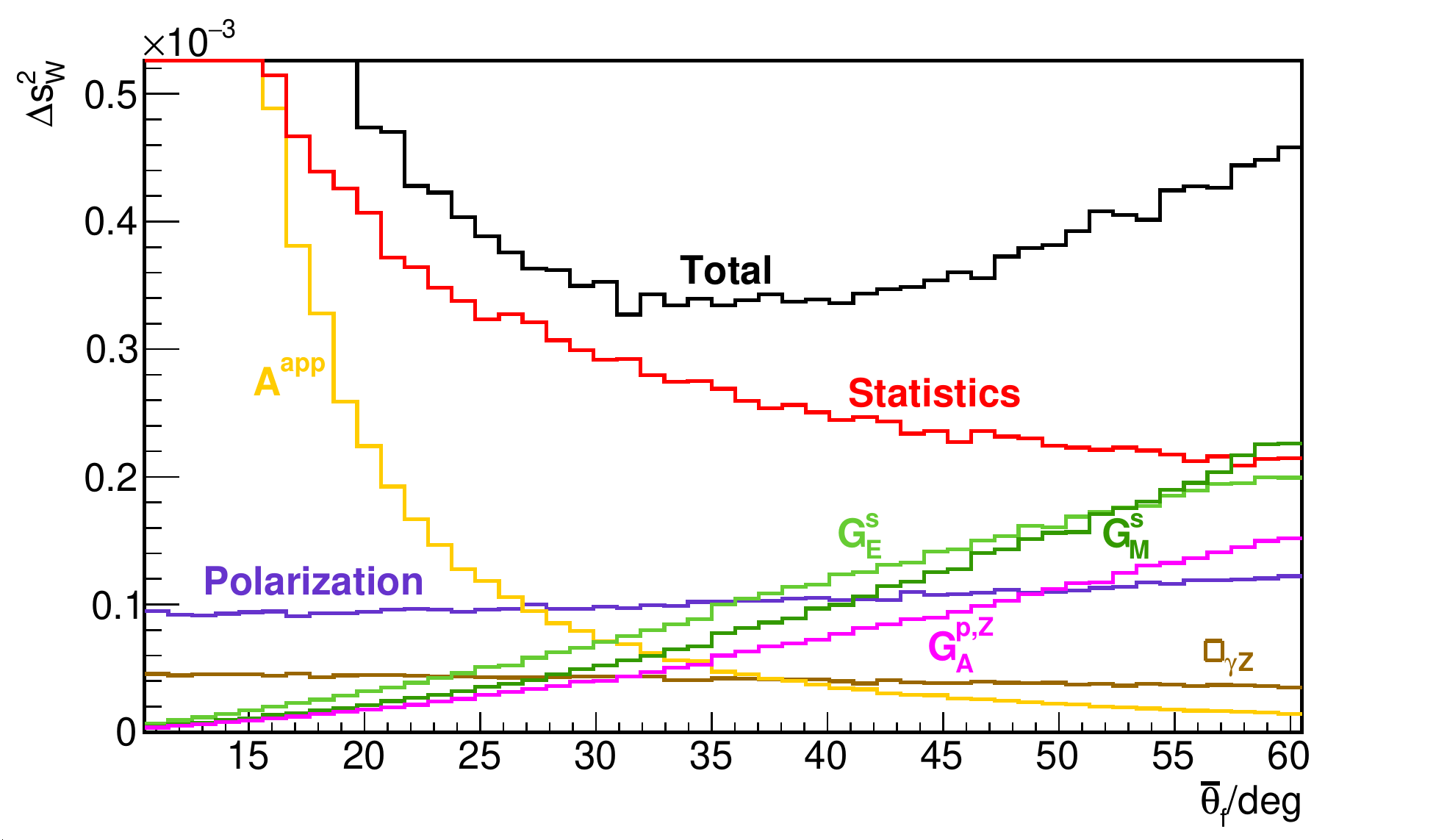}}
  \caption{Dependence of $\Delta s_\text{W}^2$ on the central scattering angle
  $\bar{\theta}_\text{f}$ for $E_\text{beam} = \SI{155}{MeV}$ and
  $\delta\theta_\text{f} = \ang{20}$. The total uncertainty $\Delta
  s_\text{W}^2$ of the electroweak mixing angle is shown in black and 
  other dominating error contributions in color.}
  \label{fig:DsZ2_vs_theta}
\end{figure}

Figure~\ref{fig:DsZ2_vs_theta} shows the dependence of $\Delta s_\text{W}^2$ on
the central electron scattering angle $\bar{\theta}_\text{f}$ for $E_\text{beam} =
\SI{155}{MeV}$ and $\delta\theta_\text{f} = \ang{20}$. For $\ang{17} \leq
\bar{\theta}_\text{f} \leq \ang{55}$, the total uncertainty is dominated by the
statistical uncertainty of the measured asymmetry $A^\text{exp}$. For scattering
angles $\bar{\theta}_\text{f} \geq \ang{40}$ the contributions from
$G^\text{s}_\text{E,M}$ and $G^\text{p,Z}_\text{A}$ become more significant,
because the form factors' contribution to the asymmetry increases with $Q^2$.
The increase of the form factor contributions and the decrease of the
statistical error and the contribution stemming from $A^\text{app}$ with
increasing $\bar{\theta}_\text{f}$ lead to a minimum of $\Delta s_\text{W}^2$ at
$\bar{\theta}_\text{f} \approx \ang{35}$, where $\Delta s_\text{W}^2 \approx
\SI{3.4e-4}{}$.
 \\

\begin{figure}[b]
  \centering
  \resizebox{0.5\textwidth}{!}{\includegraphics{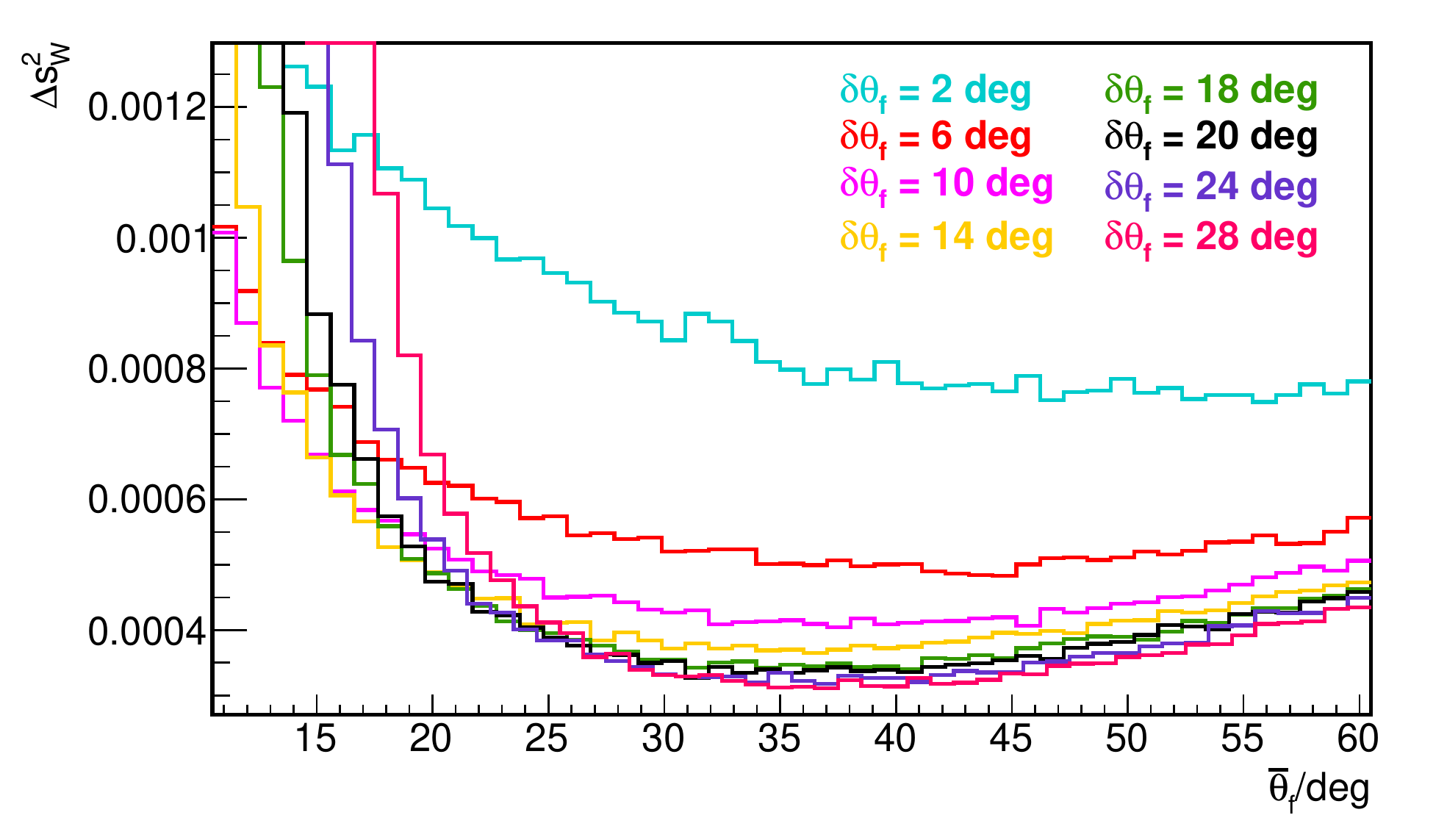}}
  \caption{Dependence of $\Delta s_\text{W}^2$ on $\bar{\theta}_\text{f}$ for
  $E_\text{beam} = \SI{155}{MeV}$ and several values of $\delta\theta_\text{f}$.
  The black curve represents an acceptance of $\delta\theta_f = \ang{20}$, which
  will be used in the P2 experiment.}
  \label{fig:DsZ2_vs_theta_variable_deltaTheta}
\end{figure}

Figure~\ref{fig:DsZ2_vs_theta_variable_deltaTheta} shows the dependence of
$\Delta s_\text{W}^2$ on $\bar{\theta}_\text{f}$ for $E_\text{beam} =
\SI{155}{MeV}$ and different choices of $\delta\theta_\text{f}$. In general, a
larger value of $\delta\theta_\text{f}$ leads to a larger $N$ and therefore to a
smaller statistical uncertainty of $A^\text{exp}$. Since the statistical
uncertainty of $A^\text{exp}$ is the dominant contribution to $\Delta
s_\text{W}^2$, the achievable uncertainty in the electroweak mixing angle
decreases with rising $\delta\theta_\text{f}$. The larger the acceptance, 
the smaller is the effect of increasing $\delta\theta_\text{f}$ on $\Delta
s_\text{W}^2$, because contributions by the nucleon form factors become more 
significant at larger scattering angles. To keep the nucleon form factors'
contributions reasonably small, we have decided to use $\delta\theta_\text{f}
\leq \ang{20}$.
 \\

\begin{figure}[htb]
  \centering
  \resizebox{0.5\textwidth}{!}{\includegraphics{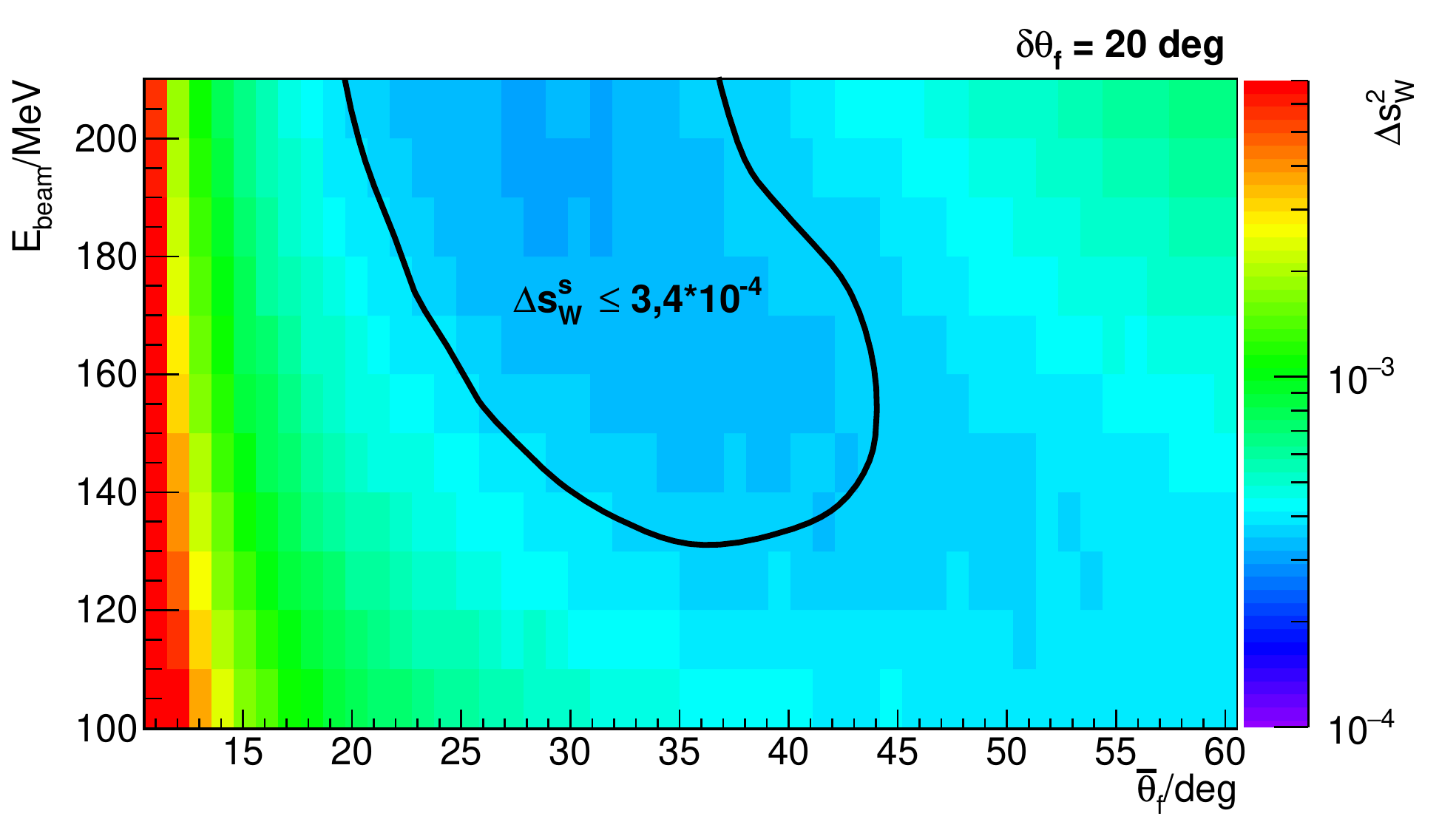}}
  \caption{Dependence of $\Delta s_\text{W}^2$ on $\bar{\theta}_\text{f}$ and
  $E_\text{beam}$ for $\delta\theta_\text{f} = \ang{20}$. In the region marked
  by the black curve, values of $\Delta s_\text{W}^2 \leq \SI{3.4e-4}{}$ are
  achievable.}
  \label{fig:DsZ2_vs_E_and_theta}
\end{figure}

Figure~\ref{fig:DsZ2_vs_E_and_theta} shows the dependence of $\Delta
s_\text{W}^2$ on $E_\text{beam}$ and $\bar{\theta}_\text{f}$ for
$\delta\theta_\text{f} = \ang{20}$. Values of $\Delta s_\text{W}^2 \leq 
\SI{3.4e-4}{}$ can be achieved in the region marked by a black curve. 
 \\

To carry out the P2 experiment within the envisaged measurement time of $T =
\SI{10e4}{h}$, we have decided to use a beam energy of $E_\text{beam} =
\SI{155}{MeV}$, a central scattering angle of $\bar{\theta}_\text{f} = \ang{35}$
and a detector acceptance $\delta\theta_\text{f} = \ang{20}$.
Table~\ref{tab:achievable_precision} lists the results of an error propagation
calculation for this choice of kinematic parameters along with the error
contributions stemming from the statistical uncertainty of $A^\text{exp}$, the
contribution of the beam polarization as well as the contribution from
helicity correlated beam fluctuations. In order to extract the electroweak
mixing angle from the measured uncertainty, one has to take the nucleon form
factors and radiative corrections to the proton's weak charge into account. The
expected contributions to $\Delta s_\text{W}^2$ due to uncertainties of the 
form factors and of $\Box_{\gamma Z}$ are also listed in 
Tab.~\ref{tab:achievable_precision}.
\begin{table}[htb]
  \begin{center}
    \begin{tabular}{cc}
      \toprule[1.5pt]
      $E_\text{beam}$ & $\SI{155}{MeV}$ \\
      \midrule
      $\bar{\theta}_\text{f}$ & $\ang{35}$ \\
      \midrule
      $\delta\theta_\text{f}$ & $\ang{20}$ \\
      \midrule
      $\langle Q^2 \rangle_{L = \SI{600}{mm}, ~\delta\theta_\text{f} = \ang{20}}$ &
      \SI{6E-3}{(GeV/c)^2}\\
      \midrule[1.5pt]
      $\langle A^\text{exp} \rangle$ & \SI{-39.94}{ppb} \\
      \midrule
      $(\Delta A^\text{exp})_\text{Total}$ & \SI{0.56}{ppb}
      (\SI{1.40}{\percent}) \\
      \midrule
      \midrule
      $(\Delta A^\text{exp})_\text{Statistics}$ & \SI{0.51}{ppb}
      (\SI{1.28}{\percent}) \\
      \midrule
      $(\Delta A^\text{exp})_\text{Polarization}$ & \SI{0.21}{ppb}
      (\SI{0.53}{\percent}) \\
      \midrule
      $(\Delta A^\text{exp})_\text{Apparative}$ & \SI{0.10}{ppb}
      (\SI{0.25}{\percent}) \\
      \midrule[1.5pt]
      \midrule[1.5pt]
      $\langle s_\text{W}^2 \rangle$ & \num{0.23116} \\
      \midrule
      $(\Delta s_\text{W}^2)_\text{Total}$ & \num{3.3E-4} (\SI{0.14}{\percent}) \\
      \midrule
      \midrule
      $(\Delta s_\text{W}^2)_\text{Statistics}$ & \num{2.7E-4}
      (\SI{0.12}{\percent}) \\
      \midrule
      $(\Delta s_\text{W}^2)_\text{Polarization}$ & \num{1.0E-4}
      (\SI{0.04}{\percent}) \\
      \midrule
      $(\Delta s_\text{W}^2)_\text{Apparative}$ & \num{0.5E-4}
      (\SI{0.02}{\percent}) \\
      \midrule
      \midrule
      $(\Delta s_\text{W}^2)_{\Box_{\gamma Z}}$ & \num{0.4E-4}
      (\SI{0.02}{\percent}) \\
      \midrule
      $(\Delta s_\text{W}^2)_\text{nucl. FF}$ & \num{1.2E-4}
      (\SI{0.05}{\percent}) \\
      \midrule[1.5pt]
      \midrule[1.5pt]
      $\langle Q^2 \rangle_\text{Cherenkov}$ & \SI{4.57e-3}{(GeV/c)^2} \\
      \midrule
      $\langle A^\text{exp} \rangle_\text{Cherenkov}$ & \SI{-28.77}{ppb} \\
      \bottomrule[1.5pt]
    \end{tabular}
  \end{center}
  \caption{Results of the error propagation calculation performed for the design
  parameters of the P2 experiment. $\langle Q^2 \rangle_{L = \SI{600}{mm},
  ~\delta\theta_\text{f} = \ang{20}}$ is the expected value of $Q^2$ after
  averaging over the target's length $L$ and the acceptance in the electron
  scattering angle $\theta_\text{f}$ and has been calculated in analogy to
  Eq.~(\ref{eq:Apv_averaged}). The values given in round brackets are the
  relative errors with regard to the expected value. $\langle Q^2
  \rangle_\text{Cherenkov}$ and $\langle A^\text{exp} \rangle_\text{Cherenkov}$
  are the expected values obtained if electrons scattered with 
  $\theta_\text{f} < \bar{\theta}_\text{f} - \delta\theta_\text{f}/2$ and hitting 
  the Cherenkov detector are taken into account 
  (see Sect.~\ref{sec:Simulation} for details). 
  }
  \label{tab:achievable_precision}
\end{table} \\
The expected value of the parity-violating asymmetry is
\begin{equation}
  \langle A^\text{exp} \rangle = \SI{-39.94}{ppb} 
\end{equation}
with an uncertainty of
\begin{equation}
  \Delta A^\text{exp} = \SI{0.56}{ppb}
\end{equation}
in \SI{1e4}{h} of measurement time. This corresponds to a relative uncertainty
of
\begin{equation}
  \frac{\Delta A^\text{exp}}{\langle A^\text{exp} \rangle} = \SI{1.40}{\percent}. 
\end{equation}
The expected uncertainty for the weak mixing angle is 
\begin{equation}
  \Delta s_\text{W}^2 = \num{3.3e-4}
\end{equation}
corresponding to a relative uncertainty of
\begin{equation}
  \frac{\Delta s_\text{W}^2}{\langle s_\text{W}^2 \rangle} = \SI{0.14}{\percent}
\end{equation}
for $s_\text{W}^2$ and
\begin{equation}
  \frac{\Delta Q_\text{W}(\text{p})}{Q_\text{W}(\text{p})} = \SI{1.83}{\percent}
\end{equation}
for the proton's weak charge.

\subsubsection{Scattering off the target entry and exit windows}
\label{sec:Scattering off the target entry and exit windows}

Beam electrons which scatter off the windows of the target cell are an 
additional source of
uncertainty. This effect is briefly discussed here and will be
included in the error propagation calculation (see Sect.~\ref{sec:Results
of the error propagation calculations}). 

Omitting all other sources of background and beam polarization, the measured
asymmetry consists of two contributions:
\begin{equation}
A^\text{exp}=(1-f)\cdot\langle A^\text{PV}\rangle+f\cdot \langle A^\text{Alu}\rangle
\end{equation}
where $A^\text{Alu}$ is the parity-violating asymmetry in eAl scattering and
$f$ is the dilution factor
\begin{equation}
f=\frac{Y_\text{eAl}}{Y_\text{ep}+Y_\text{eAl}}
\end{equation}
$Y_\text{eAl}$ and $Y_\text{ep}$ is the yield of scattering events off
aluminum and of elastic scattering off the proton, respectively. This equation
can be solved for $A^\text{PV}$. Gaussian error propagation yields
\begin{equation}
\begin{split}
& \Delta\langle A^\text{PV}\rangle
= 
\\ &
\sqrt{\left(\frac{\Delta A^\text{exp}}{1-R}\right)^2+\left(\frac{f\Delta
\langle A^\text{Alu}\rangle}{1-f}\right)^2+\left(\frac{A^\text{exp}-\langle A^\text{Alu}\rangle}{(1-f)^2}\Delta f\right)^2}.
\end{split}
\end{equation}
Both the asymmetry of the aluminum scattering $\langle A^\text{Alu}\rangle$
and the dilution factor $f$ need to be measured in the P2 experiment. For an
estimation of the size of the effect we use measurements performed by the A4
experiment \cite{Maas:2004ta} and the QWeak experiment \cite{Androic:2013rhu}.
From the A4 measurements with an aluminum target which where performed at 
the same central scattering
angle that is foreseen for P2, $\bar{\theta}_\text{f}=\ang{35}$, but with higher electron
energies from \SI{570}{MeV} up to \SI{854}{MeV}, we find a dilution factor of
$f=0.010$ for a P2 target with a \SI{60}{cm} liquid hydrogen volume along the beam
axis and a total thickness of $d_0=\SI{250}{\micro m}$ for the aluminum entry 
and exit windows. Other window thicknesses $d_\text{Alu}$ can be calculated 
by simply applying the factor $d_\text{Alu}/d_0$ to this dilution factor. 
We plan to determine this dilution factor by measuring the detector yield 
with an empty target cell with an uncertainty of $\Delta f/f\leq0.05$. The 
QWeak measurements with an aluminum target show that the asymmetry 
$\langle A^\text{Alu}\rangle$ is about one order of magnitude larger than 
the asymmetry $\langle A^\text{PV}\rangle$. We use this result to estimate
$\langle A^\text{Alu}\rangle=\SI{400}{ppb}$ in our case. We plan to perform a 
\SI{500}{hour} measurement with a \SI{3}{mm} thick aluminum target. Based on our rate 
estimation, the expected uncertainty of this measurement is $\Delta\langle 
A^\text{Alu}\rangle=\SI{6}{ppb}$.
\begin{figure}[htb]
  \centering
  \resizebox{0.48\textwidth}{!}{\includegraphics{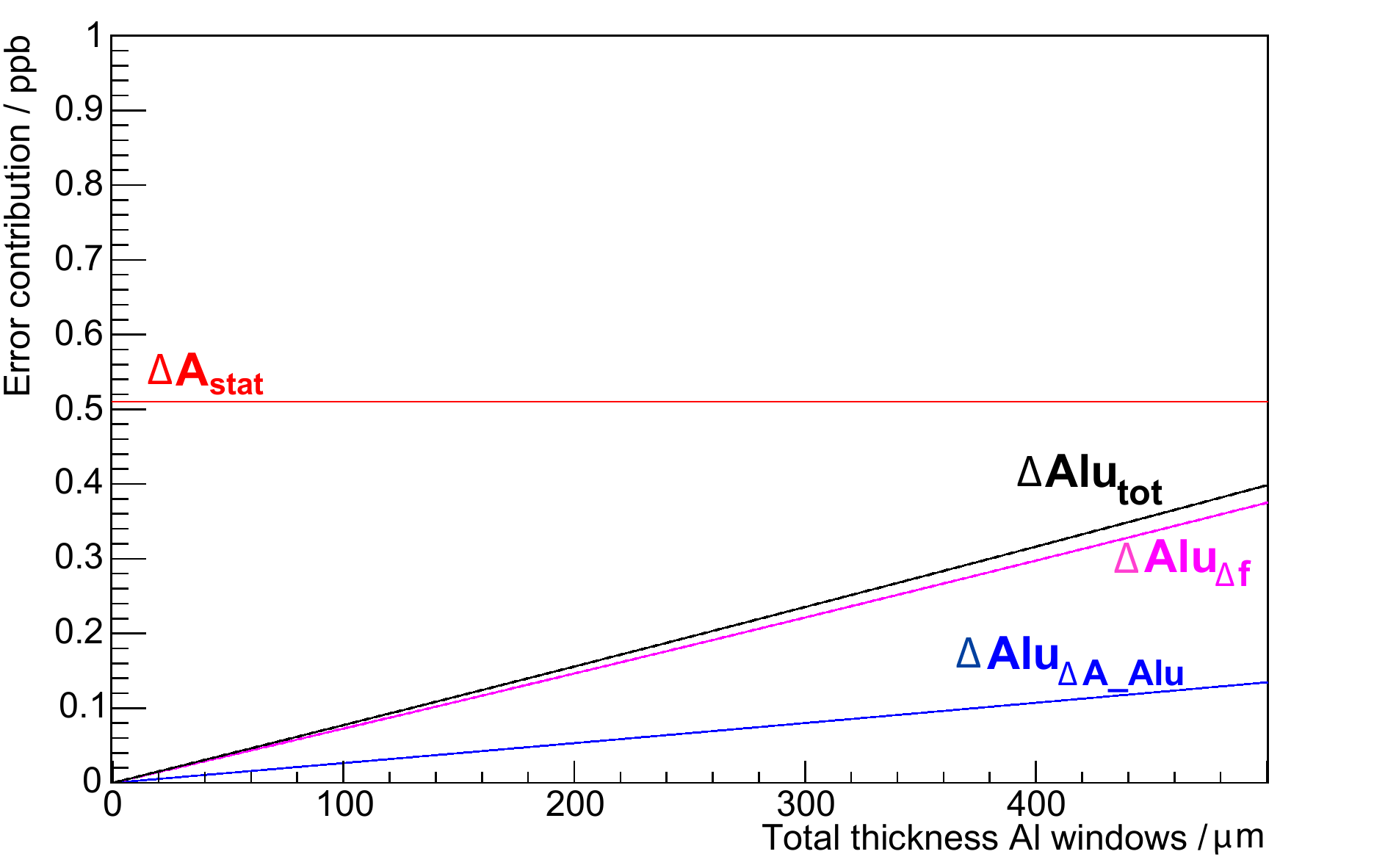}}
  \caption{
  Error contribution from scattering events of the aluminum windows
  of the target cell as a function of the total thickness of the entry and
  exit window. Shown is in blue the contribution arising from the uncertainty
  of the asymmetry in the eAl scattering $\Delta\langle A^\text{Alu}\rangle$ 
  (in ppb) and in purple the contribution arising from the uncertainty in the 
  dilution factor $f$. Here w assume $\Delta\langle
  A^\text{Alu}\rangle=\SI{6}{ppb}$ and $\Delta f/f=0.05$. The black line shows
  the quadratic sum of both terms. For comparison, the expected statistical 
  uncertainty of the P2 measurement $\Delta A^\text{exp}=\SI{0.51}{ppb}$ is 
	shown in red.
  }
  \label{fig:error_Al_windows}
\end{figure}
Based on these numbers we can perform the error calculation. 
Figure~\ref{fig:error_Al_windows} shows the contribution from the target window 
scattering to the uncertainty of $\langle A^\text{PV}\rangle$ as a
function of the total thickness of the target windows. For example, with a
total thickness of $d_0=\SI{250}{\micro m}$ we obtain an error contribution of 
\SI{0.20}{ppb}. It is of similar size as the error from the electron beam 
polarization and well below the anticipated statistical uncertainty.

%% file: accelerator.tex
\begin{figure*}
\begin{center}
\includegraphics[width=0.7\textwidth]{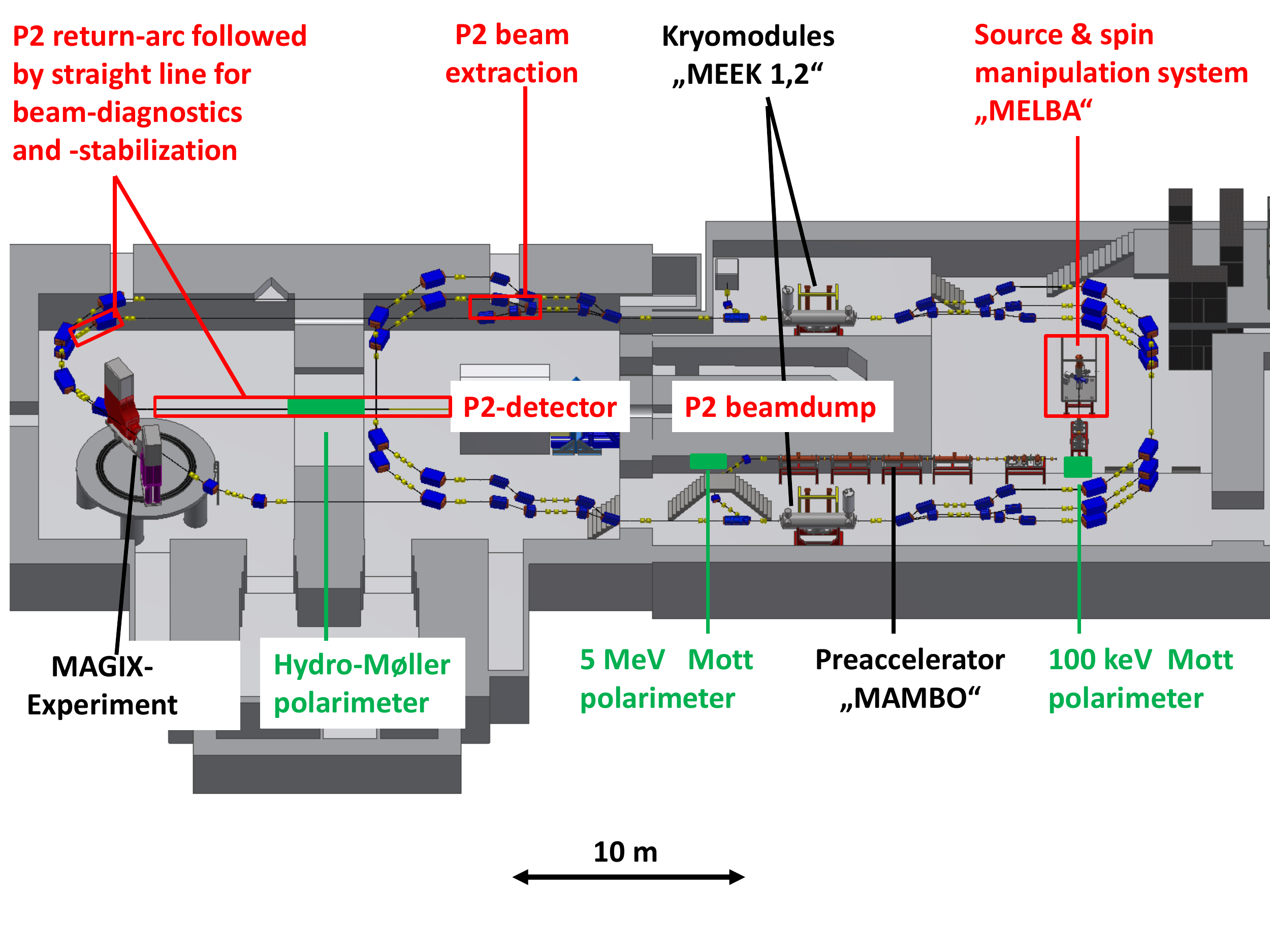}
\end{center}
\caption{View of the MESA accelerator. Areas of specific importance for P2 are indicated. Electron optical components are labelled in black colour, beam and spin control  in red and polarimeters in green.}
\label{fig:MESA_floorplan}
\end{figure*}

The increased demand for experiments at the MESA accelerator has 
necessitated  a new experimental hall which is currently being 
erected. The civil construction work will be finalized in 2020. 
Figure \ref{fig:MESA_floorplan} shows  the MESA accelerator layout 
with several components that are especially relevant  for the P2 
experiment. In P2, the beam will be extracted from the accelerator, 
directed towards the experiment and  will be stopped afterwards in the 
heavily shielded beam dump building. The beam energy gain per pass is 
given by the acceleration capacity of the two cryomodules which are 
designed for \SI{25}{MeV} each. After  three passages through the 
cryomodules the beam energy at P2 reaches \SI{155}{MeV}, 
(\SI{50}{MeV} per pass + \SI{5}{MeV} from injector), lower energies 
are possible.

An \SI{85}{\percent} spin-polarized beam of \SI{150}{\micro A} intensity is 
generated in a polarized source (described in Sect.~\ref{sec:Source}) and then injected 
into the MEsa Low-energy Beam Ap\-paratus, MELBA. One of the main purposes of 
MELBA is spin manipulation. Due to $g-2$ precession, the spin angle at P2 would 
not appear exactly longitudinal if no compensation by an additional spin 
rotation is provided. The additional spin angle can be achieved by a Wien-filter 
spin rotator of the type that is presently installed in MAMI \cite{Tioukine:2006}. 
A second Wien filter will be installed  upstream of the compensating filter. 
The purpose of this installation is to rotate the (initially longitudinal) spin 
by \ang{90} out of the accelerator plane. This spin rotation is kept fixed.  
Then a longitudinal magnetic field $B_{long}$ that is created by a solenoid will 
rotate the spin towards transverse orientation in the accelerator plane and the 
second Wien filter compensates the $g-2$ precession. By reversing the current of  
the solenoid, a \ang{180} spin flip can be obtained with no first order change 
of electron optical properties of the solenoid - the focusing strength is 
$\propto \int B_{long}^2dl$.  This spin reversal is independent from the optical 
spin reversal that is generated by helicity switching at the source. The 
procedure serves to provide a further check for beam systematics. This method 
was developed for the parity experiments at JLab \cite{Grames:2011}. After spin 
manipulation the beam is directed via an $\alpha$-magnet to the beam bunching 
and collimation system which prepares the beam for RF-acceleration in the 
pre-accelerator, the so-called MilliAMpere BOoster (MAMBO).

Space is available behind the  $\alpha$-magnet in order to install a double 
scattering Mott polarimeter. It  operates at the source energy of \SI{100}{keV} 
and will be reached by the beam after switching off the magnet. 
Behind MAMBO we will set-up a single scattering Mott polarimeter which uses a 
beam energy of \SI{5}{MeV}. These polarimeters require transverse spin 
polarization which is provided by switching off the second Wien filter. The last 
polarimeter in the system, the Hydro-M{\o}ller, requires longitudinal 
polarisation which coincides with the experimental requirements. More details 
concerning polarimetry can be found in Sect.~\ref{sec:Polarimetry}. Most of 
the components of MELBA, including the two Wien filters, are either ready for 
commissioning or already in operation in different test set-ups.

After three passages through the cryomodules the beam is extracted via a  
magnetic chicane towards the P2 beamline. It should be noted that this part of 
MESA (together with two of the recirculation arcs) is located in the 
P2 experimental hall.  Handling the radiation levels coming from the target 
seems feasible since only relatively robust accelerator components are installed 
in this region. More demanding, but also feasible, is the task to shield all 
beamline components from the magnetic fringe field of the P2 solenoid, which has 
a large aperture. 

After extraction, the beam is directed away from he P2 experiment into the 
MAGIX hall. The main idea is to obtain a long straight line in front of the 
experiment for beam diagnostic and stabilization purposes - see 
Sect.~\ref{sec:Beam control} for details. The final \ang{180} bend  in front 
of the experiment can be used to create a large longitudinal dispersion which is 
needed for energy stabilization. The straight line will also contain the 
Hydro-M{\o}ller polarimeter that occupies $\approx$ \SI{2.5}{m} of beam line 
by its cryostat. The cryostat can be installed in a rectangular opening 
(\SI{5}{m} width, \SI{0.8}{m} height) in the wall between the MAGIX and the P2 
hall, see Fig.~\ref{fig:MESA_floorplan}.

%% file: source.tex
Though a source has recently been put into operation that is capable to achieve 
the anticipated maximum beam currents of MESA of \SI{10}{mA} 
\cite{Friederich:2015}, such a device is not mandatory and maybe not even 
advisable for P2, since it incorporates increased technological risks. On the 
other hand, the source developed for MAMI \cite{Aulenbacher:1997} represents a 
sufficient basis for the P2 experiment. 
This source has been operated for decades at the MAMI accelerator 
\cite{Aulenbacher:2011} and has produced beams with nearly \SI{90}{\percent} of 
polarization and currents well above the level needed for P2. An important 
factor is the operational lifetime that can be expected. The ability of the 
photocathode to convert light quanta into electrons, the so-called quantum 
efficiency, decays due to radiation damage which in turn causes a finite 
operational lifetime of the source.  Operation of the MAMI source at 
\SI{200}{\micro A} \cite{Aulenbacher:2006} has revealed a charge lifetime of 
\SI{200}{C} - that is \SI{200}{Coulombs} of charge can be produced while the 
initial quantum efficiency drops to 1/e, i.e.~to about \SI{37}{\percent}, of 
its initial value. This can be handled even with the existing laser system of 
the polarized source at MAMI which is able to deliver \SI{300}{mW}.  
A moderate quantum efficiency of \SI{0.5}{\percent} (corresponding to a 
photo-sensitivity of \SI{3}{mA/Watt} at the operational wavelength of 
\SI{778}{nm}) is assumed for the GaAs/GaAsP superlattice photocathode that has 
to be employed to achieve polarizations surpassing \SI{85}{\percent}. Then, 
after one lifetime, less than \SI{200}{mW} of laser power on the cathode will be 
necessary. At this power level some improvement of heat transfer from the 
photocathode is needed in order to limit a temperature increase which reduces 
the lifetime. We consider this feasible, which will allow to maintain the 
\SI{200}{C} of lifetime -- this value  was observed with a cathode of much higher 
initial quantum efficiency. Our estimation of the continuous operation during P2  
is therefore more than \SI{12}{days}. This is matching the run-time schedule 
foreseen for MESA which has a period of \SI{2}{weeks} composed of \SI{12}{days} 
of run-time + \SI{2}{days} of maintenance. Since the photocathodes can be 
regenerated within a very short time \cite{Aulenbacher:1997}, the long term 
operation of P2 with a design based on the MAMI source is feasible. The source
has been built with  several improvements concerning, in particular, a better 
vacuum system.  It has been thoroughly tested and has indeed obtained improved 
lifetime parameters \cite{Aulenbacher:2011zzb}. The source is currently used in 
a separate lab to investigate the double scattering Mott polarimeter foreseen 
for P2 and will be installed at the MESA site when the civil construction work 
for the experimental halls of MESA is finalized.

%% file: polarimetry.tex
The concept of polarimetry for P2 was sketched in \cite{Aulenbacher:2012}. 
A chain of three polarimeters will provide for three independent polarization 
measurements achieving an absolute accuracy of better than $\Delta P/P =0.01$ 
with each polarimeter. The chain consists of a double Mott polarimeter at 
\SI{100}{keV} beam energy, a single scattering Mott polarimeter at \SI{5}{MeV} 
and a M{\o}ller polarimeter which is located directly in front of the P2 
experiment. The latter, the so-called ``Hydro-M{\o}ller'' will have online 
capability. The others are invasive devices but will allow very fast 
measurements. Laser-Compton back-scattering which is a very expedient way of 
polarimetry at energies above \SI{1}{GeV} is not promising in our case due to 
the small asymmetries caused by the low beam energy. We discuss the individual 
features and the status of the polarimeters in the following subsections.

\subsubsection{Double scattering Mott polarimeter}

An intriguing approach towards high precision polarimetry is to use double scattering.
The idea is to measure the effective analyzing power $S_{\mathit{eff}}$ of a scattering
experiment. All other types of polarimeters have to rely on (i) a theoretical 
determination of the analyzing power of the elementary scattering process and 
(ii) a careful determination of the deviations from the elementary process in the 
real experiment. The double scattering experiment does away completely with 
the first and offers considerable advantages with respect to the second aspect. 

The method  was thoroughly analyzed
in a series of articles by the group of Prof.~Kessler at University of M{\"u}nster 
\cite{Kessler:1990,Gellrich:1991,Mayer:1993}. The measurement works in the following
way. A first elastic scattering of an unpolarized beam produces a polarized scattered
beam with a vertical polarization $P_{\mathit{Scat}}=S_{\mathit{eff}}$. This polarization is in general
lower than the theoretical analyzing power of the process $S_0$ due to the spin diffusion
in the target of finite thickness. This creates several of the systematic errors in
conventional Mott polarimetry whereas it is -- at least in principle -- not important in
this case. The secondary beam is directed under a given solid angle to an identical target where a scattering
asymmetry is observed under the same solid angle. Provided that
the two scatterings -- notably the targets and the solid angles -- can be made identical, 
the observed asymmetry $A_{\mathit{obs}}$ is given by     

\begin{equation}\label{eq:double_scat}
	A_{\mathit{obs}}=S^2_{\mathit{eff}}
\end{equation}
This is due to the equivalence of analyzing power and polarization power in 
elastic scattering. Then, $S_{\mathit{eff}}$ is  obtained by taking the square 
root of the observed asymmetry, the sign of the analyzing power is known from 
theory. A challenge in such an experiment is the control of false asymmetries 
since the spin in the second scattering cannot easily be reversed (the initial 
beam is unpolarized!). However, an  ingenious scheme was designed 
in \cite{Kessler:1990} from which a control of such asymmetries in the range 
of \SI{0.1}{\percent} can be achieved.  

After this procedure  the targets are calibrated and each of them can be used to 
analyze a polarized beam with the effective analyzing power 
$S_{\mathit{eff}}$. At \SI{120}{keV}, an accuracy of 
$\Delta S_{\mathit{eff}}/S_{\mathit{eff}}=0.24\%$ was claimed for the secondary 
target \cite{Gellrich:1991}.  

Kessler's group used primary beams of up to \SI{120}{keV}. This scheme cannot be 
extended to much higher energies due to the rapidly falling  elastic cross 
section. The method is therefore restricted to energies typical for polarized 
sources and is, of course, invasive. The MESA source will be operated at 
\SI{100}{keV} which is well suited to the application of the method. 
Once $S_{\mathit{eff}}$ is determined,  the polarized beam will be analyzed in 
single scattering, where measurements with statistical accuracies of 
$<$\SI{1}{\percent} are possible within minutes. The beam current in such 
measurements is several orders of magnitude smaller than needed for P2. To make 
sure that the spin polarization of the electron beam does not change, we will 
make use of the large dynamic range of the \SI{5}{MeV} Mott polarimeter 
discussed in the next section. 

The apparatus of the M{\"u}nster group was transferred to Mainz \cite{Aulenbacher:2012}
where its applicability for the P2 experiment at the MESA accelerator is tested. 
It has been demonstrated that the mechanically complicated apparatus can be 
operated very reliably together with the \SI{100}{keV} polarized source of MESA.  
Statistical errors of $\Delta S_{\mathit{eff,stat.}}/S_{\mathit{eff}} <0.5\%$ 
have been achieved, no indications of drifts of the apparatus during the 
several days long calibration procedure were observed. The extracted effective 
analyzing power is in satisfactory agreement with the values observed by the 
M{\"u}nster group \cite{Gellrich:1991} but not with $S_{\mathit{eff}}$ 
measurements by us and other groups that are based on polarized single 
scattering \cite{Steigerwald1994}. We will address this issue by a direct 
comparison of the two methods in the near future \cite{Molitor2018}. 

Double scattering offers another attractive feature that may allow to reduce the
systematic error further. It was
observed by Hopster and Abraham \cite{Hopster:1987} that additional observables 
can be gained if a polarized primary beam is employed, provided that vertical 
beam polarization can be flipped while maintaining its absolute value. For the 
polarized source at MESA this is done by switching the circular polarization 
($\sigma^\pm$) of laser light that excites the photocathode of the source. It is 
assumed that under this helicity flip the following condition holds for the 
vector of the electron beam polarization:
\begin{equation}\label{eq:helicity_flip_symmetry}
    \vec{P}_{\sigma^+} \rightarrow -\vec{P}_{\sigma^-}.
\end{equation}
The primary target is
considered as an auxiliary target which does no longer have to have the same effective
analyzing power as the second one ($S_{\mathit{eff}}$), but has a value $S_T$ instead.  
The double scattering experiment with
unpolarized beam now yields 
\begin{equation}\label{eq:double_scat1}
	A_{1}=S_TS_{\mathit{eff}}.
\end{equation}
One can also move the second target including its detection system into the 
primary beam path, then observing  
\begin{equation}\label{eq:double_scat2}
	A_{2}=P_{0}S_{\mathit{eff}}.
\end{equation}

Scattering on the first target with the two input polarizations $\pm P_0$ yields different
secondary beam polarizations $P_{\uparrow, \downarrow}$ which depend on $S_T$ and 
on the depolarization factor of the auxiliary target $\alpha$, a fact that was 
observed in Ref.~\cite{Mayer:1993}: 
\begin{eqnarray}\label{eq:double_scat_aux}
	P_{\uparrow} = \frac{S_T+\alpha P_0}{1+P_0S_T}, \\
	P_{\downarrow} = \frac{S_T-\alpha P_0}{1-P_0S_T} \nonumber. 
\end{eqnarray}
Taking this into account, more asymmetries can be measured by double scattering: 
\begin{eqnarray}\label{eq:double_scat3}
	A_{3} &=& P_{\uparrow}S_{\mathit{eff}} , 
    \\
    \label{eq:double_scat4}
	A_{4} &=& P_{\downarrow}S_{\mathit{eff}} . 
\end{eqnarray}
While measuring $A_3$, $A_4$ one can also monitor the scattered beam current from the
auxiliary target which is nothing else than the single scattering asymmetry:
\begin{equation}\label{eq:double_scat5}
	A_{5}=P_{0}S_{T} .
\end{equation} 

One finds that the extension proposed in \cite{Hopster:1987} implies considerable advantages: 
\begin{itemize}
	\item The five observations $A_1$, $\ldots$, $A_5$ depend on the four unknowns $S_T$, 
	$S_{\mathit{eff}}$, $P_0,$ and $\alpha$. They result in an over-determined system 
	of equations hence allowing to
	extract the unknowns in five independent ways, providing systematic cross-checks.
	\item The condition of identical targets is revoked, but it is replaced by the
	symmetry condition expressed in Eq.~(\ref{eq:helicity_flip_symmetry}). 
	The equality of circular polarizations of the photo-exciting light can be 
	determined very accurately ($<$\SI{0.1}{\percent}).  In principle a solid 
	state effect that would lead to significantly unequal spin densities after 
	photo-absorption cannot be ruled out, but does not seem very likely. 
	\item The set of double scattering observables also can help to check the 
	validity of Eq.~(\ref{eq:helicity_flip_symmetry}) for our experimental set-up. 
	Note that the validity is implicitly assumed in virtually all polarized 
	scattering experiments, including P2.  		
\end{itemize}

It was shown in \cite{Mayer:1993} that the auxiliary target thickness can be 
varied by a factor eight without any observable influence on the extracted 
analyzing power of the second target at a level $<$\SI{0.4}{\percent}. 
The hard to achieve condition of identical target thicknesses is therefore not 
mandatory for precision polarimetry in double scattering.

\subsubsection{Single scattering Mott polarimeter at \SI{5}{MeV}}

\begin{figure}
	\centering
	\includegraphics[width=0.5\textwidth]{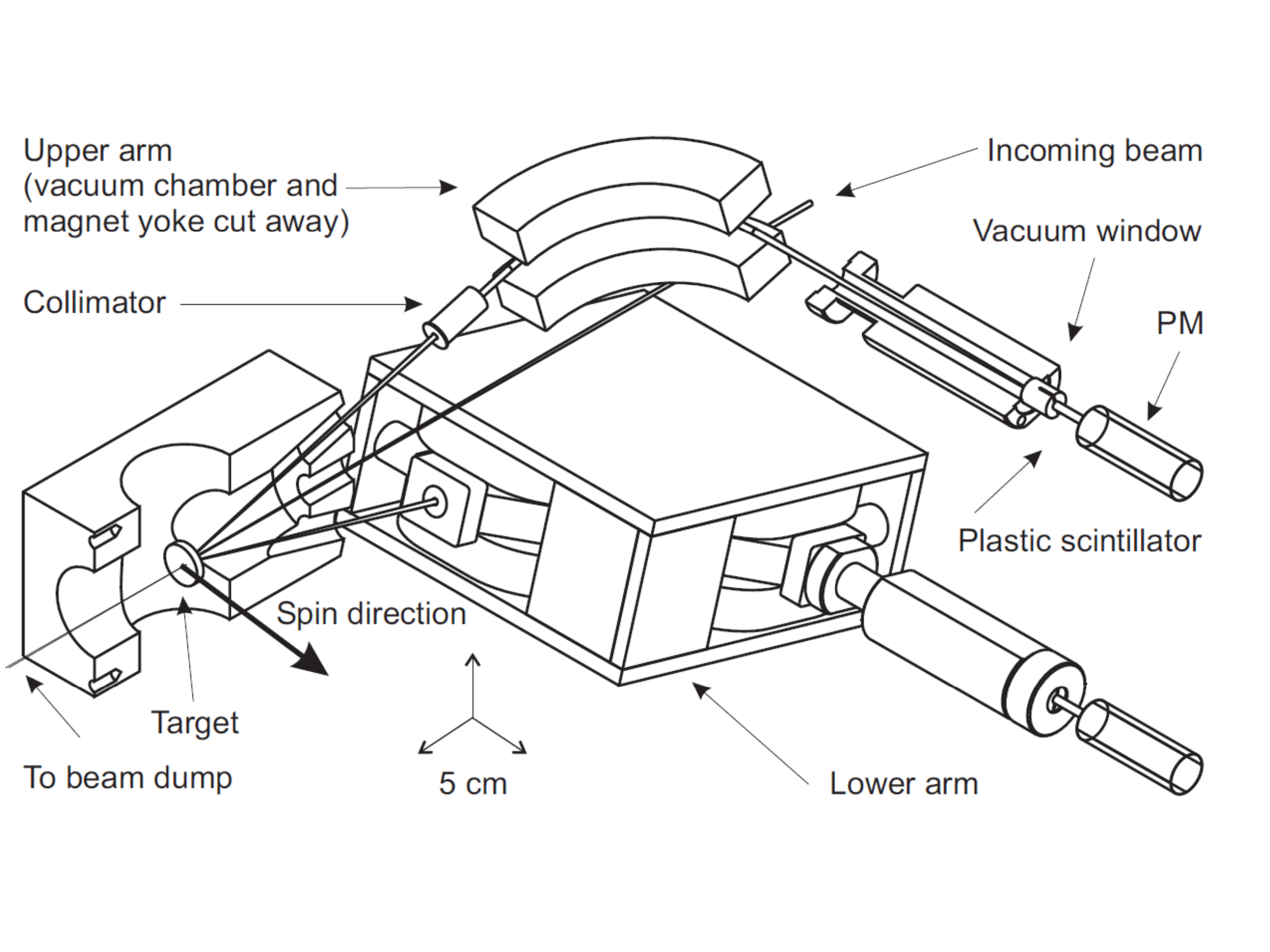}
	\caption{Schematic set-up  of the MAMI \SI{3.5}{MeV} Mott polarimeter 
	\cite{Tioukine:2011}. Backgrounds are suppressed by double-focusing magnetic 
	spectrometers.}
	\label{fig:MAINZ_MOTT_scheme}
\end{figure}

The design of the \SI{3.5}{MeV} Mott polarimeter \cite{Tioukine:2011} at MAMI 
can be used at MESA with small adaptations since the beam energy will be 
similar. An advantage of the polarimeter is its compact size, see 
Fig.~\ref{fig:MAINZ_MOTT_scheme}. 

The calibration of the analyzer requires measuring the scattering asymmetry of
targets of different thicknesses. The polarization is then determined by 
extrapolating the asymmetry to foil thickness zero ($A_0$) and then 
obtaining the beam polarization by using $P=A_0/S_0$, with $S_0$ being the 
theoretically calculated analyzing power of Mott scattering on the single atom.  
The three major sources of uncertainty are (i) background (ii) the uncertainty 
of extrapolation and (iii) the theoretical uncertainty of $S_0$.  
Based on the results from the \SI{3.5}{MeV} polarimeter we can state that  
\begin{itemize}
\item The background contribution can be reduced below \SI{0.5}{\percent} by 
dedicated experiments and improved Monte Carlo simulation. 
\item The spin diffusion 
 which reduces the effective analyzing power with target thickness can nowadays  
be modelled quite accurately starting from first principles 
\cite{Khakoo:2001,Aulenbacher:2008}. It is therefore no longer necessary to rely 
on fits to the data which have always caused significant uncertainties since 
there is no good motivation to prefer a specific fit function. It is therefore 
possible to suppress the uncertainty associated with the extrapolation below 
\SI{0.5}{\percent}. 
\item The remaining factor is the theoretical uncertainty of the analyzing 
power (called Sherman-function, $S_0$). An energy of several MeV is well suited 
for precision calculations since neither the nuclear nor the electronic details
of the atom contribute significantly \cite{Tioukine:2011,Roca:2017}. However, 
at multi-MeV energies and for high-$Z$ targets, a possible contribution of 
radiative effects has to be addressed. Though our findings and those at 
Jefferson Lab \cite{Steigerwald:2001} give reasons to assume that radiative 
effects are not important at the \SI{1}{\percent} level, no comprehensive 
theoretical treatment exists. Due to the revived interest in such calculations 
one can hope for improvement in the future. Progress has already been achieved 
in \cite{Roca:2017}, where the contribution of vacuum polarization has been 
calculated to be less than \SI{0.5}{\percent}. 
\end{itemize}
Overall one can expect that the absolute accuracy of our single-scattering 
device will be pushed to below \SI{1}{\percent}. 

\begin{figure}[ht]
\begin{center}
\includegraphics[width=0.5\textwidth]{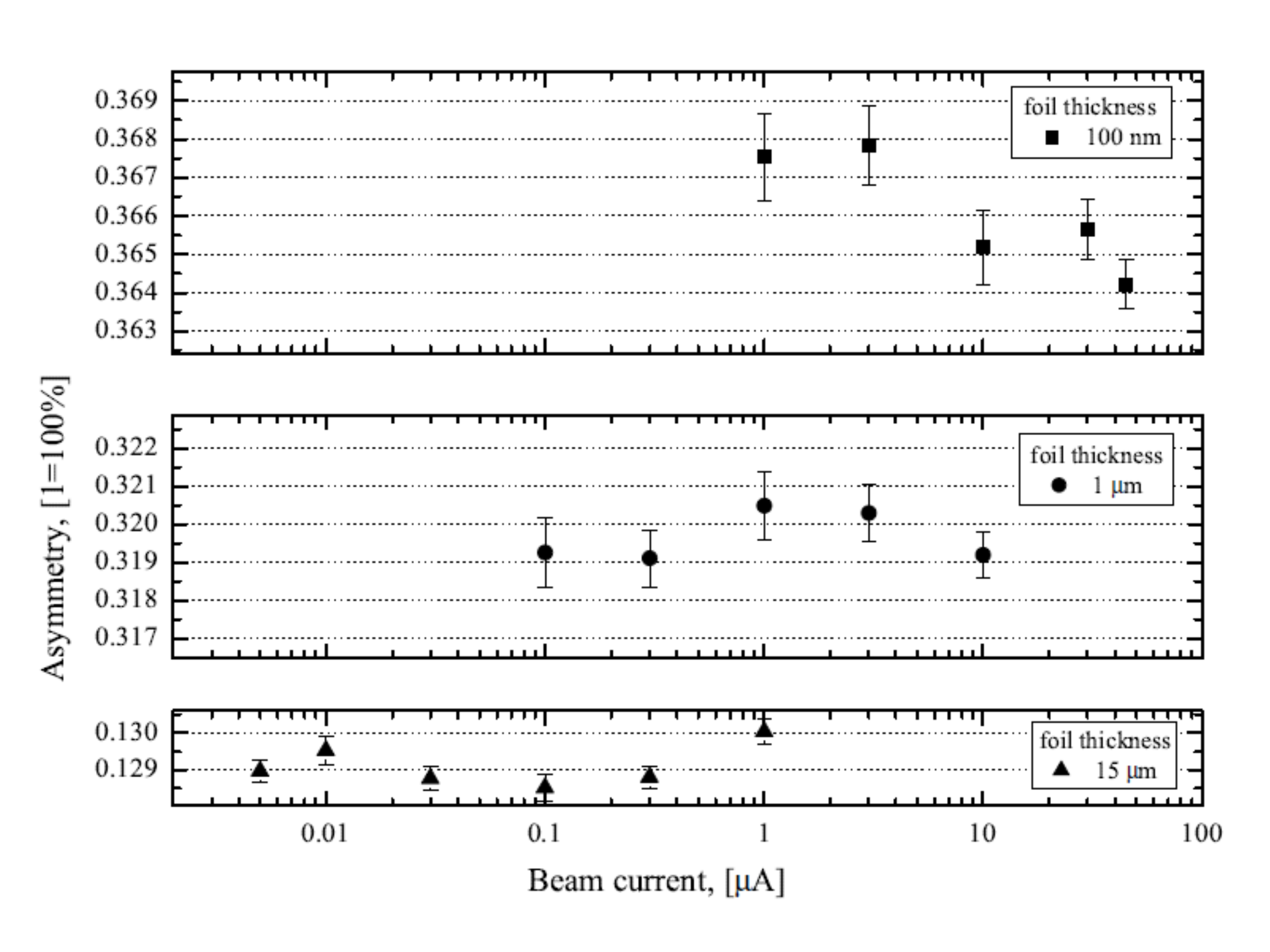}
\end{center}
\caption{Asymmetries in the MAMI 3.5\,MeV Mott polarimeter as a function of the 
primary current. Variable target thicknesses are used to limit the count rate. 
(Figure from Ref.~\cite{Tioukine:2011}).}
\label{mott:MAINZ_current_variation}
\end{figure}

A virtue of the multi-MeV Mott is its dynamic range and good reproducibility. 
Figure~\ref{mott:MAINZ_current_variation} demonstrates that consistent results 
at the sub-percent level can be achieved while varying the current between 
\SI{6}{nA} (order of magnitude intensity that is used for the double Mott) and 
\SI{50}{\micro A} which  approaches the design current of P2. Note that targets 
were changed to thinner ones while increasing the current to avoid dead time 
corrections. Such measurements will serve to connect the results of the double 
Mott with the Hydro-M{\o}ller polarimeter which is discussed in the next section. 
Efficient measurements with this polarimeter are only possible at currents 
exceeding \SI{20}{\micro A} and, of course, current-dependent effects cannot be 
excluded also in this device, see for instance \cite{Chudakov:2004de}. An 
illustration of the problem is the slight change of asymmetry at the largest 
beam currents (Fig.~\ref{mott:MAINZ_current_variation}) that could be caused by 
heating up of the photocathode due to the high laser intensity but might also 
be related to changes of the local background due to space charge effects in the 
\SI{3.5}{MeV} pre-accelerator. By comparing two polarimeters with high current 
capabilities such effects can be identified. 

\subsubsection{Hydro-M{\o}ller target}

We want  to replace the ferromagnetic target of a conventional 
M{\o}ller polarimeter by trapped polarized hydrogen atoms, an idea that was 
proposed by Chudakov and Luppov \cite{Chudakov:2004de}. Hydrogen atoms are 
injected into the fringe region of a solenoid with a magnetic field $\vec{B}$ 
pointing along the beam direction. The hydrogen atoms carry a magnetic moment 
$\vec{\mu}$ which makes them experience a force 
$\vec{F}=- \nabla \left( \vec{\mu} \cdot \vec{B} \right )$ which pushes the high 
field seeking hyperfine states $|a\rangle$ and $|b\rangle$ (see Eq.~\ref{eq:HFS}) towards 
the homogeneous field region. One of the injected states enters with both electron 
spin orientations, hence diluting the electronic spin polarization, 
\begin{eqnarray}\label{eq:HFS} 
	|a\rangle &=& \alpha |\Downarrow, \uparrow\rangle + \beta |\Uparrow, \downarrow \rangle ,
	\\
	|b\rangle &=& |\Downarrow, \downarrow \rangle .
\end{eqnarray} 
Here double arrows denote electron spins and the single-line arrows denote nuclear 
spin. The parameters $ \alpha = \sin\theta$, $\beta = \cos\theta$ can be calculated 
via $\tan(2 \theta) = 0.05/B[T]$. Note that $\theta$ varies between $\pi/4$ and 
$\pi/2$ when $B$ changes from  $0 \rightarrow \infty$. Since the relative 
density of particles in the 'wrong' state $|\Uparrow \downarrow \rangle$ 
is $\propto \beta^2$, it is strongly reduced by the magnetic field. A large field 
will thus lead to an electron polarization of $|P_e|= 1 - 10^{-5}$ at \SI{8}{Tesla}. 
This high-purity electronic spin ensemble represents the main advantage with respect 
to existing M{\o}ller polarimeters.

\begin{figure}[bht]
\begin{center}
\includegraphics[width=0.5\textwidth]{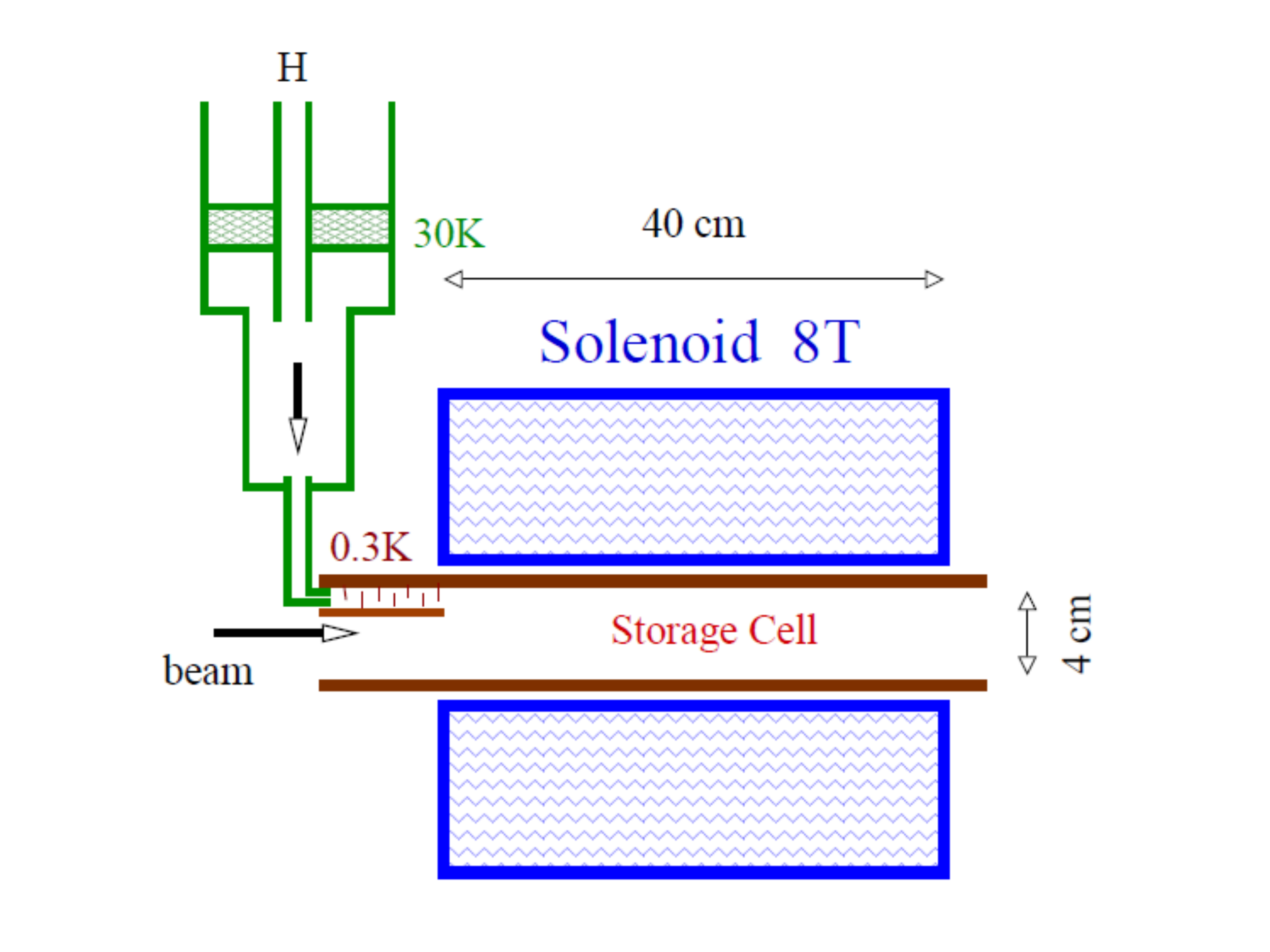}
\end{center}
\caption{Schematic of the Hydro-M{\o}ller atomic trap \cite{Chudakov:2004de}.}
\label{fig:Hydro}
\end{figure}

Figure \ref{fig:Hydro} sketches the working principle. Hydrogen atoms are 
created, for instance,  by a thermal dissociator and are injected into the 
fringe field of a strong solenoid. While entering the homogeneous part of the 
solenoid they are cooled by wall collisions to a temperature of 
$\approx$\SI{0.3}{K} so that they cannot escape in the axial direction anymore.  
Normally, the wall collisions would lead to adsorbtion and/or recombination 
of the hydrogen atoms, but these effects are strongly suppressed when the wall 
is covered with a superfluid film of helium. It is for the same reason that 
hydrogen is efficiently stored  since the particles are enclosed radially by the 
wall. A very low temperature of the helium film is required to keep the vapour 
pressure of helium low enough since helium adds unpolarized electrons to the 
target. 

Traps of this type have been built and achieved 
target densities of nearly \SI{3E15}{cm^{-3}} in  volumes of 
\SI{100}{cm^3} \cite{Mertig:1991}. They have however never been operated with 
a high power electron beam so far. 

The calculations in \cite{Chudakov:2004de} lead to the conclusion that even 
when interacting with the \SI{150}{\micro A} beam foreseen for P2, a target 
area density of \SI{E16}{cm^{-2}}, and high spin purity can be maintained. To 
achieve this, a first task is  the control of depolarizing resonances due to 
interaction with time varying magnetic fields from the beam time structure. This 
effect can be controlled by tuning the magnetic field in a way that the 
frequency of hyperfine transitions does not coincide with the beam harmonics.  
A more challenging issue is linked to the production of slow electrons or 
charged ions with at least one electron  (H$^-$, H$_2^-$, H$_2^+$) within the 
beam profile. Scattering reactions from such electrons will dilute the 
M{\o}ller signal. In contrast to neutral hydrogen their diffusion out of the 
beam area is suppressed since they are bound within the cyclotron radius which 
is smaller than the beam radius.  A possible solution is to add a small transverse 
electrical field ($E \approx \SI{1}{V/cm}$) in order to cause an $E \times B$ 
drift which removes the charged particles from the beam. It is technically 
challenging to provide the $E$-field, since insulated electrodes must be 
maintained at a temperature of \SI{0.3}{K} while also covering them with 
superfluid helium. If this can be achieved, a contamination of unpolarized 
electrons of less than \SI{0.01}{\percent} is expected. 

The low area density allows online operation since the beam is not significantly 
deteriorated by the target atoms. When the trap is irradiated with the full 
current of P2, a statistical precision of $<$\SI{0.5}{\percent} within less 
than \SI{1}{hour} could be achieved, if the solid angle of conventional 
M{\o}ller polarimeters is assumed. 

The completely polarized target eliminates two main error sources of existing 
M{\o}ller polarimeters, namely the uncertainty of the target polarisation 
\cite{Hauger:2001} and the enhancement of valence electrons in the detected 
sample of electron pairs by the Levchuk effect \cite{Levchuk:1994}. These 
represent about \SI{80}{\percent} of the error budget of the best existing 
M{\o}ller polarimeters \cite{Hauger:2001}, which opens the perspective to 
achieve accuracies well below $\Delta P/P = \SI{0.5}{\percent}$.

\subsubsection{Status of the Hydro-M{\o}ller target}

Generating the cryogenic environment  for the Hydro-M{\o}ller is not easy for 
several reasons, some of the most important ones are:   

\begin{itemize}
\item The trap must be open to let the beam pass which will expose the 
\SI{0.3}{K} surfaces to thermal radiation. The apertures cannot be made 
arbitrarily small in order to avoid beam losses.  

\item Though the direct deposition of heat by the beam in the target is not a 
major issue \cite{Chudakov:2004de}, the main heat source  is recombination of 
hydrogen  atoms to H$_2$ (\SI{4.5}{eV} per molecule). Taking into account the 
planned densities and leaving some headroom for inefficiencies of the 
refrigerator and limited heat conduction towards the trap surface, we find that 
a cooling power of \SI{60}{mW} at \SI{0.25}{K} is needed.  

\item The hydrogen dissociator and the solenoid must be incorporated.

\item The height of the refrigerator  must not exceed \SI{60}{cm} since, according 
to the present design of the P2 beamline, it has to be installed in a 
slit-shaped breakthrough between experimental halls 
(see Fig.~\ref{fig:MESA_floorplan}). The horizontal orientation of the 
refrigerator components is especially challenging.  

\item The polarimeter will have to run for very long times continuously. 
Since maintenance and repair will be difficult once the polarimeter is 
integrated in  the beamline, only very reliable components and procedures are 
recommendable. Glued interconnections and indium seals should be avoided. 
This leads to a large effort in prototyping of the many individual components. 
\end{itemize}
 
\begin{figure}[b]
\begin{center}
\includegraphics[width=0.5\textwidth]{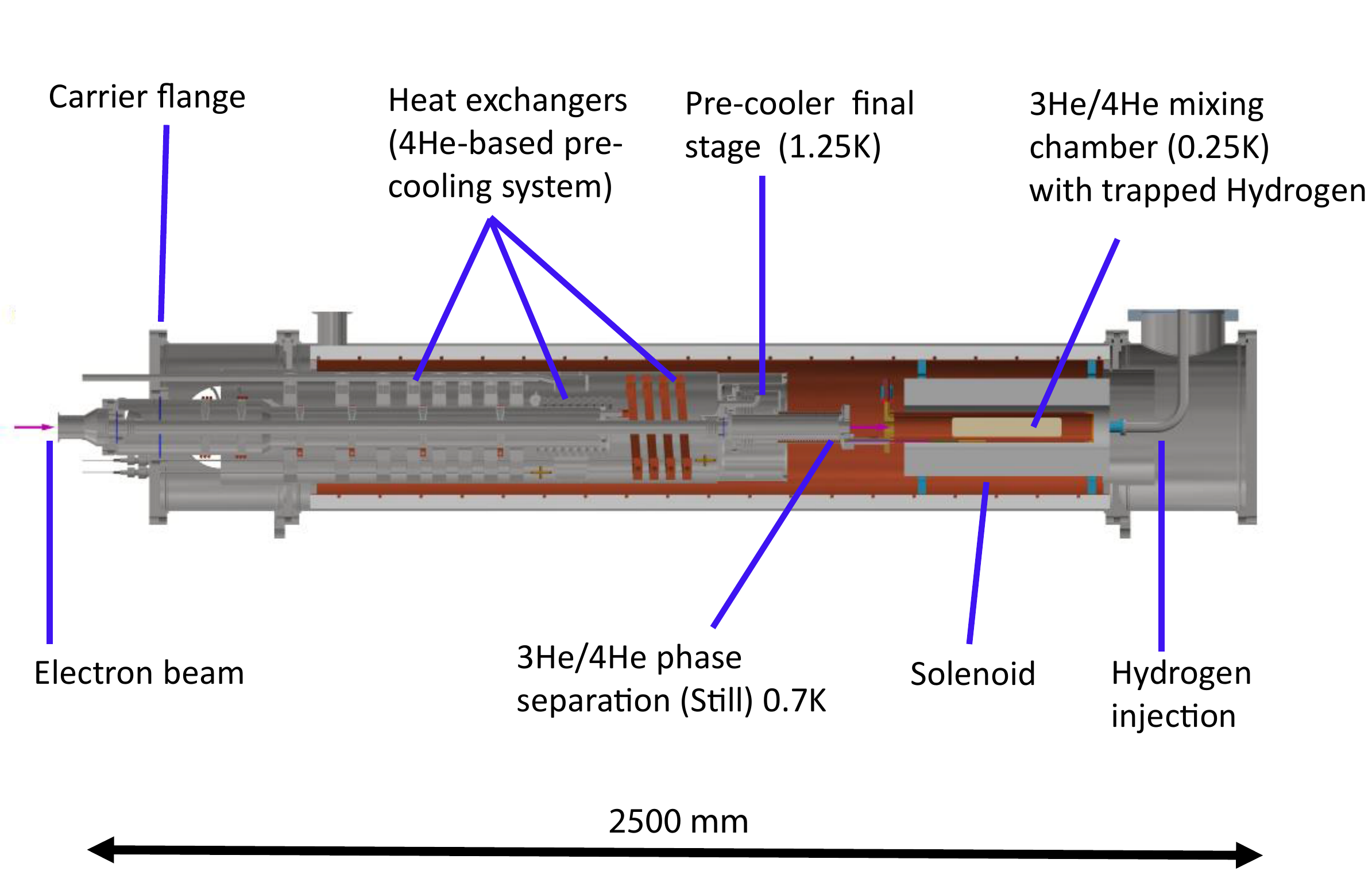}
\end{center}
\caption{Schematic of the refrigerator for the Mainz Hydro-M{\o}ller. 
Only positions of  major components are indicated.}
\label{fig:refrigerator}
\end{figure}

Figure~\ref{fig:refrigerator} shows a schematic of the device which is currently 
under construction. The design has been developed after a long series of 
consultations with cryostat experts from Dubna, JLab and CERN.  

\begin{figure}[ht]
\begin{center}
\includegraphics[width=0.48\textwidth]{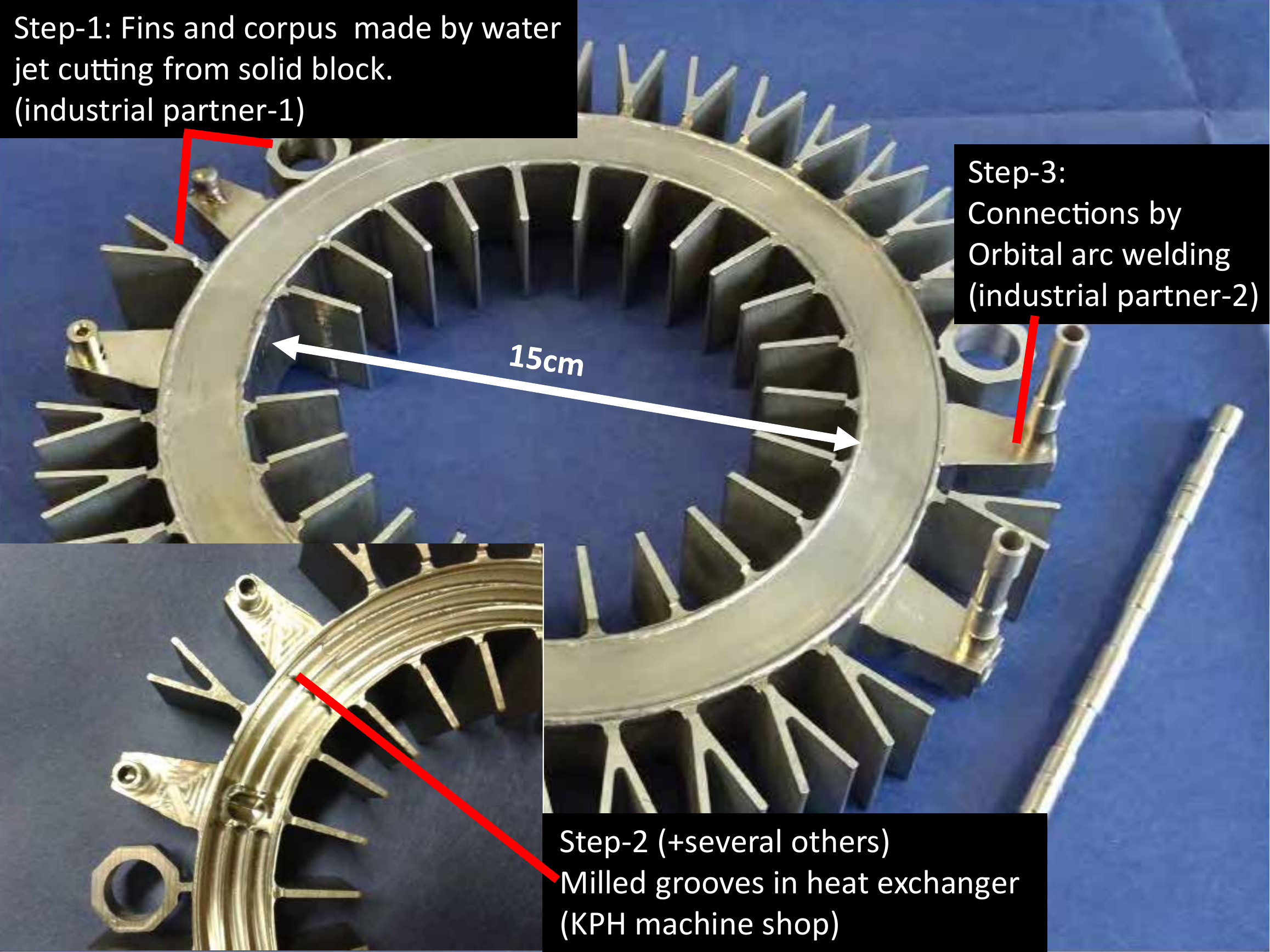}
\end{center}
\caption{All welded stainless steel heat exchanger disk for the pre-cooling circuit of the Hydro-M{\o}ller refrigerator. We indicate several techniques necessary to accomplish this component, illustrating the collaboration between external  companies  and our institute during the fabrication.}
\label{fig:heat_exchanger}
\end{figure}

An optimized pre-cooling circuit based on evaporation of $^4$He at \SI{1.25}{K} 
will minimize the amount of liquid helium needed for operation to less than 
\SI{10}{l/hour}. Concerning  the mixing circuit, the desired cooling power 
requires a mass flow of \SI{25}{mmol/s} of $^3$He, the total amount of 
liquid $^3$He will be $\approx$\SI{100}{cm^3}. Several parts have 
already been manufactured (see Fig.~\ref{fig:heat_exchanger}) and, 
based on these successful examples, we believe that assembly of the refrigerator 
can start early 2019. First test runs of the refrigerator are foreseen end of 
2019 with the final objective to obtain a ready-to-operate target by end of 2020.  

\subsubsection{Hydro-M{\o}ller spectrometer}

The Hydro-M{\o}ller target described above provides the opportunity to monitor the 
electron beam polarization in-situ with fractional accuracy well below \SI{1}{\percent}. 
The Hydro-M{\o}ller spectrometer must detect the electrons that result from 
the longitudinally polarized beam electrons scattering from the practically \SI{100}{\percent} 
longitudinally polarized target electrons. One can then monitor the electron
beam polarization by constructing the asymmetry $A_{m}$
\begin{equation}
A_m = \frac{\sigma^{\uparrow\uparrow}-\sigma{\uparrow\downarrow}}{\sigma^{\uparrow\uparrow}-\sigma{\uparrow\downarrow}} = P_TP_B\frac{\sin^2\theta(7+\cos^2\theta)}{ (3+\cos^2(\theta))^2}
\end{equation}
where $\sigma$ is the rate of scattered electrons. The first and second 
superscripts depict the sign of the beam and target electron helicities, 
respectively. $P_T$ is the $\sim 100\%$ target electron polarization, $P_B$ the
beam polarization we seek to monitor and $\theta$ is the scattering angle
in the electron-electron scattering center-of-mass frame. 

In the following, we describe a conceptual design for the spectrometer. The
analyzing power is maximum at a center-of-mass scattering angle of \ang{90},
corresponding to a scattered electron momentum of \SI{77.5}{MeV} and a scattering
angle of approximately \ang{5} in the laboratory frame. Due to
the large field integral of the Hydro-M{\o}ller solenoid 
experienced by the scattered electrons, the traditional method of picking up
a small solid angle byte around the \ang{5} lab scattering angle followed by two 
dipoles is insufficient to sample the M\o ller electrons from the full length of 
the target. As we describe below, we have found that a focusing quadrupole 
centered on the beam axis, followed by a
dipole chicane, will allow us to isolate the M\o ller electrons of interest 
in a region that can be instrumented with segmented detectors while allowing the 
primary beam to pass to the liquid hydrogen target in front of the main P2 
solenoid.

\paragraph{Optics concept.}
An optics simulation was performed using a simulated superconducting solenoidal 
magnet with a simplified geometry and a maximum field of approximately \SI{8}{T}.  
Azimuthal symmetry was assumed with a coil radius of \SI{5}{cm} and axial length 
of \SI{20}{cm} with no magnetic field-return components.  
The resulting 3-D field map was included into a Geant4 simulation which 
included an electron pair generator for a \SI{140}{MeV} electron beam and a 
\SI{20}{cm} long target.  Events were generated uniformly along the target 
length with an energy and angles corresponding to M{\o}ller kinematics over the 
full center-of-mass phase space. It was empirically observed that there is an 
approximate linear correlation between the incident radial position at a 
transverse plane downstream of the target and the direction tangent to the 
radial direction for particles near the center-of-mass scattering angle of 
\ang{90}. It was also observed that the azimuthal momentum component was small 
relative to the other components.

The particles therefore appear as if they were emanating from a point with no 
subsequent magnetic field interaction. This situation implies that to first 
order, a simple single-element optics designed for \SI{77.5}{MeV} charged 
particles can be used to focus these particles into a narrow detector region.  
Particles which do not match these characteristics, such as lower or higher 
energy M{\o}ller electrons or electrons from non-radiative elastic proton 
scattering are naturally separated.  

A quadrupole magnet with a focal length equal to its position downstream of the 
target center transforms the apparent point-emanating electron 
trajectories to a beam-parallel transport.   
These beam-parallel electrons can then be separated by dipole magnets and 
transported to a detector off-axis of the beam. 
A quadrupole doublet might possibly extend the azimuthal coverage beyond a 
single plane, though this remains to be studied.

\begin{figure}[ht]
    \includegraphics[width=0.5\textwidth]{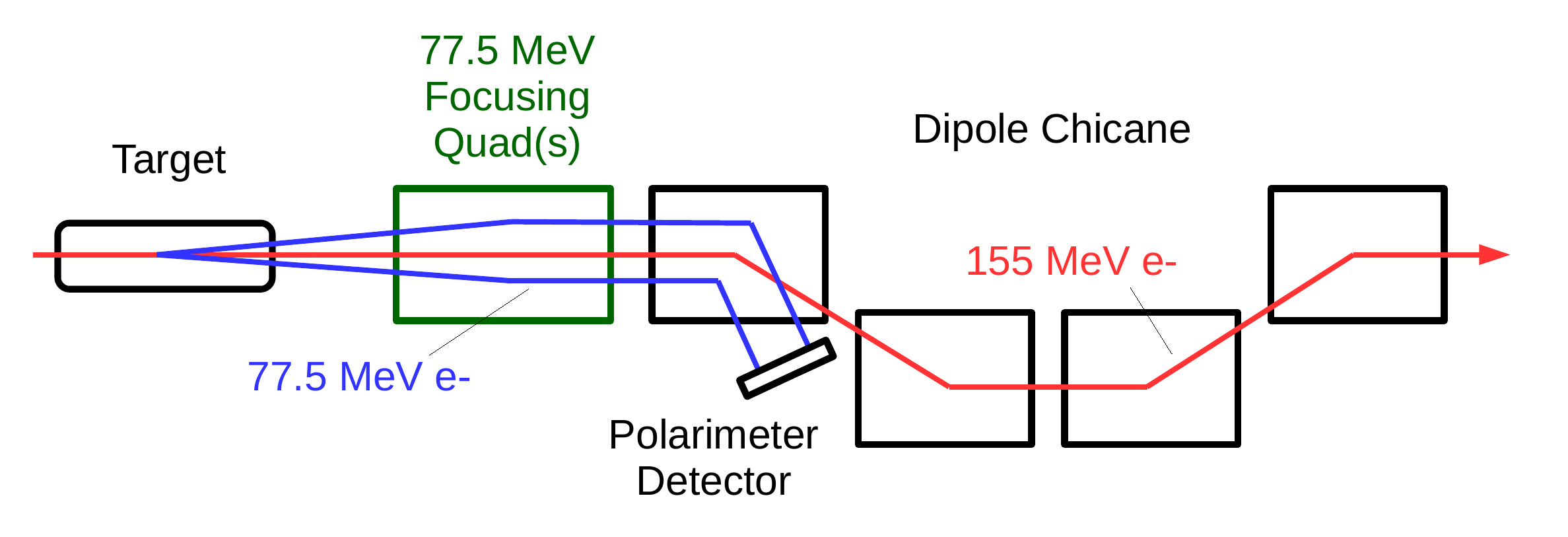}
    \caption{An optics concept for separating and focusing \ang{90} 
        center-of-mass angle M{\o}ller electrons. For \SI{77.5}{MeV} M\o ller-scattered 
		electrons, the target's \SI{8}{T} solenoidal field induces a linear 
		correlation (see text) that results in trajectories that appear 
		as if originating from a point target.}
    \label{fig:hydro_chicane}
\end{figure}

Due to the small scattering angles, the particle envelope of interest will 
naturally only span several centimeters transverse to the beam and should not 
require significant additional focusing.  Such a concept can likely be adapted 
to a system with a dipole upstream of the quadrupole using a chicane configuration.
The above ideas are incorporated to the conceptual schematic in Fig.~\ref{fig:hydro_chicane}

Additional studies are ongoing to determine ideal field values and their 
tolerances, the appropriate collimation system, effects of radiative events 
from M{\o}ller and elastic proton scattering which contribute irreducible 
backgrounds, the detection of coincidence events, and anticipated rates and 
analyzing power. Preliminary studies indicate that, given the size 
of $A_m$ is of the order of 0.5 to 0.6, it should be feasible to
design a spectrometer, collimator, and detector system that will obtain a rate 
for the M\o ller electrons of interest $\sim 0.1 - \SI{1}{kHz}$ during 
production running.  A fractional statistical uncertainty on $A_m$ at the 
level of \SI{0.1}{\percent} then could be obtained within one hour. This will make 
feasible detailed systematic studies and careful monitoring of the variations in 
beam polarization so that a determination of the beam polarization integrated 
over the running period with a total systematic error $\sim \SI{0.5}{\percent}$ 
should be possible.

%% file: beamcontrol.tex


An apparative asymmetry $A^\text{app}$ will arise from helicity correlations of 
the six beam parameters position $x$, $y$, angle $x'$, $y'$, intensity $I$,
and energy $E$ at the P2 target. Therefore an accurate, continuous
measurement of the beam parameters is mandatory to determine $A^\text{app}$
and correct for it. Such a correction should not exceed a certain
fraction of the physics asymmetry $A^\text{exp}$ and its uncertainty must
not exceed $\Delta A^\text{app}=$\SI{0.1}{ppb} as given in
Tab.~\ref{tab:input_parameters_ep}.

There are two possibilities to keep the correction small. First, 
beam parameter fluctuations can be actively suppressed by feedback
systems. This was done successfully in the A4
experiment \cite{BaunackPhD} at MAMI, using analog feedback loops
for position, angle, intensity and a digital loop to stabilize the
beam energy. However, this approach removes helicity-correlated as
well as non-helicity correlated beam fluctuations. Since the latter
are needed to decorrelate the individual contributions $A^\text{app}_\text{i}$ 
to the total apparative asymmetry, suppressing them can increase the
uncertainty $\Delta A^\text{app}$ of the correction.

A second possibility is to measure the helicity-correla\-tions of all
beam parameters online. Then the helicity signal can be used for a
feed-forward suppression of the helicity-correlated fluctuations.

In the following we will describe some of the technology and expertise
that already exists from running the existing MAMI accelerator and the
recent A4 parity violation experiment. We will then discuss dedicated
tests with a new digital data acquisition and control system. From
these results we will see that we can be confident to fulfil the
requirements of the P2 experiment on the beam parameter stability and
measurement.

\subsection{Proposed beam control system}

The operation of the existing MAMI accelerator and the former A4
parity-violation experiment rely on measuring the beam parameters with
cavity beam monitors: beam current monitors (BCM) can be used for beam
intensity and also for beam phase measurement, while beam position
monitors (BPM) provide measurements of beam position and angle (from
differences between two BPMs). Due to our experience and the good
performance and reliability with cavity monitors we will also use them
for instrumentation of the P2 beamline at MESA. The cavities for MESA
will be based on the design of the cavities used at MAMI, but with
resonance frequencies adapted to MESA (\SI{1.3}{GHz} or
\SI{2.6}{GHz}) and with enhanced vacuum properties (bakeable design)
and better tuning capabilities.

Due to the tiny physics asymmetry $A^\text{exp}$ we believe that a digital
system for the beam monitors is mandatory. Such a system provides
flexibility and can be adapted to the exact needs of the
experiment. If all beam monitors of the accelerator and experiment
beamline are read out digitally, all of these data will be available
for diagnostics to both, the accelerator operator as well as to the
experiment. This will be of great importance in the commissioning
phase of the experiment.

\subsection{Beam monitor tests at MAMI}

\begin{figure}[b]\centering 
  \includegraphics[width=0.95\linewidth]{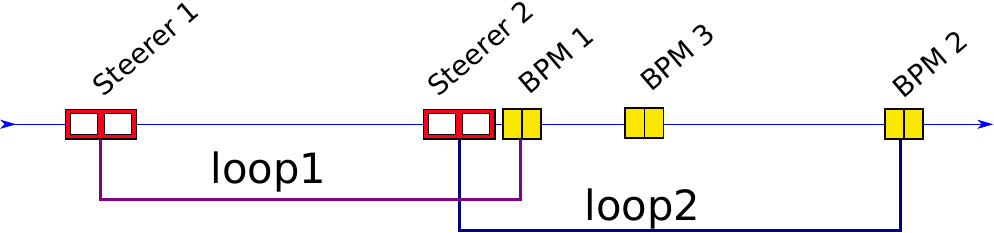}
  \vspace*{2mm}
  \caption{Control system components in the beamline. Steerer\,1 and BPM\,1 form a
    first loop, which controls the position on BPM\,1. Steerer\,2 and BPM\,2 form
    the second loop, which controls the position on BPM\,2. This, in fact, controls
  the angle of the beam. BPM\,3 is used as an unbiased observer.}
  \label{fig:beamline}
\end{figure}

In order to develop and test a new, digital data acquisition and control system
for MESA we instrumented \SI{20}{m} of beamline at the existing MAMI
accelerator with additional beam monitors and steering magnets. For
feedback tests digital control loops have been arranged as shown in
Fig.~\ref{fig:beamline}. The beam has an energy
of \SI{180}{MeV} which is close to the \SI{155}{MeV} planned
for the P2 experiment. 
We explored different techniques for down-conversion of the cavity monitor RF
signals to baseband, employing direct down-conversion as well as IQ
demodulation with and without employing an intermediate frequency. One
advantage of IQ demodulation via an intermediate frequency would be
the decrease of baseband noise/interference contributions collected on
the signal path.

To gain significantly higher accuracy we use fast ADCs and DACs
(both \SI{125}{MSa/s}, \SI{14}{bits}) as part of a control system that
is established on an FPGA. This approach enabled us to control the
beam in the classical feedback as well as in feed-forward loop. The
flexibility of such a digital system allows for small remote
modifications as well as for the implementation of new features in the
system without the need for a redesign of electronics. To avoid the
tedious development of the hardware consisting of FGPAs, ADCs, DACs,
and periphery for communications (possibly even a CPU), we decided to
use a commercially available board (RedPitaya) for the moment to carry
out our measurements.

\subsubsection{Accuracy results}
A key issue is the achievable accuracy of our data acquisition system. 
A digital signal always carries information as well as noise
stemming from variation in voltage supply, clock jitter, and the quantization error.
The ADCs show \SI{2.5}{bits} of noise and \SI{11.5}{bits} of ENOB
(effective number of bits) which corresponds to the effective resolution.


For the P2 experiment a beam position or beam current data rate of the
order of few times the helicity flip rate is sufficient, far below
the \SI{125}{MHz} our ADCs provide. Therefore we can average over $N$
samples, improving the effective resolution, and record only these
average values with \SI{16}{bits} resolution, of course, at a data
rate decimated by the factor $N$. With \SI{125}{MHz} sampling rate, a
decimation factor of 8192 would give a data rate of
about \SI{15.2}{kHz}, which was used in our test measurements. The
gain in effective resolution can also be calculated. For example to
reduce the error $\sigma$ by half one needs to increase the number of
measurements $N$ four times to $N'$:
\begin{align}
  \frac{\sigma}{\sqrt{N'}} = \frac{1}{2}\frac{\sigma}{\sqrt{N}} =
  \frac{\sigma}{\sqrt{4N}}
  \label{eq:resolution}
\end{align}
under the condition that the noise is white noise.


In the following paragraphs we discuss the different sources of noise
contributing to the overall noise budget. We refer to beam monitor
BPM 2 which revealed a sensitivity of
640\,\si{\milli\volt\per\milli\meter} at \SI{10}{\micro A} beam
current. At \SI{150}{\micro A} beam current the sensitivity
would be 9.6\,V\si{\per\milli\meter}. From the sensitivities one can
convert each contribution to beam position uncertainty.

\paragraph{ADC effective resolution.}
We carried out measurements of the ADC noise without any further
electronics attached. This allows to find out what the theoretically
achievable resolution limit was, if none of the other components would
add any noise. Table~\ref{tab:resolution} shows the effective
resolution of data acquisition in bits, as well as converted to the
expected effective beam position resolution in nm at
\SI{150}{\micro A} beam current.





\begin{table}[hbt]
  \centering
  \begin{tabular}{ccc}
    \toprule[1.5pt]
    bits & W/$\si{\micro\volt}$  & W/$\si{\nano\meter}$
    @$150\,\si{\micro\ampere}$  \\
    \midrule
    0.42 & 40.8 & 4.3 \\
    \bottomrule[1.5pt]
  \end{tabular}
  \vspace*{2mm}
  \caption{Maximal possible resolution for a decimation of 8192 in number of bits
    and \SI{}{\micro V}. W stands for
    signal width. The measurement was taken with unplugged BPM electronics. The width in
    nm is the extrapolated contribution to the beam's uncertainty stemming from the effective
  resolution, if measured at \SI{150}{\micro A}.}
  \label{tab:resolution}
\end{table}

\paragraph{Signal width without beam.}
Noise measurements with cavity beam monitors were performed without
beam to determine the total noise of the full acquisition system. The
signal width contains contributions coming from the electronic components of the
IQ-demodulation, such as mixers, amplifiers, splitters, and
connections. Reducing the distortions collected on the transfer path
through the hall, we have carried the signals
differentially. Table~\ref{tab:signal_wo_beam} summarizes the signal
width without beam.

%

\begin{table}[hbt]
  \centering
  \begin{tabular}{ccc}
    \toprule[1.5pt]
    bits & W/$\si{\micro\volt}$  & 
  \begin{tabular}{@{}l@{}}W/$\si{\nano\meter}$\\@$150\,\si{\micro\ampere}$\end{tabular}

    \\
    \midrule
    0.97 & 59.8 & 6.2 \\
    \bottomrule[1.5pt]
  \end{tabular}
  \vspace*{2mm}
  \caption{Signal width without beam for decimation of 8192. This data derives
    from a measurement with fully plugged electronics but switched off beam.
    Similar to the effective resolution the signal width in nm represents the
    extrapolated
    contribution to the beam width caused by electronic noise, if measured with
  \SI{150}{\micro A}.  }
  \label{tab:signal_wo_beam}
\end{table}


\paragraph{Signal width with beam.}
Finally we discuss measurements of the beam position with a beam current
of $5\,\si{\micro\ampere}$. During these measurements we also
employed our digital feedback loop to stabilize the beam position and
angle in our test beamline.


\begin{table}[hbt]
  \centering
  \begin{tabular}{cccc}
    \toprule[1.5pt]
    bits & \begin{tabular}{@{}l@{}}
  W/$\si{\milli\volt}$\\@5\,$\si{\micro\ampere}$\end{tabular}
  & W/$\si{\micro\meter}$
  & 
\begin{tabular}{@{}l@{}}W/$\si{\nano\meter}$\\@10\,kh\end{tabular}\\
  \midrule
  3.15 & 6.25 & 54.3 & 0.29\\
  \bottomrule[1.5pt]
\end{tabular}
\vspace*{2mm}
\caption{Signal width and beam width of a stabilized \SI{5}{\micro A}
  beam measured with BPM\,3. The table shows results from measurements
  with a beam current of \SI{5}{\micro A} in V and nm as well as the estimated
  mean value of the
  width after 10\,000\,h measuring time. The control unit was the PID element of
the FPGA. }
\label{tab:signal_w_beam}
\end{table}

Evidently the beam position fluctuation creates a width that is 100
times larger than the electronic noise. One can therefore extrapolate
the uncertainty of the beam position (or helicity-correlated beam
position difference) to be of the order \SI{0.3}{nm}
after \SI{10000}{h} of data taking. For a perfectly symmetric P2
detector one can estimate the resulting systematic error from this
uncertainty taking into account only the geometry effect from the
beam position fluctuations, neglecting magnetic field and detector
properties. This gives an extremely small value,
about \SI{0.003}{ppb}. It is now necessary to fully model 
the experiment, including magnetic field, alignment
errors of the beamline, beam monitors, and detector modules. This will
allow to determine the sensitivity of the experiment on beam position
fluctuations.

\paragraph{Beam current.}
Also a beam current monitor was read out with the new DAQ system. A Short
Term Asymmetry (STA) can be determined straight from the beam current. STAs are calculated from
quadruplets, a pattern of helicity states (either positive or negative) of 1 and
\SI{2}{ms} length in the following order: +\,--\,--\,+ and 
--\,+\,+\,--. Each of our quadruplets is \SI{4}{ms} long and $9\times 10^9$ quadruplets
would correspond to 10\,000 hours measuring time.

Table~\ref{tab:current} shows the results for the uncertainty of the
STAs. The measurements were carried out at a beam current of
$10\,\si{\micro\ampere}$ and the expected width for a current of
$150\,\si{\micro\ampere}$ was calculated. The validity of such an
extrapolation is motivated by our evaluation of beam current monitor
calibration data from the A4 experiment. The results in Tab. \ref{tab:current}
show that the beam current stability at MESA has to be improved compared to
MAMI.

From Tab.~\ref{tab:current} one can read a projected precision of the
beam current measurement for \SI{10000}{h} of \SI{0.3}{ppb} at
MAMI. For P2 at MESA this still needs to be improved by about a factor
of ten. We believe that this is possible, because at MAMI the beamline
between polarized electron source and injector LINAC contains a number
of critical apertures. Therefore even small beam position fluctuations
already cause significant beam current fluctuations. The design of the
MESA low energy beam apparatus (MELBA) avoids such aperture limitations. In
addition, for MESA a more sophisticated laser optics system for the
polarized electron source will allow to keep beam position
fluctuations much smaller in this critical part of the beamline.

\begin{table}[hbt]
  \centering
  \begin{tabular}{cccc}
    \toprule[1.5pt]
  \begin{tabular}{@{}l@{}}W/\si{\milli\volt}\\@$10\si{\micro\ampere}$ \end{tabular}
    & W/\si{\nano\ampere}
    &
  \begin{tabular}{@{}l@{}}$\Delta$STA/ppm\\@$150\si{\micro\ampere}$ \end{tabular}
    & 
  \begin{tabular}{@{}l@{}}$\Delta$STA/ppb\\@$10\si{\kilo\hour}$ \end{tabular}
    \\
    \midrule
    6.0 & 5.9 & 28 & 0.29\\
    \bottomrule
  \end{tabular}
  \vspace*{2mm}
  \caption{Width of the beam current and extrapolation of the expected STA at
  150\si{\micro\ampere} beam current and the average after 10\,000\,h measuring
time. }
  \label{tab:current}
\end{table}

\paragraph{Beam energy.}

For the beam energy stabilization the 180$^\circ$ P2 return-arc 
(see Fig.~\ref{fig:MESA_floorplan}) with maximal
longitudinal dispersion, along with two beam phase cavity monitors
will be used. A similar concept has proven successful at MAMI for
A4. So far, our measurements at MAMI did not consider beam energy yet
(we have  not positioned  our monitors at positions with large longitudinal
dispersion so far). However, we
have estimated the effect of helicity correlated beam energy
fluctuations in terms of apparative asymmetry in comparison to
$A^\text{exp}$ as shown in Tab.~\ref{tab:beamenergy}. In the A4 experiment,
for instance, the beam energy was very stable (with a feedback system)
and the corresponding contribution to $\Delta A^\text{app}$ in 10\,000 hours
would be safely below \SI{0.1}{ppb}. However, we believe this is due to the
intrinsic stability and longitudinal self-focussing properties of
MAMI, which can not be expected for MESA. Therefore any possible
source of beam energy fluctuation (noise) in MESA should be kept as
small as possible. One can see from Tab.~\ref{tab:beamenergy} that energy stability at
MAMI was not an issue for A4, while even a helicity correlated change
(or its uncertainty when correcting for it) of 1~eV at MESA would lead
to a 23\% effect at P2. Beam energy (helicity correlation or even
noise) will probably be the most critical beam parameter at MESA.

However, an uncertainty of \SI{0.1}{ppb} would correspond to
\SI{14}{meV} in \SI{10000}{h}, which is more than $10^{10}$
helicity gates. This gives an upper limit on the uncertainty of a
single energy measurement of about \SI{1400}{eV} or about $10^{-5}$
relative uncertainty at \SI{155}{MeV}. A longitudinal dispersion
of \SI{10}{mm/10^{-3}} can be routinely achieved in
a \SI{180}{\degree} arc in the beamline. Such an arc then leads to a
change in the longitudinal path length of \SI{0.1}{mm} for $10^{-5}$
relative energy change. This \SI{0.1}{mm} corresponds to a RF phase
difference of \SI{0.14}{\degree} at \SI{1.3}{GHz} to be detected in a helicity window
of \SI{0.5}{ms} which seems feasible. Alternatively the transverse
dispersion of a \SI{90}{\degree} arc of about \SI{3}{mm/10^{-3}} would
require measuring a beam displacement of \SI{0.3}{nm}
in \SI{10000}{h}, which is possible as can be seen from
Tab.~\ref{tab:signal_w_beam}.

\begin{table}[hbt]
  \centering
  \begin{tabular}{lll}
    \toprule[1.5pt]
    E$_\text{beam}$ / MeV & A$^\text{app}$ / ppb/eV & A$^\text{PV}$ / ppb \\
    \midrule
    855                  & 2.0 & $\approx$ 5000\\
    155                  & 6.8 & $\approx$ 30 \\
    \bottomrule
  \end{tabular}
  \vspace*{2mm}
  \caption{Estimated apparative asymmetries from helicity correlated
  beam energy fluctuations for P2 at MESA.}
  \label{tab:beamenergy}
\end{table}

\paragraph{Expected asymmetry uncertainties for P2 from beam position.}

Table~\ref{tab:asymmetries} shows the summarized results for asymmetry
uncertainties expected after \SI{10000}{h} of measuring time. We expect improvement of
the results for the beam position by further increasing the degree of averaging.
Also increasing the beam intensity of MESA compared to MAMI will lead to
higher monitor sensitivities. 

\begin{table}[hbt]
  \centering
  \begin{tabular}{llll}
    \toprule[1.5pt]
    & Width/nm & 
  \begin{tabular}{@{}l@{}}$\Delta$STA/ppb\\per quadr. \end{tabular}
    & 
  \begin{tabular}{@{}l@{}}$\Delta$STA/ppb\\@\SI{10}{kh} \end{tabular}
    \\
    \cmidrule(r){1-4}
    eff. resolution & 4.3 & 0.83 & 8.7$\times10^{-6}$ \\
    electronics & 6.2 & 1.2 & 1.3$\times10^{-5}$ \\
    stabilized beam & 5.4$\times 10^4$ & 1.0$\times10^4$ & 0.11 \\
    \bottomrule
  \end{tabular}
  \vspace*{2mm}
  \caption{Uncertainties of the asymmetry projected from A4 to P2 for the data
  with a decimation of 8192.}
  \label{tab:asymmetries}
\end{table}

\subsection{BPM cavity design}
Cavity BPMs have been chosen as the measuring element of the control system
because of reliable and long experienced operation at MAMI as well as the fact
that they fulfil the requirements for a high precision experiment with high
current.
The most appealing properties of cavity BPMs are the non-invasive measuring 
technique and the possibility to measure very low currents. This
comprises the possibility of increasing precision with increasing beam current.
The measuring principle is based on amplification via
resonance: A bunch of electrons flying through excites a spectrum in the cavity that
depends on the length of the bunch. The shorter the bunch, the more higher
frequencies are excited. Due to the repetition of the bunches, only modes with a
frequency of multiples of the bunch rate frequency are excited positively and build
standing waves inside the cavity. The inner design of the cavity determines
which modes will survive, so that different purposes, such as phase and energy
measurements can be pursued. For a BPM the TM$_{110}$-mode is of interest which
can be sensed with antennas coupled to the cavity. In our case, the cavity is
designed for a resonance frequency of 2.6\,GHz (first harmonic) to keep its
transversal dimensions reasonably small.

Combining a BPM with a feedback system, one has to take into account the
characteristics of the whole loop, when designing the cavity. A BPM resembles a
low pass with a cut-off frequency, where its bandwidth ends, and at which the 
phase shift is \ang{45}. This
cut-off frequency is determined by the quality factor of the cavity. It is the
target of the control loop designer to keep amplification below one at frequencies
at which the \ang{180} shift of the whole loop is hit. Phase shifts of
all components in the loop sum up. In simple terms, the threshold comes closer to lower
frequencies the more low passes you find in the loop and the lower their cut-off
frequencies are. Therefore, the design of the cavity has to make a compromise between
high signal and high bandwidth. We decided to confine the bandwidth to \SI{250}{kHz}. 
This is equal to a loaded $Q$ factor of 5200. The loaded $Q$ factor is determined
by the $Q$ factor and the coupling $\beta$ of the antennas:
\begin{align}
  Q = (\beta+1) Q_L.
\end{align}
Simulations on a copper cavity show that the (unloaded) $Q$ factor will be above
20\,000. The often used critical coupling of $\beta$=1, ensuring a high signal and high $Q_L$,
is not sufficient here, instead the required bandwidth has to be approached by elongation of
the
antennas. As longer antennas decrease the resonance frequency, this has to
be compensated with a tuning piston.

\begin{figure}[t]\centering 
  \includegraphics[width=0.95\linewidth, trim={6cm 0cm 20cm 5cm}, clip]{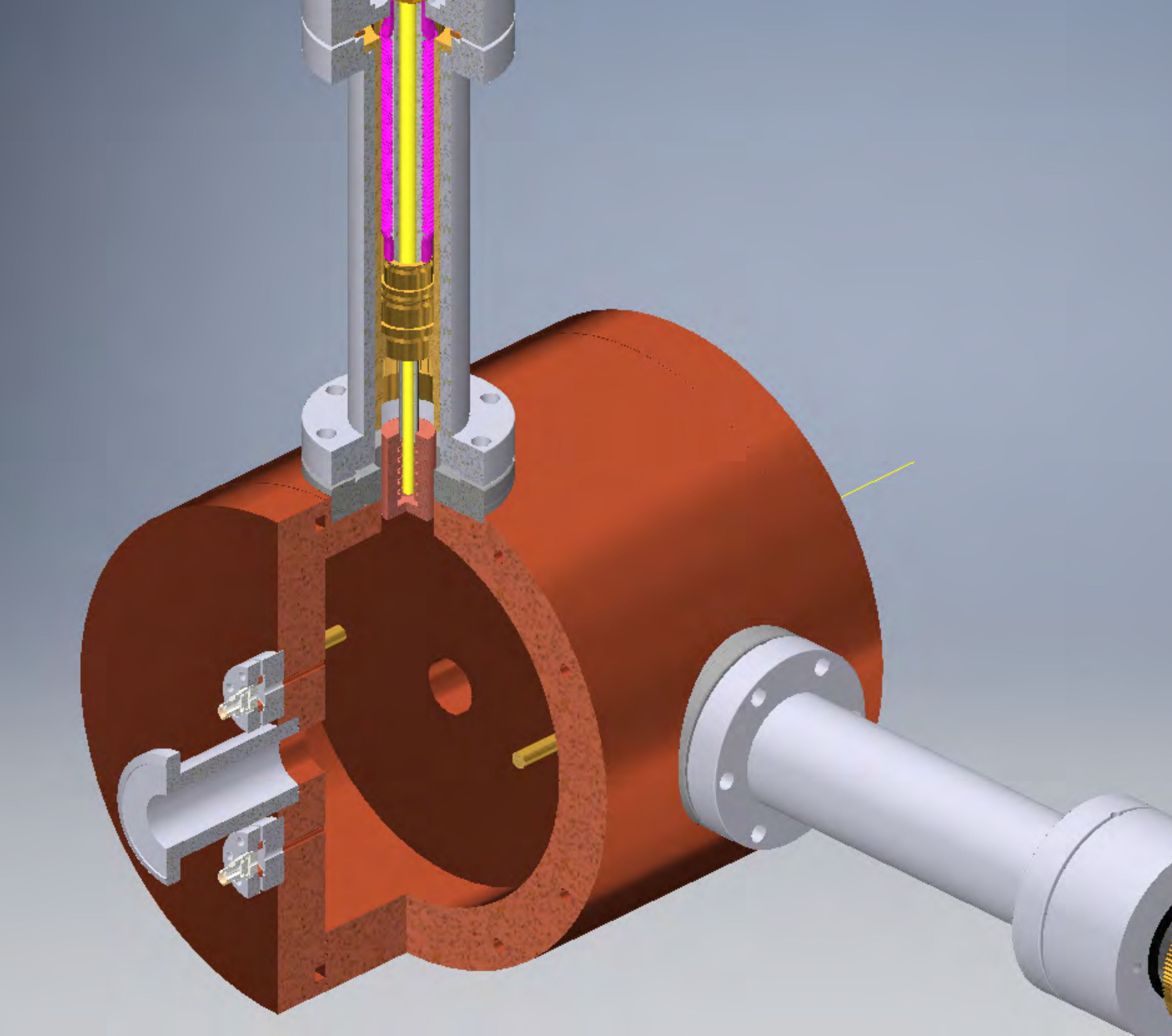}
  \vspace*{2mm}
  \caption{Cavity design. The corpus is made of copper and all connections are 
		of Conflat Standard, as the cavity
    must be bakeable to guarantee an ultra-high vacuum environment. The caps on each
    side are soldered to the corpus.
    The tuning pistons can be moved during operation to adjust the
    sensitivity of the BPM. Mode isolators inside the cavity suppress the
    TM$_{110}$-mode in perpendicular direction. Flanges with the antenna
    feedthroughs can be seen in
  front above and below the beam pipe. The outer diameter of the cylinder is
\SI{184}{mm}.}
  \label{fig:cavity}
\end{figure}

A picture of the current state of the design is shown in
Fig.~\ref{fig:cavity}.
The tuning pistons are attached perpendicular to the cavity and can change its
volume. They can shift the resonance frequency within about 8\,MHz. This can be
used firstly to hit the exact frequency of 2.6\,GHz and secondly to detune the
frequency and damp the signal in order to save electronics, if high
displacements are expected.
The antennas are designed in pairs, as it is useful to subtract the two
signals in order to double the outcome and annihilate remains of TM$_{0\text{x}0}$
modes.

%% file: target.tex
The P2 experiment requires a high luminosity liquid 
hydrogen ($\ell$H$_2$) target to measure a very small asymmetry with a very high 
precision. The cell geometry has to accommodate a full azimuthal angle about the 
electron beam direction and a polar angle range between \ang{25} and \ang{45} 
along its full length with minimal materials in the path of scattered particles.

\begin{table}
\caption{P2 target design parameters}
\begin{center}
\begin{tabular}{cc}
\toprule[1.5pt]
Pressure/temperature                &  \SI{2.4}{bar} / \SI{20}{K}      \\  
Cell length                         &  \SI{60}{cm}      \\ 
$\dot{m}$                           &  $<\SI{2}{kg/s}$        \\ 
$\ell$H$_2$ pump head               &  $<\SI{0.1}{bar}$       \\
Beam area on target                 &  \SI{25}{mm\squared}        \\
HX cooling power                    &  \SI{4}{kW}             \\
Target thickness                    &  \SI{4.3}{g/cm\squared}     \\ 
\midrule
$\ell$H$_2$ ($\Delta \rho/\rho$)    &  $<\SI{2}{\percent}$           \\
$\ell$H$_2$ ($\delta \rho/\rho$) at \SI{1}{kHz}   &  $<\SI{10}{ppm}$        \\ 
\bottomrule[1.5pt]
\end{tabular}
\end{center}
\label{tab:tgtparams}
\end{table}

\begin{table}
\caption{P2 target heat load}
\begin{center}
\begin{tabular}{cc}
\toprule[1.5pt]
Source                          &  Value (W)    \\  \midrule
Beam power in $\ell$H$_2$       & 3100          \\ 
Beam power in cell windows      & 35            \\ 
Viscous heating                 & 275           \\
Radiative losses                & 200           \\
Pump motor                      & 150           \\
Reserve heater power            & 240           \\ 
\midrule
Total heat load                 & 4000          \\ 
\bottomrule[1.5pt]
\end{tabular}
\end{center}
\label{tab:tgtheat}
\end{table}

The total heat load of the target is estimated to be \SI{4000}{W}, 
see Tab.~\ref{tab:tgtheat}. The 
heat deposited by the electron beam in the target cell materials can be 
calculated with the formula: $P = I\rho L(dE/dx)$, where $I$ is the beam 
current in $\mu$A, $\rho$ is the material density in beam in g/cm${}^3$, $L$ 
is the material thickness in beam in cm, $dE/dx$ is the collisional 
energy loss of electrons in the material in beam in MeV/(g/cm${}^2$) and $P$ 
is the heating power in W. The electron beam will deposit 
\SI{3135}{W} in the P2 target cell materials, which will make the P2 target the 
highest power $\ell$H$_2$ target in the world. A fluid target in a parity violation 
experiment produces two systematic effects that affect the PV asymmetry 
uncertainty: density reduction, denoted $\Delta\rho/\rho$ in Tab.~\ref{tab:tgtparams}, 
and density fluctuation, denoted $\delta\rho/\rho$ in Tab.~\ref{tab:tgtparams}. 
The density reduction effect is caused by the electron beam heating of the target 
fluid in the beam illuminated volume in the cell, which increases the temperature 
of the fluid and decreases its density. A \SI{5}{\percent} $\ell$H$_2$ density 
reduction produces a \SI{5}{\percent} luminosity reduction, which means that the 
experiment will have to run \SI{5}{\percent} longer to achieve its proposed 
statistical uncertainty. The density fluctuation is a time dependent effect 
caused by the $\ell$H$_2$ density fluctuation due to the electron beam
heating over time periods of the electron beam helicity. The $\ell$H$_2$ density
fluctuation noise produces an enlargement of the PV asymmetry
width~\cite{Covrig:2005vx}, which increases the PV asymmetry systematic
uncertainty. A \SI{10}{\percent} increase in the PV asymmetry width due to $\ell$H$_2$ 
density fluctuation means a \SI{20}{\percent} increase in the experimental running 
time to achieve the same systematic uncertainty as from a noiseless target. The 
P2 PV asymmetry is the smallest one that will have been measured in all completed
and proposed PVES experiments and with the smallest relative uncertainty too.
In order to achieve this goal the P2 experiment requires a very high performance 
$\ell$H$_2$ target with no more than \SI{2}{\percent} $\ell$H$_2$ density 
reduction and no more than \SI{2}{\percent} PV asymmetry width enlargement or 
less than \SI{10}{ppm} $\ell$H$_2$ density fluctuation noise over the time 
period of electron helicity reversal.

\begin{figure}
    \begin{center}
        \includegraphics[width=0.47\textwidth]{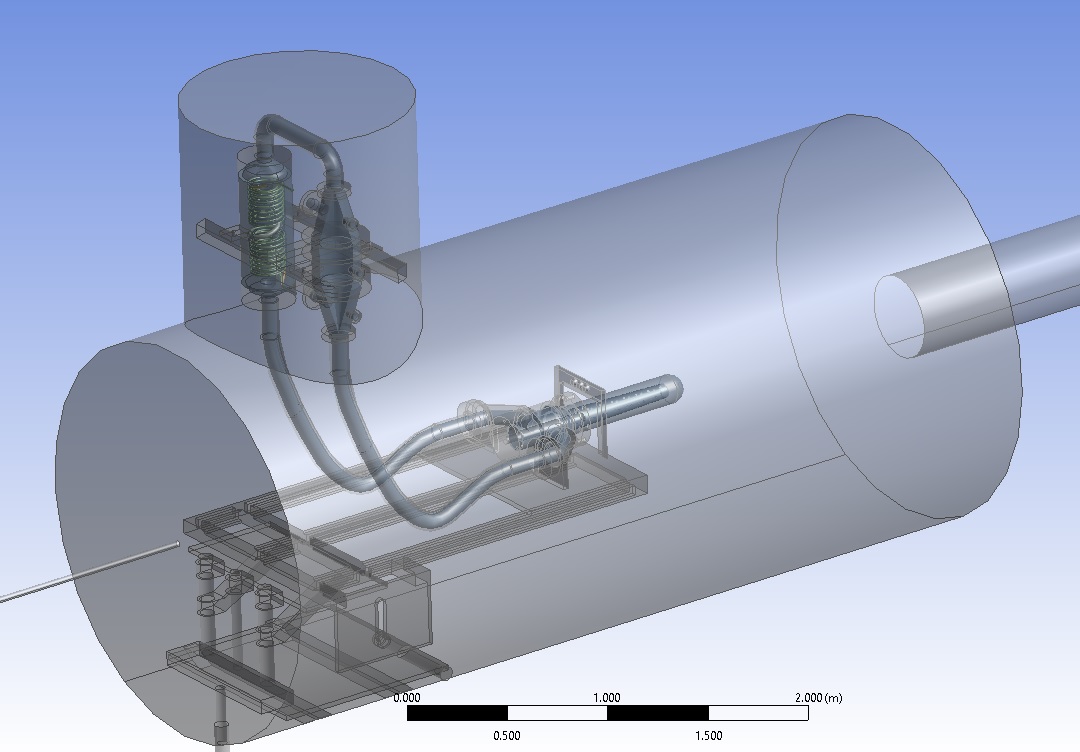}
    \end{center}
\caption{Configuration of the P2 target loop}
\label{fig:tgtloop}
\end{figure}

In order to satisfy the requirements of the P2 experiment a closed recirculating
cryogenic $\ell$H$_2$ target loop will be designed, built and tested. The target
will incorporate a control system to monitor, safely operate and control it
during experimental commissioning and data acquisition periods. Hydrogen 
is a highly flammable gas in atmosphere in concentrations between \SI{4}{\percent}
and \SI{74}{\percent} by volume and can produce explosions in concentrations
between \SI{18}{\percent} and \SI{54}{\percent} by volume. Special safety measures 
will be taken in the design and control of the target system to mitigate hazards.
The main components of the target loop are the $\ell$H$_2$ centrifugal pump, the heat
exchanger (HX), the high power heater (HPH), and the target cell in beam. A
model of the proposed $\ell$H$_2$ cryogenic P2 target loop can be seen in Fig.~\ref{fig:tgtloop}. 
In this configuration the \SI{60}{cm} long $\ell$H$_2$ cell and its upstream manifold 
are supported on a table and placed inside the P2 solenoidal spectrometer. The
supporting table will be instrumented with an all-metal 6 degrees of freedom alignment 
mechanism that will place the upstream and downstream windows of the long
target cell within \SI{0.5}{mm} each of their ideal positions respectively. The
table will also be connected to a motion mechanism that will move the target cell
in and out of the beam line and place solid targets in beam or no target in beam.
The target cell manifold will connect with the rest of the target loop through
stainless steel pipes that will have flexible sections in order to accommodate
the target cell alignment in beam and the range of its motion mechanism. The HX, 
the HPH and the $\ell$H$_2$ pump will be supported inside of a vertical vacuum 
chamber but outside of the bore of the P2 magnet. These components of the cryogenic 
target loop will be thermally insulated from the shell of the vacuum chamber, 
which together with the vacuum in the chamber will reduce heat leaks and mitigate
the target system cooling load. The target loop chamber and the P2 magnet will 
share vacuum space. 

\begin{table*}
\centering
\caption{Parity-violation electron scattering $\ell$H$_2$ target parameters}
\begin{tabular}{cccccc}
\toprule[1.5pt]
Experiment      & Length & $P/I/E$      & $\Delta \rho/\rho$ & $\delta \rho/\rho$  \\ 
                & $cm$  & $W/\mu A/GeV$  & \%        & ppm               \\
\midrule
SAMPLE          & 40    & 700/40/0.2    & 1         &  $<1000$ @ \SI{60}{Hz}  \\ 
HAPPEX          & 20    & 500/35-55/3   & -         & 100 @ \SI{30}{Hz}       \\ 
PV-A4           & 10    & 250/20/0.854  & 0.1       & 392 @ \SI{50}{Hz}       \\
E158            & 150   & 700/12/48     & 1.5       & $<65$ @ \SI{120}{Hz}    \\
G0              & 20    & 500/40-60/3   & $<1.5$    & $<238$ @ \SI{30}{Hz}    \\
QWeak           & 35    & 2500/180/1.1  & $<1.6$    &$< 50$ @ \SI{960}{Hz}    \\ 
\midrule
P2              & 60    & 4000/150/0.155 & $<2$     &  10 @ \SI{1000}{Hz}\\ 
MOLLER          & 150   & 4500/75/11    & $<2$      & 25 @ \SI{1920}{Hz} \\ 
\bottomrule[1.5pt]
\multicolumn{5}{p{14cm}}{$P$: beam power on target, $I$: maximum beam
current, $E$: beam energy, $\Delta \rho/\rho$: target bulk density reduction,
$\delta\rho/\rho$: size of target density fluctuation at given helicity
reversal frequency.}\\
\end{tabular}
\label{tab:PVESLH2Targets}
\end{table*}

All previous PVES experiments or series of such experiments have designed,
build, commissioned and operated their own target systems. Table~\ref{tab:PVESLH2Targets} 
shows the design parameters of various PVES $\ell$H$_2$ targets. Before P2 the 
highest power $\ell$H$_2$ PVES target in the world was the QWeak target. The QWeak 
$\ell$H$_2$ target at Jefferson Lab was the first such target to be designed with 
Computational Fluid Dynamics (CFD). The QWeak target achieved all its design goals 
with a measured $\ell$H$_2$ density reduction of \SI{0.8}{\percent} and $\ell$H$_2$ 
density fluctuation of less than \SI{50}{ppm} at \SI{960}{Hz}. CFD-driven target 
design has been validated as the critical tool in the performance of the QWeak 
target. Currently the QWeak target is the highest power $\ell$H$_2$ target in 
the world and with the smallest noise figure. The P2 target will be \SI{50}{\percent} 
more powerful while being required to have five times less noise from $\ell$H$_2$ 
density fluctuation than the QWeak target. The design of the P2 target cryogenic 
loop components will be modelled after the successful targets for the QWeak and 
G0 experiments that ran at the Jefferson Lab. In addition, the performance of 
each component of the P2 target will be assessed with CFD. 

\paragraph{Cooling requirement.}
The total heat load on the target is \SI{4}{kW}, see Tab.~\ref{tab:tgtheat} for 
a list of contributions to the target power load. Therefore, the target will 
require at least \SI{4}{kW} of helium cooling from the MESA refrigerator in order
to support \SI{150}{\micro A} electron beam operations on target. The flow of He cooling
needed can be calculated with the formula $\dot{m} = P/\Delta H$, where $P$ is
the total cooling power and $\Delta H$ is the He enthalpy variation between the
inlet and the outlet of the HX. If the He inlet to the HX were at 10~atm, 14~K 
and the outlet were at 3~atm, 20~K, then 101~g/s He flow would be needed to provide
4~kW of cooling power. The same amount of cooling power could be provided by 40~g/s 
He delivered to the P2 HX at 3~atm, 4.5~K and returned from the HX at 1.5~atm, 20~K.

\begin{figure*}
\centering
\includegraphics[width=0.8\textwidth]{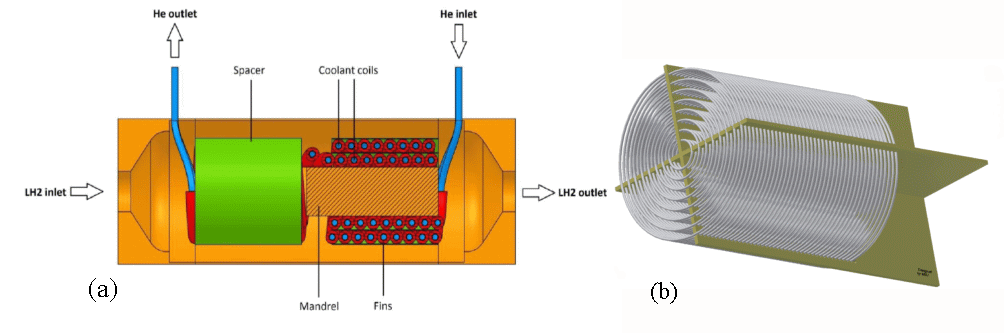}
\caption{(a) CAD drawing of the P2 HX.  (b) CAD drawing of the 8 layer P2 HPH}
\label{fig:HX}
\end{figure*}

\paragraph{Heat exchanger.}
The heat exchanger will be used to liquefy the hydrogen gas and then provide
enough cooling power to remove the total heat load from the target loop in order to 
keep the target stable. It will be modelled after the QWeak counter-flow HX, however, 
it will only require two layers of copper finned tube. The two layers will be 
wrapped around a central baffle to provide maximum heat transfer. At the midpoint 
of the heat exchanger the two layers of tubing will swap to equalize the pressure 
loss in each layer, see Fig.~\ref{fig:HX}(a). The HX design will be assessed with 
CFD. The HX will be cold shocked and tested for leaks with a sensitive He leak 
detector under both vacuum and pressure in both the H$_2$ and He circuits.

\paragraph{High power heater.}

To mitigate the $\ell$H$_2$ density variations and relaxation time with electron 
beam trips the P2 target will be operated in constant heat load. In this mode of
operation the HPH power supply works in a feedback loop with the read-back of
a temperature sensor embedded into the target fluid to keep the fluid
temperature constant regardless of the presence of the electron beam. The heater 
is used to replace the beam heat load and regulate the loop
temperature. The heater is designed to have eight layers of 18 Awg Nichrome
wire wrapped around crossed G10 boards. The heater will reside in a section of
loop pipe with conflat flanges. Heat transfer calculations were done assuming
one can treat the heater as an array of cylinders or tubes in 
a flow. This assumption was largely confirmed for the QWeak heater by comparing
the heat transfer calculations with CFD simulations. One major difference
between this heater design and the QWeak design is that we plan to use a power 
supply with higher DC voltage and, of course, the heat load will be \SI{50}{\percent} 
more than the value used for QWeak, hence $P = \SI{4000}{W}$. Therefore, the 
design calls for a greater than \SI{4}{kW}, \SI{250}{V} DC power supply which 
has a maximum current of \SI{20}{A}. The ideal resistive load for this power 
supplies is \SI{12.5}{\ohm}, which provides a design constraint for the heater. 
A design of the HPH coils is shown in Fig.~\ref{fig:HX}(b).

\paragraph{$\boldsymbol\ell$H$_2$ pump.}
Comparing the P2 target design parameters with the QWeak target design parameters, 
it is expected that a centrifugal pump will be required for the P2 target. The QWeak 
$\ell$H$_2$ pump was designed in-house at Jefferson Lab for a volume displacement 
of \SI{16}{l/s} at a rotational motor shaft frequency of \SI{30}{Hz}. The QWeak 
$\ell$H$_2$ pump head was estimated with engineering calculations and corroborated 
with CFD simulations to be \SI{0.077}{bar}. During commissioning of the QWeak target 
pump with $\ell$H$_2$ the pump head was measured to be $\SI[separate-uncertainty=true]{0.076\pm0.007}{bar}$ 
at a rotational frequency of \SI{30}{Hz}. The QWeak $\ell$H$_2$ pump achieved its 
design goals validating the engineering calculations and CFD simulations. Before 
$\ell$H$_2$ operations the QWeak pump underwent a series of tests at Jefferson Lab 
fully immersed in a liquid nitrogen ($\ell$N$_2$) bath. $\ell$N$_2$ is not 
expensive, is not flammable and its density is eleven times higher than $\ell$H$_2$ 
which makes it very useful in testing the performance of a cryogenic pump. The 
$\ell$N$_2$ tests resulted in several improvements to the pump system: optimized 
spacing on the motor shaft to accommodate the thermal contraction of various pump 
parts in cryogenic conditions without compromising performance, adequate bearings 
for cryogenic conditions and optimized pump controls. The P2 $\ell$H$_2$ pump will be 
designed to have less than \SI{2}{kg/s} $\ell$H$_2$ mass flow or a volume rate of less 
than \SI{28}{l/s} with a pump head of less than \SI{0.1}{bar}, which is \SI{36}{\percent} 
higher than the QWeak $\ell$H$_2$ pump head. The impeller shape can be characterized
using the specific speed ($N_s = N\sqrt{Q}/H^{0.75}$ where $N$ is the rotational 
frequency, $Q$  is the flow rate and $H$ is the total head in the loop). Since 
the viscous heating is proportional to the 1.5th power of the pump head, the target
loop geometry will need to be optimized while also trying to minimize the inventory 
of $\ell$H$_2$ in the target. Depending on the final geometry of the loop, it may 
be possible to have the pump motor outside the loop, hence reducing the overall 
heat load and mitigating the risk associated with servicing the motor. The P2 
pump will be assessed in $\ell$N$_2$ to optimize it. The pump housing will be 
pressure tested and leak checked.

\begin{figure} 
\centering
\includegraphics[width=0.49\textwidth]{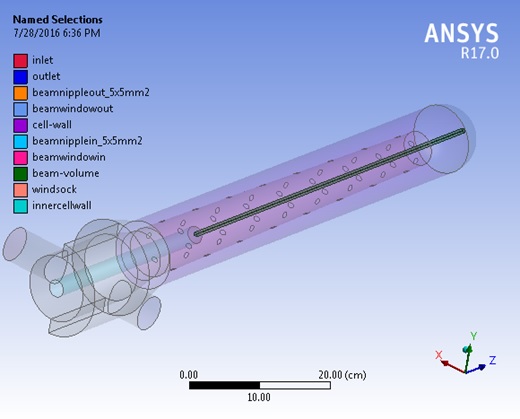}%
\caption{G0 type cell design.}
\label{fig:cell} 
\end{figure}

\begin{figure} 
\centering
\includegraphics[width=0.49\textwidth]{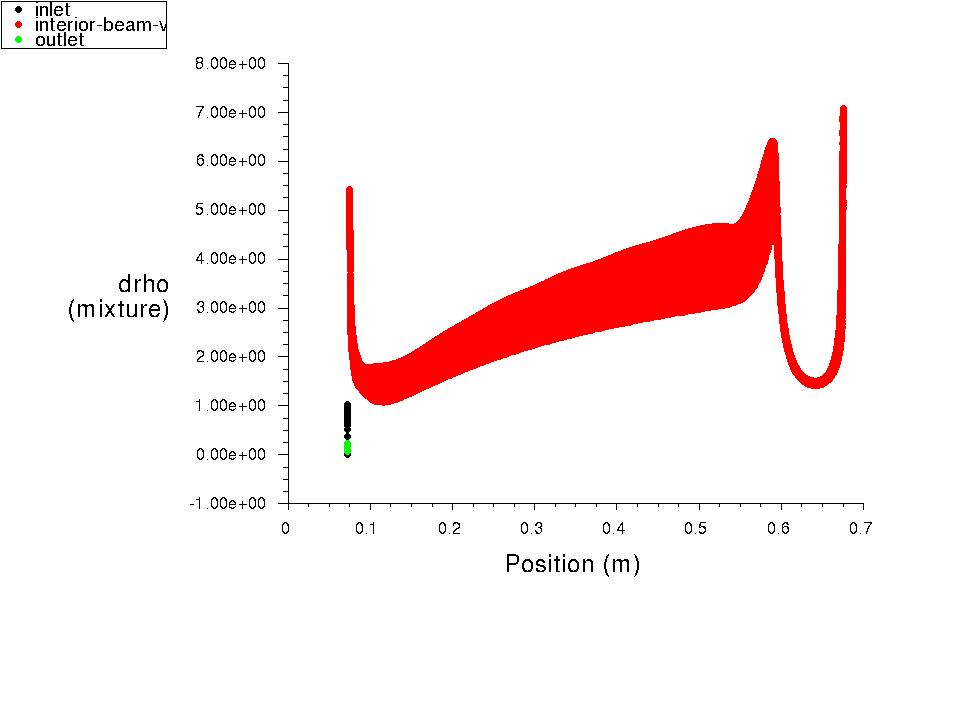} 
\caption{Preliminary CFD simulations of the $\ell$H$_2$ density loss.
The vertical axis is the absolute relative $\ell$H$_2$
density loss in (\%) and the horizontal axis is the distance along the beam
path from the beam entrance window of the cell to the downstream beam exit
window. 
}
\label{fig:drho} 
\end{figure}

\paragraph{$\boldsymbol\ell$H$_2$ cell.}
To accommodate the experimental acceptance for scattered particles the target
cell design will start with a G0-type cell geometry, shown in Fig.~\ref{fig:cell}. The
cell in this figure is \SI{10}{cm} in diameter with a hemispherical window at the
downstream end. The cell has an internal conical flow diverter made of \SI{0.075}{mm}
thick aluminum foil with holes on its wall. The internal flow diverter, also called
windsock, will be e-beam welded on a baffle that will separate the inlet $\ell$H$_2$
flow into the cell from the outlet flow. The \SI{60}{cm} $\ell$H$_2$ length in beam will be
defined between an upstream thin aluminum window and the downstream aluminum hemispherical
cell window. The upstream aluminum window will be a vacuum window with a diameter of
\SI{12}{mm} and thickness of \SI{0.125}{mm}. This window will be supported at the end of an
aluminum tube manufactured on an aluminum conflat flange that will mate on the target
manifold. This configuration keeps the asymmetric cell manifold completely
outside of the experimental acceptance. The cell connects with the rest of the
cryogenic loop through its manifold. The target cell and manifold will be made
from an aluminum alloy. The cell will be manufactured on a conflat flange that will
mate with the aluminum manifold. Preliminary structural engineering calculations show
that a cell wall thickness of \SI{0.25}{mm} will be sufficient to self support such
a long cell. $\ell$H$_2$ enters the cell through the inner flow diverter and is
accelerated and jetted at the aluminum beam exit window where it turns around and
flows upstream in the annular space between the cell wall and the flow diverter
wall towards the cell manifold, which directs it back into the target loop.

Figure~\ref{fig:drho} shows the $\ell$H$_2$ density loss from 
CFD simulations of the cell geometry in Fig.~\ref{fig:cell}~\cite{Covrig:P2CFD}. 
The CFD simulations were done in steady state assuming the design 
parameters from Tab.~\ref{tab:tgtparams} and the electron beam heating into 
$\ell$H$_2$ and aluminum windows from Tab.~\ref{tab:tgtheat}. 
These CFD simulations cover a temperature range of \SI{20}{K} to {300}{K} for
the heated fluid and the cell walls, properly accounting for $\ell$H$_2$ boiling and
treating the fluid as a liquid-gas mixture wherever it undergoes phase change
in the geometry. Turbulence is accounted for in these simulations through a
model. The vertical spread at a specific location along the beam path shows
the absolute relative $\ell$H$_2$ density loss in the square beam spot area of
\SI{25}{mm\squared}. If the $\ell$H$_2$ density decreases by \SI{6.7}{\percent} 
the liquid reaches saturation
at \SI{23.7}{K} and is susceptible of boiling. The average $\ell$H$_2$ density loss
over the beam volume for this model is predicted with CFD to be \SI{3}{\percent}. 
The QWeak target was the first PVES $\ell$H$_2$ target designed with CFD. The QWeak
target group originally proposed using a G0-type cell, extended to \SI{35}{cm} long,
CFD modelling changed that design from a fully longitudinal flow to a fully
transverse $\ell$H$_2$ flow to the beam path. Computational fluid dynamics 
technologies will be used to drive the design of the P2 target cell, starting 
from a G0-type cell, extended to \SI{60}{cm} length aiming for a luminosity 
loss due to the target boiling of less than \SI{2}{\percent}. The thin-walled 
aluminum cells are structurally the weakest part of the cryogenic target loop. 
A safety testing protocol will be developed to verify that the $\ell$H$_2$ 
cells are safe for operations (pressure tests, leak checks etc.).

\paragraph{Density fluctuations.}
The parity-violation asymmetry will be measured experimentally by using quartet 
helicity-flip structures $+ - - +$ and $- + + -$. 
The time period of an asymmetry quartet is \SI{4}{ms}. The quartet asymmetry is 
defined as $A_{m} = (N_+-N_-)/(N_++N_-)$, where $N_+$ and $N_-$ are the total 
number of scattered electrons, normalized to the electron beam, in the two 
positive and the two negative helicity states in the quartet respectively. 
In the absence of other noise sources, $A_m$ has a width, the counting 
statistics~\cite{Covrig:2005vx}, given by $\sigma_0 = (N_+ + N_-)^{-0.5} = 
(f/4R)^{0.5}$, where $f$ is the electron helicity frequency and $R$ is the 
expected electron rate in the P2 main detector. The P2 counting statistics 
is estimated to be \SI{50}{ppm}, four times smaller than QWeak's. The 
$\ell$H$_2$ target density fluctuation noise on the time period of the 
electron helicity is called boiling noise and adds in quadrature to the 
counting statistics to yield the measured asymmetry width $\sigma_m^2 = 
\sigma_0^2 + \sigma_b^2$. If the P2 target boiling noise contribution to 
the measured asymmetry width $\sigma_m$ is capped at \SI{2}{\percent}, 
then $\sigma_b< $\SI{10}{ppm} or five times smaller than the QWeak target 
boiling noise and twice smaller than the projected MOLLER $\ell$H$_2$ target 
boiling noise at Jefferson Lab. This design parameter makes the P2 target the 
most challenging PV $\ell$H$_2$ target in the world today. The QWeak $\ell$H$_2$ 
target underwent a thorough assessment of its performance in beam. 
The QWeak target noise was also measured at various electron beam helicity 
frequencies and found to vary like $\sigma_b\propto f^{-0.38}$. CFD simulations 
can reliably predict the $\ell$H$_2$ density reduction. A Facility for 
Computational Fluid Dynamics (CFDFAC) is being used to develop state of the art 
time-dependent simulations that aim to capture a $\ell$H$_2$ target cell's noise 
over various time scales. These simulations will be critical in the design of 
the MOLLER target cell~\cite{Benesch:2014bas}. The P2 target cell design
will benefit from these design technologies. A fine tuning of the CFD P2 
target cell design along with fine tuning of operational parameters like 
beam current, beam spot size, $\ell$H$_2$ pump frequency, $\ell$H$_2$ 
temperature and helicity frequency will achieve the design noise figure 
of the target of \SI{10}{ppm}.


\paragraph{Solid targets.}
The solid target ladder will be attached to the hydrogen cell and will contain
thick aluminum alloy foils (the same alloy as the cell) at the same positions along the
beam axis as the cell windows. In addition, it will contain centering targets
for beam steering/alignment and several carbon targets for optics studies. CFD
will be used to assess the beam heating of these targets and establish beam
current limits for safe operations.

\paragraph{Controls.}
Target controls include instrumentation, hardware and software to monitor, 
operate and control the target system. The target loop will be instrumented
with electrical feed-throughs for temperature sensors, $\ell$H$_2$ pump motor and HPH.
We plan to place temperature sensors at six locations around the loop, with two
sensors at each location for redundancy, across the target cell manifold, across
the HPH and across the HX. We also plan to place temperature sensors in the He
circuit of the HX at inlet and outlet to the HX. All temperature sensors will be
immersed in the fluid and calibrated. The locations of the sensors will allow
us to measure the $\ell$H$_2$ pump performance, the HX cooling power and efficiency and
measure and monitor the beam heating power. Redundancy of the sensors will
mitigate the risk that some of them will die in radiation. The solid targets ladder 
will be instrumented with up to six resistance temperature detectors (RTDs).
The target electronics will be made rack mountable. These will include the
temperature sensors monitors/controllers, the pump motor controller, the HPH
power supplies (two identical supplies connected in parallel, one active, one
redundant), vacuum gauge controllers, pressure monitors/transmitters, ADCs, DAQ
cards, low level data acquisition computer or PLC. MESA plans to use
EPICS as a software environment. The underlying software to
monitor, control, operate and archive target parameters data will be written in
EPICS. As appropriate other software, Python, C++ etc. will be used to control
and monitor target parts. 

\paragraph{Safety.}
The G0 target contained \SI{6}{liters} of liquid hydrogen and ran safely for over
five years at Jefferson Lab. The QWeak target contained \SI{55}{liters} of liquid
hydrogen and ran safely over two years at the same lab. The safety experience
accumulated by previous PVES liquid hydrogen targets will be leveraged in the
design, manufacturing and operations of the P2 target.

%% file: spectrometer.tex

The P2 spectrometer will use a large superconducting solenoid and is designed
to perform different tasks. The first one is the detection of the elastically scattered
electrons. Fast and radiation-resistant detectors are needed for this. Second, background processes such as Moller
scattering and bremsstrahlung has to be suppressed. Here, a careful design of the detector
layout performing Geant4 simulations is performed. Third, a measurement of the
momentum transfer $Q^2$ has to be provided. A dedicated detector system for track reconstruction
in the magnetic field is developed for this task. All these aspects are
discussed in the following sections.

%% file: simulation.tex


\begin{figure*}[htb]
  \centering
  \resizebox{0.8\textwidth}{!}{\includegraphics{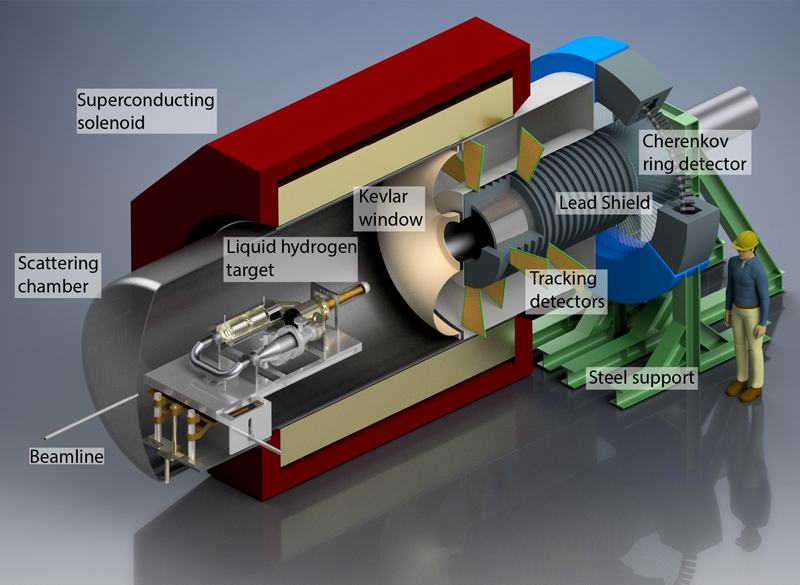}}
  \caption{CAD drawing of the experimental setup which has been implemented in
  the Geant4 simulation using CADMesh.}
  \label{fig:Geometry_P2Sim}
\end{figure*}

In order to simulate the P2 experiment, a Geant4
\cite{Agostinelli:2002hh,Allison:2006ve,Allison:2016lfl} application has been
developed. Geant4 is a software framework that allows to simulate the passage of
particles through matter with a computer. The simulation of the physics
processes involved is based on Monte Carlo methods, where the differential cross
sections are interpreted as probability density distributions, which are used to
sample the relevant kinematic variables of the particles. 

The purpose of the experiment's simulation is to ensure the feasibility of the
$Q_\text{W}(\text{p})$ measurement with the foreseen apparatus. In this section,
the main aspects of the Geant4 application will be discussed and results
presented.

\subsubsection{Geometry definition}
\label{sec:Geometrydefinition}

The application employs an interface to Computer-Aided Design (CAD) software for
defining the geometrical objects the experimental apparatus is comprised of. CAD
software is a widely used designing and analyzing tool in engineering science.
The simulation of the P2 experiment uses CADMesh \cite{Poole2012} to import
geometrical objects created with CAD software into Geant4. For this purpose, the
surfaces of the objects under consideration are first parametrized by applying a
tessellation procedure and then converted into a Geant4-native geometrical
object. The big advantage of this procedure is that engineering studies can be
performed using CAD applications and the resulting geometrical shapes may be
directly imported into Geant4. Furthermore, implementing new and altering
existing parts of the apparatus using realistic, complex geometrical shapes is
possible with a minimum of programming effort this way. The downside of using
CADMesh as compared to Geant4's standard method of defining geometry directly in
the source code is that the runtime of the application is slightly increased due
to the higher number of surfaces resulting from the tessellation procedure.
However, the prolongation of runtime is a minor effect and easily outweighted by
the benefits of the CAD interface, especially when using multiple CPU cores in
parallel to perform the simulation. 

Figure~\ref{fig:Geometry_P2Sim} shows a CAD drawing of the experimental setup,
which has been implemented in the simulation using CADMesh. The beam electrons
enter the scattering chamber's vacuum through the final part of the beamline and
interact with the $\ell\mathrm{H}_2$ target. Both target and scattering chamber
are contained within a superconducting solenoid that generates a magnetic field
of $B_\text{z} \approx \SI{0.6}{T}$ along the beam axis. The beam electrons,
which have been scattered off protons in the target, pass a Kevlar window which
separates the vacuum of the scattering chamber from the helium filled chamber
that contains the tracking detectors. The tracking detectors will be used to
reconstruct the $Q^2$ of the detected electrons and are described in section
\ref{sec:TrackingDetectors}. After passing the tracking system, the electrons
are detected in a Cherenkov ring detector for the measurement of the
parity-violating asymmetry.

\subsubsection{Event generation}
\label{sec:Eventgeneration}

One of the simulation's central aspects is the realistic simulation of the
interaction between the electron beam and the \SI{600}{mm} long
$\ell\mathrm{H}_2$ target. Since the beam energy $E_\text{beam} = \SI{155}{MeV}$
is rather small, energy loss and angular straggling of the beam in the target
material due to collisions and bremsstrahlung cannot be neglected. While Geant4
is an excellent tool to simulate these processes, the simulation of elastic
electron-proton scattering under large scattering angles is not foreseen in a
manner that is coherent with the simulation of the energy loss processes. The
reason for this is that the probability for scattering an electron elastically
off a proton with $\theta_\text{f} \sim \ang{35}$ is in the order
$\mathcal{O}(10^{-4})$ and therefore too low to simulate the process with it's
actual probability in an efficient way. 

\begin{figure}[htb]
  \centering
  \resizebox{0.5\textwidth}{!}{\includegraphics{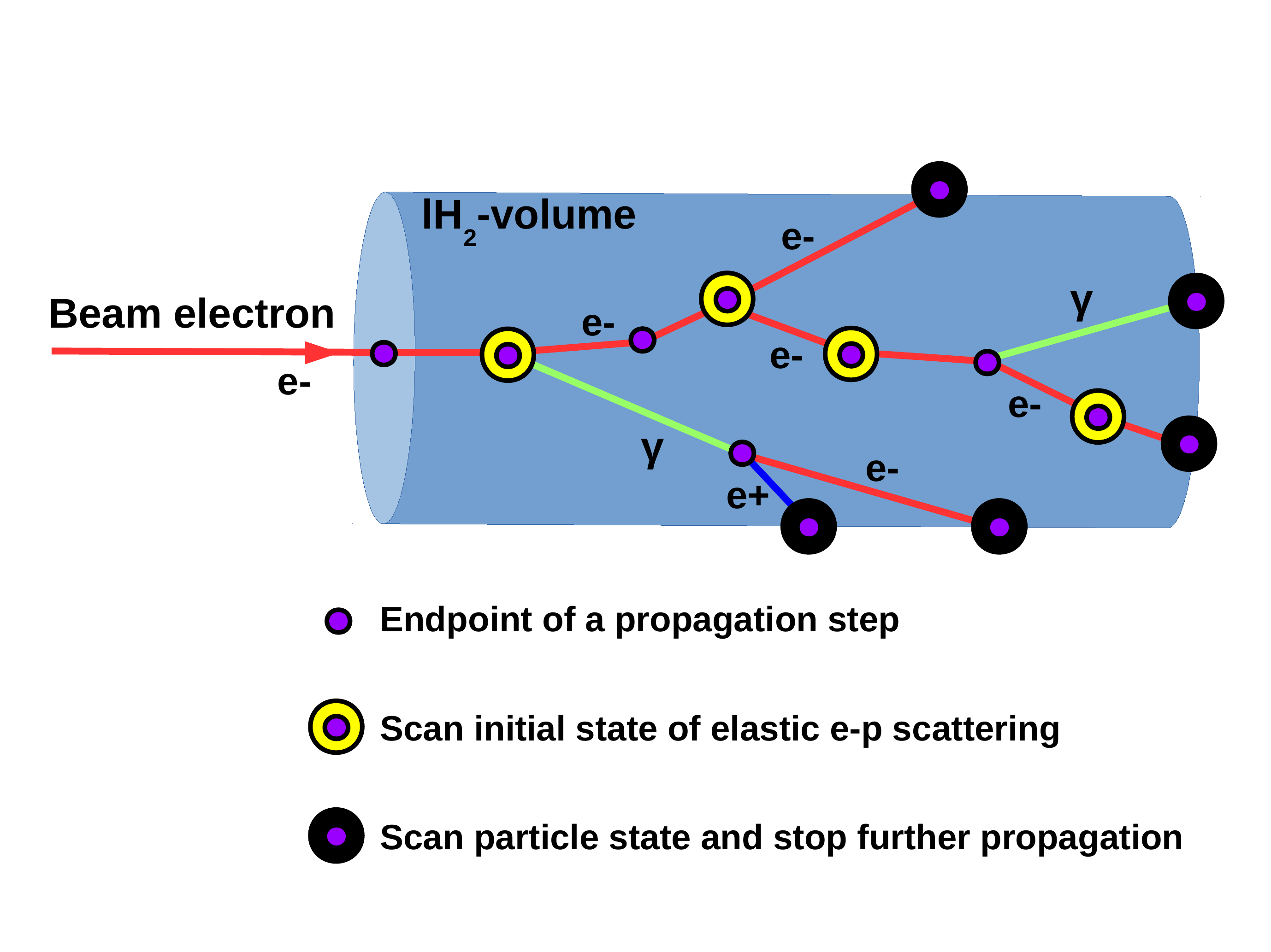}}
  \caption{Principle of generating ensembles of initial states of elastic e-p
  scattering and background particle states. The beam electrons are impinged
  upon the $\ell\mathrm{H}_2$-volume. In Geant4, all particles are propagated in
  spatial steps of finite length. The soft energy loss processes are simulated
  using the process models implemented in Geant4. Initial states of elastic e-p
  scattering are scanned at random positions along the beam electrons'
  trajectories without interfering with the simulation of the other physics
  processes. When a particle reaches the surface of the
  $\ell\mathrm{H}_2$-volume from it's inside, the particle's state is scanned
  and the simulation of the particle's trajectory is terminated.}
  \label{fig:event_gen_principle}
\end{figure}

\begin{figure}[b]
  \centering
  \resizebox{0.5\textwidth}{!}{\includegraphics{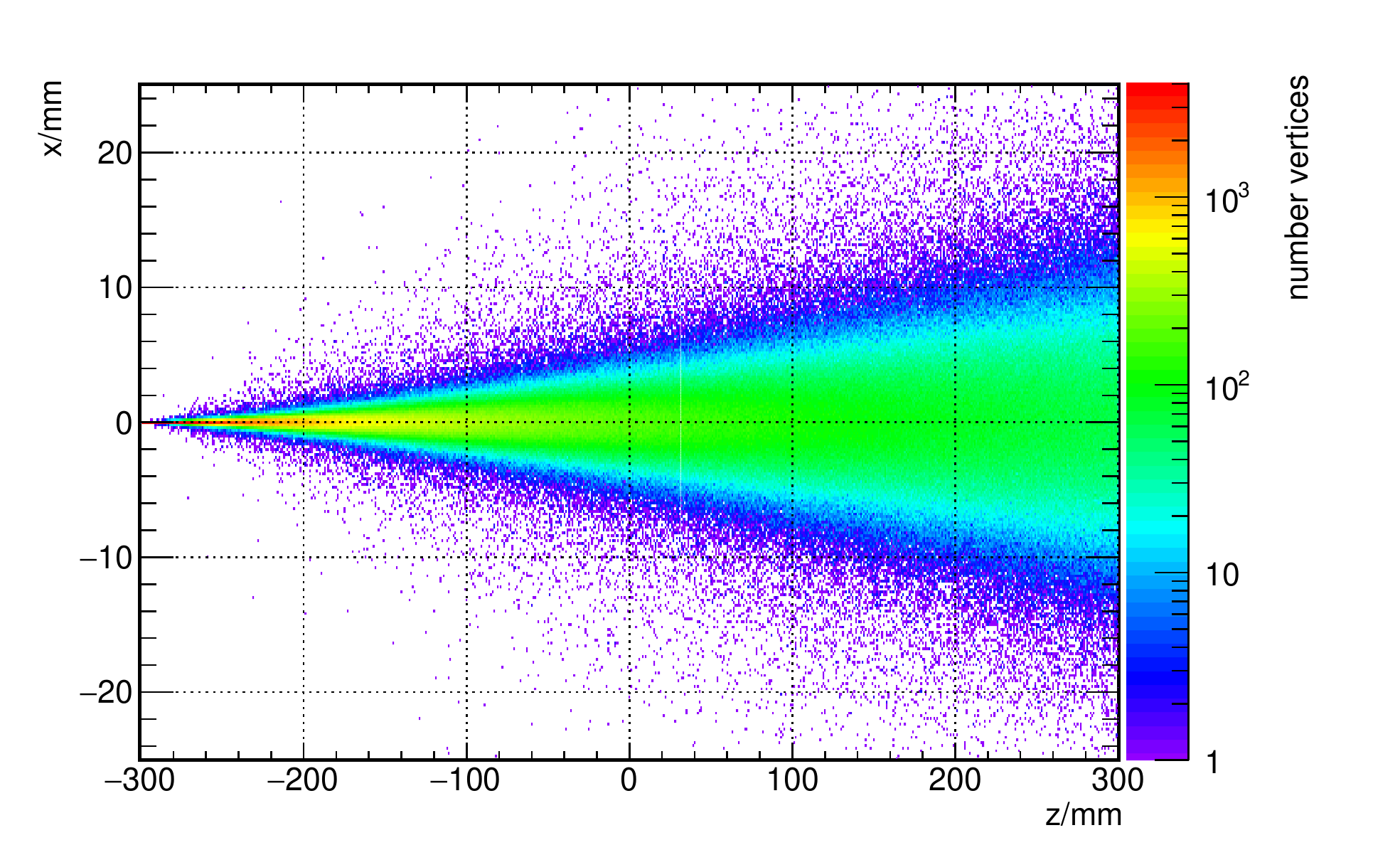}}
  \caption{Projection of the spatial distribution of sampled interaction
  vertices for the simulation of elastic electron-proton scattering to the
  $x$-$z$-plane. The cylindrical $\ell\mathrm{H}_2$ volume stretches from $z =
  \SI{-300}{mm}$ to $z = \SI{+300}{mm}$ and has a radius of \SI{25}{mm}. The
  electron beam enters the target volume from the left. One recognizes the
  widening of the beam profile due to collision processes with increasing
  $z$-coordinate.}
  \label{fig:Initial_state_distro_ep}
\end{figure}

In order to enable an efficient simulation of the ep scattering process, a
dedicated event generator has been developed. Initially, the passage of the beam
electrons through the target volume is simulated by impinging electrons with
$E_\text{beam} = \SI{155}{MeV}$ upon the target volume. The beam electrons are
tracked inside the $\ell\mathrm{H}_2$ volume, while the soft energy loss
processes are simulated using Geant4 built-in process models. As the beam
electrons travel through the target volume, initial states of the elastic ep
scattering process are scanned at random positions along their trajectories
without interfering with the simulation of the other physics processes.
Figure~\ref{fig:event_gen_principle} illustrates the principle.
An initial state of elastic electron-proton scattering is defined by:
\begin{itemize}
  \item The position of the vertex inside the target volume; 
  \item The initial state energy $E_\text{i}$ of the beam electron; 
  \item The 3-momentum vector of the beam electron. 
\end{itemize}
This method of sampling an ensemble of initial states of the ep scattering
process is valid, because the beam electrons undergo very similar processes as
they travel through the target volume so that each of the beam electrons'
trajectories may be regarded as the mean of an ensemble of similar trajectories.
This mean trajectory can be used to scan several initial states of elastic ep
scattering. Figure~\ref{fig:Initial_state_distro_ep} shows a sample distribution
of initial states of the ep scattering process.
As the beam electrons are propagated through the $\ell\mathrm{H}_2$ volume,
Geant4 generates secondary particles in the course of the simulation of the
collision and bremsstrahlung processes. All of these particles are tracked
through the target volume as well until they leave the volume. Once at this
point, the particles' state is scanned, saved, and the particle is stopped and
terminated in order to save CPU time. This leads to an ensemble of particle
states stemming from background processes in the target volume, which are
located on the target volume's surface. Such a state is defined by:
\begin{itemize}
  \item The particle's type; 
  \item The position of the particle on the target's surface; 
  \item The 4-momentum vector of the particle. 
\end{itemize}
Once calculated for a specific target geometry, both the initial state ensemble
of the ep scattering process and the ensemble of background particle states may
be re-used an arbitrary number of times to generate final state ensembles for the
detector simulation. In order to be able to predict event rate distributions
expected in the real experiment with the method described above, one has to
normalize the simulated events properly. 

\begin{figure}[htb]
  \begin{minipage}[t]{1\linewidth}
    \centering
    \resizebox{1\textwidth}{!}{\includegraphics{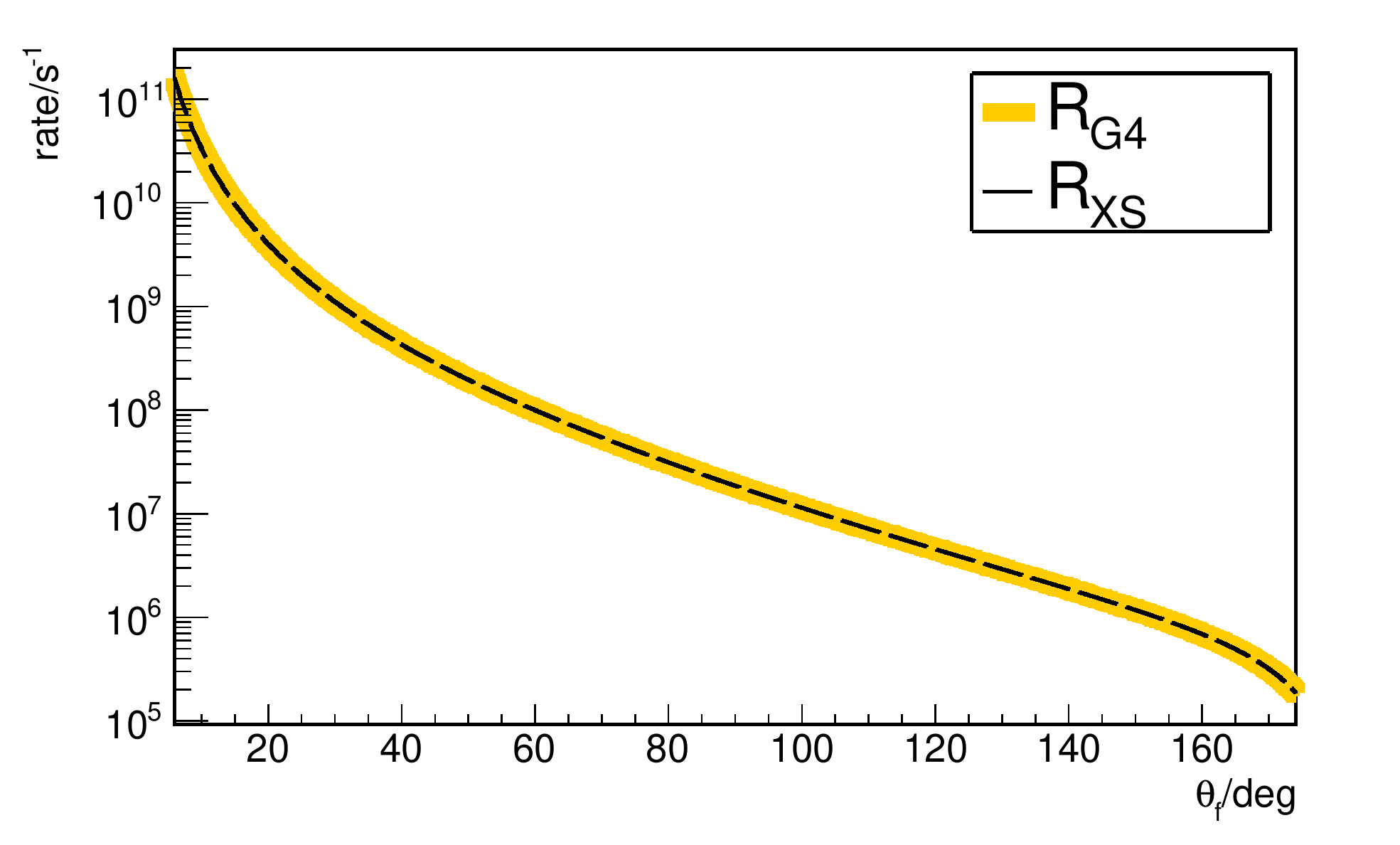}}
  \end{minipage}
  \hfill
  \begin{minipage}[t]{1\linewidth}
    \centering
    \resizebox{1\textwidth}{!}{\includegraphics{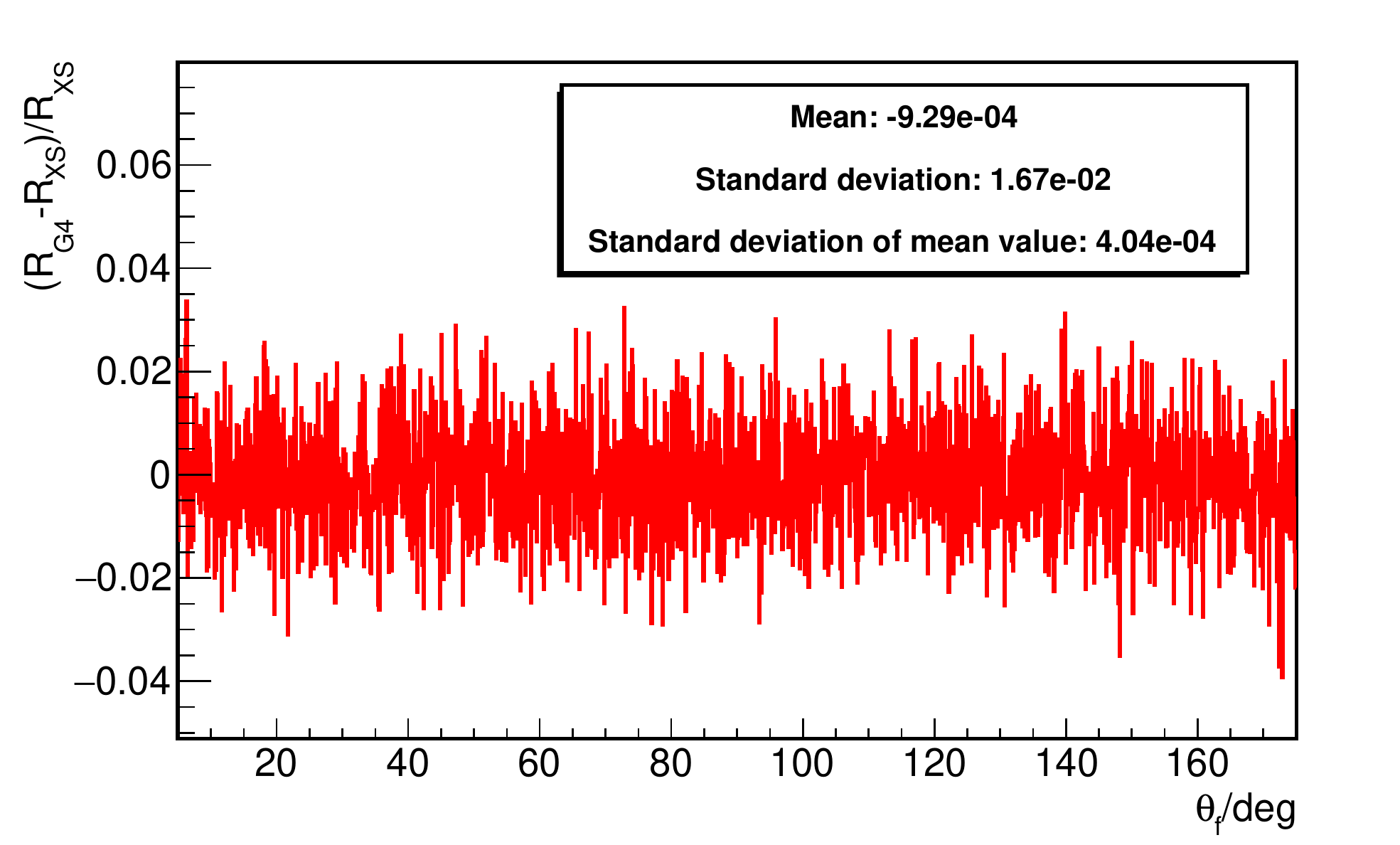}}
  \end{minipage}
  \caption{Comparison of two predictions of the scattering rates expected in the
  P2 experiment at design luminosity. $R_\text{G4}$ is the prediction made using
  the event generator implemented in the Geant4 simulation, and $R_\text{XS}$ is
  the rate prediction based on a numerical integration of the Rosenbluth
  formula. The upper picture shows the dependence of $R_\text{G4}$ (yellow 
  thick line) and $R_\text{XS}$ (thin black dashed line) on $\theta_\text{f}$.
  The two curves shown in the upper picture do overlap each other. The lower
  picture shows the relative deviations between the two rates as a function of
  $\theta_\text{f}$. The relative deviations scatter statistically around $0$
  for the considered values of $\theta_\text{f} \in [\ang{5}, ~\ang{175}]$,
  which indicates that the event generator reproduces the expected rate
  distribution in $\theta_\text{f}$ correctly.}
  \label{fig:Rosi_test}
\end{figure}

For each of the initial states of elastic ep scattering, one final state is
generated. For this purpose, a final state generator has been developed. The
generator uses elastic kinematics and creates an electron and a proton in the
final state of the scattering process. For this, the electron's scattering
angles $\theta_\text{f}$ and $\phi_\text{f}$ are sampled using flat probability
density distributions. The Rosenbluth formula
(Eq.~(\ref{eq:Rosenbluth_formula})) is used as a weighting factor for the
sampled event. Figure~\ref{fig:Rosi_test} shows a comparison between the rate
prediction of the event generator and one that is based on a numerical
integration of the differential cross section. The event generator reproduces
the rate prediction of the independent numerical integration to high accuracy.
Furthermore, a prototype of a final state generator has been developed, which
allows for a real photon in the final state of the scattering process. This
final state generator will be available in addition to the currently used
generator in the near future and make it possible to take the shifts in $Q^2$
into account, which are caused by the radiation of photons (see section
\ref{sec:QED}).

The states of the beam electrons and background particles that have been scanned
during the passage of the beam electrons through the target volume are simply
recreated after the simulation of elastic ep scattering has been completed.
After their creation in the target volume, all particles' trajectories in the
magnetic field of the superconducting solenoid are simulated.

\subsubsection{Simulation of trajectories in the magnetic field}
\label{sec:Simulation of trajectories in the magnetic field}

\begin{figure}[htb]
  \begin{minipage}[t]{1\linewidth}
    \centering
    \resizebox{1\textwidth}{!}{\includegraphics{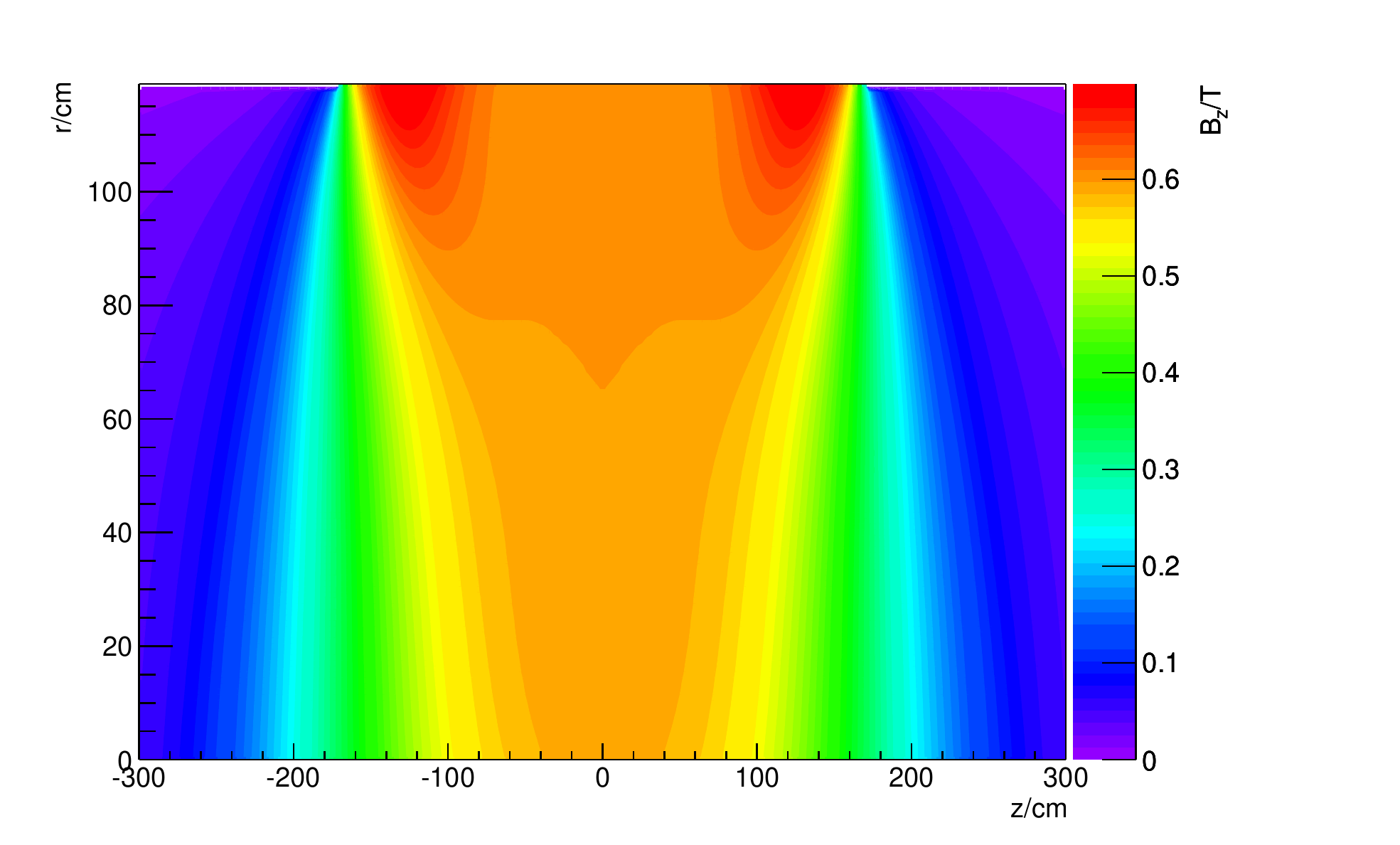}}
  \end{minipage}
  \hfill
  \begin{minipage}[t]{1\linewidth}
    \centering
    \resizebox{1\textwidth}{!}{\includegraphics{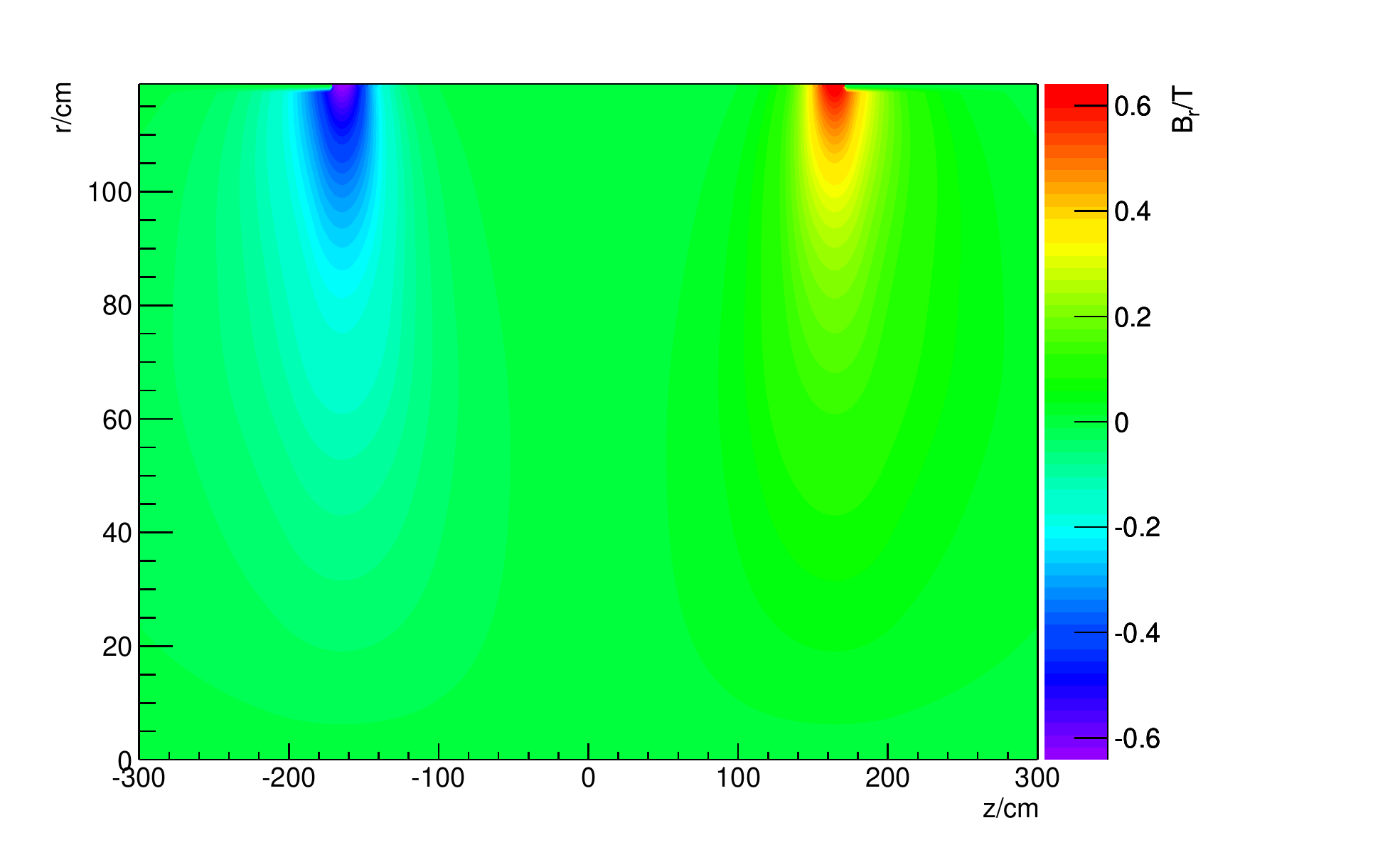}}
  \end{minipage}
  \caption{Magnetic field-map of the FOPI-solenoid. Courtesy of Y. Leiffels (FOPI
  Collaboration). The upper picture shows the $z$-component $B_\text{z}$ of the
  magnetic field, which is parallel to the beam axis, as a function of the
  $z$-coordinate and the distance $r$ from the beam axis. The lower picture
  shows the radial component $B_\text{r}$ of the magnetic field, which is
  perpendicular to the beam axis, as a function of $z$ and $r$.}
  \label{fig:fopi_fieldmap}
\end{figure}

In order to be able to properly determine the positions of the tracking system,
the Cherenkov ring detector and the lead shielding, a realistic simulation of
charged particles' trajectories in the magnetic field of the superconducting
solenoid is indispensable. For this reason, the simulation enables the usage of
realistic magnetic field maps to calculate the trajectories. In particular, the
P2 Collaboration has studied the usability of the superconducting solenoid that
has been used in the FOPI \cite{Ritman:1995td} experiment. The field-map of this
magnet has been provided to the P2 Collaboration courtesy of the FOPI
Collaboration. It is shown in Fig.~\ref{fig:fopi_fieldmap}.

For the calculation of charged particles' trajectories in the magnetic field, an
implicit Euler method is used to numerically solve the equations of motion. This
method makes use of the fact that in a time-independent magnetic field, which is
locally homogeneous, the trajectories may be regarded as a concatenation of
helix-shaped chords. 

\begin{figure}[htb]
  \centering
	\includegraphics[width=0.5\textwidth]{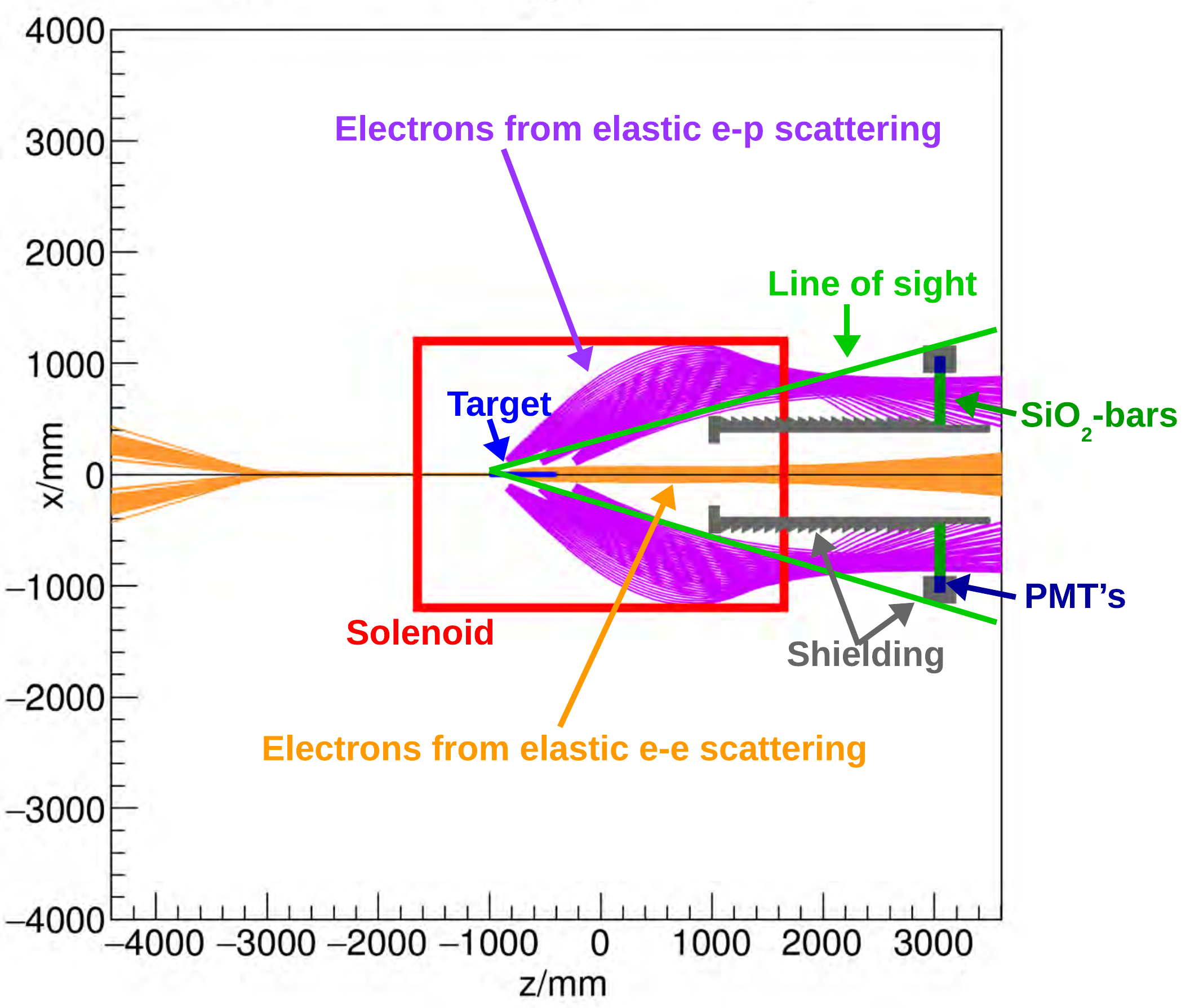}
  \caption{Simulation of electron trajectories in the magnetic field of the FOPI
  solenoid. Shown are the projections to the $x$-$z$-plane of the trajectories
  of electrons in the final states of elastic ep scattering (purple), which are
  to be detected in the P2 experiment, and electrons in the final state of the
  M{\o}ller scattering process (orange) for $E_\text{beam} = \SI{155}{MeV}$.}
  \label{fig:raytraces}
\end{figure}

Figure~\ref{fig:raytraces} shows projections of simulated electron trajectories
in the magnetic field of the FOPI solenoid. For this specific calculation, all
physics processes implemented in Geant4 have been disabled so that there is no
angular straggling or energy loss of the particles as they travel through the
materials of the apparatus. To illustrate the effect of the lead shielding, all
particles which hit the shields were stopped instantly. The picture illustrates
the purpose of the magnetic field, which is to separate electrons in the final
state of the elastic ep scattering process from electrons in the final state of
the M{\o}ller scattering process. The function of the lead shielding is also
illustrated: It covers the lines of sight between the Cherenkov detector and the
target volume in order to prevent photons emerging from the target from hitting
the detector's active volume and photomultiplier tubes.

\subsubsection{Simulation of the Cherenkov ring detector}
\label{SimulationoftheCherenkovringdetector}

The simulation of the Cherenkov detector is another central aspect of the Geant4
application. A detailed description of the Cherenkov ring detector can be found
in section \ref{sec:IntegratingDetectors}. The main goal of the Monte Carlo
simulation is to predict the distribution of particles which hit the detector
as well as the detector response. 

\begin{figure}[tb]
  \centering
  \resizebox{0.5\textwidth}{!}{\includegraphics{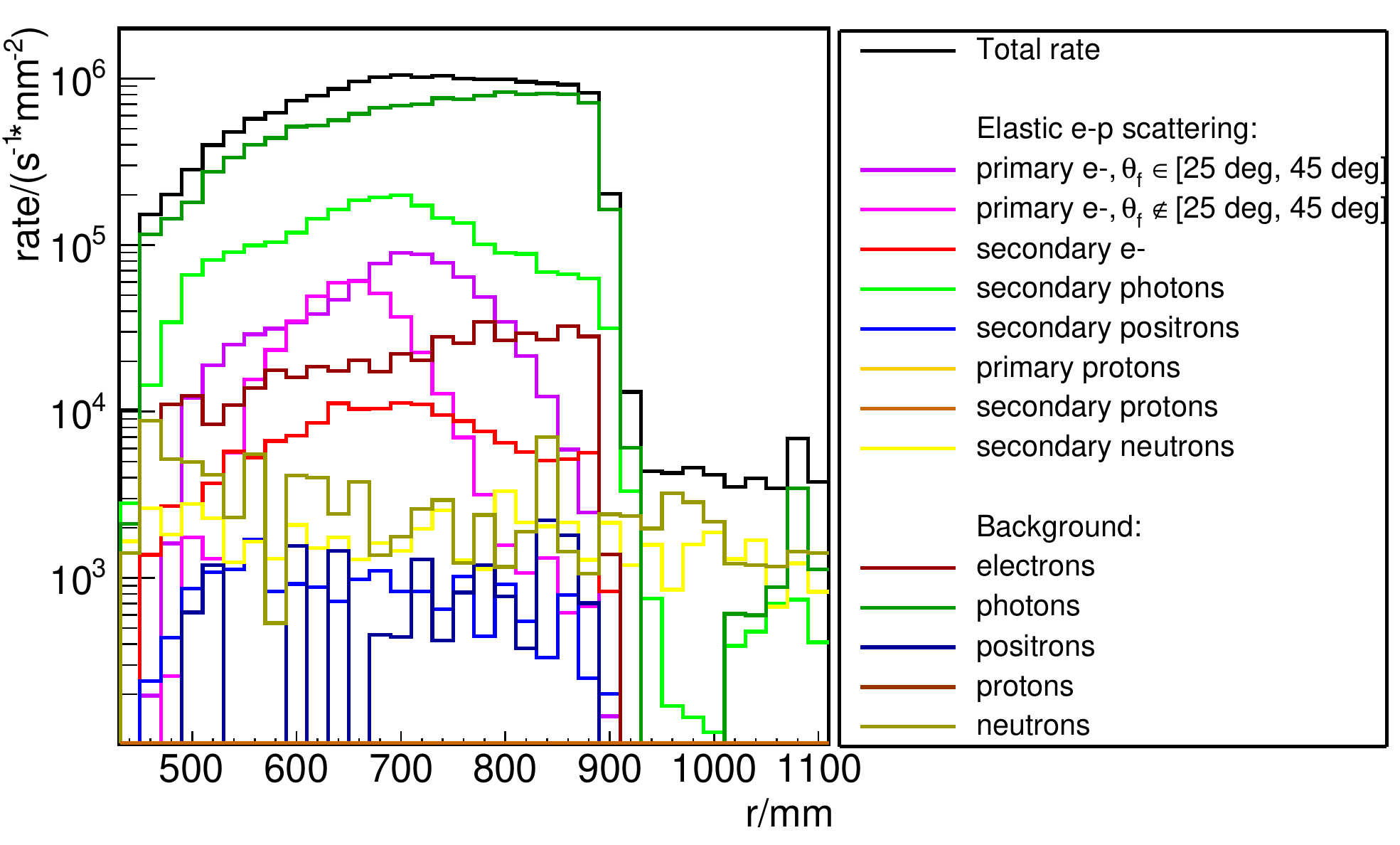}}
  \caption{Rate distributions on the surface of the Cherenkov ring detector as a
  function of the distance $r$ from the beam axis. The rates have been
  calculated assuming the experimental conditions listed in
  Tab.~\ref{tab:input_parameters_ep} and Tab.~\ref{tab:achievable_precision} of
  Sect.~\ref{sec:SinThetaW}. The rates have been normalized to the areas of
  the ring segments which correspond to the width of the histograms' bins in
  $r$-direction. Further explanations and the discussion of the distributions
  can be found in the text.}
  \label{fig:detector_hit_distribution}
\end{figure}

The calculation of the particle distribution that is incident upon the
detector's active volume, which will consist of $\mathrm{SiO}_2$ bars wrapped in
a reflective foil (see Fig.~\ref{fig:detectorelement}), is done by scanning the
particles' states when they reach the surface of a $\mathrm{SiO}_2$ bar in the
simulation. The information gathered includes
\begin{itemize}
  \item the particle's type, 
  \item the impact position of the particle on the active volume's surface, 
  \item and the momentum $4$-vector of the particle at impact position
\end{itemize}
for all particles. Figure~\ref{fig:detector_hit_distribution} shows the
simulated rate distribution on the surface of the ring detector for the particle
types considered in the calculation. In the figure legend, the particle types
are sorted into two categories according to the way the event generator operates
(see section \ref{sec:Eventgeneration}): The first category comprises
particles which reach the detector as a consequence of an elastic ep scattering
in the $\ell\mathrm{H}_2$ target, the second category includes all particles
which hit the detector as a consequence of a background process in the target
volume. For the first category, a distinction is made between ``primary'' and
``secondary'' particles. A ``primary'' particle is a particle that has been
generated by the final state generator, so it can be either an electron or a
proton in the final state of the ep scattering process. All particles labeled
``secondary'' have emerged from the interaction of primary particles with the
materials of the experimental setup. The total rate distribution in
Fig.~\ref{fig:detector_hit_distribution} is dominated by photons stemming from
background processes in the target volume and secondary photons created during
the simulation of elastic ep scattering. As described in
\ref{sec:IntegratingDetectors}, the $\mathrm{SiO}_2$ bars reach from $r =
\SI{450}{mm}$ to $r = \SI{1100}{mm}$, but only the parts with $r \in
[\SI{450}{mm}, ~\SI{900}{mm}]$ serve as active volumes for
particle detection. The sections with $r > \SI{900}{mm}$ guide the
Cherenkov light to the photomultiplier tubes, which will be located at $r >
\SI{1100}{mm}$. The parts used as light-guides will be surrounded by a
$\SI{100}{mm}$ thick lead shield, which is the reason for the reduction of the
total rate in the region with $r \geq \SI{900}{mm}$.
Table~\ref{tab:rates} lists the total rates expected in the P2 experiment on the
surface of the full $\mathrm{SiO}_2$ ring.

\begin{table}
  \centering
  \begin{tabular}{cc}
    \toprule[1.5pt]
    Contribution & Hit rate/$s^{-1}$ \\
    \midrule[1.5pt]
    Total & $\SI{1.54e+12}{}$ \\
    \midrule[1.5pt]
    Elastic ep scattering in the target: & \\
    \midrule
    Primary electrons, $\theta_\text{f} \in [\ang{25}, \ang{45}]$ & $\SI{7.10e+10}{}$ \\
    \midrule
    Primary electrons, $\theta_\text{f} \notin [\ang{25}, \ang{45}]$ & $\SI{3.21e+10}{}$ \\
    \midrule
    Primary protons & $\SI{0.00e+00}{}$ \\
    \midrule
    Secondary electrons & $\SI{1.33e+10}{}$ \\
    \midrule
    Secondary positrons & $\SI{1.47e+09}{}$ \\
    \midrule
    Secondary photons & $\SI{2.12e+11}{}$ \\
    \midrule
    Secondary protons & $\SI{5.11e+05}{}$ \\
    \midrule
    Secondary neutrons & $\SI{5.41e+09}{}$ \\
    \midrule[1.5pt]
    Background processes in the target: & \\
    \midrule
    Electrons & $\SI{4.05e+10}{}$ \\
    \midrule
    Photons & $\SI{1.14e+12}{}$ \\
    \midrule
    Positrons & $\SI{1.40e+09}{}$ \\
    \midrule
    Protons & $\SI{0.00e+00}{}$ \\
    \midrule
    Neutrons & $\SI{8.31e+09}{}$ \\
    \bottomrule[1.5pt]
  \end{tabular}
  \caption{Survey of the hit rates expected on the full $\mathrm{SiO}_2$ ring of
  the Cherenkov detector for the projected experimental conditions.}
  \label{tab:rates}
\end{table} 

It is obvious from Fig. \ref{fig:detector_hit_distribution} that a good
understanding of the Cherenkov detector's response to the incident particles is
needed, since the photon rate is much higher than the rate of the primary
electrons from elastic ep scattering. In order to be able to predict the
response of the detector, one needs to know the number of photo-electrons
emitted from the photomultiplier's cathode in consequence of a particle hitting
the detector. Since the reproduction of the Cherenkov effect in the simulation of
the full experiment is not feasible due to the CPU time required for a coherent 
simulation of the effect, a parametrization of the detector modules' response
has been created. The parametrization is described in detail in 
Sect.~\ref{sec:IntegratingDetectors}. It allows to calculate the mean number of 
photo-electrons expected from a particle passing through a $\mathrm{SiO}_2$ bar based
on:
\begin{itemize}
  \item The particle's type; 
  \item The particle's total energy; 
  \item The particle's momentum direction at impact position on the detector
  module. 
\end{itemize}

\begin{figure}[b]
  \centering
  \resizebox{0.5\textwidth}{!}{\includegraphics{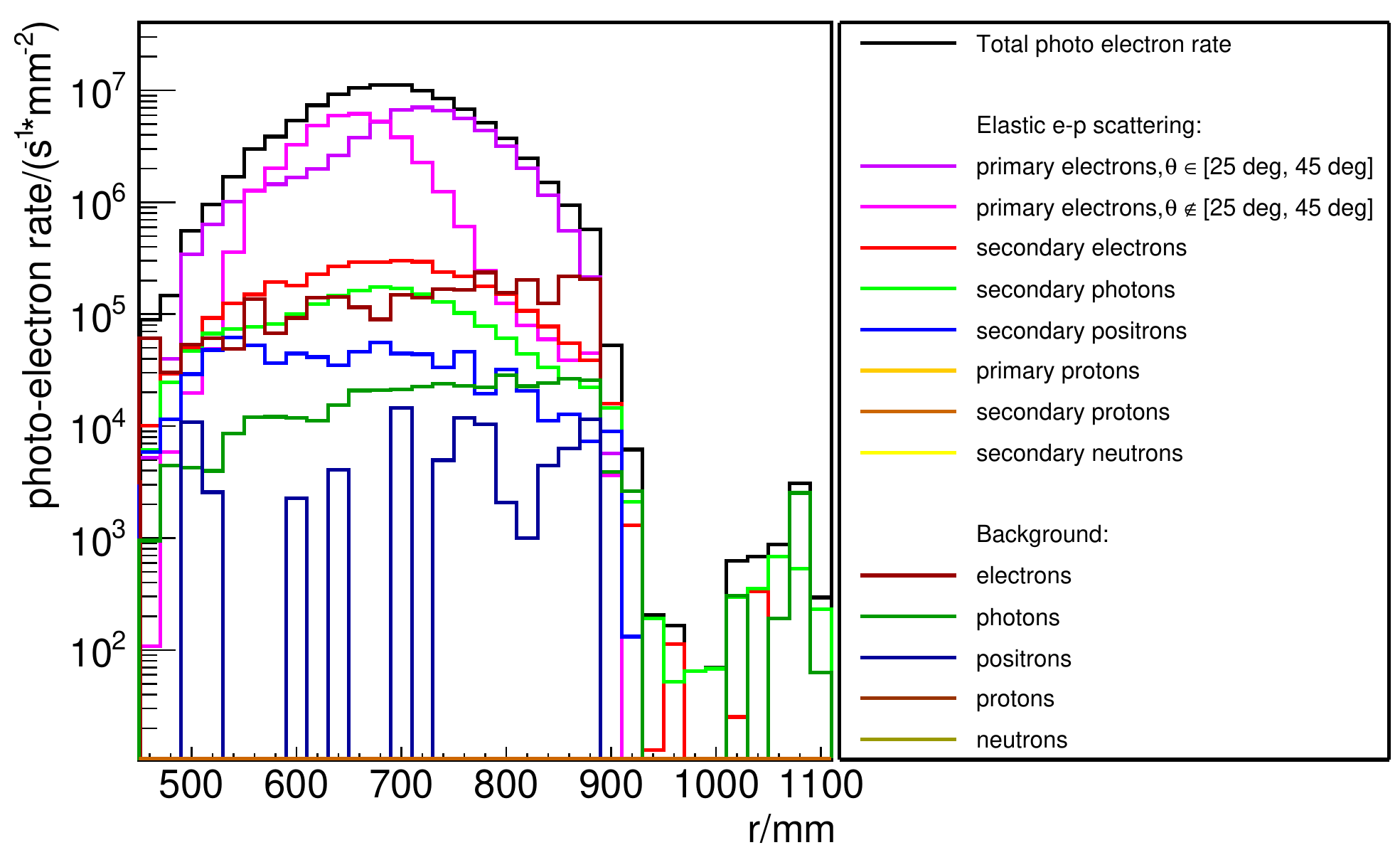}}
  \caption{Photo-electron rate distributions as a function of the distance $r$
  from the beam axis at which the particles hit the surface of the
  $\mathrm{SiO}_2$ ring. Like the distributions shown in
  Fig.~\ref{fig:detector_hit_distribution}, the photo-electron distributions are
  normalized to the areas of the ring segments associated with the bin widths of
  the histograms. Further discussion in the text.}
  \label{fig:pe_rate_distro}
\end{figure}

Utilizing the detector response parametrization allows to convert the hit rate
distributions shown in Fig.~\ref{fig:detector_hit_distribution} into the 
photo-electron rate distributions shown in Fig.~\ref{fig:pe_rate_distro}. Here, the
photo-electron rates emitted from the photocathodes are shown as functions of
the radius $r$ at which the particles have hit the detector. The total 
photo-electron rate is clearly dominated by the contributions stemming from electrons,
which have been scattered elastically off protons in the target volume. The
parametrization of the detector response currently includes only electrons,
positrons and photons. All other particle types' contributions to the spectrum
have been set to $0$ manually.
The expected photo-electron rates are listed in Tab.~\ref{tab:pe_rates}.

\begin{table}
  \centering
  \begin{tabular}{cc}
    \toprule[1.5pt]
    Contribution & Photo electron rate/$s^{-1}$ \\
    \midrule[1.5pt]
    Total & $\SI{9.06e+12}{}$ \\
    \midrule[1.5pt]
    \multicolumn{2}{l}{Elastic ep scattering in the target:} \\
    \midrule
    Primary electrons, $\theta_\text{f} \in [\ang{25}, \ang{45}]$ &
    $\SI{5.12e+12}{}$ (\SI{56.5}{\%}) \\
    \midrule
    Primary electrons, $\theta_\text{f} \notin [\ang{25}, \ang{45}]$ &
    $\SI{3.11e+12}{}$ (\SI{34.3}{\%}) \\
    \midrule
    Secondary electrons & $\SI{3.07e+11}{}$ (\SI{3.4}{\%}) \\
    \midrule
    Secondary positrons & $\SI{6.17e+10}{}$ (\SI{0.7}{\%}) \\
    \midrule
    Secondary photons & $\SI{1.63e+11}{}$ (\SI{1.8}{\%}) \\
    \midrule[1.5pt]
    \multicolumn{2}{l}{Background processes in the target:} \\
    \midrule
    Electrons & $\SI{2.54e+11}{}$ (\SI{2.8}{\%}) \\
    \midrule
    Photons & $\SI{3.48e+10}{}$ (\SI{0.4}{\%}) \\
    \midrule
    Positrons & $\SI{7.97e+09}{}$ (\SI{0.1}{\%}) \\
    \bottomrule[1.5pt]
  \end{tabular}
  \caption{Survey of the photo-electron rates expected in the Cherenkov ring
  detector for the projected experimental conditions.}
  \label{tab:pe_rates}
\end{table} 

\begin{figure}[tb]
  \centering
  \resizebox{0.5\textwidth}{!}{\includegraphics{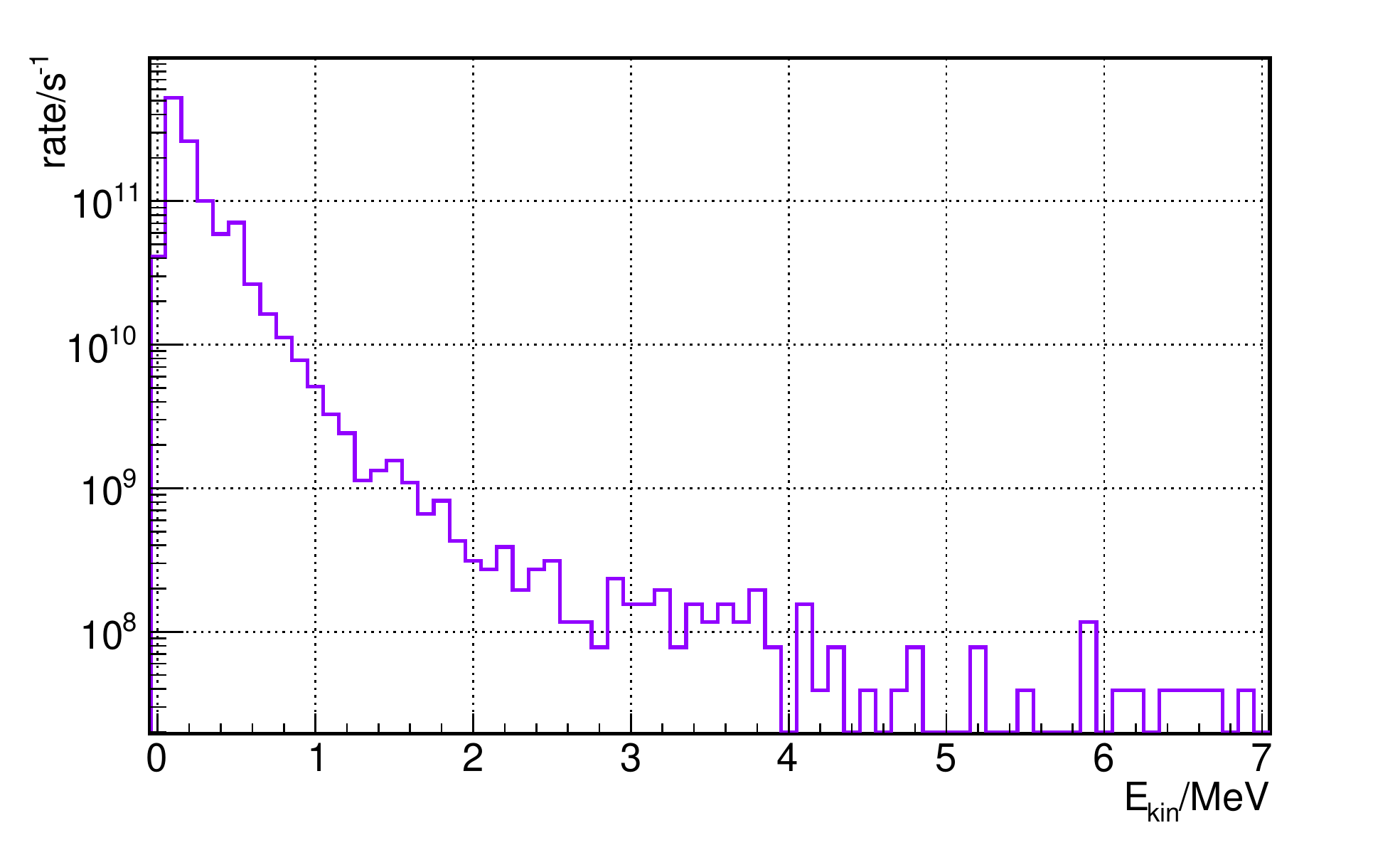}}
  \caption{Energy distribution of the photons stemming from background processes
  in the target. The majority of the photons have energies below $E_\text{th}
  \approx \SI{0.7}{MeV}$.}
  \label{fig:energy_distro_photons}
\end{figure}

By comparing the spectra shown in Fig.~\ref{fig:detector_hit_distribution} and
Fig.~\ref{fig:pe_rate_distro}, one observes that the relative contribution of
photons to the total photo-electron rate is drastically reduced compared to the
photons' relative contribution to the total hit rate distribution. The
explanation for this is that many photons do not lead to a detector signal,
because they cannot produce Cherenkov light inside the $\mathrm{SiO}_2$ bars
directly. The photons have to interact with the detector material first leading
to charged particles with mass $m$ travelling through the $\mathrm{SiO}_2$ bars,
whose energies are above the threshold energy
\begin{equation}
  E_\text{th} = \frac{m}{\sqrt{1 - (1 / n^2)}}
  \label{eq:threshold_energy_cherenkov}
\end{equation}
for the production of Cherenkov light. In 
Eq.~(\ref{eq:threshold_energy_cherenkov}), $n$ is the refractive index. For
electrons traveling through $\mathrm{SiO}_2$, it is $E_\text{th} \approx
\SI{0.7}{MeV}$. Most of the photons which hit the detector have energies smaller
than $E_\text{th}$, as can be seen in Fig.~\ref{fig:energy_distro_photons}. This
circumstance enhances the suppression of photons in the Cherenkov ring detector.
 \\

\begin{figure}[htb]
  \centering
  \resizebox{0.5\textwidth}{!}{\includegraphics{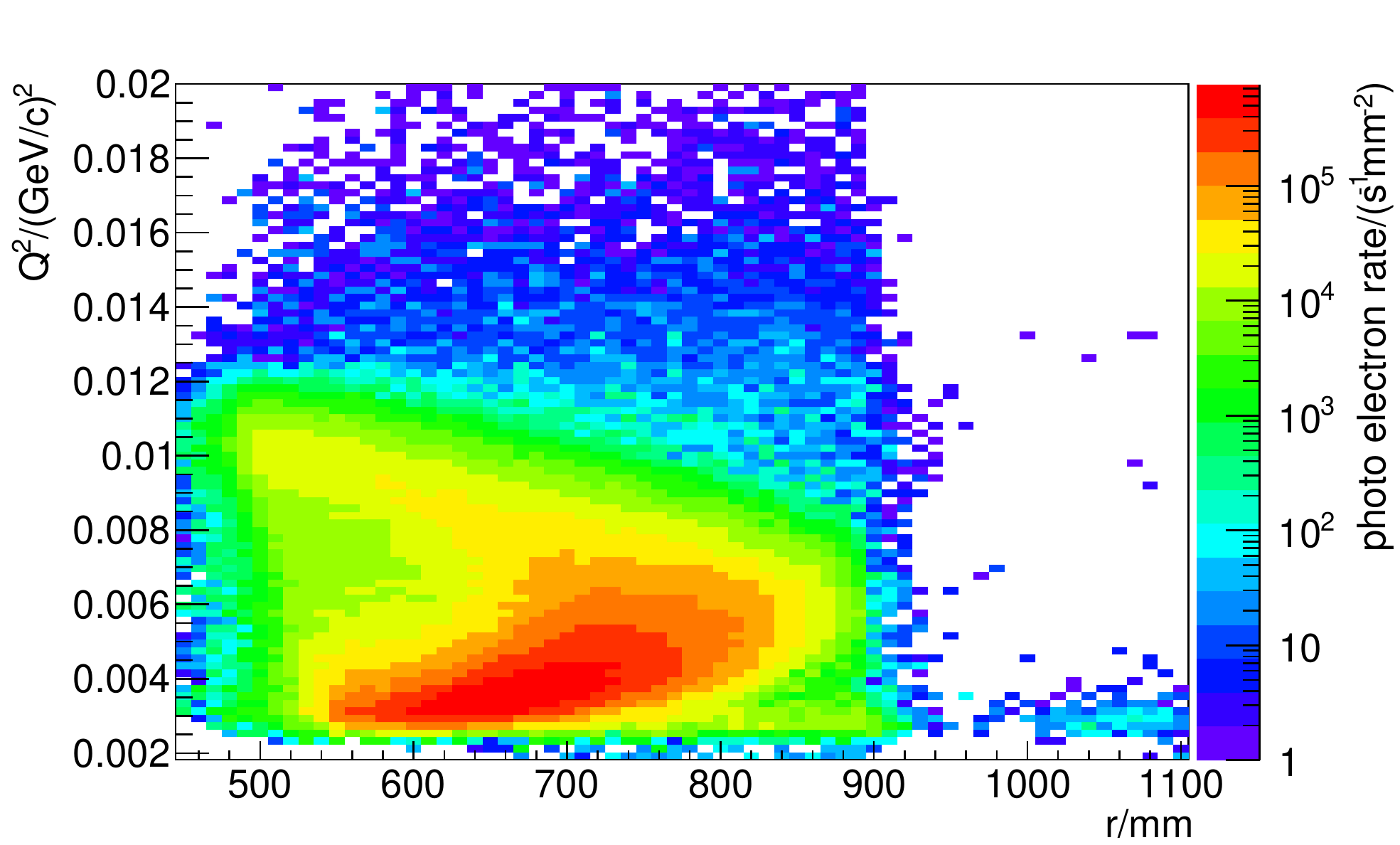}}
  \caption{Photo-electron rate distribution of all particles hitting the
  detector in consequence of an elastic ep scattering in the target volume.}
  \label{fig:distro_photons_qsquare}
\end{figure}

Figure~\ref{fig:distro_photons_qsquare} shows the photo-electron rate
distribution of all particles hitting the detector in consequence of an elastic
ep scattering in the target volume in dependence of the distance $r$ from the
beam axis and the $Q^2$ value of the scattering event. From this distribution
the average value of $Q^2$ can be calculated:
\begin{equation}
  \langle Q^2 \rangle_\text{Cherenkov} = \SI{4.57e-3}{(GeV/c)^2}.
\end{equation}
This corresponds to 
\begin{equation}
  \langle A^\text{exp} \rangle_\text{Cherenkov} = \SI{-28.77}{ppb}
\end{equation}
for the asymmetry, which is to be measured. The reduction of the absolute value
of the asymmetry by \SI{28}{\%} compared to the value of \SI{-39.94}{ppb}, which
has been predicted by the error propagation calculation in section
\ref{sec:Results of the error propagation calculations} is due to the dilution
of the asymmetry caused by the background and the admixture of electrons from
elastic ep scattering with smaller values of $Q^2$.

%% file: cherenkovdetector.tex
A high-precision, high-intensity electron scattering experiment such as P2 
imposes substantial demands on the detector system. In order to reach the 
precision goal by measuring the tiny 
parity-violating asymmetry in electron-proton scattering of only 
$A^\text{exp}=\SI{28.77}{ppb}$, we need to collect very high statistics in a manageable 
run-time. The design of the detector was optimized with regard to speed, 
radiation hardness and optimal coverage of polar and azimuth angles as well as 
a strong signal from elastically scattered electrons and a suppression of 
background particles.

\subsubsection{General detector concept}
\label{sssec:generaldetectorconcept}
\begin{figure}
	\centering
	\includegraphics[width=0.45\textwidth]{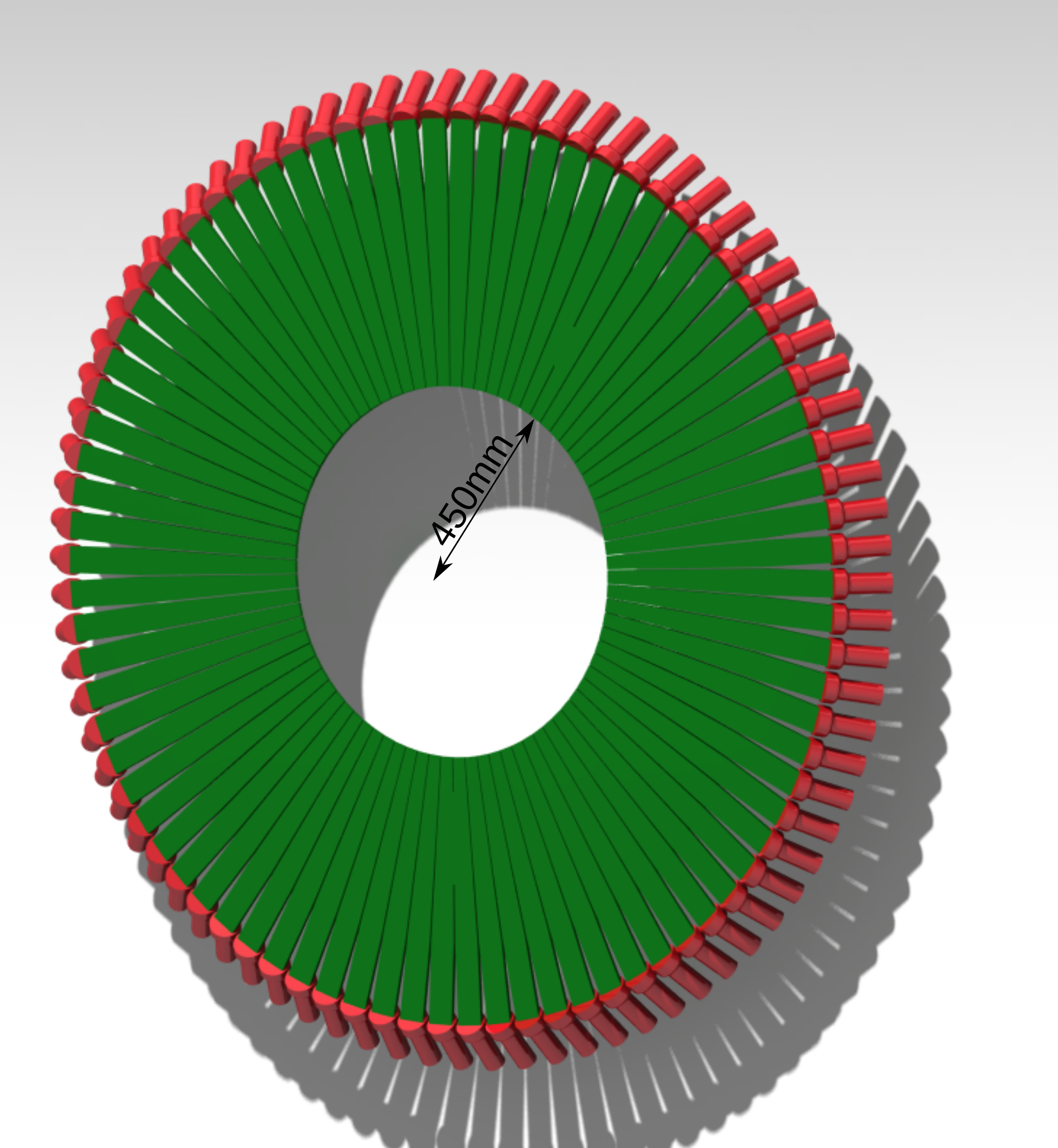}
	\caption{Detector ring consisting of 82 fused silica bars.}
	\label{fig:detring}
\end{figure}

The P2 detector will detect high energy electrons via the Cherenkov effect. 
As shown in Fig.~\ref{fig:detring} it is going to consist of 82 wedged fused 
silica bars (also referred to as quartz bars) that will cover the whole azimuth 
angle aside from very small gaps between the single detector elements occupied 
by wrapping and mounting material. Each quartz bar is wrapped in Alanod 4300UP, 
a highly reflective aluminum foil, which is not anodized to avoid wave 
interference caused by a coating. A photomultiplier tube (PMT) is attached to 
each bar.

\begin{figure}
	\centering
	\includegraphics[width=0.4\textwidth]{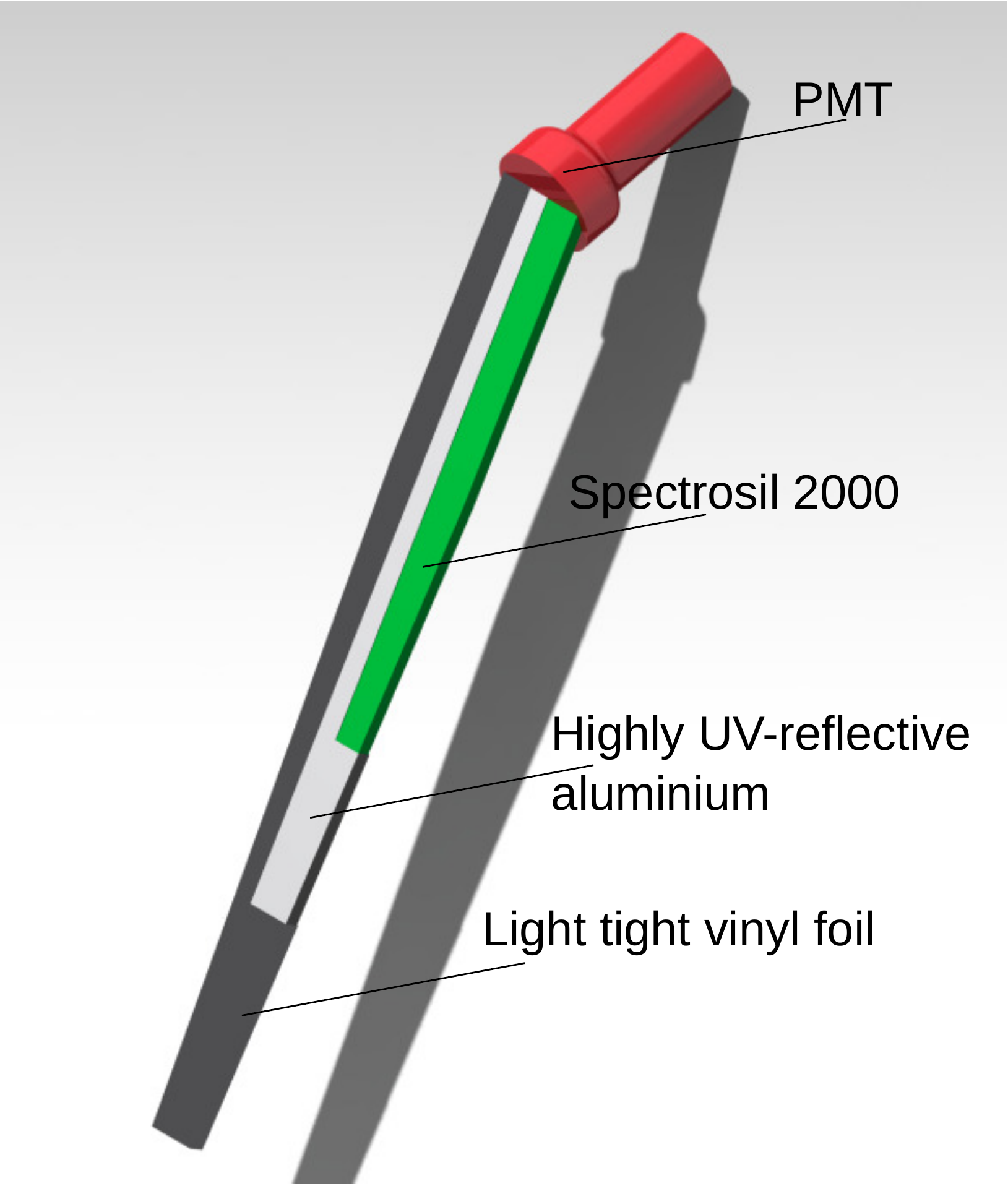}
	\caption{Detector element consisting of quartz bar with optical outlet, PMT, UV reflective wrapping, and light-tight vinyl.}
	\label{fig:detectorelement}
	\end{figure}

\begin{figure}
	\centering
	\includegraphics[width=0.3\textwidth]{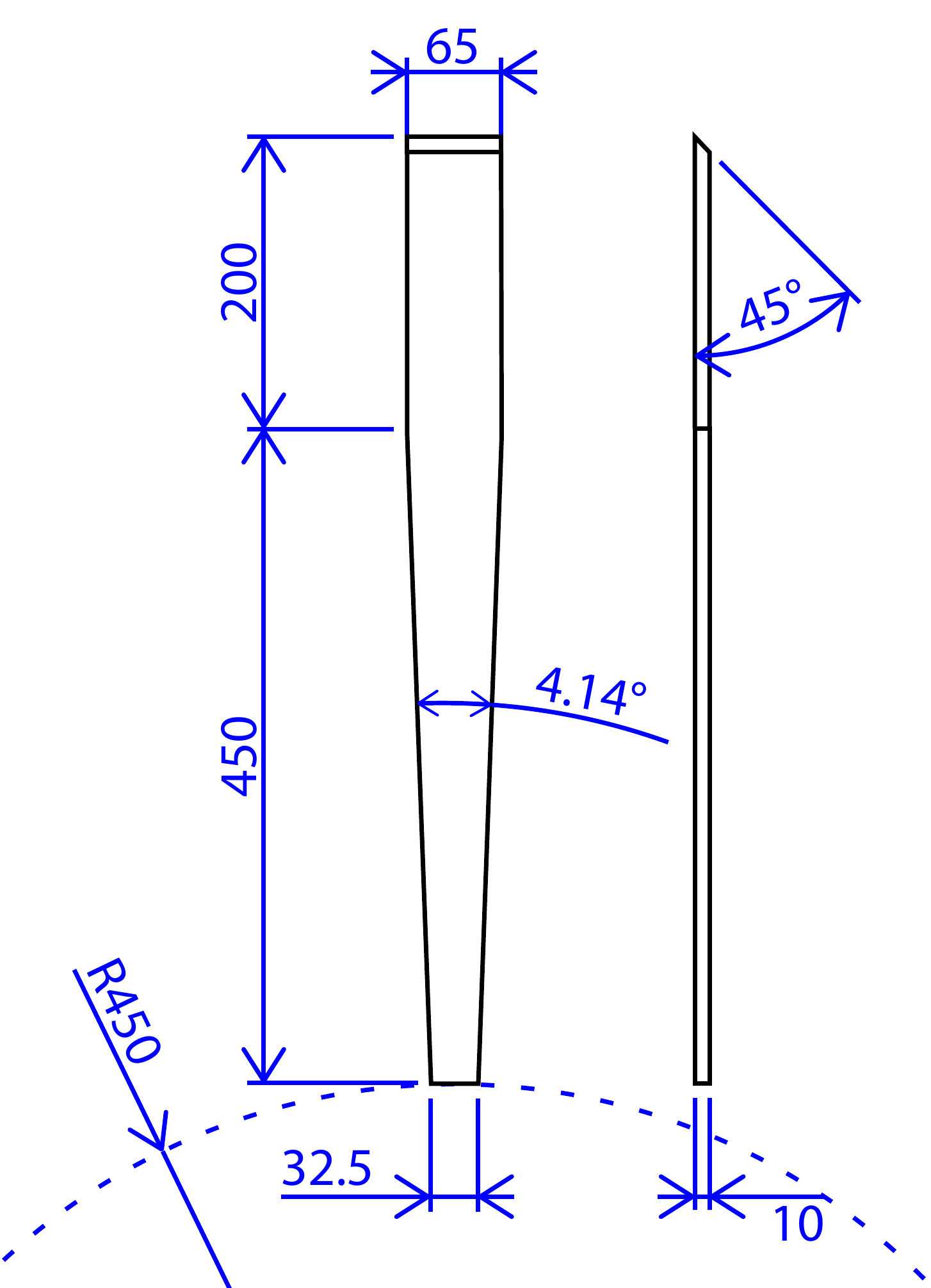}
	\caption{Technical drawing of fused silica bars.}
	\label{fig:tecquartz}
\end{figure}

The active area of the bars will cover the radial range of \SI{450}{mm} to 
\SI{900}{mm} from the beam line. The quartz bars will have an additional 
\SI{200}{mm} inactive part which will be shielded by \SI{100}{mm} of lead and 
will serve as a light-guide for the Cherenkov light to the photomultipliers.

\subsubsection{Requirements}
\label{requirements}

The hit rates during the data taking runs onto the P2 detector are going to 
be in excess of \SI{E12}{Hz}. They will be recorded in charge-integrating mode 
(see Sect.~\ref{sec:HighResolutionADCs}).

Operation in integrating mode precludes event mode cuts based on pulse shape, 
and true counting statistics is not achievable. Therefore the two primary 
concerns for the detector design are to reduce background sensitivity as much as 
possible and to increase the signal-to-excess noise ratio to a level that allows 
the detectors to operate as close to counting statistics as possible.

The asymmetry in the $i_{th}$ helicity pair is
\begin{equation}
A_i = \frac{N_i ^+ - N_i ^-}{N_i ^+ + N_i ^-}~.
\end{equation}
Using simple error propagation and the approximation $N_i ^+ = N_i ^- =N_i$ 
this gives:
\begin{equation}
\sigma_{A_i}  = \frac{1}{2N_i ^2} \sqrt{2N_i ^2 \sigma_N ^2}~.
\end{equation}
If one assumed Poisson statistics ($\sigma_{N_{stat}}=\sqrt{N}$), the asymmetry 
error for a given detector in the $i_{th}$ helicity pair would be
\begin{equation}
\sigma_{A_i} = \frac{1}{\sqrt{2N_i}}~.
\end{equation}
However, the production of showers inside the quartz, suboptimal geometry, and 
poor light collection efficiency typically lead to excess noise (because they 
produce additional variation in photo-electron count), expressed by an additional 
term:
\begin{equation}
\sigma_{A_i} = \sqrt{1+\alpha^2}/\sqrt{2N_i}
\end{equation}
where $\alpha \equiv \sigma_{PE}/n_{PE}$, $N_i$ is the number of primary 
electrons detected within a single helicity window, and $n_{PE}$ and 
$\sigma_{PE}$ are the mean photo-electron yield and standard deviation in the 
yield respectively. A higher average number of photo-electrons per event leads to 
a decrease in excess noise.

We demand that excess noise be less than 1\%:
\begin{equation}
\sqrt{1+\alpha^2} < 1.01 \, .
\end{equation}
Approximating the standard deviation to be $\sigma_{PE} = \sqrt{n_{PE}}$, we 
find that the number of photo-electrons per electron incidence onto the quartz 
bar, $n_{PE}$, needs to be larger than 50:
\begin{equation}
\label{eq:npe50}
n_{PE} > 50 \, .
\end{equation}
These issues are intimately connected to the detector material and geometry, 
which has been optimized using simulations verified by experiment. The final 
choice for the geometry and orientation between the quartz pieces, the beam, 
and the PMTs is being studied using simulations and prototype tests at the MAMI 
electron beam, both of which are described in more detail in 
Sect.~\ref{sssec:simandtests}.

The amount and spectral distribution of Cherenkov photons, produced per 
\SI{1}{cm} of trajectory in the radiator, are well described by the formula 
\begin{equation}
\frac{dN}{d\lambda} = \frac{2\pi z^2\alpha}{\lambda ^2} \left( 1- \frac{1}{\beta ^2 n(\lambda)^2} \right)
\end{equation}
where $z$ is the particle charge expressed in units of $e$, $\alpha$ is the 
fine structure constant, $\beta$ the velocity as a fraction of the speed of light, 
and $n(\lambda)$ the wavelength-dependent refractive index of the material. 
Thus, a high energy electron generates a number of approximately 900 Cherenkov 
photons when travelling through a \SI{1}{cm} piece of quartz.
\begin{figure}
	\centering
	\includegraphics[width=0.46\textwidth]{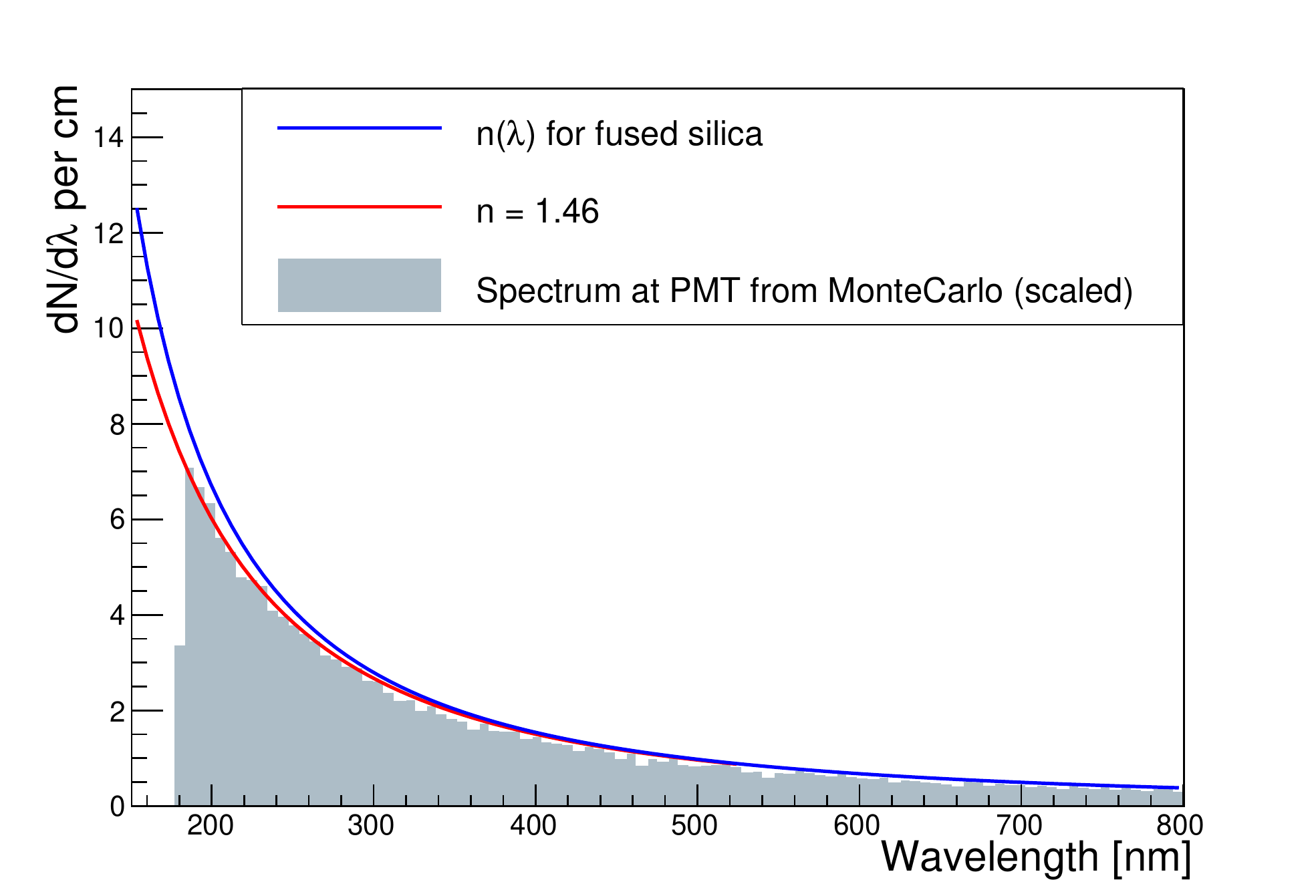}
	\caption{Number of photons produced by the Cherenkov effect per \SI{1}{cm} track length and nm wavelength in fused silica and photon spectrum detected at the location of the PMT in the Geant4 simulation (scaled for comparison purposes).}
	\label{fig:cherenkovspectrum}
\end{figure}
It is evident from the plot of this spectrum in Fig.~\ref{fig:cherenkovspectrum} 
that a large portion of this light is in the deep UV region. Special care is 
taken to collect these photons in the PMT, where they will be transformed into 
an electric signal. The PMT will be located at the end of the quartz bar facing 
away from the beam axis. In order to minimize the light losses on the way to the 
PMT and maximize the number of electrons from the PMT cathode, all the material 
used for the detector elements has to be suitable for UV applications.

\begin{figure}
	\centering
	\includegraphics[width=0.46\textwidth]{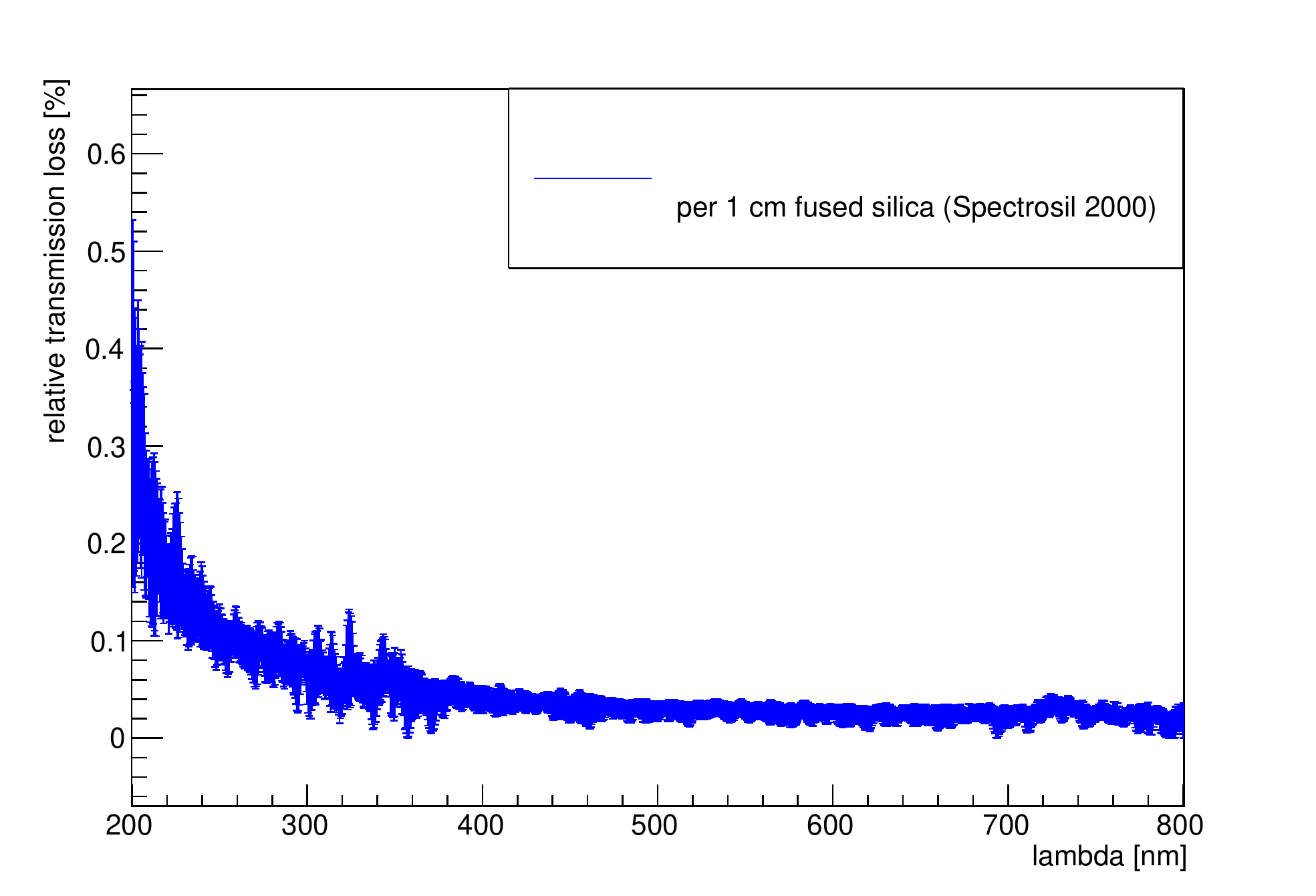}
	\caption{Spectral transmission losses per \SI{1}{cm} of fused silica.}
	\label{fig:quartztransm}
\end{figure}

The fused silica bars are made of Heraeus Spectrosil 2000, a very pure form of 
amorphous silicon dioxide, which is extremely well transmissive to UV light. 
As shown in Fig.~\ref{fig:quartztransm} the light intensity losses per 
\SI{1}{cm} are very low even at \SI{200}{nm} wavelength.

\begin{figure}
	\centering
	\includegraphics[width=0.46\textwidth]{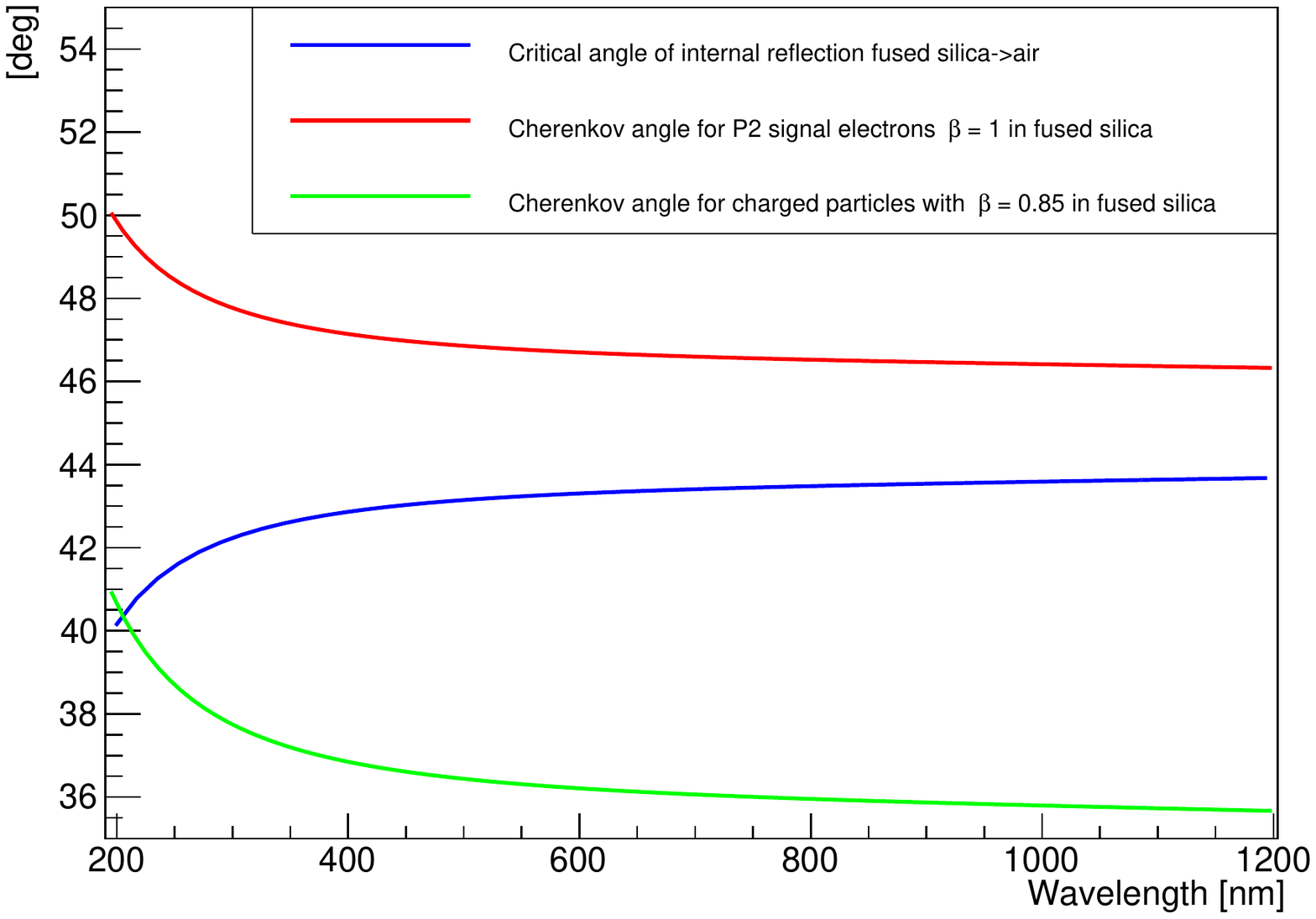}
	\caption{Cherenkov angle and critical angle of total internal reflection as a function of the wavelength for fused silica.}
	\label{fig:AngleCC}
\end{figure}

Figure~\ref{fig:AngleCC} shows that for fused silica both the Cherenkov angle 
for ultra-relativistic charged particles with $\beta \approx 1$ and the critical 
angle of total internal reflection at the boundary to air are approximately 
\ang{45}, the Cherenkov angle being slightly larger over the whole spectrum.

\begin{figure}
	\centering
	\includegraphics[width=0.46\textwidth]{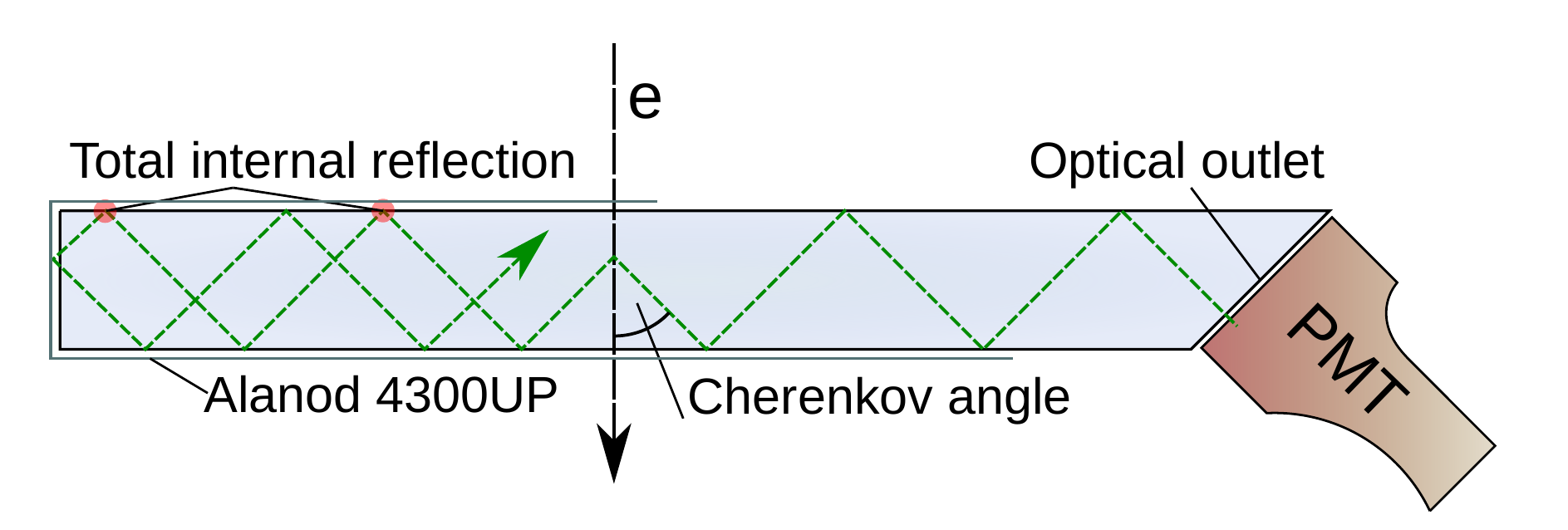}
	\caption{Quartz as Cherenkov medium serves as an effective light-guide at the same time.}
	\label{fig:quartzinternalref}
\end{figure}

\begin{figure}
	\centering
	\includegraphics[width=0.46\textwidth]{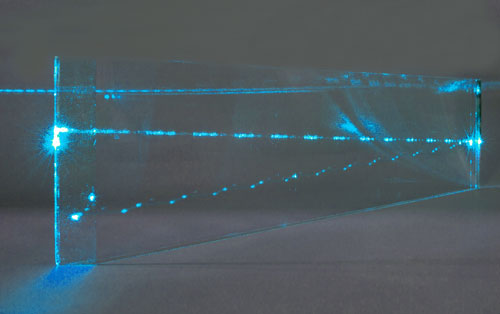}
	\caption{Internal reflection of light in fused silica demonstrated with a laser pointer.}
	\label{fig:quartzbl}
\end{figure}

We chose the geometry of the quartz bars in such a way as to enhance the light 
collection by the effect of total internal reflection. It is ensured that 
Cherenkov light emerging from a perpendicular electron incidence is contained 
within the material and guided to the outer end, where an optical outlet allows 
it to exit and reach the photomultiplier tube cathode. 
Figure~\ref{fig:quartzinternalref} schematically shows this concept. 
For potential background particles with $\beta < 0.85$ the Cherenkov angle is 
smaller, resulting in more losses of light from these events and thus improving 
the signal-to-background ratio.
In addition, the bar is wrapped in a highly reflective aluminum foil with a 
thin layer of air (\SI{0.2}{mm}) in between. The foil helps to contain the light 
within the detector element and thus enhances the signal yield.
The photograph in Fig.~\ref{fig:quartzbl} visualizes the effect of total internal 
reflection with a blue laser pointer.

\subsubsection{Photomultipliers}

\begin{figure}
	\centering
	\includegraphics[width=0.48\textwidth]{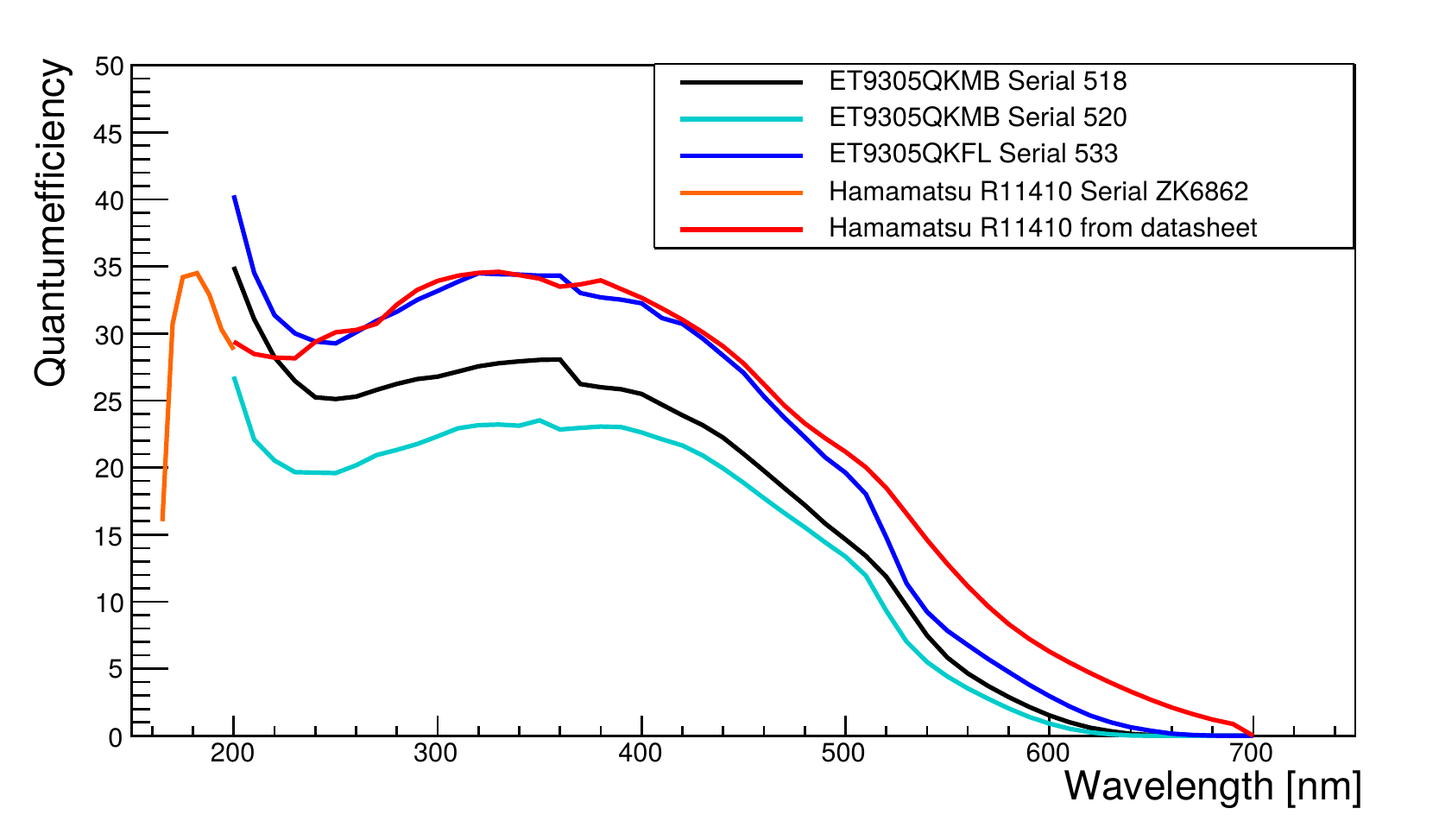}
	\caption{Spectral responses of three different models of PMTs which are suitable for Cherenkov light detection.}
	\label{fig:pmtsqe}
\end{figure}

The photomultipliers for the integrating detectors have to satisfy five main 
criteria: (1) high efficiency in the UV, (2) support for high cathode currents, 
(3) uniform sheet resistivity across the photocathode, (4) good linearity at 
relatively low bias voltages for current mode, and (5) fast charge collection 
at high bias voltages for tracking mode operation. The last two criteria are as 
much dependent on the base design as they are on the PMT itself. Three models of 
photomultipliers have been investigated, all especially developed for UV usage:
\begin{itemize}
\item ElectronTubes 9305QKMB; 
\item ElectronTubes 9305QKFL; 
\item Hamamatsu R11410. 
\end{itemize}
All three are \SI{78}{mm} in diameter with an effective area of \SI{64}{mm}. 
They all have UV sensitive bi-alkali cathodes and quartz windows for good 
transmission of short wavelengths. Their spectral responses are plotted in 
Fig~\ref{fig:pmtsqe}. The two specimen of the ET9305QKMB as well as the 
ET9305QKFL were individually calibrated for wavelengths ranging from 
\SI{200}{nm} to \SI{800}{nm}. The Hamamatsu PMT was only calibrated from 
\SI{165}{nm} to \SI{200}{nm}. The data for longer wavelengths was taken from 
the manufacturer data sheet.

\begin{figure}
	\centering
	\includegraphics[width=0.48\textwidth]{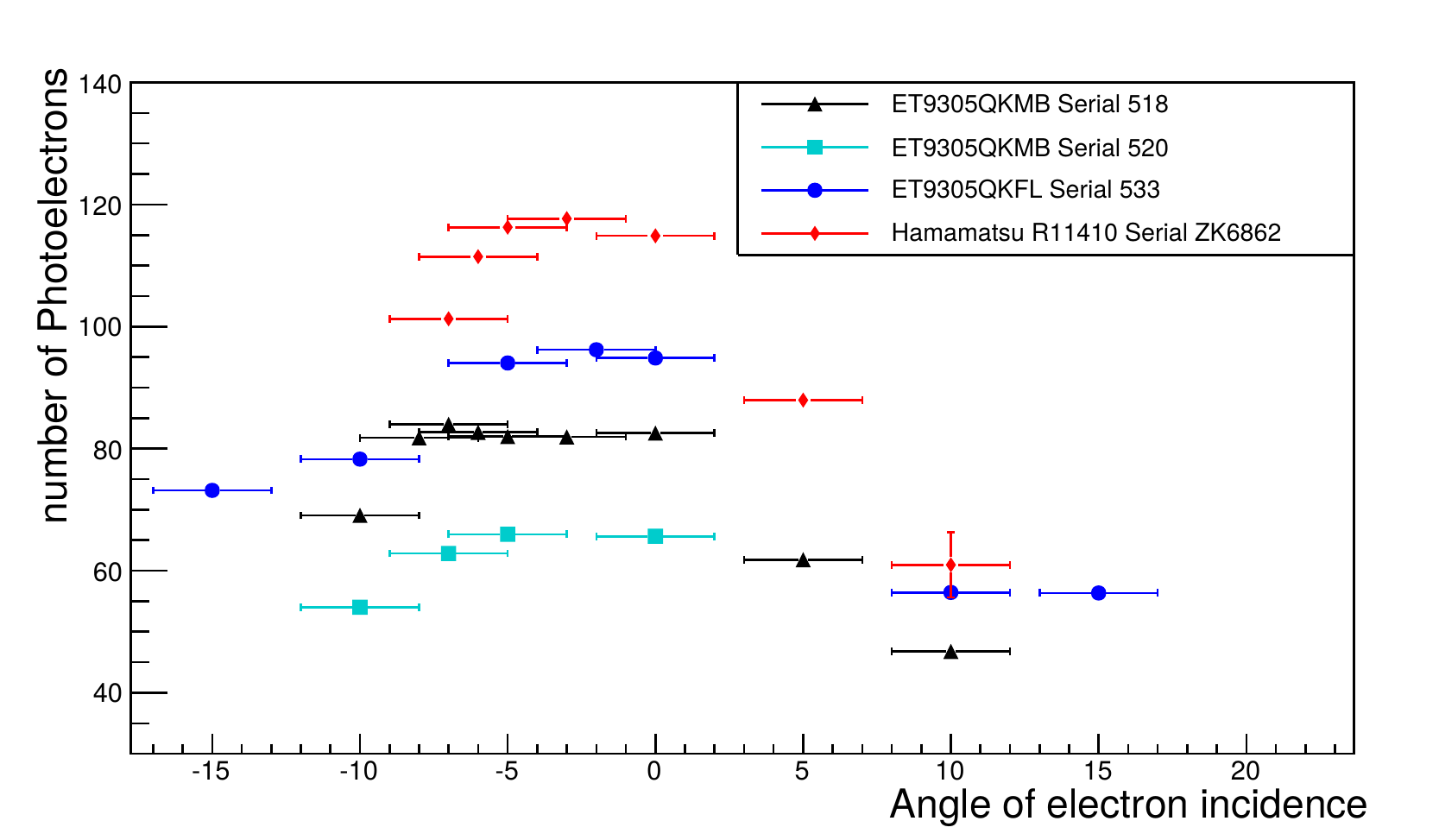}
	\caption{MAMI test beam data from detector element prototypes with the different PMTs.}
	\label{fig:pmtsmeasurement}
\end{figure}

The data in Fig.~\ref{fig:pmtsmeasurement} was taken during a test beam at MAMI, 
where we used a prototype detector element with exchangeable PMTs. Each assembly 
was then irradiated with beam electrons of \SI{195}{MeV} and the number of 
photo-electrons from the PMT cathode per electron incidence onto the detector 
was determined for different angles between beam and quartz surface, 
\ang{0} being perpendicular. The requirement of Eq.~(\ref{eq:npe50}) is met with 
each one of those PMTs.

\begin{figure*}
\centering
	\includegraphics[width=0.8\textwidth]{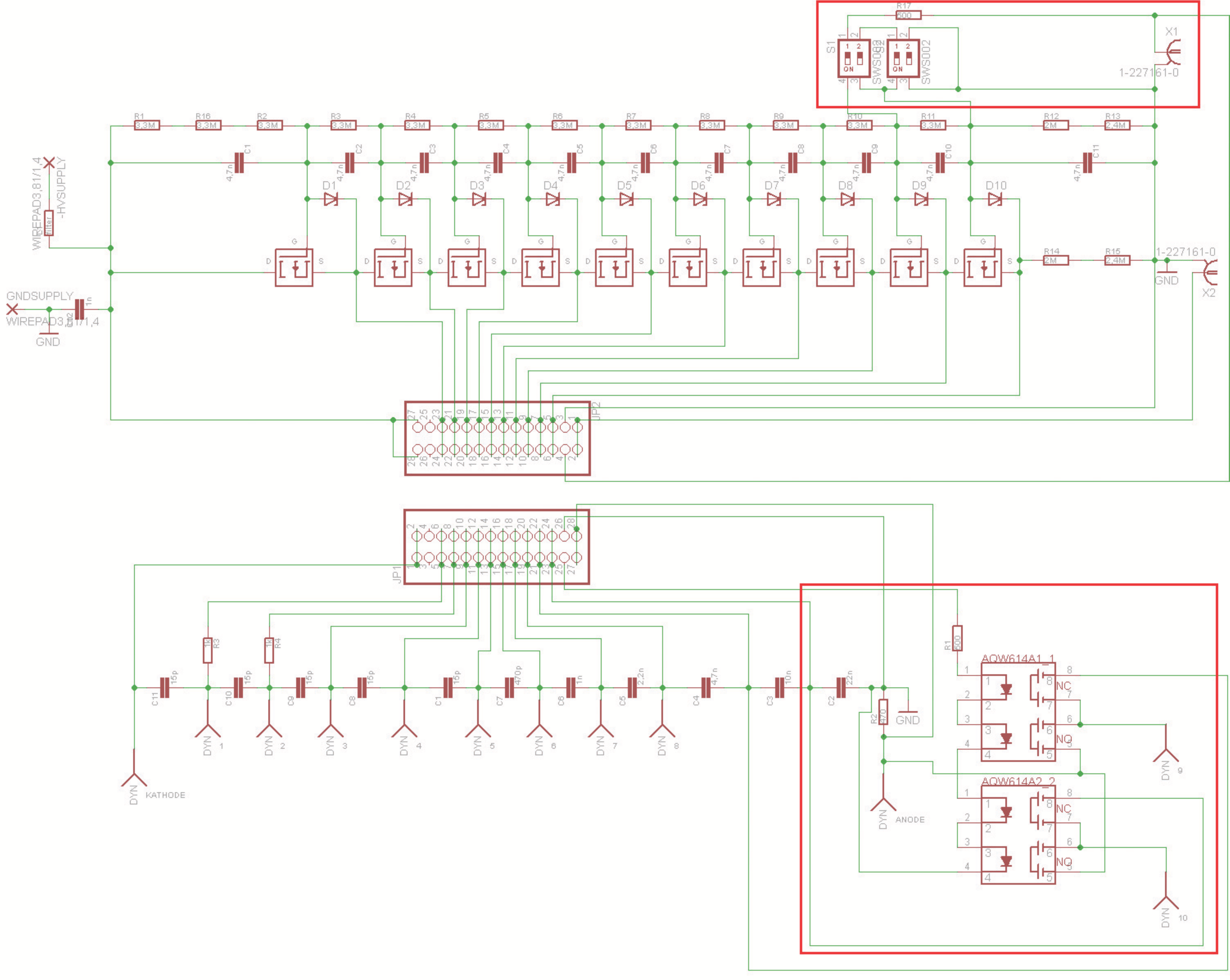}
	\caption{Schematic of the PMT base prototype for the main Cherenkov detectors. This is an active base design, using FETs and diodes, to stabilize the gain behavior at high event rates. The base implements the switching mechanism, using photomos relays.}
	\label{fig:pmtbaseschem}
\end{figure*}

Since the detectors are to be operated both in current mode (low gain), and 
tracking or event mode (high gain), we require a base that is remotely 
switchable between these two modes. A prototype base has been developed and 
tested with the prototype detectors at the MAMI facility. The schematic for the 
base is shown in Fig.~\ref{fig:pmtbaseschem}. This is an active base design, 
using FETs and diodes to stabilize the gain behavior of the PMT at high event 
rates. The base implements the switching mechanism using photomos relays, the 
viability of which still requires radiation hardness testing. Alternatively, we 
can implement the switching mechanism with standard radiation-hard relays.

\begin{figure}
	\centering
	\includegraphics[width=0.48\textwidth]{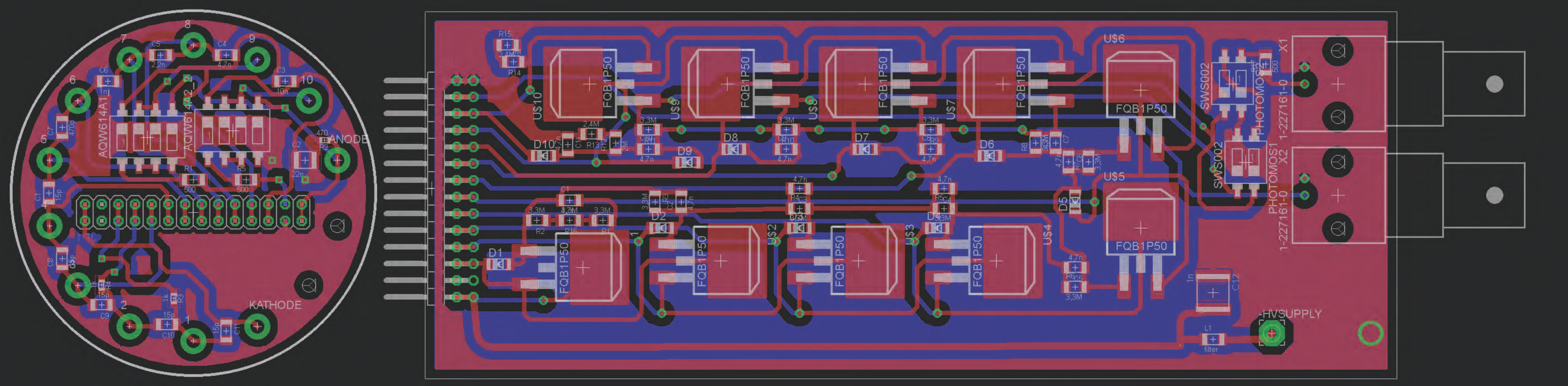}
	\caption{PCB layout of the PMT base prototype for the main Cherenkov detectors.}
	\label{fig:pmtbasepcb}
\end{figure}

\begin{figure*}
    \centering
	\includegraphics[width=0.8\textwidth]{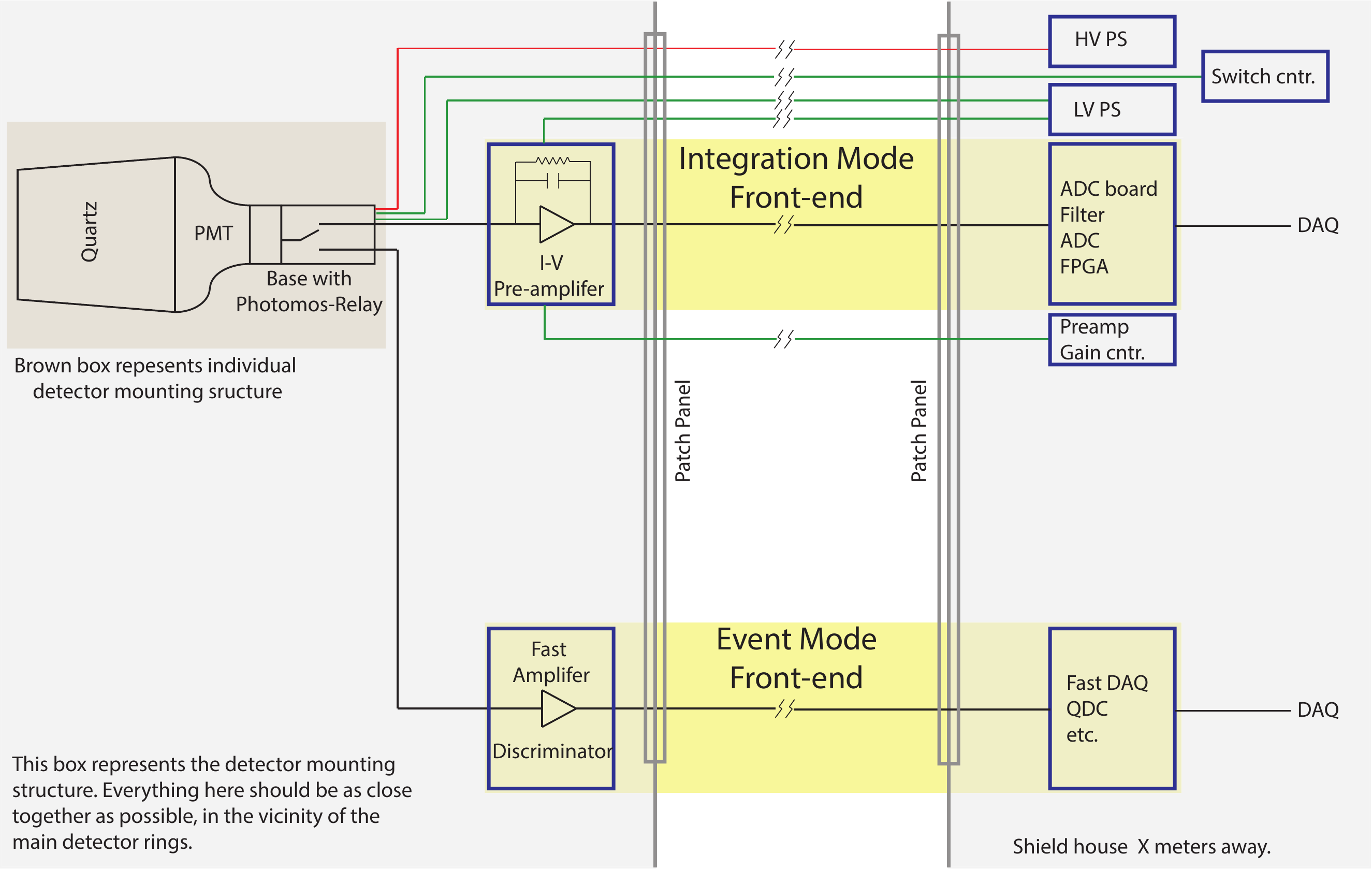}
	\caption{Schematic layout of the main detector system, including the separate 
	integration mode and event (tracking) mode signal chains. Also shown are the
    modules that control the PMT base switching and the gain adjustment for the 
		preamplifiers. The low voltage power supplies needed for the base and 
		preamplifier circuits are also shown, along with the high-voltage power 
		supplies. The details of the integration mode front-end electronics system 
		are discussed in Sect.~\ref{sec:HighResolutionADCs}.}
	\label{fig:IDSystemLayout}
\end{figure*}

It is very important that the PMT and base system design interfaces well with 
the constraints placed on the integrating and event mode front-end electronics. 
For the integration mode measurement, the interdependencies that need to be 
optimized with respect to each other are summarized in the list given below and 
discussed in detail in Sect.~\ref{sec:HighResolutionADCs}. 
The complete main detector system layout is shown in Fig~\ref{fig:IDSystemLayout}.

\begin{enumerate}
  \item The PMT base design and the corresponding gain are set by the amount of 
	charge that can be drawn from the PMT cathode and dynodes. The event rate, 
	experiment running time and the number of photo-electrons per event determine 
	the former, while the base design determines the latter.
  \item The linearity of the PMT-base system deteriorates at lower gain 
	(lower bias voltage), but higher gain means a larger anode current and higher 
	charge drawn on the dynodes.
  \item The integrating measurement noise bandwidth requirements 
	(see Sect.~\ref{sec:HighResolutionADCs}) limit the available gain range for 
	the preamplifier, which influences the minimum (and maximum) anode current 
	that can be accommodated by the rest of the front-end electronics.
\end{enumerate}

\subsubsection{Prototype tests and Geant4 simulation of the detector response}
\label{sssec:simandtests}

Extensive material studies and prototype tests for the P2 Cherenkov detector 
have been performed between 2013 and 2016, examining different Cherenkov medium 
materials, photosensors, wrapping material, reflectors and detector geometries. 
Following a period of material and geometry considerations and studies, several 
detector elements have been tested at the MAMI beam with low currents of 
\SI{195}{MeV} electrons in order to measure the signal yield of single events. 

\begin{figure}
	\centering
	\includegraphics[width=0.49\textwidth]{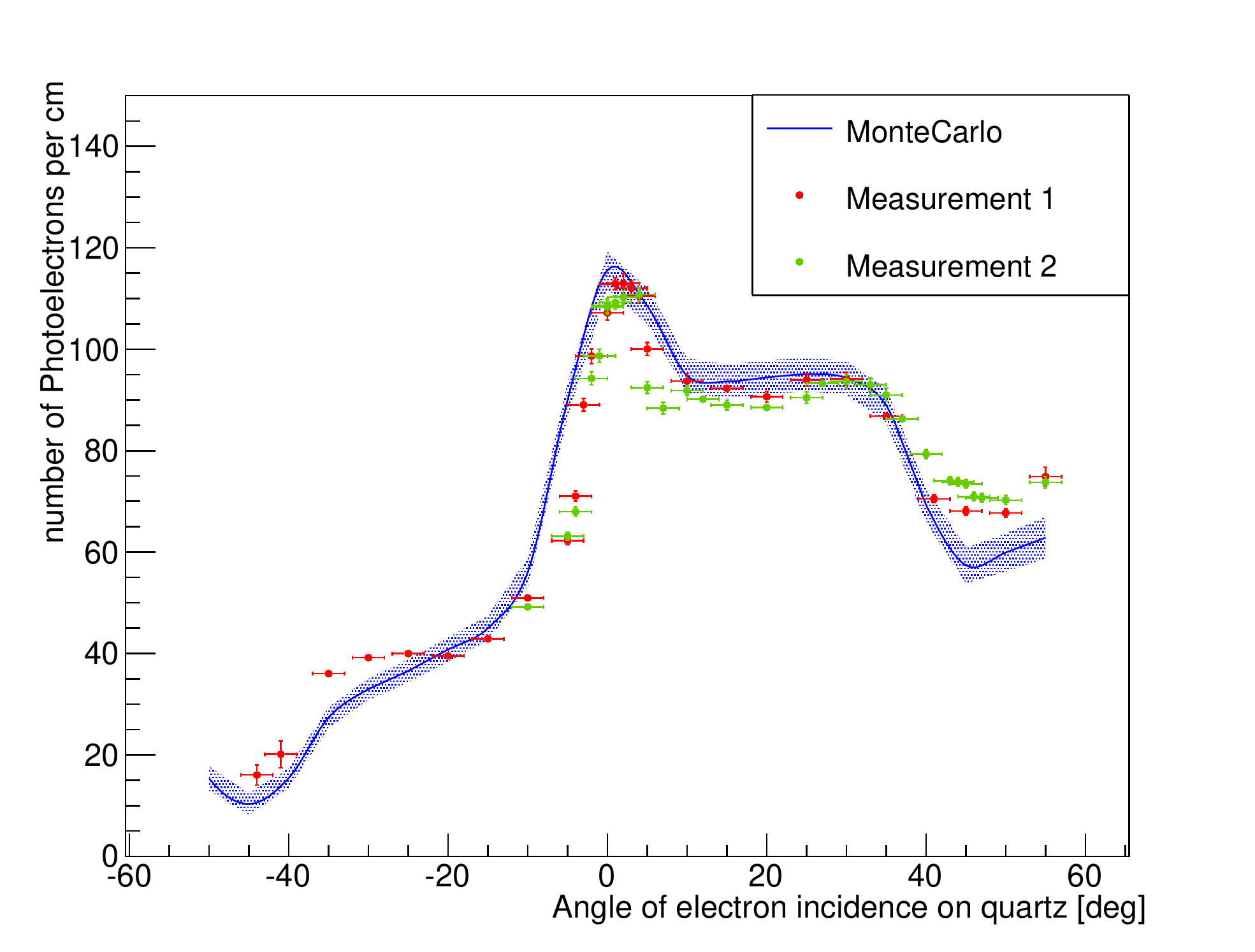}
	\caption{Measurement results and Monte Carlo simulation of the photo-electron 
	yield per electron event onto the detector element for angles of electron 
	incidence ranging from \ang{-50} to \ang{55}, the maximum signal being at 
	perpendicular electron incidence onto the quartz surface.}
	\label{fig:churchplot}
\end{figure}

As an example, Fig.~\ref{fig:churchplot} shows the result of a run series with 
the detector design described in Sect.~\ref{sssec:generaldetectorconcept} and 
listed in Tab.~\ref{tab:runconditions}. In order to measure the signal 
dependence on the angle of the particle incidence onto the detector surface, 
the prototype was rotated with respect to the MAMI beam line. Due to the effect 
of total internal reflection the number of photocathode electrons is highest for 
perpendicular impact.

\begin{table}
    \begin{tabular}{ll}
\toprule[1.5pt]
Electron energy & \SI{195}{MeV}\\ \hline
Beam rate & \SI{3}{KHz}\\ \hline
Cherenkov medium & Spectrosil 2000\\ \hline
Polish & Optical polish\\ \hline
Geometry & \SI{650}{mm} long wedged (Fig.~\ref{fig:tecquartz})\\ \hline
Photomultiplier & Hamamatsu R11410 ZK6862 Assy\\ \hline
Reflective wrapping & Alanod 4300UP\\ \hline
Light tight wrapping & \SI{0.3}{mm} Vinyl\\ 
\bottomrule[1.5pt]
\end{tabular}
\caption{Run conditions and prototype setup for measurements and simulation shown in 
Fig.~\ref{fig:churchplot}.}
\label{tab:runconditions}
\end{table}

These measurements were used to benchmark a Geant4 simulation with optical 
photon processes. Within this simulation, several particle types with 
properties expected for P2 signal electrons and background were shot onto 
different positions of the detector bars at different impact angles.
As primary photon production processes, the Cherenkov process, and
scintillation processes (not relevant in quartz) are implemented. 
Additional particles resulting into signal contributions can be 
created by pair production, Compton scattering, and bremsstrahlung.

\begin{figure}
	\centering
	\includegraphics[width=0.46\textwidth]{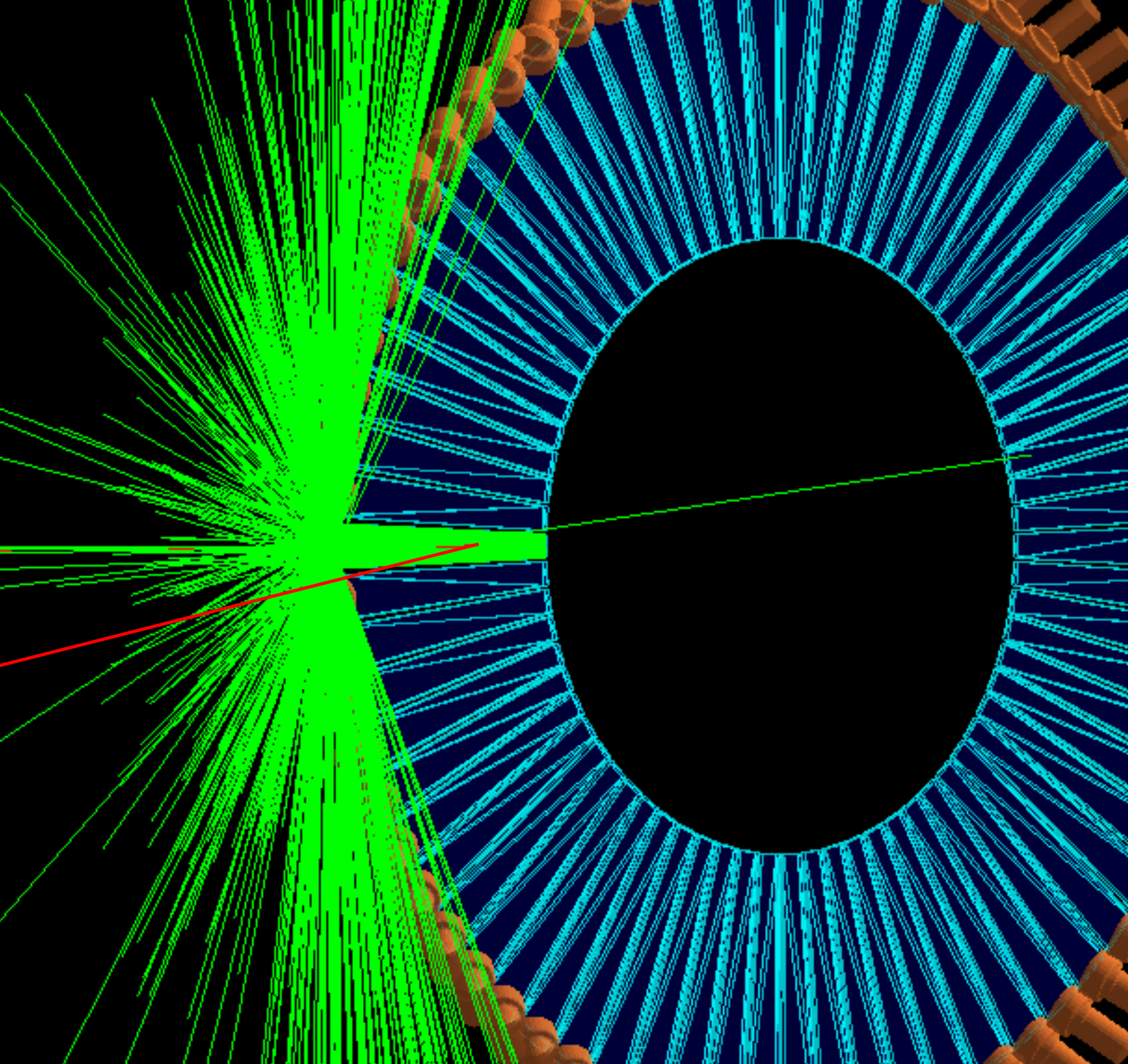}
	\caption{Visualization of electron incidence on detector element. In red: electron, green: photons}
	\label{fig:simvis}
\end{figure}

The detector geometry and the materials were defined along with their optical 
properties as functions of the photon energies: refractive indices 
($n(\lambda)$), light transmittance ($T(\lambda)$), and surface properties. 
The photons are then subject to wavelength dependent processes, namely
refraction, reflection, absorption, and Rayleigh scattering.
A typical simulated electron impact event onto one of the detector elements is 
visualized in Fig.~\ref{fig:simvis}.

At the position of the PMT cathode, the simulation contains a sensitive volume, 
which detects all particles passing through including their type, total energy,
momentum direction, creation process, and place of origin.
Using this information along with the cathode's quantum efficiency, the 
simulation can determine the number of photo-electrons per event.
Figure~\ref{fig:churchplot} contains test beam results as well as simulated data 
generated using the same detector geometry, material properties, and primary 
particle characteristics. Thus, measured and simulated data can be directly 
compared. Both share the same characteristic angle dependence and agree well in 
the magnitude of the photo-electron signal.

\begin{figure}
	\centering
	\includegraphics[width=0.46\textwidth]{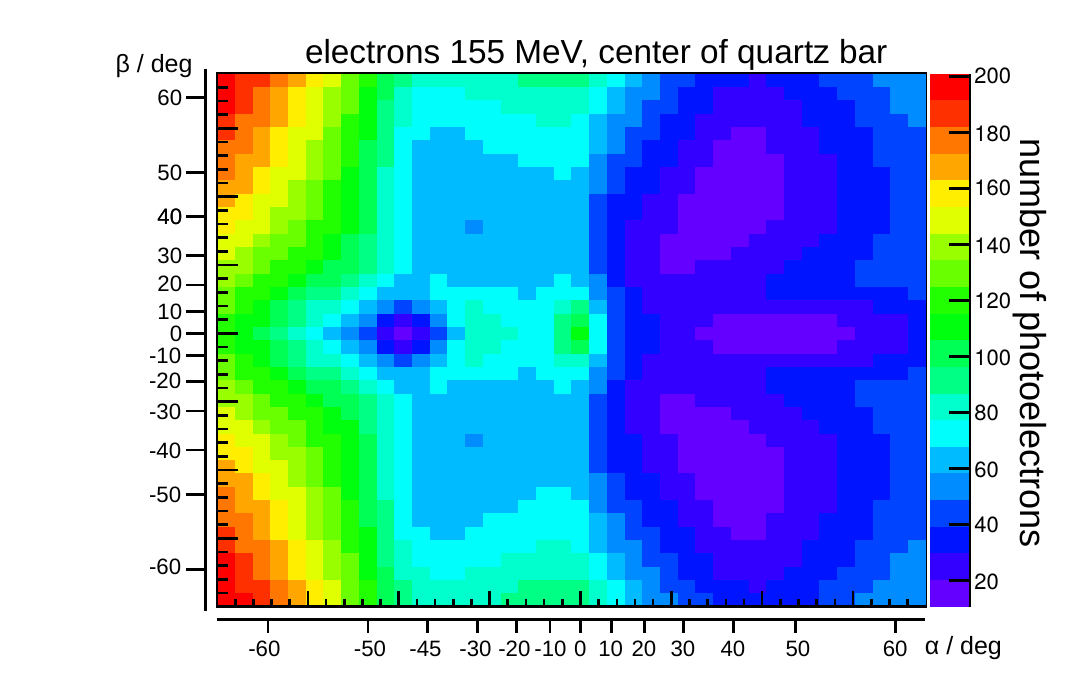}
	\caption{A small fraction of the detector response database as an example. 
	The number of photoelectrons is plotted for electrons of 
	\SI{155}{MeV} hitting a detector module at the center of the active at a 
	range of different angles to the detector surface.}.
	\label{fig:detresp}
\end{figure}

The Monte Carlo simulation described in Sect.~\ref{sec:Simulation} delivers a 
rate distribution of particles hitting the Cherenkov detector plane. 
Figure~\ref{fig:detector_hit_distribution} shows the particle rates incident on 
the detector ring for different particle types as a function of the radial 
distance from the $z$-axis. Particles are divided into two groups: the ones 
originating from an elastic ep scattering process and background particles. 
We see a very prominent photon background over-powering the signal electrons by 
one to two orders of magnitude.

The detector response simulation has been used to compile a comprehensive 
database of signal yields of particles hitting the detector. It includes the 
mean number of photo-electrons expected for electrons, positrons, and photons 
with energies ranging from \SI{0}{MeV} to \SI{155}{MeV}, various incidence 
positions, and incidence angles. Some example data is shown in 
Fig.~\ref{fig:detresp}.

Along with the particle information from the ray tracer simulation 
(Sect.~\ref{sec:Simulation}) the event rates on the detector can now be
converted into photo-electron rates which are plotted in 
Fig.~\ref{fig:pe_rate_distro} of Sect.~\ref{sec:Simulation}. The background 
signal is suppressed by the small detector response to photons since they do not 
directly produce Cherenkov light. They only lead to a detector signal if they 
first convert into charged particles (e.g., by pair production or Compton 
scattering) inside the detector, and if the resulting particles are above the 
Cherenkov threshold (\SI{0.71}{MeV} for electrons traveling through the quartz 
bars). The photon background in P2 is mainly below this threshold.

\subsubsection{Radiation hardness}

The energy deposition in \SI{1}{cm} of quartz 
at \SI{150}{MeV} is \SI{15.5}{MeV}. The detector bars are going to cover an angle 
range of \ang{25} to \ang{45}, and we expect an electron rate of 
$\sim$\SI{5}{GHz} onto each detector element, corresponding to an average 
of $\sim$\SI{20}{MHz/cm^2} and a deposited energy of 
\SI[inter-unit-product =\ensuremath{\cdot}, per-mode=symbol]{5e-5}{\joule\per\second \per\centi\meter\cubed}. 
This leads to an absorbed radiation dose of $\sim$\SI{80}{Mrad} over the data 
taking time of \SI{10000}{h}. The material used for the detector components 
must be able to endure this radiation. It has been shown that for quartz, low 
contamination levels lead to low susceptibility to radiation damage. 
The PANDA DIRC group has done extensive studies with Spectrosil 2000 by 
irradiating the material with photons up to a dose of \SI{100}{krad} 
and --- in another study --- with protons up to 
\SI{10}{Mrad} \cite{Hoek:2011zz,Hoek:2008zz}. 
The research has demonstrated the extraordinary radiation hardness of synthetic 
fused silica.

%% file: highresolutionadc.tex
The small asymmetries and the precision goal for P2 require very high
statistics, meaning very high event rates in the detectors. Given the needed
accuracy of the proposed experiment, background susceptibility, linearity, noise
behavior, and radiation hardness are major issues. For these reasons, P2 has
adopted radiation hard, highly linear, and relatively large active area
($O(\SI{100}{cm^2})$) fused silica (quartz) Cherenkov detectors.
The anticipated total rate in a given detector is on the order of one to
several GHz. At these detector rates the counting of individual pulses is no
longer possible, requiring integration mode operation, in which individual
pulses overlap to such a degree, that they produce a continuous current at the
detector output (therefore this is also sometimes referred to as current mode
detection, as opposed to pulse mode). The electronics discussed in this section
refers to the integrating mode electronics for these detectors, as well as the
beam monitors, which are used to normalize the detector signal and should
therefore, ideally, have the same digitization scheme, as further discussed
below. Due to the similarities in the detection and measurement methodologies
with the QWeak experiment~\cite{Allison:2014tpu,Androic:2013rhu}, the P2
design introduced here is based on the design of and experience with the
front-end electronics for the QWeak integrating detectors.

\begin{figure*}
	\centering
		\includegraphics[width=0.98\textwidth]{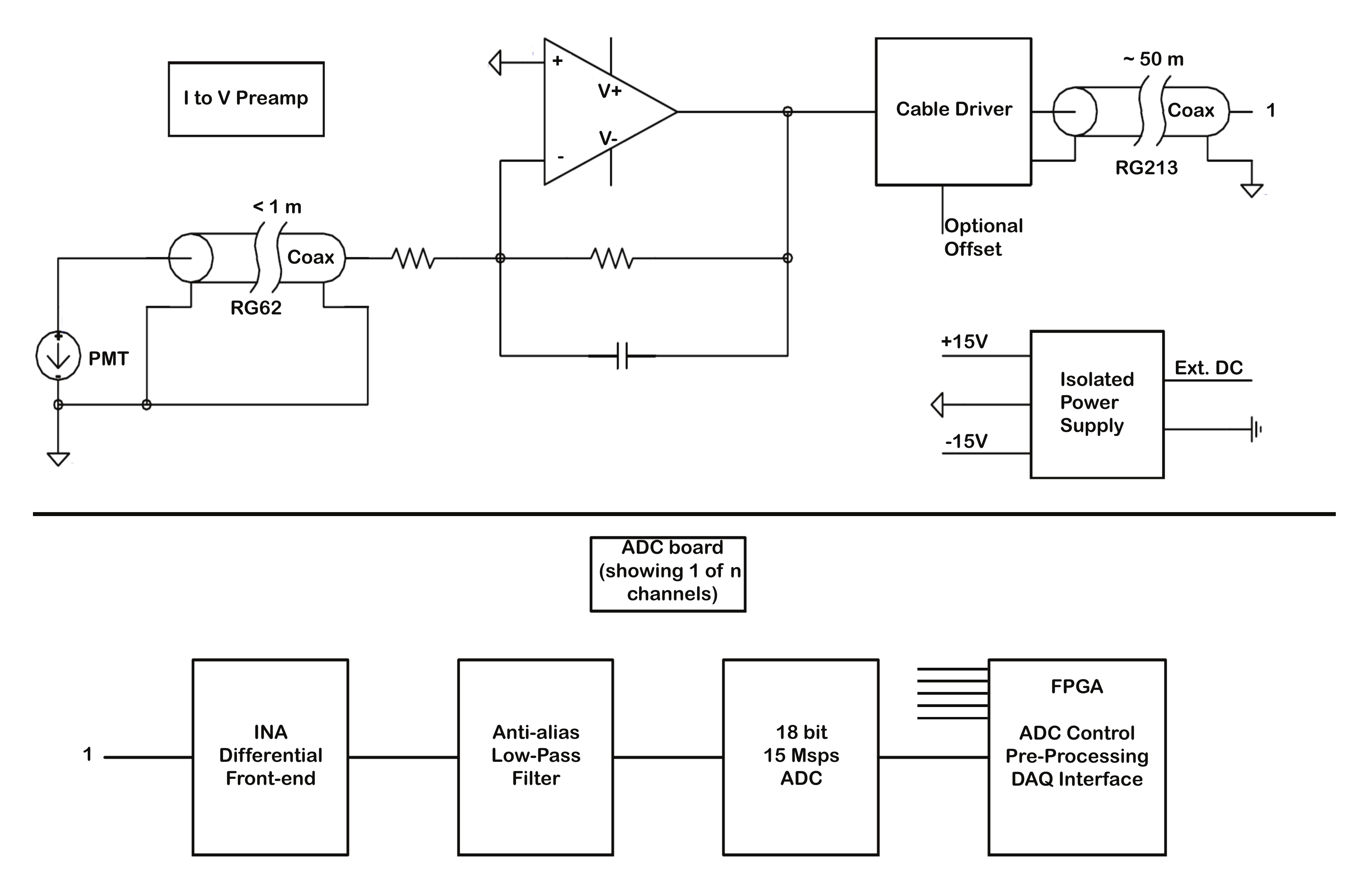}
	\caption{Schematic of the front-end signal path for integration mode measurements.
	The continuous current signal from the PMT will be converted to a voltage
  signal using a trans-impedance amplifier. A cable driver will maintain the
	signal strength from the amplifier to the ADC board, which is located away
	from the detector behind a shielding wall. The filtered amplifier signal will
	be digitized by a high resolution ADC and the digitized data will be
	pre-processed and transmitted to the DAQ system using an FPGA.}
	\label{fig:BFeSchem}
\end{figure*}

\subsubsection{Integrating detector signal chain}

The proposed P2 integration or current mode measurement front-end signal chain
is shown in Fig.~\ref{fig:BFeSchem}. The light from the quartz is converted to
a current using quartz glass photomultiplier tubes (PMT) with a high quantum
efficiency (QE) in the UV. At a nominal rate of 1 GHz and for a given detector
geometry with a photo-electron yield of \SI{50}{pe} per primary electron, a mean
current of roughly \SI{8}{\micro A} is produced at the anode of a PMT with a
gain of about 1000. A trans-impedance preamplifier is used to convert the
current signal to a voltage and to provide the primary filtering stage for the
signal. The voltage signal is then further filtered and digitized by an ADC.
Several channels of the filter and ADC chain are implemented on a single board,
together with an FPGA, which queries the ADCs, collects (meaning in this case
integrates) the data and facilitates the readout. The preamplifier also
implements a line driver to sustain the signal over the longer cable distance
between the detectors and the ADC. The preamplifiers and the integrating ADC
boards are the main subject of this section.


The design of the electronics is, of course, dictated by the properties of the
signal, including the mean amplitude, the RMS width, noise sources, and expected
variation as a function of helicity. For the purposes of discussing the
integrating electronics, the electronic signal from a single detector $(i)$ can
be written as
\begin{equation}\label{eqn:Current1}
  I^{\pm}_i(t) = I^{\pm}_{i_{_P}}(t) + \displaystyle\sum_{B}I^{\pm}_{i_{_B}}(t) + I_{i_{_D}}(t) \, .
\end{equation}
This is the current that is present at the anode of the PMT, where
$I^{\pm}_{i_{_P}}(t)$ is the contribution to the current from the physics
process of interest (elastically scattered electrons and the secondaries they
produce, indicated by the subscript $P$), the sum over $I^{\pm}_{i_{_B}}(t)$
is the contribution from the various background processes (indicated by the
subscript $B$), and $I_{i_{_D}}(t)$ is the dark current (indicated by the
subscript $D$). The current is converted to a voltage by an I-to-V amplifier
with a gain of $g^{\pm}_{_{Amp_{i}}}(t)$, in units of Ohms, so that the
corresponding voltage yield is
\begin{equation}\label{eqn:Yield1}
  V^{\pm}_i(t) = g^{\pm}_{_{Amp_{i}}}(t) I^{\pm}_i(t) + V^{\pm}_{_{i_E}}(t) \, .
\end{equation}
Equation~(\ref{eqn:Yield1}) includes a voltage contribution
$(V^{\pm}_{_{i_E}}(t))$ from a possible amplifier offset and the voltage signal
corresponding to the dark current can contain helicity dependence due to
possible electronic coupling of the helicity signal in the preamplifer. The
latter should be suppressed or eliminated if possible through the use of
appropriate signal isolation (avoidance of ground loops, signal isolation, the
use of fiber optics, etc.).

For measurements in integration-mode, the counting statistics manifests itself
primarily in the root-mean-square (RMS) width of the shot noise at the PMT
anode, which has contributions from all sources that generate an anode current,
including background and dark current. To get the total RMS width in the signal,
as it is sampled by the ADC, one has to add all sources of electronic noise
$\sigma^2_{i_{_E}}$, which includes resistive (Johnson) noise, as well as noise
introduced by active components and filters. The total squared RMS noise density
in the signal is then given by
\begin{equation}
\label{eqn:RMS1}
\sigma^2_i = 
2 Q_{i_{_P}} I_{i_{_P}} 
+ \displaystyle\sum_{B}2 Q_{i_{_B}} I_{i_{_B}} 
+ 2 Q_{i_{_D}} I_{i_{_D}} 
+ \sigma^2_{i_{_E}} 
\end{equation}
where  $I_{i_{_{P,B,D}}} = \langle I^{\pm}_{i_{_{P,B,D}}}(t)\rangle$ is the mean
anode current and $Q_{i_{_{P,B,D}}} = \langle Q_{i_{_{P,B,D}}}(t) \rangle $ is
the mean quantum of charge at the anode. For the anode current this is simply a
time average, but for the charge quantum the assumption is that
this is the mean of a Poisson distribution and that the time integration for
the current is done over a time period for which the quantum efficiency was
stable. Note that, under ideal operating conditions, the dark current is
dominated by thermionic emission of single electrons at the cathode, so that
$Q_{i_{_{D}}}(t) = g_{_{PMT_{i}}}(t)e^{-}$ and the corresponding dark current
is given by $I_{i_{_D}}(t) = AT^2(t)e^{-(W)/kT(t)}$, where $A$ is constant
depending on the cathode material and size, $T$ is the temperature in Kelvin,
$W$ is the (field modified) work function of the cathode material, and $k$ is
the Boltzmann constant.

In the expressions above, a possible, explicit dependence on helicity state of
a given parameter is indicated by the $\pm$ superscript, while explicit time
dependence (both slow and fast), that is generally not correlated with the
helicity state is indicated by the continuous time parametrization $(t)$.
Aside from the physics asymmetries, the variation of various parameters with the
helicity state of the beam may be a result of either electronic coupling of the
helicity signal or direct variations in the beam as a function of helicity
state. Examples of the former would be the coupling of the helicity gate into
the signal that comes from the detector preamplifiers $g^{\pm}_{_{Amp_{i}}}(t)$
or the ADC electronics $V^{\pm}_{_{i_E}}(t)$. Helicity correlated changes in
beam conditions result in explicit changes of the rate seen in the detectors,
even in the absence of any asymmetry that results from the interaction of the
primary beam in the target or the rest of the experiment. Continuous time
dependence in the detector signal is primarily due to variations in the beam
current, target conditions, electronic drifts, temperature fluctuations, and
PMT ageing, but also due to electronic drifts. A great deal of effort goes into
the overall experimental design and analysis methods to remove or mitigate the
dependencies on these unwanted experimental factors and many of the solutions
are explicitly related to or influence the design of the integrating electronics.

\subsubsection{Design criteria for the integrating electronics}

The list below briefly describes the requirements the integrating electronics
has to satisfy.

\begin{enumerate}
  \item Helicity correlated changes in the beam
	(current, energy, position, and angle) are unavoidable, but they can be
	minimized to some degree and they can be measured using charge and position
	monitors. The experiment will be designed to measure the sensitivity of the
	detectors to these beam changes. The sensitivity measurements are made by
	measuring the correlation between the changes measured with the beam monitors
	to those measured with the detectors, which requires that the electronics
	chain used to process the measurements from the beam monitors is identical to
	that of the integrating detectors. This is particularly important in the case
	of the beam current monitors, since the integrating detector signal is
	normalized to the current monitor signal to remove random drifts,
	fast fluctuations, and helicity correlated changes, in the beam current.
  \item Random and systematic drifts in the detector signal
	$V^{\pm}_i(t) \rightarrow V^{\pm}_i$: The detector signal contains random
	changes and systematic drifts, both fast and slow. The slow drifts emerge due
	to variations in beam current, target conditions (target boiling and slower
	density changes), electronic drifts, temperature fluctuations, and PMT ageing.
	Almost all of these can be controlled to some degree (for example by
	operating the PMTs at a reasonable HV bias, in a reasonably stable temperature
	environment, and removed from or shielded from direct radiation exposure),
	but the primary way to deal with possible false asymmetries due to drifts and
	unwanted noise above shot noise, due to large amplitude random drifts, is to
	perform each asymmetry measurement on a timescale that is short compared to
	the timescale of the drifts. This is achieved by running the experiments with
	a fast helicity reversal rate (currently planned $\simeq \SI{1}{kHz}$) and a
	suitable choice of helicity patterns. Helicity quartets such as +$-$$-$+ and 
	$-$++$-$
	remove linear drifts and the introduction of pseudo-random initial state
	($\pm$) of the quartet removes quadratic drifts. The faster the helicity
	reversal rate, the more the signal variation with respect to time is well
	approximated as constant on the timescale of an asymmetry calculation:
	$$A_{i_{_{raw}}} = \frac{V^{+}_i(t_+) - V^{-}_i(t_-)}{V^{+}_i(t_+) + V^{-}_i(t_-)} \rightarrow \frac{V^{+}_i - V^{-}_i}{V^{+}_i + V^{-}_i}.$$
	Random variations that are faster than the helicity reversal rate are dealt
	with and used to advantage in the electronics chain, as described in detail
	below. The choice of helicity reversal rate influences or determines almost
	every detail of the electronics design.
  \item Minimization of electronics noise contribution: $$\sigma^2_{i_{_E}} \ll 2 Q_{i_{_P}} I_{i_{_P}} + \displaystyle\sum_{B}2 Q_{i_{_B}} I_{i_{_B}} + 2 Q_{i_{_D}} I_{i_{_D}}.$$
  \item Minimization of drift effects and helicity correlated electronic
	pickup in the pedestal: $V^{\pm}_{i_{_{E}}}(t) \rightarrow V_{i_{_{E}}}$.
  \item Minimization of drift effects and helicity correlated electronic pickup
	in the amplifier: $g^{\pm}_{_{Amp_{i}}}(t) \rightarrow g_{_{Amp_{i}}}(t)$.
  \item Bandwidth matching of all electronic components and between the various
	detectors that will be read out in integrating mode.
  \item Optimization of ADC resolution and sampling rate with respect to a
	$\SI{1}{kHz}$ reversal signal and the $\SI{1}{MHz}$ input bandwidth is required to
        follow (resolve) the helicity reversal signal transition (see below).
\end{enumerate}

\subsection{Signal structure and sampling scheme}\label{sec:SamplingScheme}

\begin{figure}
  \begin{center}
  \includegraphics[width=0.9\columnwidth]{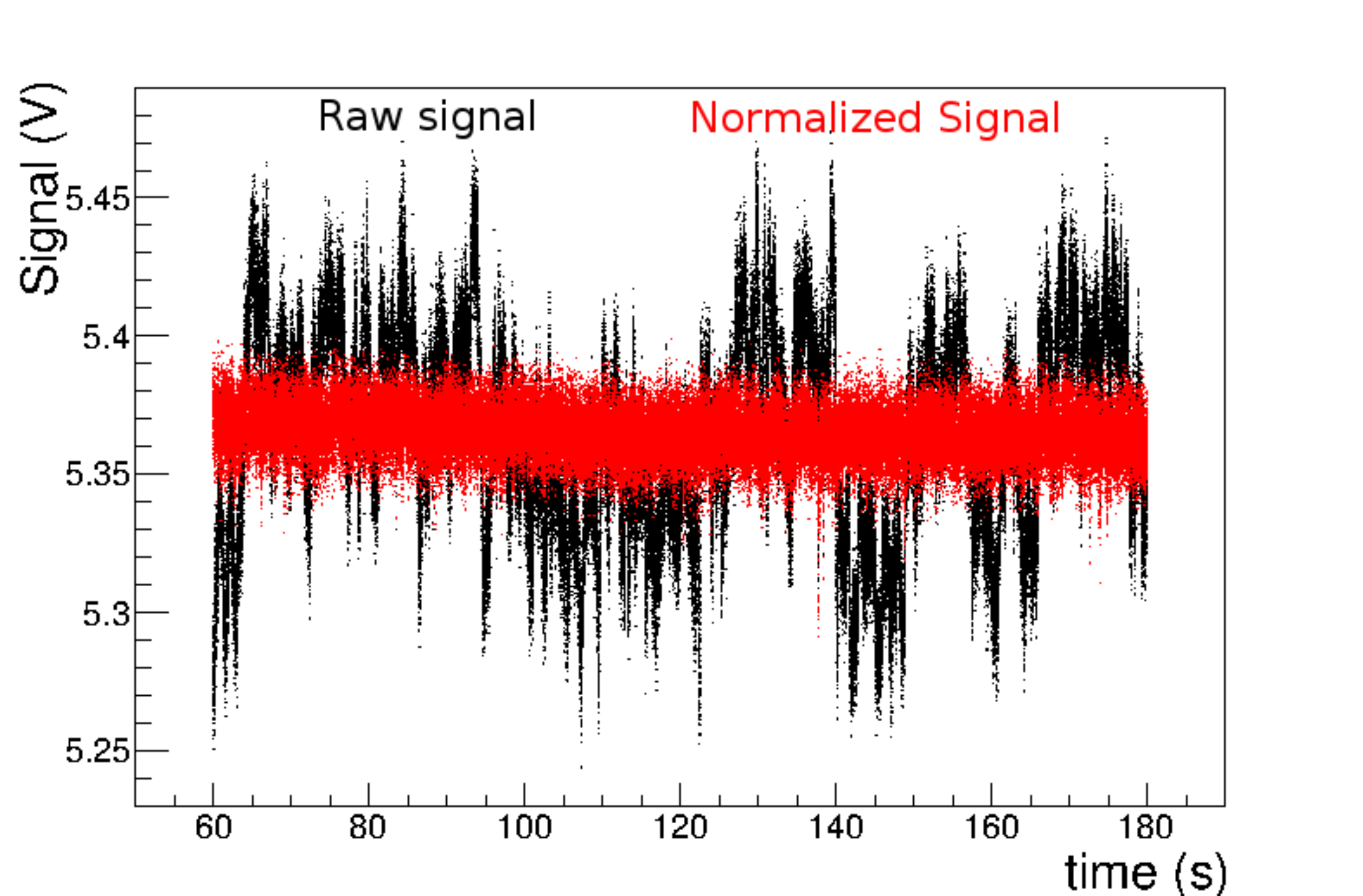}
  \caption{Example period of a signal from one of the QWeak integrating detectors.
	The black data shows the raw, unnormalized yield and the red
    data shows the beam current normalized yield.}
    \label{fig:DetSigNorm}
  \end{center}
\end{figure}

Figure~\ref{fig:DetSigNorm} shows a 2-minute period of the raw and beam current
normalized signals for one of the QWeak integrating detectors.
This data corresponds to an integration with an equivalent bandwidth of half
the helicity reversal rate ($\SI{480}{Hz}$), since the data for each helicity
window is averaged to one value per helicity window, in the FPGA, after
sampling. Even after normalization, drifts can be seen at timescales down to
seconds, along with much longer drifts, and large non-Gaussian drops in yield,
due to target boiling, can be seen at much shorter timescales (around
$\SI{30}{ms}$). A high helicity reversal rate prevents these types of drifts
from producing false asymmetries and reduces the contribution to the
RMS width in the asymmetry signal, due to these effects. For a mean signal of,
say, $\SI{4}{V}$ a physics asymmetry of $~\num{40E-9}$ produces a shift in
the mean of about $\SI{0.16}{\micro V}$, which is much smaller than the observed RMS
width of a few $\rm{mV}$. Since the bit resolution of even the best ADCs
today is not high enough to measure a signal difference at that level, in a
single sample, the combination of a suitable bandwidth selection and a high
rate of oversampling is used to increase the effective bit resolution of the ADC.

\begin{figure}
  \begin{center}
  \includegraphics[width=0.9\columnwidth]{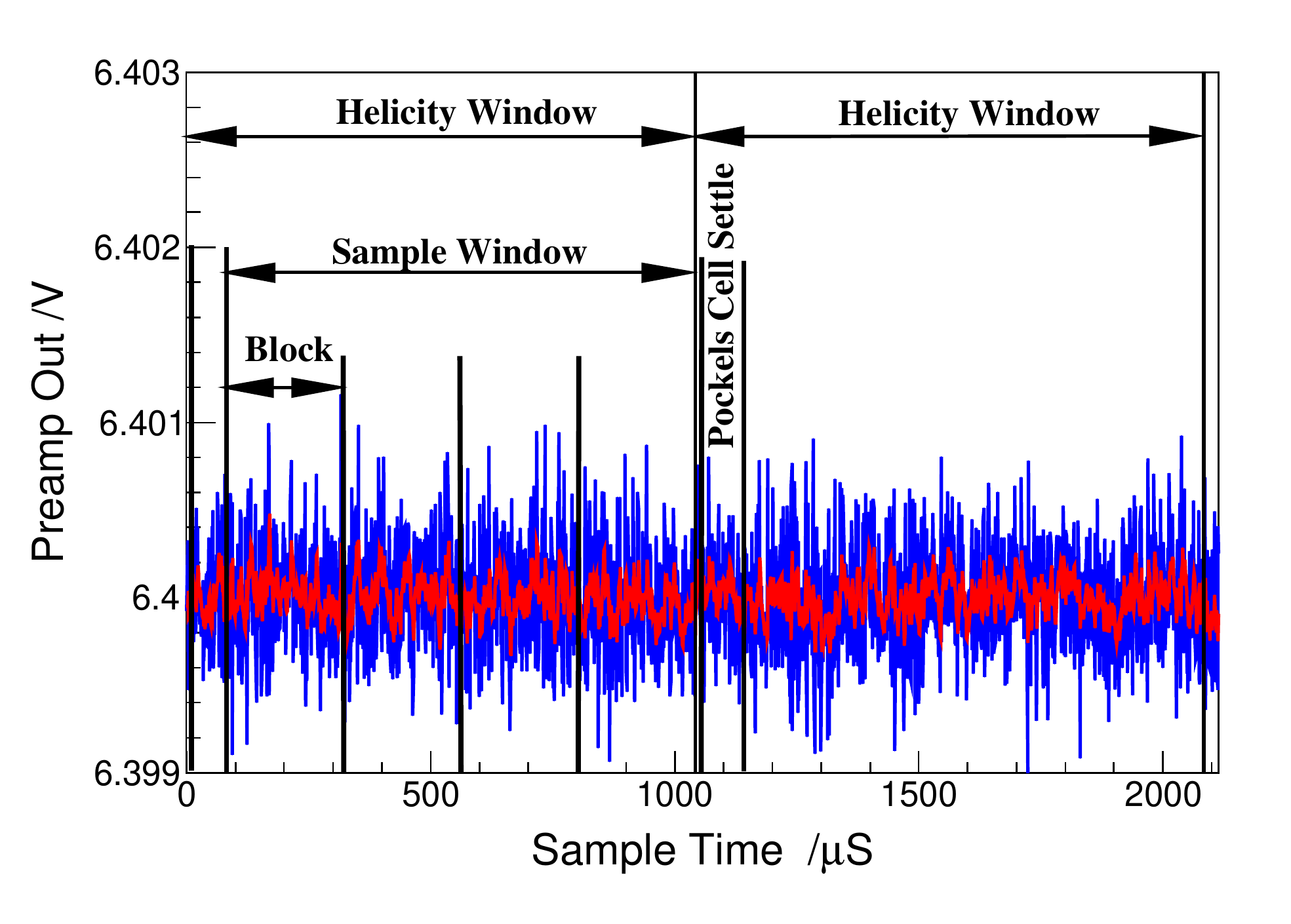}
  \caption{Integration mode detector signal (simulated Gaussian white noise) at
	the preamplifier output with a $\SI{500}{kHz}$ bandwidth (blue) and a
  $\SI{25}{kHz}$ bandwidth (red) and how this signal is divided into helicity
	windows and sampling regions. A primary goal of the integrating electronics is
	to match the ADC bit resolution and sampling rate with the detector signal
	bandwidth needed to follow the changes in the beam (related to the helicity
	change and otherwise).  }\label{fig:SigStruct}
  \end{center}
\end{figure}

Figure~\ref{fig:SigStruct} illustrates the division of the integration mode
detector signal into sampling regions. The period sampled for the
physics measurement must be stable, which means that the time periods
corresponding to the Pockels cell settling and ADC stability are
excluded. However, it is important that the electronics be designed to allow
sampling during the excluded time periods, so that the settling
time can be monitored and used for systematic checks. The example in
Fig.~\ref{fig:SigStruct} shows the time structure used for the QWeak experiment,
with a helicity reversal rate of $\SI{960}{Hz}$, Pockels cell settling time of
 $\SI{70}{\micro s}$, and an ADC delay of $\SI{42.5}{\micro s}$. For P2, the
reversal rate will be $\SI{1}{kHz}$ and the goal Pockels cell settling time is
$\SI{10}{\micro s}$. This means that the integrating electronics settling should
be correspondingly faster and operate at a higher sampling rate. The primary
goal of the integrating electronics is to match the ADC bit resolution and
sampling rate with the detector signal bandwidth needed to follow the changes in
the beam (related to the helicity change and otherwise).

\begin{figure}
  \begin{center}
  \includegraphics[width=0.99\columnwidth]{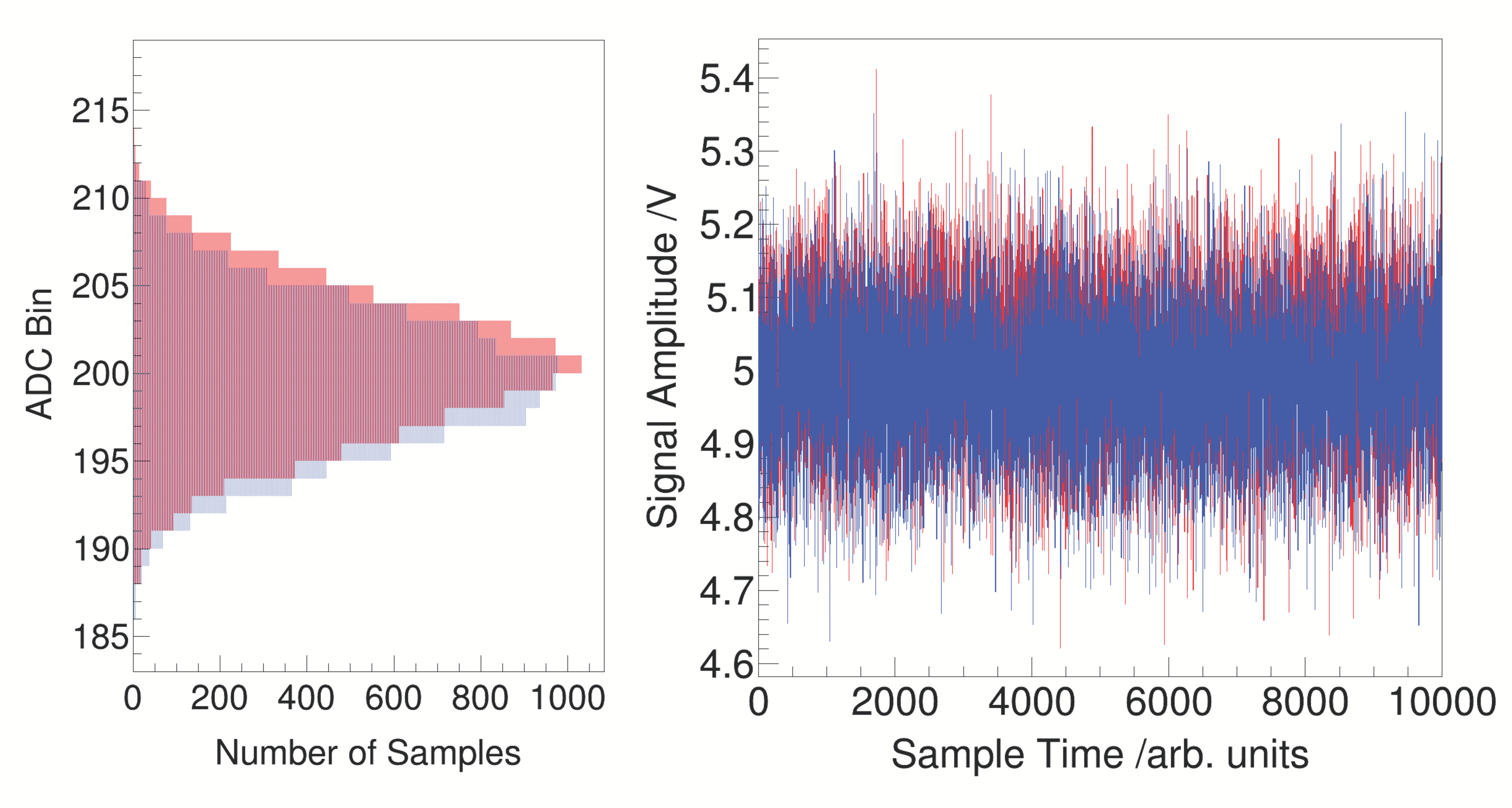}
  \caption{Illustration of the concept behind oversampling. The right-hand side
	of the figure shows two simulated data sets (red and blue) for which the mean
	value differs (e.g., as a result of an asymmetry), as a function of sample time.
	The left-hand side shows the samples accumulated by the ADC in histogram form,
	where the bin width of the histogram is synonymous with the resolution of the
	corresponding ADC. The difference is again shown in red and blue, with the
	darker region being the overlap of the two distributions. The difference in
	the mean for the two signals is $\SI{20}{mV}$, while the bin width (ADC
	resolution) is $\SI{25}{mV}$, but the oversampling, together with the large
	RMS width in the signal, allows a high precision difference measurement
	between two helicity windows.}\label{fig:OverSampCpt}
  \end{center}
\end{figure}

 Figure~\ref{fig:OverSampCpt} illustrates the concept behind the oversampling.
The right hand side of the figure shows two simulated data sets (red and blue)
for which the mean value differs (e.g., as a result of an asymmetry), as a
function of sample time. The left hand side shows the samples
 accumulated by the ADC in histogram form. In this simulated data set, the
mean difference between the two data sets is $\SI{20}{mV}$, while the bin width
 in the histogram, which represent the resolution of the ADC, corresponds to
$\SI{25}{mV}$. Both of these values are exaggerated for illustrative purposes,
but they show the general effect. The ADC resolution is too large to resolve the
difference in the signal caused by the asymmetry, but the signal is
 sampled many times within each period (i.e., the helicity window) and the mean
difference of the two distributions is clearly visible. This only works
 effectively, if the RMS width in the signal is significantly larger than the
resolution of the ADC ($\Delta = V_{ref}/2^n$ for an $n$-bit ADC), which then
 automatically exceeds the digitization error in the ADC
$\sigma_d = \Delta/\sqrt{12}$. Another benefit of the larger RMS width in the
signal is that the ADC differential non-linearity becomes less and less significant 
while increasing the number of ADC channels over which the signal is spread. The
experimental design and a suitable choice for the bandwidth in the preamplifiers
and filters at the ADC input will ensure that this requirement is satisfied. The 
final ADC has to be chosen to have an integral non-linearity that is as small as
possible, but since the difference in the mean of the signal for any pair of 
opposite helicity windows is smaller than the ADC resolution, the resulting 
distortions in the output distributions (left-hand panel in Fig.~\ref{fig:OverSampCpt}) 
for a given helicity pair will be nearly the same and the averaging over a large
number of ADC bins, inherent in the this sampling scheme, reduces the effect
further. For most good ADCs, the integral non-linearity as a function of ADC 
output code is distributed around $0\pm 0.6$ LSB.   

\subsubsection{Bandwidth requirements}

The lower limit to the system bandwidth is set by the need to follow the
helicity flip settling time. The goal for the settling is $\SI{10}{\micro s}$
 and the front-end electronics has to be fast enough to follow this transition,
to ensure that the amount of lost data within a helicity window is minimized.
This means that the preamplifier output should also settle, within a certain
accuracy relative to the mean helicity signal level, within a few $\si{\micro s}$.
Taking a rate of $\SI{100}{GHz}$ in the detectors (approximately, from primary 
scattered electrons only), counting statistics within
a single $\SI{1}{ms}$ helicity window is about $\SI{100}{ppm}$.
Assuming that the noise from ringing during helicity transitions is perfectly correlated in all detectors and
that we want the corresponding peak-to-peak variations to be much smaller than
the counting statistics noise (shot-noise), we demand that the amplifier must
settle to \SI{0.001}{\percent} ($\SI{10}{ppm}$) within a time window that is
small compared to the helicity flip settling time. A single pole filter with a
signal bandwidth cutoff of $f_{\SI{3}{db}} = \SI{1}{MHz}$ has a time constant
of $RC \simeq \SI{0.16}{\micro s}$ and reaches the desired settling in
$t_s = - \ln(0.001/100)\cdot RC \simeq \SI{2}{\micro s}$, which is reasonable
and also allows for some component variation and monitoring of additional
ringing during the helicity transition. Suitable operational amplifiers for the
preamplifiers have to be selected with these criteria in mind.

\subsubsection{The preamplifier}

The design of the P2 integrating electronics will be based on the QWeak
electronics, but will be modified to satisfy, primarily, the criteria for
higher bandwidth discussed above. The schematic for a single channel of the
QWeak preamplifier is shown below, in Fig.~\ref{fig:PreampSchem}. The inputs
are protected with a combination of resistors and diodes. Each channel has a
separate set of switches that set the gain between various values (here shown
for QWeak, between $0.5$ and \SI{4.0}{\mega\ohm}).
The bandwidth is primarily
set by the parallel C1/R9 capacitor resistor feedback circuit, which was
$\SI{25}{kHz}$ for QWeak. The displayed operational amplifier (OPA2604) can
not support the desired $\SI{1}{MHz}$ bandwidth at the needed gains, so we will have to choose a
different OpAmp, or implement a second gain stage. The constraints for the
achievable signal bandwidth are set by the competing needs for the feedback
capacitance $(C_F)$ (which controls the noise behavior), feedback resistance
$(R_{F})$ (which sets the overall gain), and the gain bandwidth product of
the OpAmp. The two relevant relations are $f_{SB} \leq 1/2\pi R_F C_F$ and
$f_{GB} > (C_{IN} + C_{F})/2\pi R_F C^2_F$, where $f_{SB}$ is the desired
signal bandwidth, $f_{GB}$ is the gain bandwidth of the OpAmp (which is
$\SI{20}{MHz}$ for the OPA2604 used for the QWeak design), and $C_{IN}$ is the
capacitance at the amplifier input, which is a combination of the OpAmp input
capacitances, the PMT capacitance, and the cable capacitance. The latter two
are, of course, dependent on the PMT base design and the cable length needed
to connect the PMTs to the preamplifiers. The QWeak amplifier also implemented
an offset control that sets an output bias voltage on each channel, to provide
an indication that the preamp is powered and connected. An external 5V source
powers an isolated DC-DC converter to supply $\pm \SI{15}{V}$ internally.

\begin{figure}
  \begin{center}
  \includegraphics[width=0.9\columnwidth]{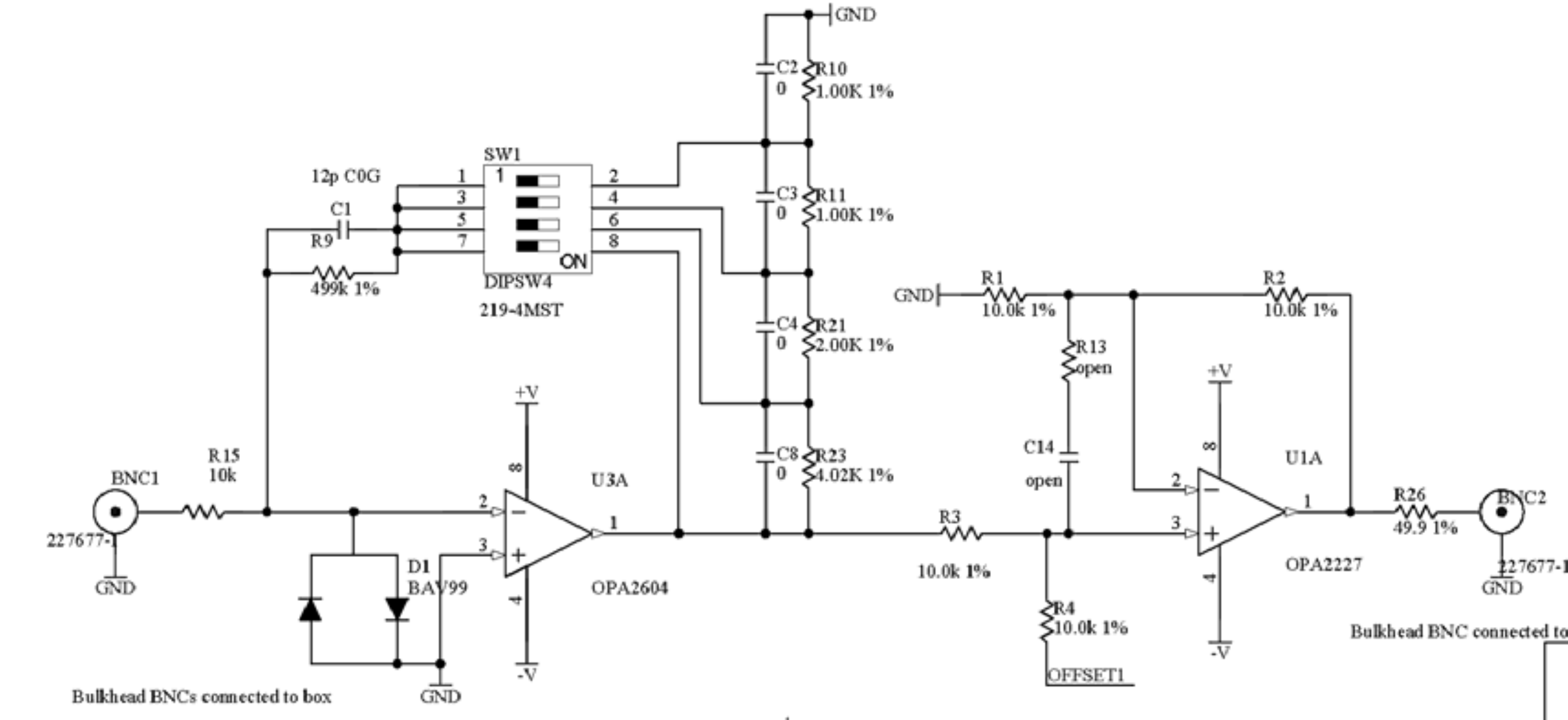}
  \caption{QWeak style trans-impedance preamplifier schematic. The gain is set
	by a selectable feedback amplifier, using a dip switch. The bandwidth is
	mainly determined by the parallel capacitor-resistor pair (C1 and R9).}\label{fig:PreampSchem}
  \end{center}
\end{figure}

\subsubsection{The integrating ADC}

As discussed in Sect.~\ref{sec:SamplingScheme}, two critically important
parameters for the integrating detector ADCs will be the sampling rate and the
amplitude or bit resolution. These are also competing parameters, in that ADCs
with a high-bit resolution tend to have lower sampling rates and vice versa.
Today, one can buy high performance, low noise, ADCs with 18 to 24 bit
resolution and sampling speeds of several Msps. For P2, \SI{18}{bit} resolution
is sufficient, given that the experiment will use the oversampling scheme
described in Sect.~\ref{sec:SamplingScheme}. A suitable ADC has been
identified\footnote{Analog Devices \url{http://www.linear.com/product/LTC2387-18}},
with a sampling speed of \SI{15}{Msps}, which is significantly above the minimum
required for the desired $\SI{1}{MHz}$ signal bandwidth.

For an event rate of $\SI{1}{GHz}$, a PMT gain of 1000, and about
50~photo-electrons at the cathode, per event, the average current from a single
detector would be about $\SI{8}{\micro A}$. Assuming also that the background
and electronic noise contributions are negligible, Eq.~(\ref{eqn:RMS1}) gives an
estimated RMS width in the detector signal ($\sqrt{2 Q_{i_{_P}} I_{i_{_P}} B}$)
of about $\SI{0.36}{\micro A}$ (for a $\SI{1}{MHz}$ signal bandwidth). A
preamplifier gain of \SI{0.2}{\mega\ohm} would then produce a signal at the
ADC, with a mean of $\SI{2}{V}$ and an RMS of about $\SI{90}{mV}$.
The \SI{18}{bit} ADC currently under consideration has a full scale voltage
range of $V_{fs} = \pm \SI{4.096}{V}$ and has a resolution of
$V_{Ref}/2^{18} = \SI{4.096}{V} / 2^{18} \simeq \SI{16}{\micro V}$, which means
that a signal with an RMS width of $\SI{90}{mV}$ would be spread over more
than \SI{5625}{ADC} channels, which is more than enough to make the digitization
noise and non-linearity negligible, as discussed above.

\subsubsection{Back-end data processing, readout, and DAQ interface}

\begin{figure}
  \centering
  \includegraphics[width=0.9\columnwidth]{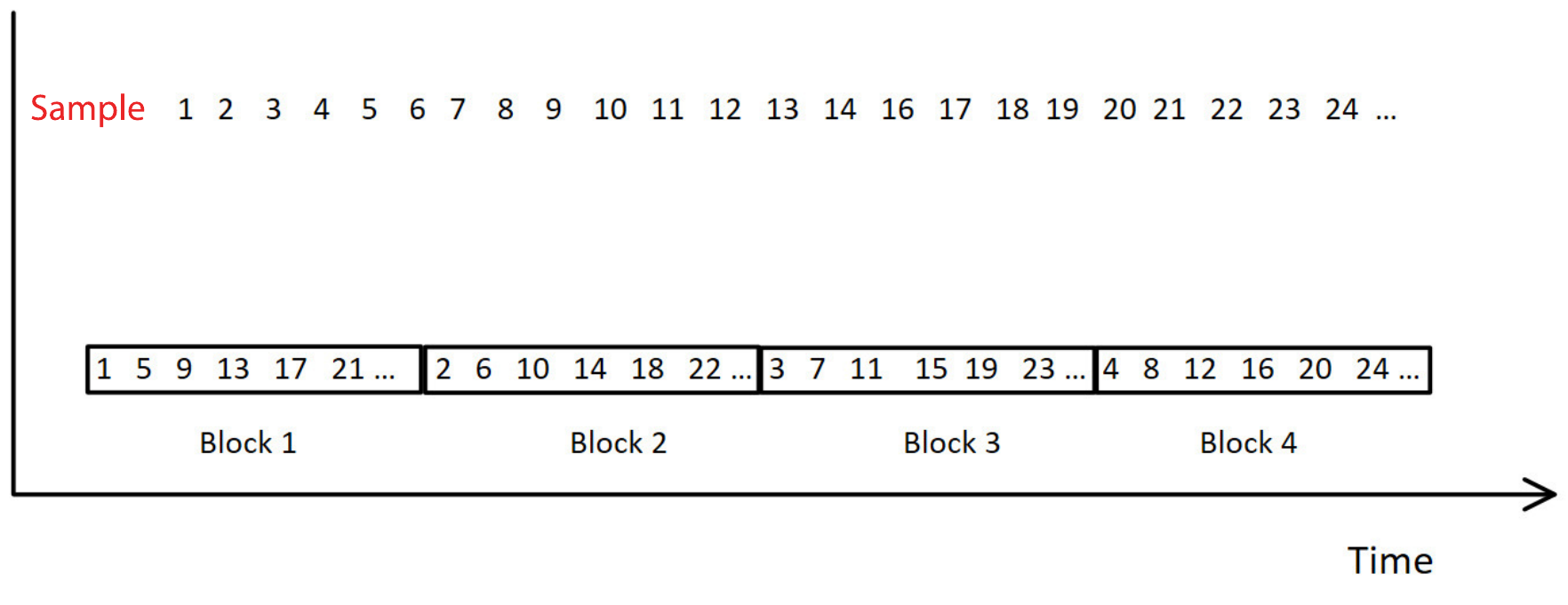}
  \caption{Each helicity period can be separated into 4 blocks which are
	normally populated by samples taken in sequence. For diagnostic purposes relating to
  possible phase slippage in detector signals (particularly between the beam
	monitors and the integrating detectors), it is desirable to have only every nth
  sample (up to $n = 4$) in a given block, as illustrated in this figure. }\label{fig:VMEBlkAccSch}
\end{figure}

As illustrated in Fig.~\ref{fig:SigStruct}, the signal readout for the QWeak
experiment included the option to split the signal into four blocks of
consecutive samples, for diagnostic purposes, and this capability should be
retained for the P2 design. In addition to this, it is desirable to implement
the possibility to accumulate only every $n$th sample in a given block, as
illustrated in Fig.~\ref{fig:VMEBlkAccSch}, primarily to monitor the effects of
possible phase slippage between the integrating detectors and the beam monitors.

The QWeak design only allowed the readout of the sample sum for each entire
helicity window and the sums for the four separate blocks within each helicity
window. It is desirable to implement the readout of additional information in
the data stream for each data block, such as the minimum sample value, maximum
sample value, and RMS of the samples in each block. In addition, for diagnostic
purposes, it is often desirable to obtain the fast Fourier transform of the
actual signal, as it is sampled by the ADC. For this to work, it is necessary
to implement the possibility of reading out every ADC sample, at a lower event
rate (tracking rates), rather than only producing integrated numbers for each
helicity window. This mode basically amounts to running the ADCs as slow
waveform digitizers. This pre-processing of the ADC data will be implemented
with an FPGA. For the purpose of commissioning, when the DAQ is being
synchronized with the rest of the experiment, it will be necessary to have
front panel outputs and data stream information about when exactly the FPGA
starts and stops accumulating ADC samples after the helicity gate. The ADC will
be controlled by an external clock that is synchronized to the helicity
reversal. The ADC data will be read out via Gbit ethernet connections for each
module.

%% file: tracker.tex

Depending on the scattering position along the target, the solenoidal field will
map different scattering angles into the acceptance of the \ICDs. A tracking
detector is needed to determine this relationship and ultimately the average
squared momentum transfer $\left\langle Q^2 \right\rangle$ of the electrons 
entering the asymmetry determination. High resolution tracking also allows for 
the study of backgrounds and position and momentum dependent systematic effects.

The tracking detector geometry and reconstruction was developed using a full 
Geant4 \cite{Agostinelli:2002hh,Allison:2006ve,Allison:2016lfl} based 
simulation of the P2 setup including a detailed field map based on 
the FOPI solenoid magnet \cite{Ritman:1995td}.

\subsubsection{Tracker operation modes}
\label{sec:TrackerOperationModes}

Two main modes of operation are foreseen for the tracker, one at low rates, 
where the Cherenkov detectors can be operated in single electron detection mode 
and coincidences with the tracker can be formed. A second mode at high rates is
used to study rate, position and momentum dependent systematic effects; here the
tracking detectors will be gated in order to keep the data acquisition rate
manageable.

The first mode requires that the tracker acceptance at the very least covers all
electrons that can reach one fused silica bar with very high efficiency, whilst 
the second mode demands a very fast, radiation tolerant detector with a geometry 
that allows for a reliable track reconstruction also at high occupancy. The low
electron momentum and the reduction of background from photons created in the
tracker put very stringent constraints on the tracker material budget.

\subsubsection{Tracker geometry}
\label{sec:TrackerGeometry}

\begin{figure}
	\centering
	\includegraphics[width=0.48\textwidth]{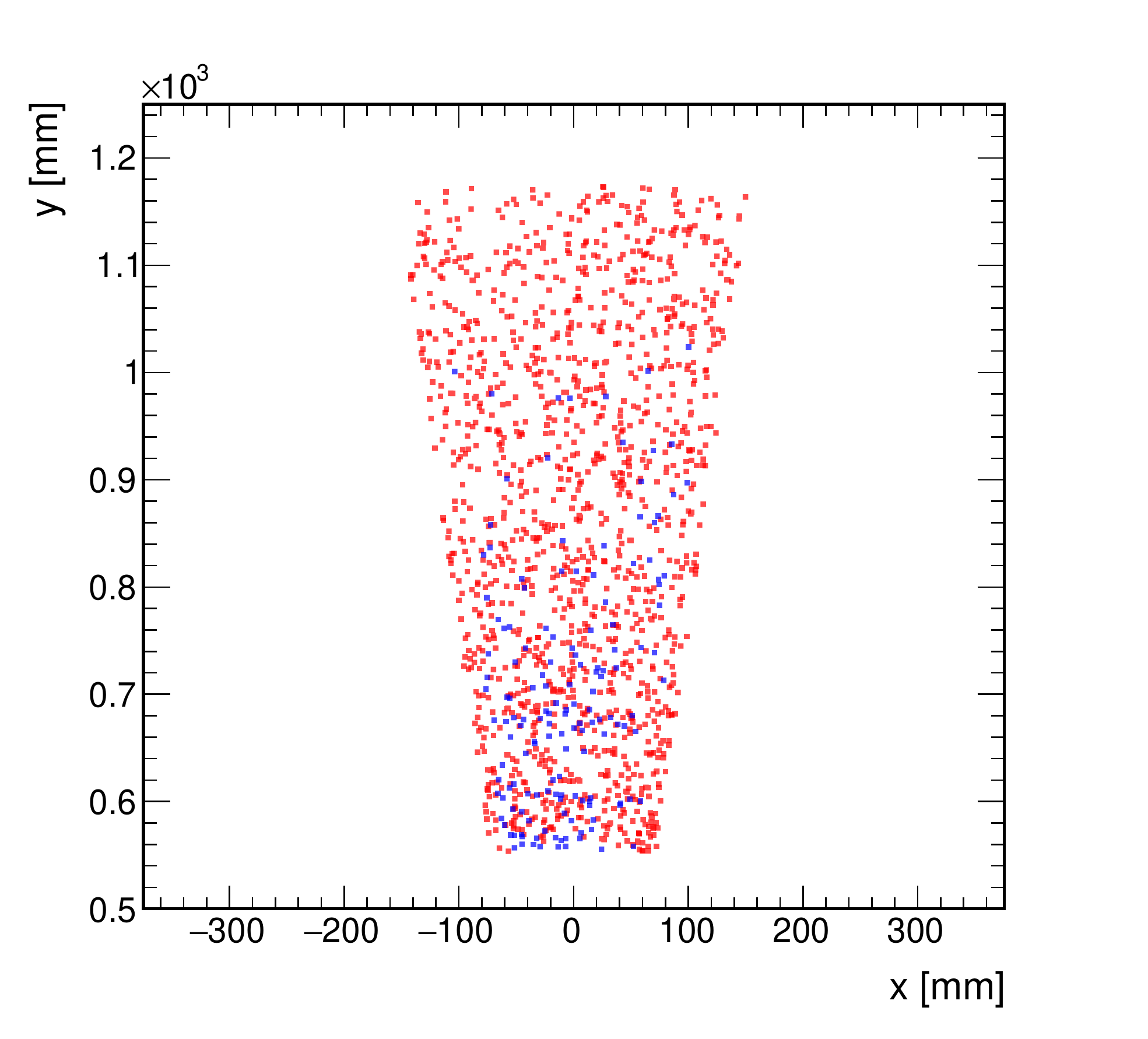}
	\caption{Simulated hit distribution in a segment of the first tracker plane at 
	full rate and with a \si{50}{ns} integration window. About one in thousand 
	bremsstrahlung photons leaves a hit (red dots). The signal electrons cause the 
	hits indicated with blue dots. A coordinate system centered in the solenoid
	center and with the $z$ axis along the beam direction is employed.}
	\label{fig:hitsinoneframeplane0}
\end{figure}

The tracking detector uses the curvature in the magnetic field for momentum 
measurements, requiring placement in the solenoid and thus prevents a complete 
geometric shielding of the bremsstrahlung photons created in the target.
Indeed, the first tracker plane is impinged by up to six orders of magnitude 
more photons than signal electrons. This in turn requires again a very thin 
tracker to minimize the photon interaction probability and a geometry that allows
for a robust track reconstruction with a very adverse signal-to-background ratio
of tracker hits (see Fig.~\ref{fig:hitsinoneframeplane0}). Such a robust track 
reconstruction can be achieved by tracking plane pairs spaced by roughly the typical 
distance of two hits on a single plane at full occupancy ($\approx$ 
\SIrange{1}{2}{cm}).

\begin{figure}
	\centering
		\includegraphics[width=0.48\textwidth]{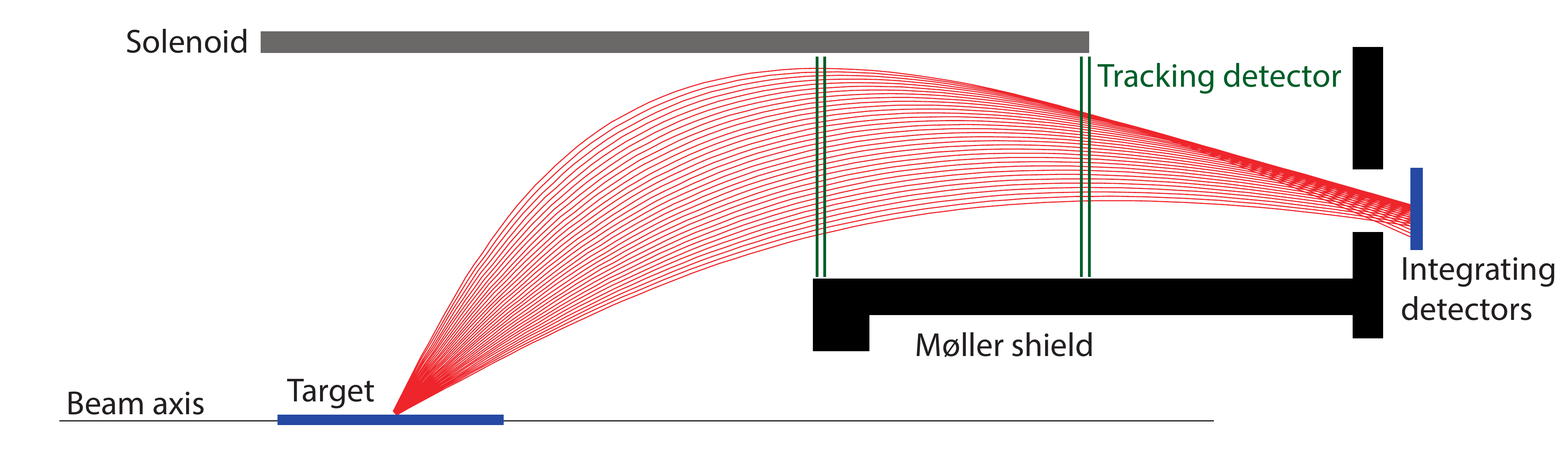}
	\caption{Schematic view of the tracking detector geometry.}
	\label{fig:TrackerSchematic}
\end{figure}

\begin{figure}
	\centering
		\includegraphics[width=0.48\textwidth]{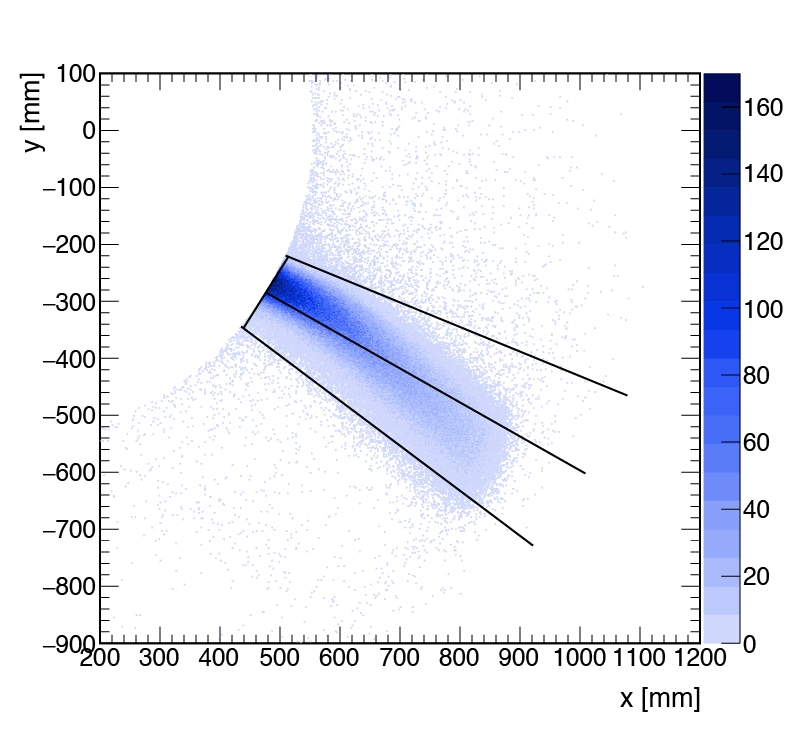}
		\includegraphics[width=0.48\textwidth]{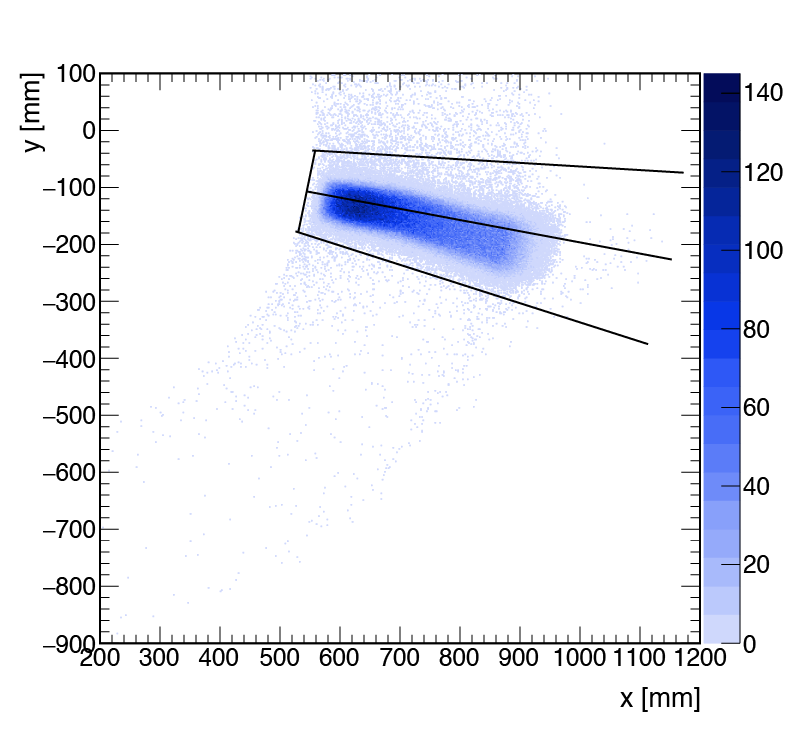}
	\caption{Simulated intercept of the electron trajectories ending in one fused silica bar
	with the first (closest to the target, top) and last (closest to the silica bar,
	bottom) tracking plane. The active area and the centerline of a tracking panel
	are indicated with the black lines.}
	\label{fig:CrystalShadowPlane0_canv}
\end{figure}

For the low momentum tracks expected in P2, multiple Coulomb scattering in
detector material is the main effect determining momentum resolution. It is thus
desirable to have a long, material-free region to obtain a good curvature 
measurement. The demands of the reconstruction and good momentum resolution can 
be met by two double planes separated by a wide, empty drift region, the 
geometry chosen for P2, see the schematic view in Fig.~\ref{fig:TrackerSchematic}.

The $\left\langle Q^2 \right\rangle$ determination does not require full
azimuthal coverage of the tracker; the active area should however be large
enough to cover virtually all electrons heading for one fused silica bar. 
This can be achieved by tracker segments covering \ang{15}, see 
Fig.~\ref{fig:CrystalShadowPlane0_canv}. 
We currently foresee four such segments to cover up-/down- and left-/right- 
asymmetries.

\subsubsection{High-Voltage Monolithic Active Pixel Sensors}
\label{sec:HighVoltageMonolithicActivePixelSensors}

The P2 tracker requires active elements which are fast, thin, radiation hard and 
highly granular in order to deal with the high rates and low momentum tracks. 
High-Vol\-tage Monolithic Active Pixel Sensors 
(HV-MAPS, see \cite{Peric:2007zz,Peric2010504,Peric2010,Peric:2012bp,Peric:2013cka})
fulfill all these requirements. A commercial high-vol\-tage CMOS process providing 
deep $n$-wells in a $p$-doped silicon substrate allows for reverse bias voltages 
of about \SI{90}{V} between wells and substrate. This creates a thin, high-field 
depletion region, from which charge is quickly collected via drift. Inside the
$n$-wells, complete CMOS electronics can be implemented, allowing for in-pixel
amplifier circuits and complete signal digitization and processing on the same 
chip. The thin active region allows for thinning of the sensors to just 
\SI{50}{\micro \meter} thickness.

\begin{figure}
	\centering
		\includegraphics[width=0.48\textwidth]{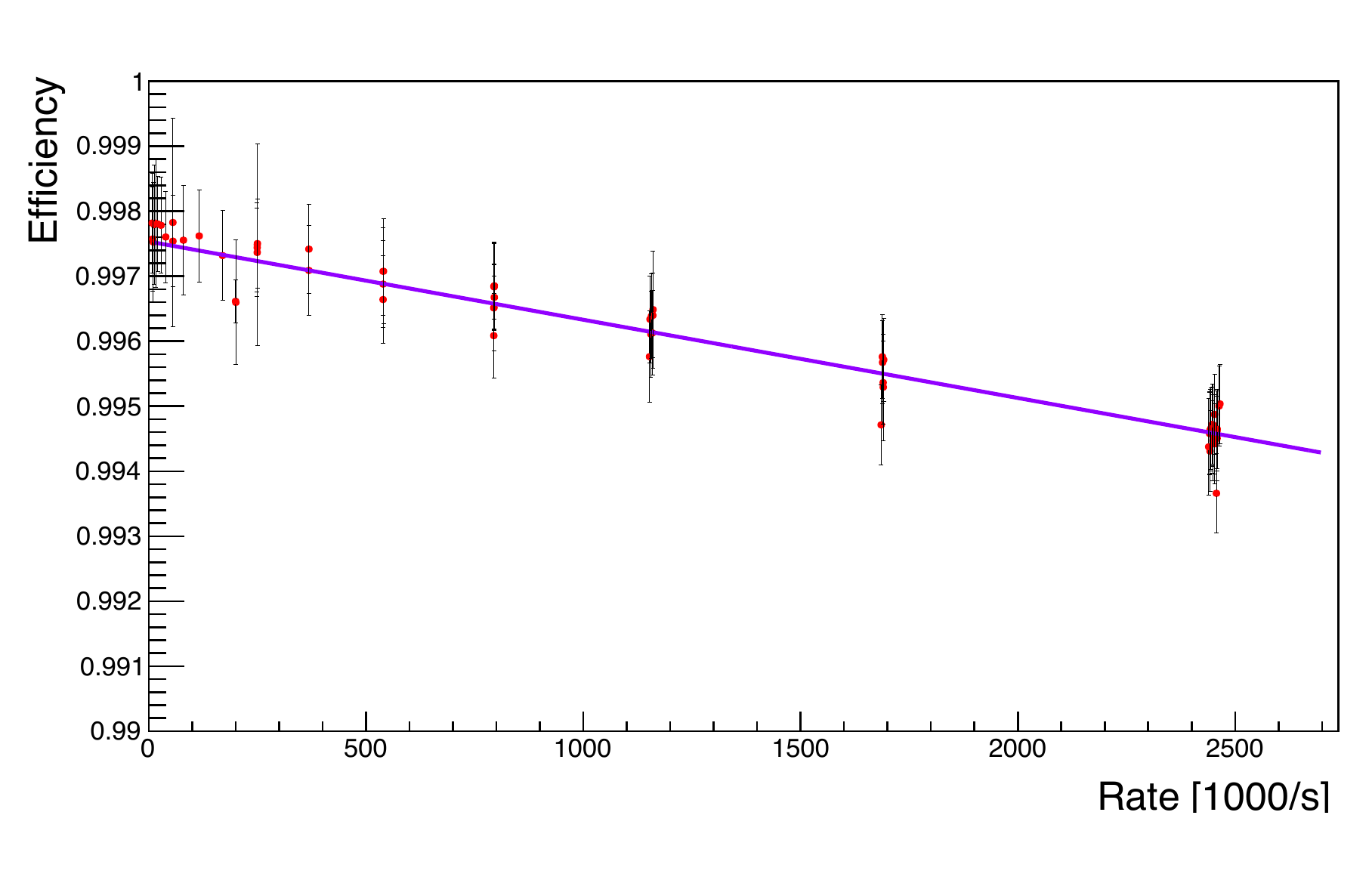}
	\caption{Rate-dependent efficiency of a MuPix7 HV-MAPS prototype illuminated 
	by a roughly \SI{500}{\micro\meter} diameter electron beam.}
	\label{fig:HighRateNew}
\end{figure}

\begin{figure}
	\centering
		\includegraphics[width=0.49\textwidth]{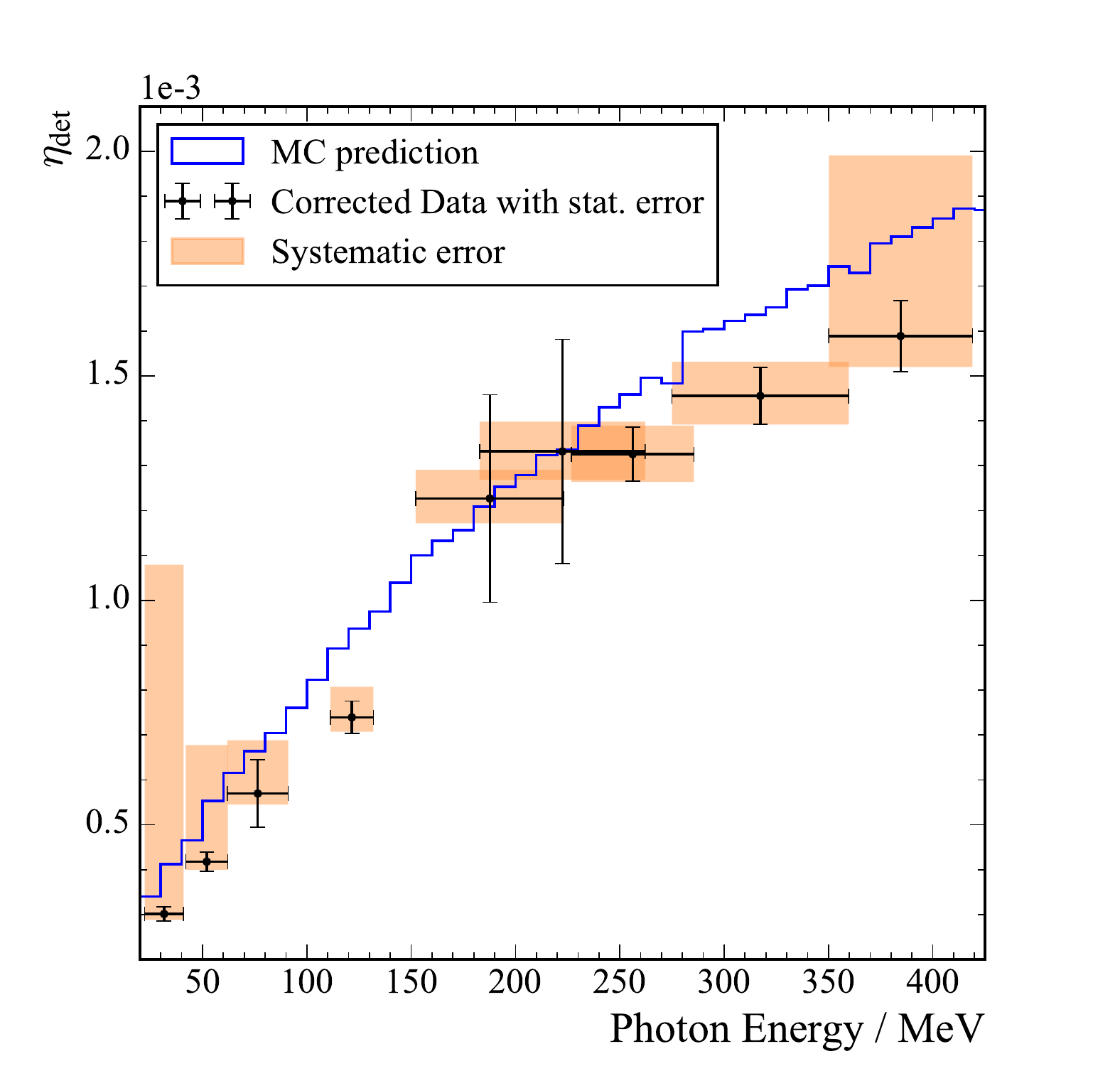}
	\caption{Efficiency of MuPix7 sensors for high energy photons compared with 
					the Geant4 based simulation.}
	\label{fig:efficiencymerged32ns}
\end{figure}

In the framework of the Mu3e collaboration \cite{Blondel:2013ia}, we have developed a series of 
HV-MAPS prototypes known as the MuPix chips \cite{Shrestha:2014oxa,Augustin:2015mqa}.
With the MuPix7 prototype \cite{Augustin:2016hzx}, we have produced a complete 
system-on-chip with a \SI{3 x 3}{mm} active pixel matrix with internal 
amplifiers and source followers, driving the signals to the chip periphery, 
where hits are detected by a tunable comparator and timestamps are assigned. 
A state machine collects and serializes the hits and sends them off-chip using a 
\SI{1.25}{Gbit/s} low-voltage differential signaling (LVDS) link. In this system, we
have measured detection efficiencies well above \SI{99}{\percent} and a time 
resolution below \SI{15}{ns} at noise rates below \SI{1}{Hz} per pixel using
a variety of beam tests at DESY, PSI, CERN and MAMI. High rate capability has 
been demonstrated by a MAMI beam test, where we illuminated a roughly 
\SI{500}{\micro\meter} large spot with more than \SI{2}{MHz} of \SI{800}{MeV}
electrons, leading to an efficiency loss consistent with the single pixel
deadtime of about \SI{1}{\micro\second}, see Fig.~\ref{fig:HighRateNew}.
Despite no special design measures being taken, the sensors turn out to be very
radiation hard \cite{Augustin:2017guc}.

Given the very large bremsstrahlung background expected in P2, it is imperative
that the sensors have a small and well understood efficiency for detecting
photons. To this end we have conducted tests with sources and at the A2 tagged
photon facility at MAMI and find detection probabilities below \num{10e-3} for
photons in the few MeV energy range, see Fig.~\ref{fig:efficiencymerged32ns}. The 
detection probability rises again towards lower energies and was found to be
$\approx \SI{30}{\percent}$ at \SI{5.9}{keV}, again in good agreement with
simulations \cite{MartinBachelor}.

We are currently testing the \SI{2 x 1}{cm} MuPix8 prototype in order to 
understand scaling effects in large sensors and prepare for the production of 
the final \SI{2 x 2}{cm} sensor.

\subsubsection{Tracker segments}
\label{sec:TrackerSegments}

\begin{figure*}
	\centering
		\includegraphics[width=0.98\textwidth]{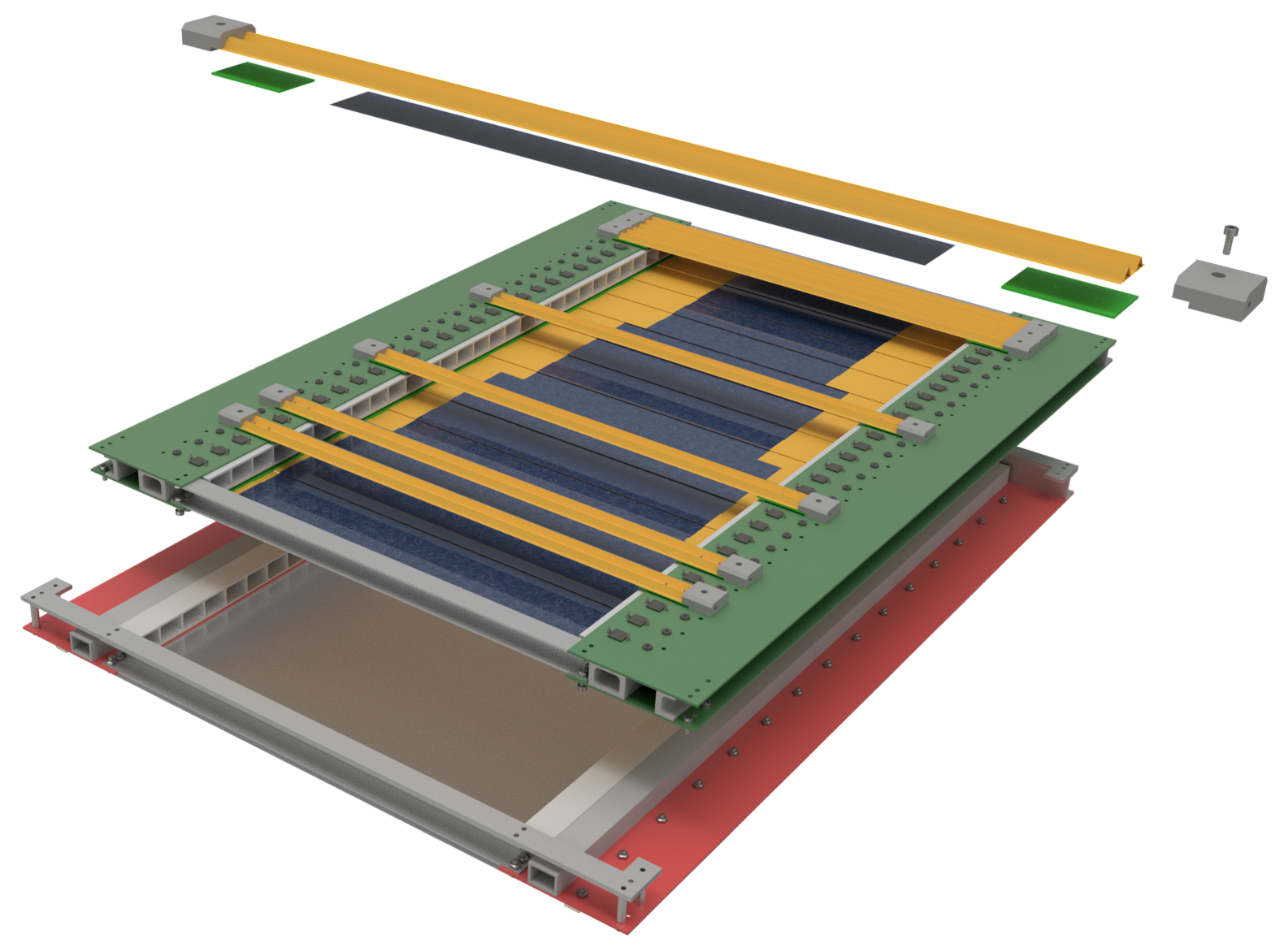}
	\caption{Rendering of a partially assembled tracker module. On top a single
	strip assembly is shown with the sensors in blue, the polymide flexprint with
	the cooling channel V-folds in yellow, the end PCBs in green and the gas 
	distribution pieces in gray. Below a module with the front cover and most of 
	the front layer strips removed. The PCB/cooling pipe frame with gas 
	distribution is visible, as are the sensors of the back layer. On the bottom 
	is the back cover with its gas distribution.}
	\label{fig:Module_explosion}
\end{figure*}

The tracker will consist of double layer segments covering \ang{15} in azimuth 
and as much as mechanically possible of the radial space between the M\o ller 
shield and the magnet inner wall (\SI{535}{mm} to \SI{1200}{mm} in radius). 
The active sensors will be cooled by gaseous helium.

\paragraph{Mechanics}
\label{sec:Mechanics}

A tracker module, see Fig.~\ref{fig:Module_explosion}, is built from 29 
staggered strips on each side with a slight overlap of the sensors along the 
radial axis to prevent ineffective areas and a frame providing mechanical 
support, cooling and electrical connectivity. 
Depending on the radius, each strip is equipped with between 8 
and 14 \SI{50}{\micro m} thin HV-MAPS sensors, the strip design is derived
from the Mu3e design \cite{Berger:2016lme}. 
About 30 single point tape-automated bond connections \cite{Oinonen:2005rm} 
to an aluminum-polymide flexprint provide the supply voltages and transfer
data and control signals. An additional polymide layer with V-shaped folds glued
to the flexprint provides mechanical stability and cooling channels. The overall 
thickness of the active part of a strip is less than \SI{0.12}{\percent} of a 
radiation length. 

The strip ends in a rigid printed 
circuit board (PCB) with a milled plastic cooling gas manifold on top; two different 
PCB thicknesses allow for the strip staggering.
At both ends of the strip, these end PCBs are connected to a frame built from 
long PCBs and cooling ducts. Signals and power are connected via a 
high-density interposer. The azimuthal sides of the frame have a sliding system
and are tensioned by springs in order to compensate for thermal expansion of the
strips. Two layers of strips are connected to the same frame such that the sensors
face inwards towards each other with a distance of roughly \SI{23}{mm}. The full
module has 632 sensors. 

The modules will be mounted on rails in their surrounding helium volume with the
possibility to remove and re-insert them in a limited amount of time. This is
in particular required for the runs with lead targets, where the radiation 
levels definitely exceed the tolerance of the tracker and gives the flexibility
to also run without tracker in hydrogen mode in order to study systematics or
perform repairs.

We are currently investigating two solutions for the powering of the HV-MAPS
sensors, which require a \SI{1.8}{V} supply voltage. Either this voltage is 
generated from an external \SI{10}{V} supply via radiation hard DC-DC converters
\cite{Feld:2013ewa} mounted on the support frame or we employ a powering 
scheme, where several sensors are connected to the external supply in series;
the required on-chip shunts are currently under test.

\paragraph{Cooling}
\label{sec:Cooling}

\begin{figure}
	\centering
		\includegraphics[width=0.49\textwidth]{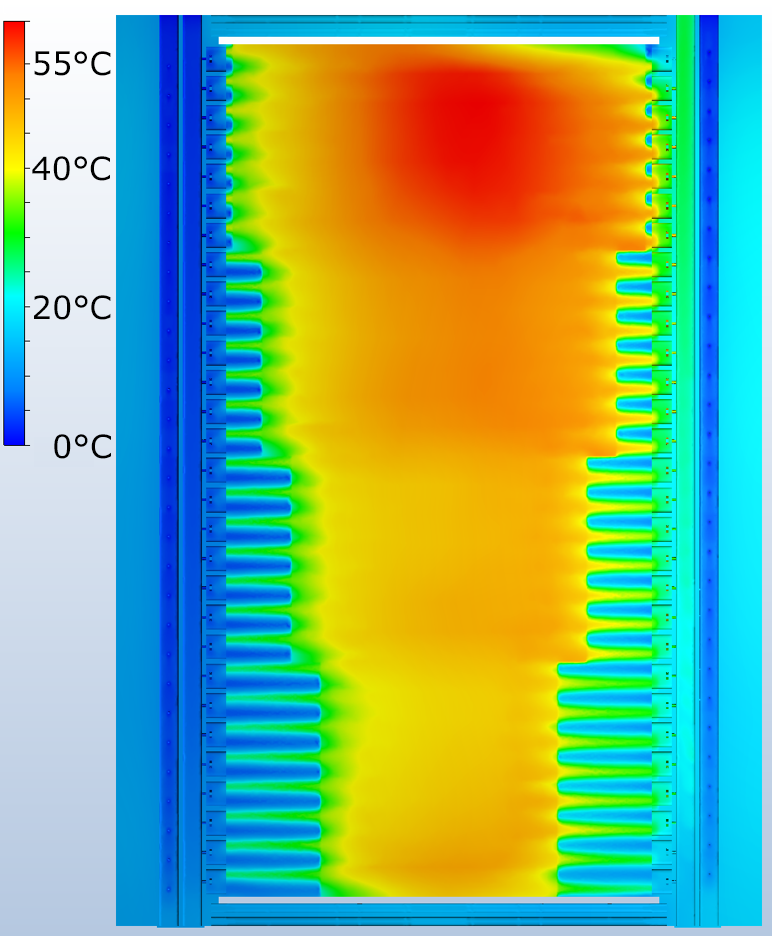}
	\caption{Computational fluid dynamics simulation of a full tracker module
	at the maximum power dissipation of the HV-MAPS of \SI{400}{mW/cm^2}.}
	\label{fig:Cooling}
\end{figure}

\begin{figure}
	\centering
		\includegraphics[width=0.35\textwidth]{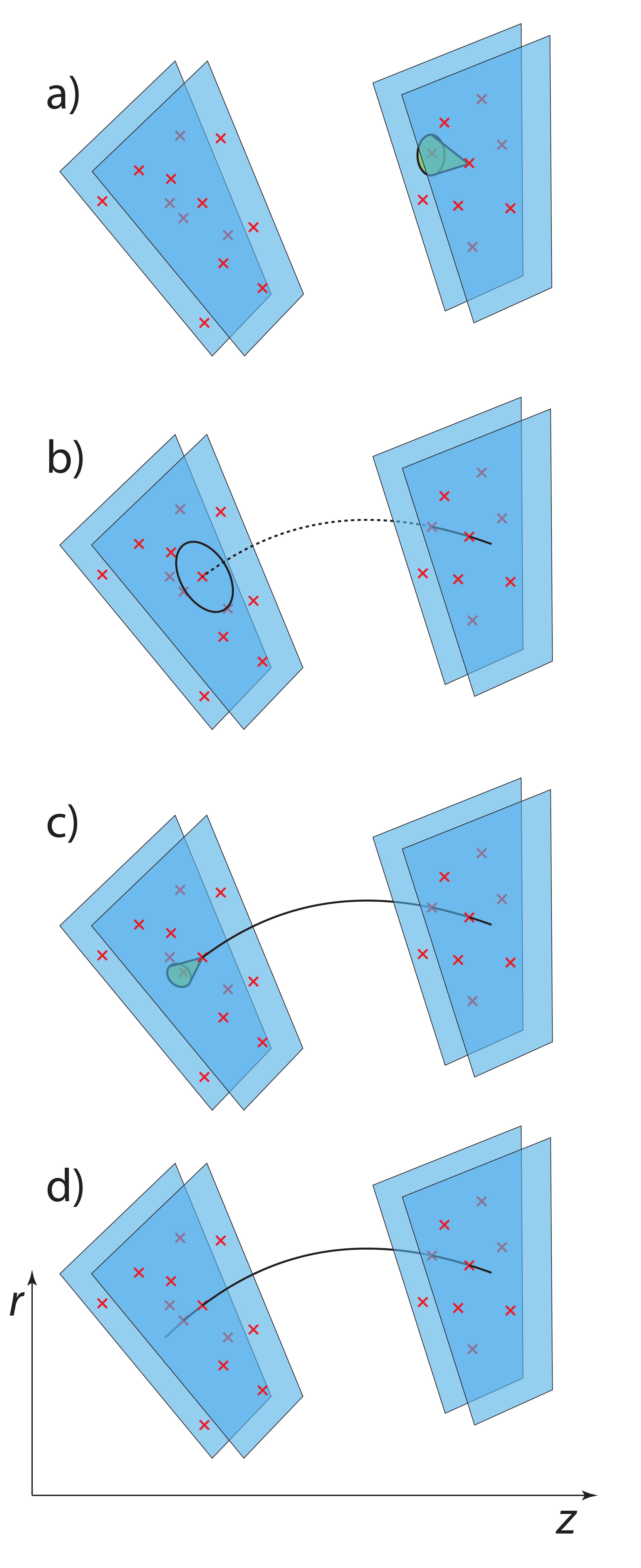}
	\caption{Schematic overview of the track reconstruction (see text for details).
	a) forming of hit pairs in the back module, b) extrapolation to the front module,
	c) validation using the first plane, d) track fit.}
	\label{fig:TrackReco}
\end{figure}

\begin{figure}
	\centering
		\includegraphics[width=0.49\textwidth]{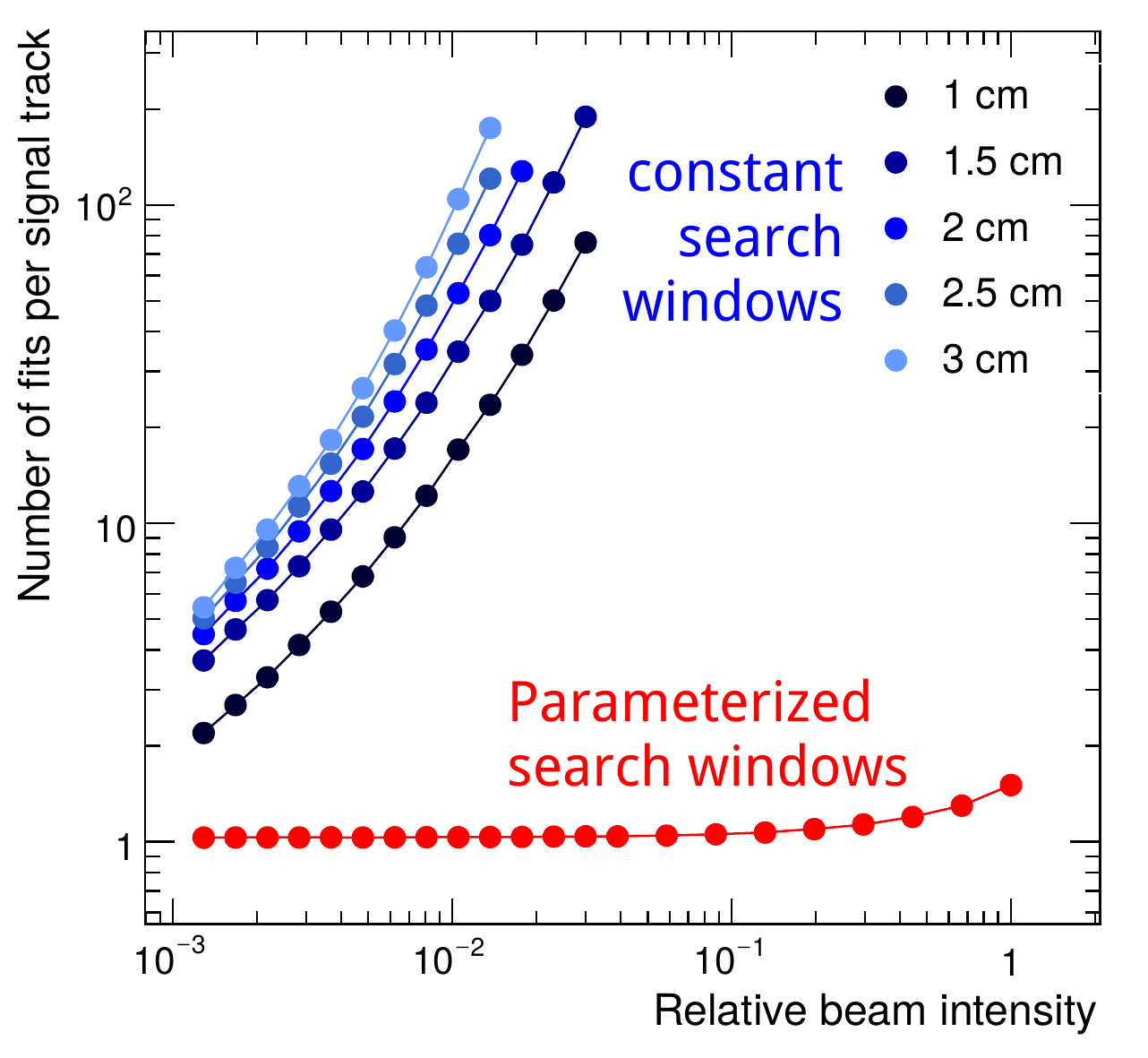}
	\caption{Number of track candidates as a function of beam intensity (relative to
	the nominal current of \SI{150}{\micro A}) using fixed size search windows and the
	parametrization technique described in the text.}
	\label{fig:tracking_n_candidates}
\end{figure}

\begin{figure}
	\centering
		\includegraphics[width=0.49\textwidth]{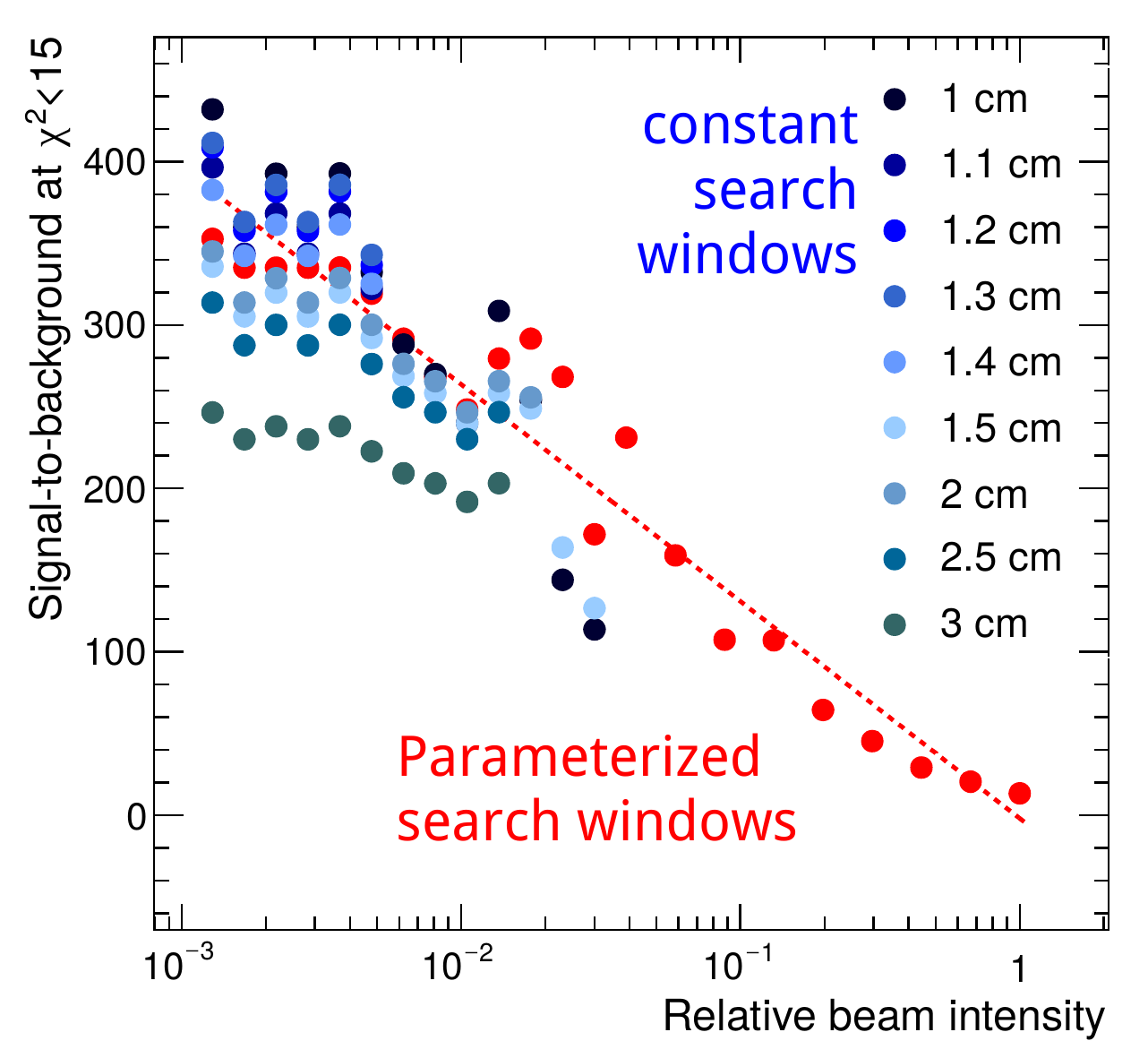}
	\caption{Signal to background ratio of track finding algorithms depending on 
	beam intensity (relative to the nominal current of \SI{150}{\micro A}).}
	\label{fig:tracking_sig_to_bg}
\end{figure}

\begin{figure}
	\centering
		\includegraphics[width=0.49\textwidth]{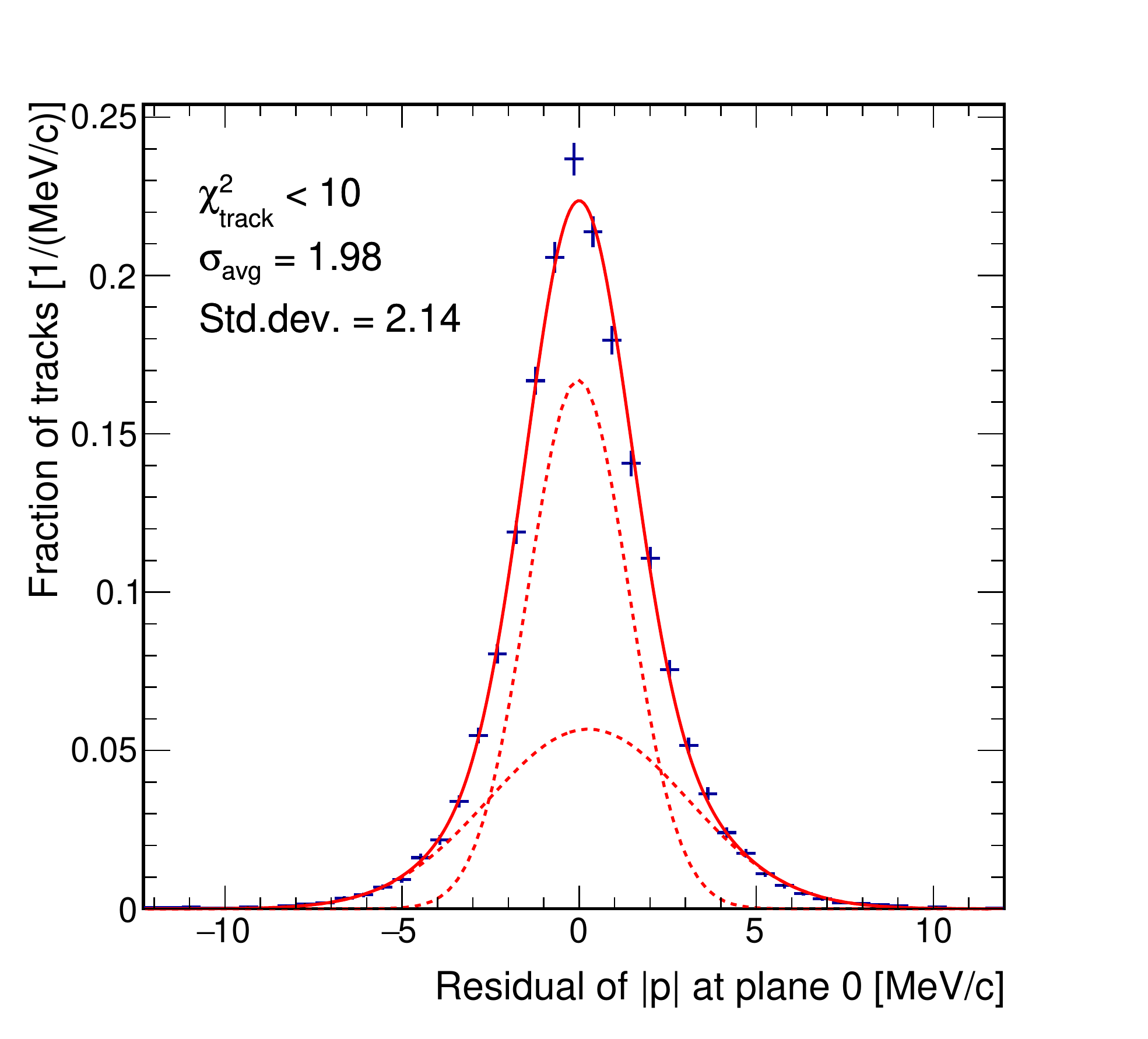}
	\caption{Reconstructed minus simulated absolute momentum of electron 
	tracks determined at the first tracking plane with requirement on the track fit
	$\chi^2$ of 10. The fit is the sum of two Gaussians, the resolution $\sigma$ is 
	the area-weighted mean.}
	\label{fig:pAbs_residuals_chi2_10}
\end{figure}

\begin{figure}
	\centering
		\includegraphics[width=0.49\textwidth]{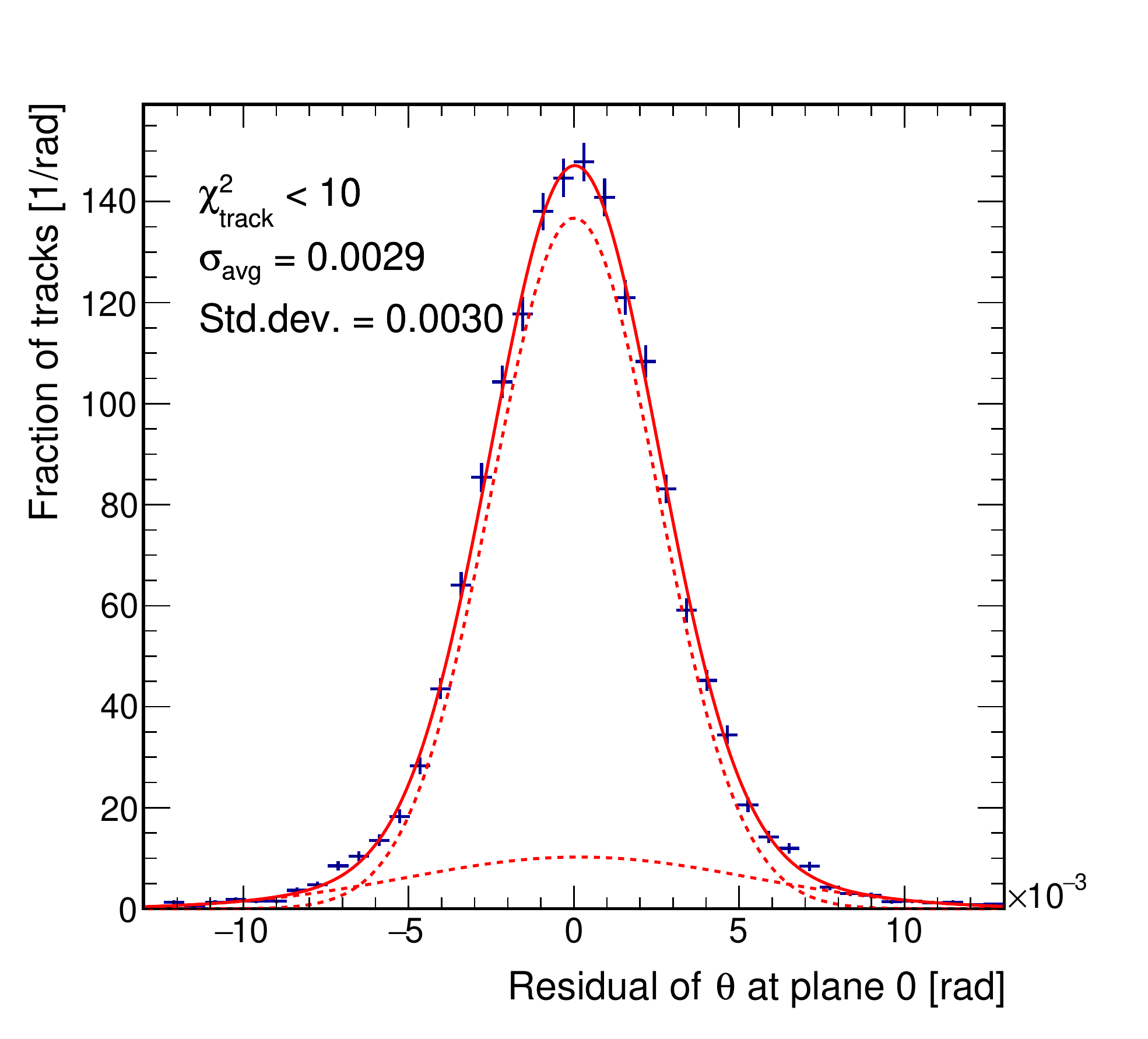}
	\caption{Reconstructed minus simulated polar angle of electron 
	tracks determined at the first tracking plane with requirement on the track fit
	$\chi^2$ of 10. The fit is the sum of two Gaussians, the resolution $\sigma$ is 
	the area-weighted mean.}
	\label{fig:pTheta_residuals_chi2_10}
\end{figure}

\begin{figure}
	\centering
		\includegraphics[width=0.49\textwidth]{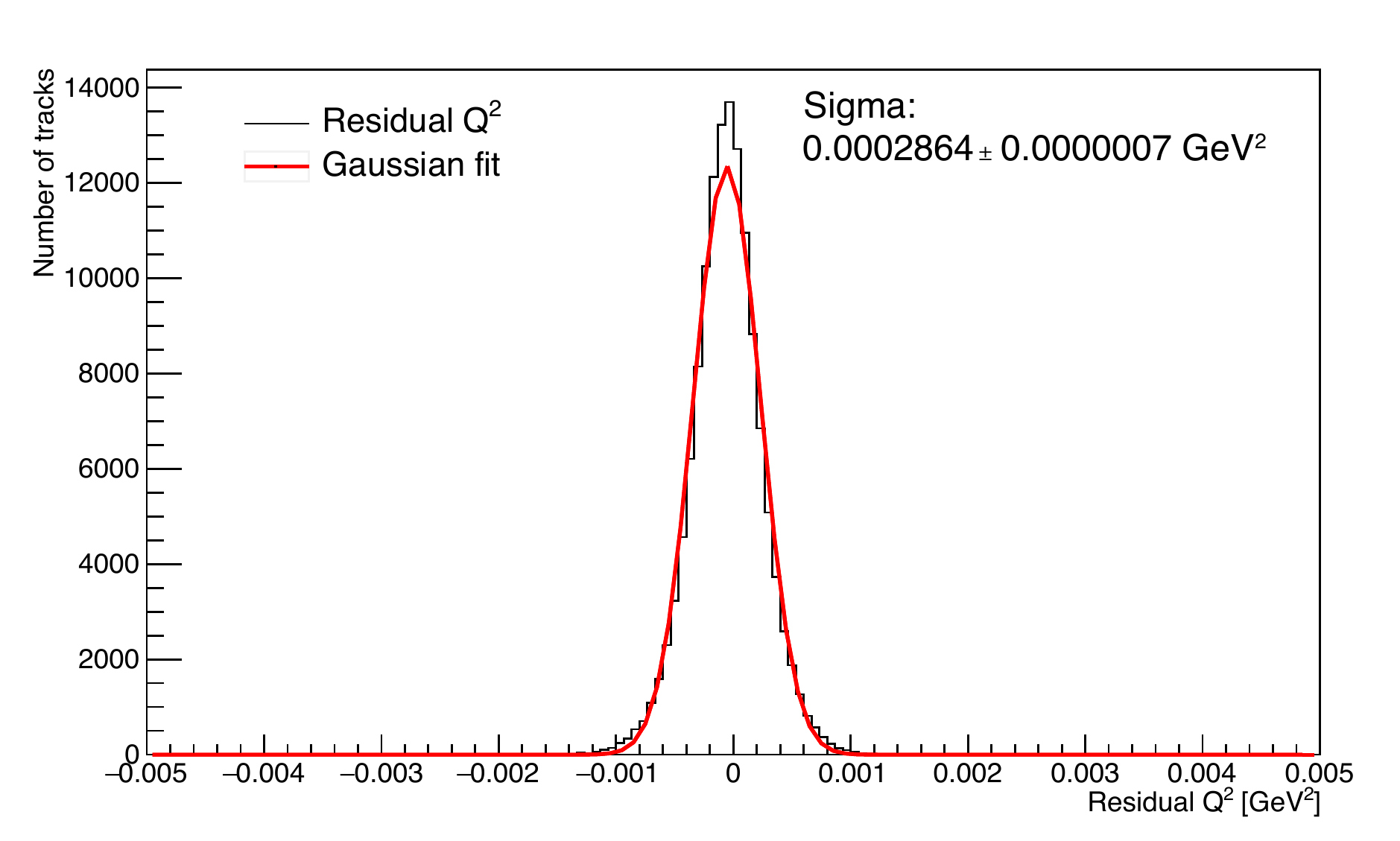}
	\caption{Reconstructed minus simulated $Q^2$ for electron tracks as determined 
	with the P2 tracking detector. }
	\label{fig:Q2ResidClean}
\end{figure}

The HV-MAPS sensors are active and, depending on their settings, dissipate 
between \SI{150}{mW/cm^2} and \SI{400}{mW/cm^2}. One module can thus produce
in excess of \SI{1}{kW} of heat, which needs to be actively cooled; any liquid
coolant would however add unacceptable amounts of material in the active region.
We therefore employ a gaseous Helium cooling with high flows in the V-folds, the
gap between the two planes in a module and over the module outside. Helium is
suitable due to its long scattering length, high mobility and high speed of sound
(allowing for large laminar flow speeds).
The piping
in the frame allows for counter-flowing helium streams in the two V-folds of 
each strip as well as on the top and bottom of a layer. We have performed 
extensive computational fluid dynamics simulations of the modules. With flow
velocities of the initially \SI{0}{\degreeCelsius} Helium of \SI{17}{m/s} in 
the V-folds and volume flows of \SI{6}{\liter/s} in between and on top of the 
layers, a mean temperature of \SI{49}{\degreeCelsius} (maximum: 
\SI{64.5}{\degreeCelsius}) can be achieved with \SI{400}{mW/cm^2} heating power, 
see Fig.~\ref{fig:Cooling}. The MuPix HV-MAPS 
sensors have been tested at temperatures above \SI{90}{\degreeCelsius} with only 
a very moderate amount of additional noise.

\subsubsection{Tracker readout}
\label{sec:TrackerReadout}

At nominal beam current, the tracker produces in the order of \SI{10}{Tbit/s} of
raw data, more than can be managed by an affordable readout system. We will thus
use a gated mode, where the sensors are only active for short time slices.

The HV-MAPS sensors send out zero suppressed hit data (column and row address,
time stamp) via a LVDS link with \SI{1.25}{Gbit/s}. The data are then sent out of the
active volume either using radiation resistant LVDS repeaters or multiplexed on
radiation hard optical links \cite{Schwemmer:2014tva,Lesma:2017oss,Soos:2017stv}.
Both options are currently under study in terms of power and space requirements
as well as signal integrity.

Outside of the radiation area, the signals are received by FPGA-based front-end
boards. The data streams are synchronized and the hits time-sorted (the MuPix
readout scheme introduces some randomization of the time-order of hits). The
double plane structure of the tracker is then used to form hit pairs and thus
reduce the data rate. These hit pairs are then forwarded to reconstruction PCs,
where track finding and fitting is performed. The geometry of the P2 
spectrometer and the limited momentum range of the tracks leads to a good 
locality of the tracking problem, reducing the need for complex data 
distribution networks.

\subsubsection{Track reconstruction}
\label{sec:TrackReconstruction}


\paragraph{Track finding.}
\label{sec:TrackFinding}

The high occupancy of the tracking detectors at the nominal beam current makes
track finding in P2 a potentially daunting task. We have developed an algorithm
progressing from regions with low-occupancy to regions with high-occupancy 
using optimized 
track-dependent search windows \cite{Sorokin:2017cyq} that elegantly manages the
combinatorial problems posed. Figure~\ref{fig:TrackReco} gives a schematic 
overview of the algorithm, which starts from the two tracking planes furthest
away from the target and finds hit pairs consistent with a track from the target. 
The target constraint and the vector connecting the two hits are then used to 
define an optimal search window on the second plane of the front module. 
Matching hits found there are then validated using the front plane. Finally the 
four hits are subjected to a track fit (see below) and the fit $\chi^2$ is used 
to accept or reject the track.
The performance of the algorithm in reducing the number of candidates reaching 
the fit state is shown in Fig.~\ref{fig:tracking_n_candidates}. The number of
wrongly combined tracks (fakes) is highly dependent on the beam rate, see
Fig.~\ref{fig:tracking_sig_to_bg} but even at the full rate, a signal to
background ratio above 10 can be achieved at a reasonable efficiency of 
\SI{85}{\percent}.

\paragraph{Track fitting.}
\label{sec:TrackFitting}

The best precision for the track parameters is obtained using a fit to the
measured hit positions.
We start by assuming that the track is a helix originating on the beam axis
and passing through the centers of gravity of the first two and the last two
tracker hits. This rough track, which assumes a constant magnetic field, is then
used as a seed in the general broken lines fit \cite{Blobel:2011az,Kleinwort:2012np}, 
which takes into account the hit position uncertainty and multiple Coulomb 
scattering at the two internal planes. The track state is propagated between 
planes using a Runge-Kutta-Nystr\"om integration of the (inhomogeneous) magnetic 
field.
From four 3D measurements, the fit
determines two global track parameters (momentum and polar angle), eight
shifts in the measurement planes and four scattering angle projections at the
internal planes. The required additional constraints are the zero expectation 
value of the scattering angle and its variance from multiple scattering theory.
In total there are 16 constraints and 14 fit parameters; the resulting track
$\chi^2$ can be used to select well reconstructed tracks with little multiple
scattering.

The achieved precision for 
the track parameters is shown in Figs.~\ref{fig:pAbs_residuals_chi2_10} and
\ref{fig:pTheta_residuals_chi2_10}: the resolution for the track momentum is
about \SI{2}{MeV/c}.

\paragraph{$\left\langle Q^2 \right\rangle$ reconstruction.}
\label{sec:Q2Reconstruction}

For elastic scattering of a beam of known energy and direction, the momentum
transfer can be determined from either the scattering angle or the outgoing
electron momentum. The very long liquid hydrogen target in P2 however leads to
energy losses and small angle scattering of both incoming and outgoing electrons,
such that even a very good knowledge of the polar angle and momentum of the
electron after leaving the target allows only a very approximate determination
of $\left\langle Q^2 \right\rangle$. The distributions for energy loss,
multiple scattering angles and detector acceptance are highly non-Gaussian and
tend to cause biases, which is also the case for the radiative corrections to
the large scattering (see Sect.~\ref{sec:QED}). The 
$\left\langle Q^2 \right\rangle$ reconstruction does thus have to rely on an 
iterative procedure using a Monte Carlo simulation including higher-order 
corrections, where not only the momentum and angle distributions, but also the 
longitudinal and radial position distributions of the point of closest approach 
need to be reproduced. 
We are still in the process of developing the respective algorithms
and study potential sources of bias; currently we achieve a resolution per track
of \SI{3e-4}{GeV^2} (\SI{4.3}{\percent}), see Fig.~\ref{fig:Q2ResidClean}.
This allows us to reach the required statistical precision with a few dozen 
tracks --- the $\left\langle Q^2 \right\rangle$ measurement will be completely
dominated by systematic uncertainties. We have started studying systematic
uncertainties arising due to detector misalignment and estimate them to be much
smaller than the required precision of \SI{1}{\percent} after a track-based
alignment. The dominating uncertainty is expected to arise from the MC 
description of multiple Coulomb scattering and energy loss in the target, which 
can only be addressed when beam data with the physical target become available.
In the meantime, we will continue our efforts to qualify simulation models using
test beam data, see e.g.~\cite{Berger:2014fsa}.

%% file: theory.tex

The interpretation of a high-precision measurement of the 
helicity asymmetry in ep scattering requires theory 
predictions with uncertainties below those of the 
experiment. In this section with describe the present 
status of the corresponding calculations. 

To leading order in the electroweak coupling constants, the 
amplitude $\mathcal{M}_\text{ep}^\pm$ for elastic scattering of 
electrons with helicity $\pm 1/2$ is given by the sum of two 
Feynman diagrams which are due to the exchange of one photon 
and one $Z$ boson, respectively,
\begin{equation}
  \mathcal{M}_\text{ep}^\pm = 
  \mathcal{M}_\gamma + \mathcal{M}_Z^\pm.
\end{equation}
While the $\gamma$ exchange is parity-conserving, the 
$Z$ exchange contains parity-violating contributions which 
flip sign depending on the helicity.  

At very small elastic momentum transfer $Q^2$ the internal 
structure of the proton is not resolved, and the two amplitudes 
only depend on the proton's electric and weak charges,  
$e_\text{p}=+1$ (in units of the positron charge) and 
$Q_\text{W}(\text{p})$, 
respectively, 
\begin{eqnarray}
\mathcal{M}_\text{ep}^\pm \sim \frac{1}{Q^2}
\mp\frac{Q_\text{W}(\text{p})}{16\sin^2\theta_W\cos^2\theta_W} 
\, \frac{1}{Q^2 + m_Z^2},
\label{eq:Mep_pm}
\end{eqnarray}
with $m_Z = \SI{91.188 \pm 0.002}{GeV}$ the mass of the 
$Z$ boson and $\theta_W$ the weak mixing angle.  The 
proton's weak charge at tree-level is related to the 
weak mixing angle, $Q_\text{W}(\text{p}) = 1 - 4\sin^2\theta_W$. 

The parity-violating asymmetry measures the difference 
between the cross sections for electrons with opposite 
helicities, 
\begin{equation}
A^\text{PV} 
= 
\frac{|\mathcal{M}_\text{ep}^+|^2-|\mathcal{M}_\text{ep}^-|^2}%
{|\mathcal{M}_\text{ep}^+|^2+|\mathcal{M}_\text{ep}^-|^2},
\end{equation}
and for $Q^2 \ll m_Z^2$ it is a small quantity which arises 
from the interference of virtual $\gamma$ and $Z$ exchange. 
Evaluating this asymmetry using Eq.~(\ref{eq:Mep_pm}) for 
very low $Q^2$ and introducing the Fermi constant 
\begin{equation}
G_{\rm F} 
= 
\frac{\pi\alpha_{\rm em}}{\sqrt2 m_W^2 \sin^2\theta_W},
\end{equation}
we obtain 
\begin{equation}
A^\text{PV}(Q^2\to0) 
= - \frac{G_{\rm F}Q^2}{4\pi\sqrt2\alpha_{\rm em}} 
Q_\text{W}(\text{p}).
\end{equation}
The direct proportionality between the PV asymmetry and the 
proton's weak charge constitutes the basis of the P2 experiment. 

To match the precision of the experimental measurement of 
$A^\text{PV}$, one has to go beyond the tree-level approximation 
and include radiative corrections described by Feynman 
diagrams with loops. They generically scale as $\alpha_{\rm em}/\pi 
= O(10^{-3})$, but may be enhanced by logarithms or large 
numerical factors, as, e.g., in the case of the $WW$ box. 
Moreover, box diagrams are in general functions of two 
kinematical variables, $E_i$ and $Q^2$, which requires 
additional caution when relating the PV asymmetry to 
the proton's weak charge.

We follow Refs.\ \cite{Gorchtein:2011mz,Erler:2013xha} to 
define the weak charge as 
\begin{equation}
Q_\text{W}^{\text{1-loop}}(\text{p}) =  \lim_{E_i \to 0} 
\lim_{Q^2 \to 0}\frac{A^\text{PV}}{A_0}
\end{equation}
with $A_0 = - G_\text{F} Q^2 / (4\sqrt{2}\pi\alpha_{\text{em}})$. 

The measurement of the PV asymmetry in PVES realized at finite 
energy $E_i$ and finite momentum transfer $Q^2$ can be cast in 
the following form that generalizes Eq.~(\ref{eq:Apv}) to 
include one-loop effects,
\begin{eqnarray}
A^\text{PV} 
&=& 
A_0 \left[Q_\text{W}^{\text{1-loop}}(\text{p}) - F(E_i,Q^2) 
\right.\nonumber\\
&&\left.\;\;\;\;\;
+ \Delta_\Box(E_i,Q^2) - \Delta_\Box(0,0) \right],
\label{eq:Apv_full}
\end{eqnarray}
where, for the sake of completeness, we keep the 
kinematically suppressed term $F(E_i,Q^2)$ introduced earlier 
in Eq.~\eqref{eq:Apv}. 

The one-loop SM result for $Q_W(p)$ has been formulated 
in Ref.\ \cite{Erler:2003yk} in the $\overline{\text{MS}}$ 
scheme, and Eq.\ (\ref{eq:QWp_tree_level}) is replaced by  
\begin{equation}
  Q_W^{\text{1-loop}}(p) =  
  (\rho_{nc} + \Delta_e) \left(1 - 4 \sin^2\hat\theta_W(\mu) +
  \Delta_e'\right) + \Delta_\Box (0,0)
  \label{eq:Qwp_with_corrections}
\end{equation}
where $\hat\theta_W(\mu)$ is the weak mixing angle defined 
in the $\overline{\text{MS}}$ scheme at scale $\mu$, where 
$\mu \simeq 0$ for the P2 experiment. In this equation, the 
Veltman parameter $\rho_{nc}$ is a universal correction which 
renormalizes the ratio of the neutral and charged current 
strengths at low energies. $\Delta_e$ and $\Delta_e'$ are 
small, non-universal corrections at the electron vertex. 
The term $\Delta_\Box$ in Eq.~(\ref{eq:Qwp_with_corrections}) 
represents the contributions to $Q_\text{W}(\text{p})$ from 
box graphs and is the subject of the next subsection. 

The scale dependence of the $\overline{\text{MS}}$ weak 
mixing angle has been studied in Ref.~\cite{Erler:2004in} 
(see also \cite{Erler:2017knj} for a recent update). The 
value of $\sin^2 \hat\theta_W(\mu)$ at low momentum transfer 
is related by 
\begin{equation}
  \sin^2\hat\theta_W(0) 
  =  
  \kappa(m_Z) \cdot \sin^2\hat\theta_W(m_Z),
\end{equation}
to its value at the $Z$ pole (with $\kappa(m_Z) = 1.0317$ 
for the present values of the SM parameters). 

\begin{figure}[t!]
  \begin{minipage}[t]{0.2\linewidth}
    \centering
    \resizebox{1\textwidth}{!}{\includegraphics{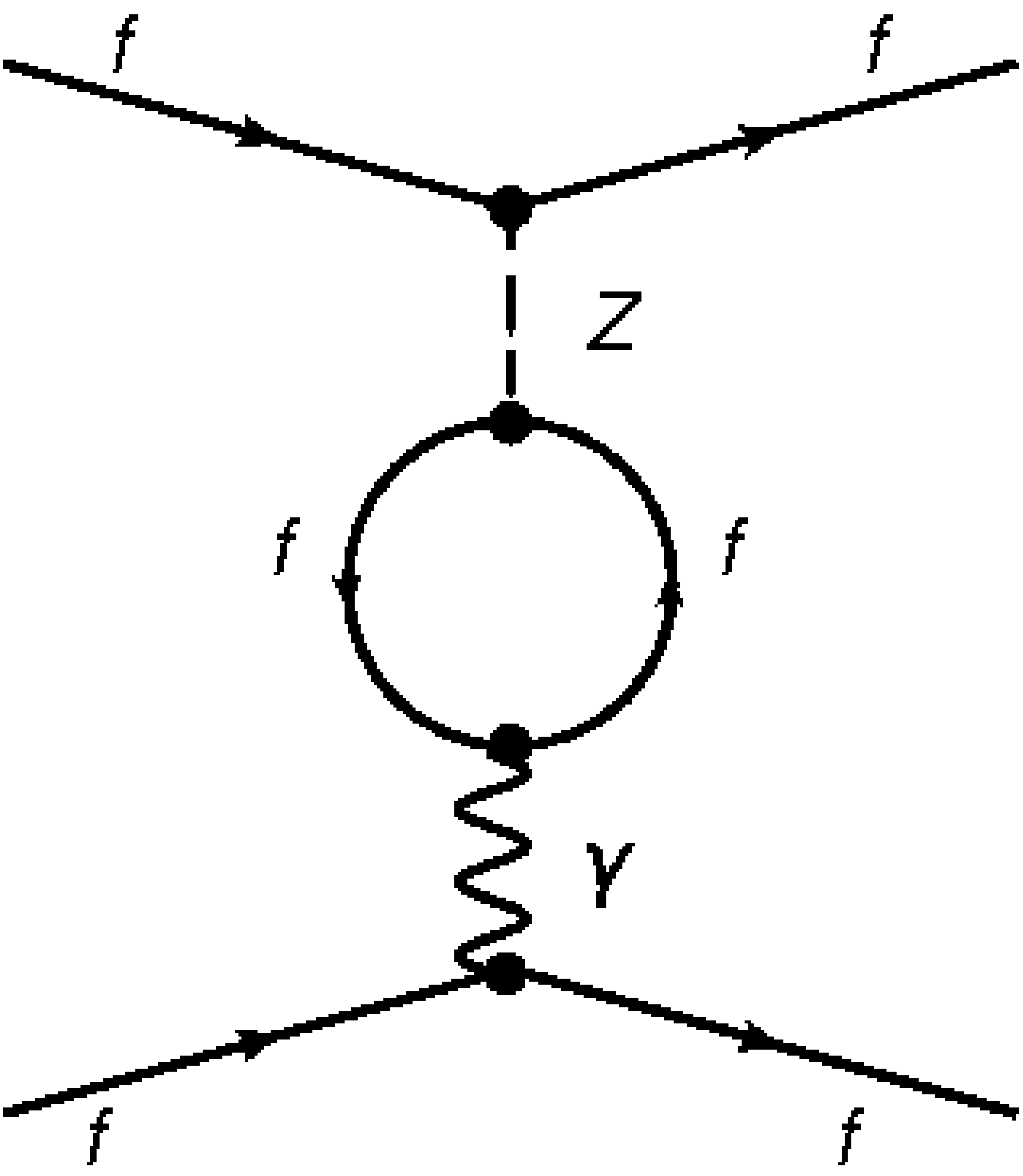}}
  \end{minipage}
  \hfill
  \begin{minipage}[t]{0.2\linewidth}
    \centering
    \resizebox{1\textwidth}{!}{\includegraphics{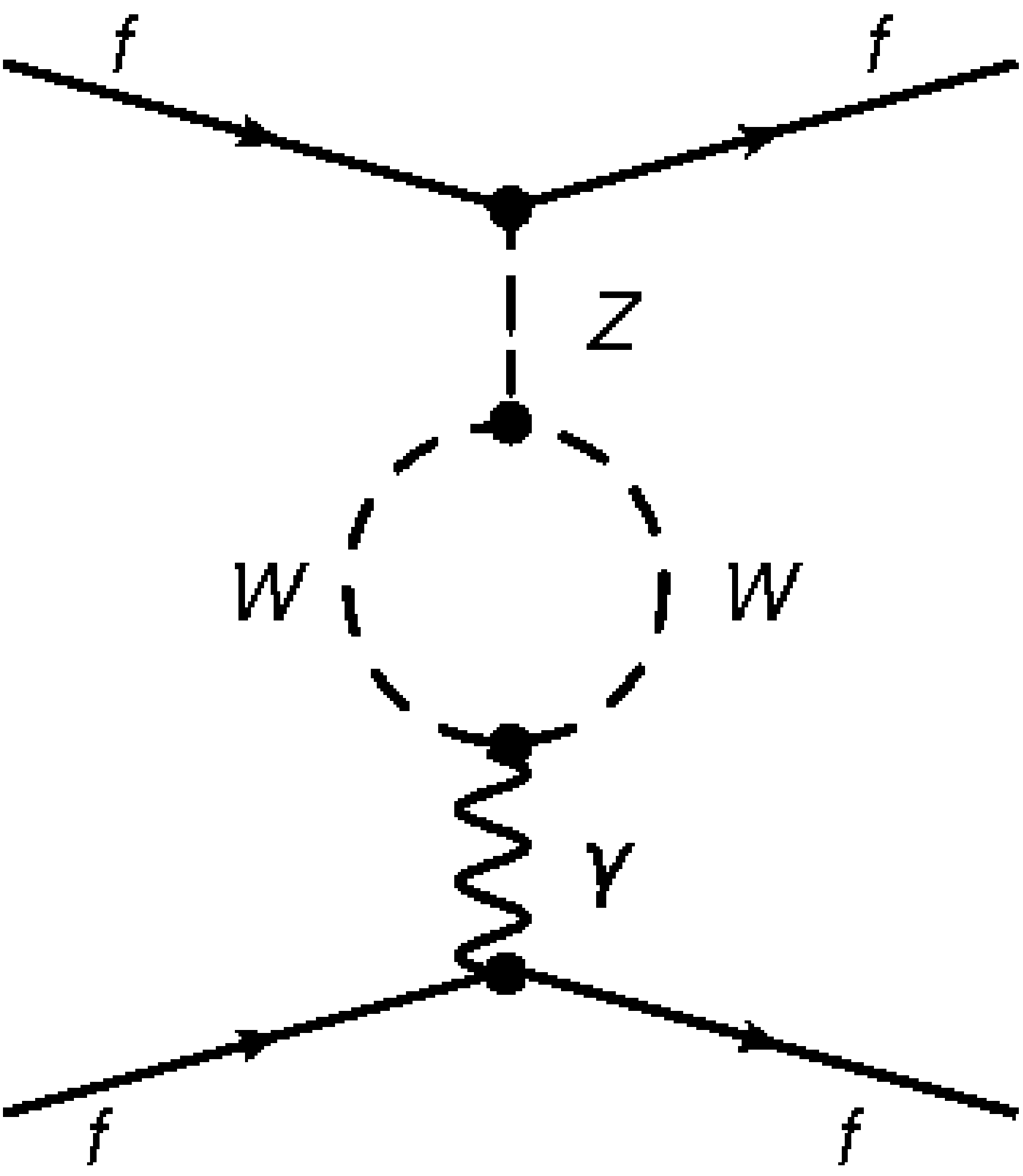}}
  \end{minipage}
  \hfill
  \begin{minipage}[t]{0.2\linewidth}
    \centering
    \resizebox{1\textwidth}{!}{\includegraphics{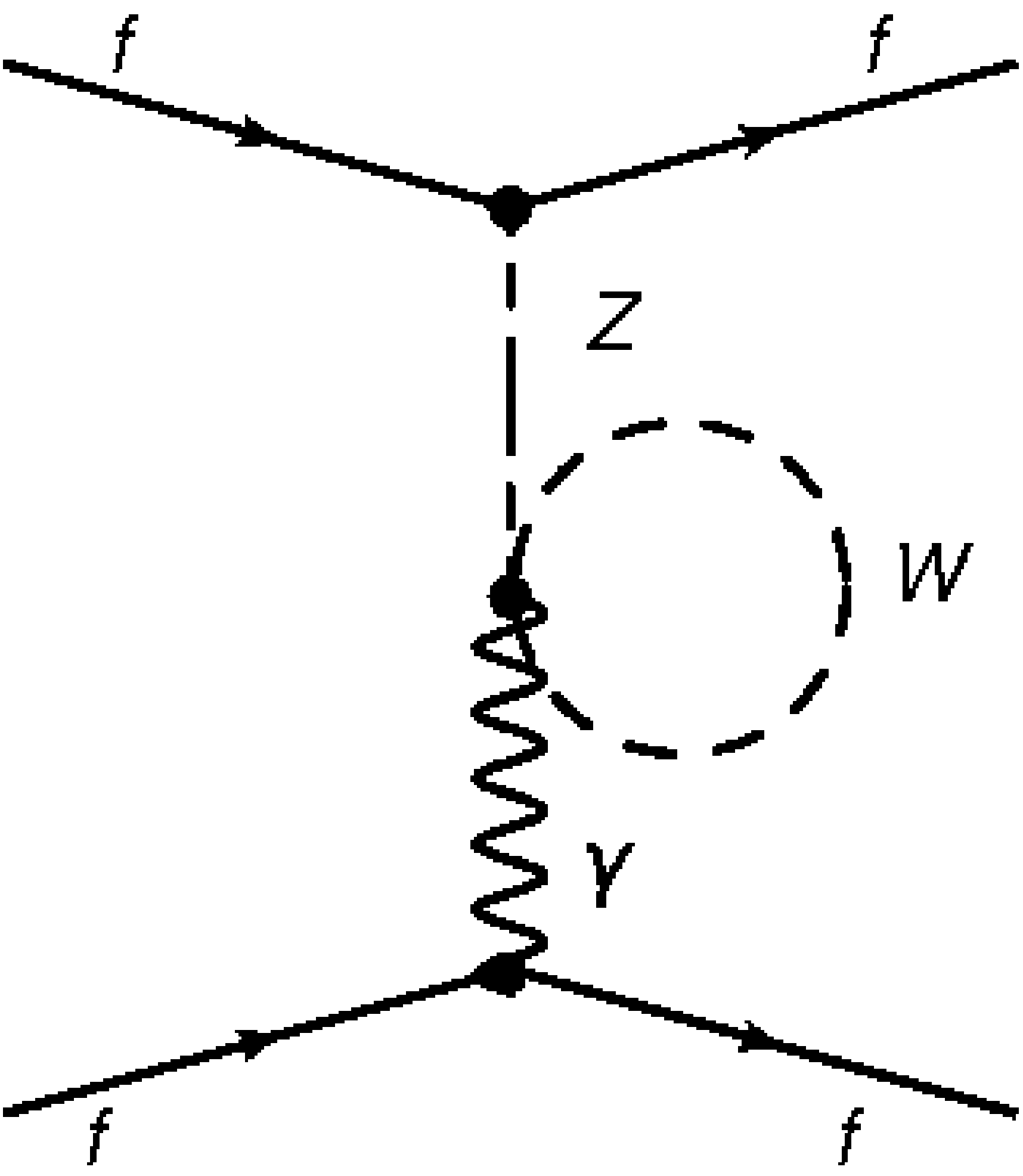}}
  \end{minipage}
  \hfill
  \begin{minipage}[t]{0.2\linewidth}
    \centering
    \resizebox{1\textwidth}{!}{\includegraphics{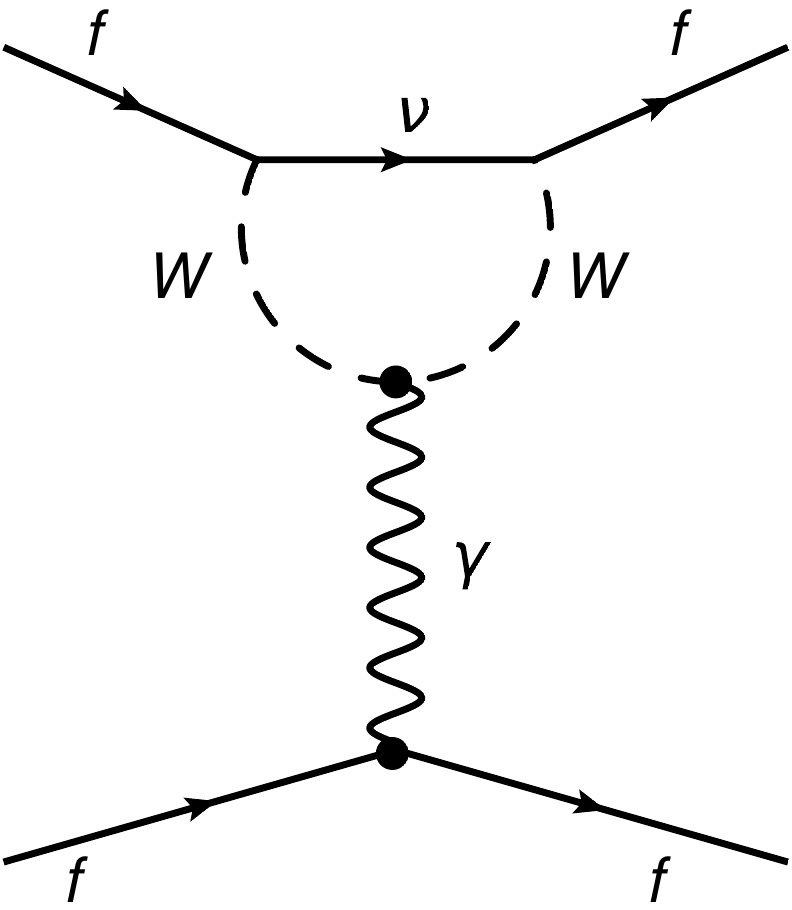}}
  \end{minipage}
  \caption{
  Feynman graphs of universal and non-universal electroweak 
  corrections which lead to an energy scale dependence of 
  $\sin\theta_{W, \text{eff}}$.}
  \label{fig:Electroweak_corrections_to_sin2ThetaW}
\end{figure}

In the $\overline{\text{MS}}$ scheme the scale dependence 
of the weak mixing angle is determined by the renormalization 
group evolution of the SM coupling constants. Other definitions 
of a scale-dependent effective weak mixing angle exist in the 
literature, see for example \cite{Czarnecki:2000ic}. They 
are based on a redefinition of $\sin\theta_W$ which absorbs 
universal and partly non-universal one-loop corrections into 
an effective weak mixing angle $\sin\theta_{W, \text{eff}}$. 
Figure~\ref{fig:Electroweak_corrections_to_sin2ThetaW} shows 
some typical Feynman diagrams contributing to the scale 
dependence of the effective weak mixing angle. 

%% file: boxgraph.tex

\begin{figure}[b!]
\includegraphics[width=8.5cm]{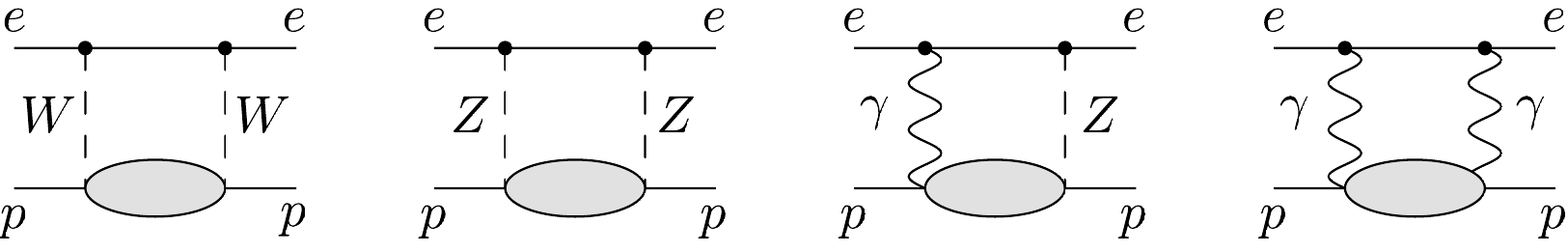}
  \caption{
  Electroweak box corrections to parity-violating 
  ep scattering. Shown are, from left to right, $WW$-, 
  $ZZ$-, $\gamma Z$- and $\gamma\gamma$-exchange diagrams, 
  respectively.
  Contributions with crossed boson lines are not displayed. 
  The grey blob at the lower part of each diagram denotes 
  inclusive hadronic intermediate states.
  }
  \label{fig:Boxgraph_corrections_to_Qw}
\end{figure}

The one-loop result of Eqs.~(\ref{eq:Apv_full}, \ref{eq:Qwp_with_corrections}) 
singles out the two-boson exchange contributions, 
\begin{equation}
  \Delta_\Box 
  \equiv 
  \Box_{WW} +
  \Box_{ZZ} + 
  \Box_{\gamma Z} + 
  \Box_{\gamma\gamma} ,
\end{equation}
representative Feynman diagrams of which are displayed in 
Fig.~\ref{fig:Boxgraph_corrections_to_Qw}. For each box 
graph, only the real part contributes and $\Box_{ab}$ is 
understood as the real part of the corresponding diagram 
here and in the following. The box graphs are specific and 
have to be added as separate contributions since they depend 
on both the 4-momentum transfer $Q^2$ and the electron 
energy. Other one-loop corrections depend on $Q^2$ only 
and can therefore be factorized and partly absorbed into 
universal correction factors as shown above in 
Eq.~(\ref{eq:Qwp_with_corrections}). 

It has been observed in 
Refs.~\cite{Marciano:1982mm,Marciano:1983ss,Musolf:1990ts} 
that the energy dependence of the heavy-boson box graphs 
associated with $WW$ and $ZZ$ exchange induces corrections of 
order $G_\text{F} E_i^2$, rather than $\sim\alpha_\text{em}/\pi$. 
For electron energies up to a few GeV these energy-dependent 
contributions can be safely neglected. The constant terms, 
however, are numerically large. Since they are dominated by 
contributions from loop momenta of the order of $m_Z$, their 
calculation in the framework of perturbation theory is safe 
with a reliable uncertainty estimate \cite{Erler:2003yk}. 

The $\gamma\gamma$ box does not contain large logarithms 
and is known to vanish at small momentum transfer as it can 
only renormalize the charge radius of the proton but not its 
charge. Since it only corrects the parity-conserving part 
of the amplitude, its effect on the PV asymmetry will also 
be multiplied by the proton's weak charge. All in all, it is 
natural to expect a correction to $A^\text{PV}$ of the order of 
$(\alpha_\text{em}/\pi) (Q^2/E_i^2)Q_\text{W}(\text{p})$ due 
to $\gamma\gamma$-box graphs. This amounts to a negligible 
correction of order $O(10^{-5})$ for the kinematical 
conditions at the P2 experiment that can be accommodated in 
the uncertainty associated with the kinematically suppressed 
correction term $F(E_i, Q^2)$. With these observations, the 
energy dependence of the boxes present in Eq.~(\ref{eq:Apv_full}) 
reduces to that of the $\gamma Z$ box,
\begin{equation}
\Delta_\Box(E_i,Q^2) - \Delta_\Box(0,0) 
= 
\Box_{\gamma Z}(E_i,Q^2)-\Box_{\gamma Z}(0,0).
\label{eq:BoxgZ-Edep}
\end{equation}

The $\gamma Z$-box graph contains a large logarithm 
$\log\frac{m_Z^2}{\Lambda^2}$ where $\Lambda\sim$ \SI{1}{GeV} 
is a typical hadronic mass scale. The coefficient in front of 
this large logarithm is energy-inde\-pen\-dent up to corrections 
$\sim G_\text{F} E_i^2$  and can be calculated precisely using quark 
sum rules \cite{Marciano:1982mm,Marciano:1983ss}. The presence 
of the hadronic mass scale $\Lambda$ signals the sensitivity 
of the $\gamma Z$ box to the hadronic structure, and this 
sensitivity was used to estimate the hadronic structure-related 
uncertainty~\cite{Erler:2003yk}. However, early studies 
described in the references given above had assumed that 
the energy dependence of the $\gamma Z$ box was negligible, 
$\sim G_\text{F} E_i^2$, following the pattern of the heavy 
boson boxes. 

\begin{figure}[t!]
  \centering
  \resizebox{0.48\textwidth}{!}{\includegraphics{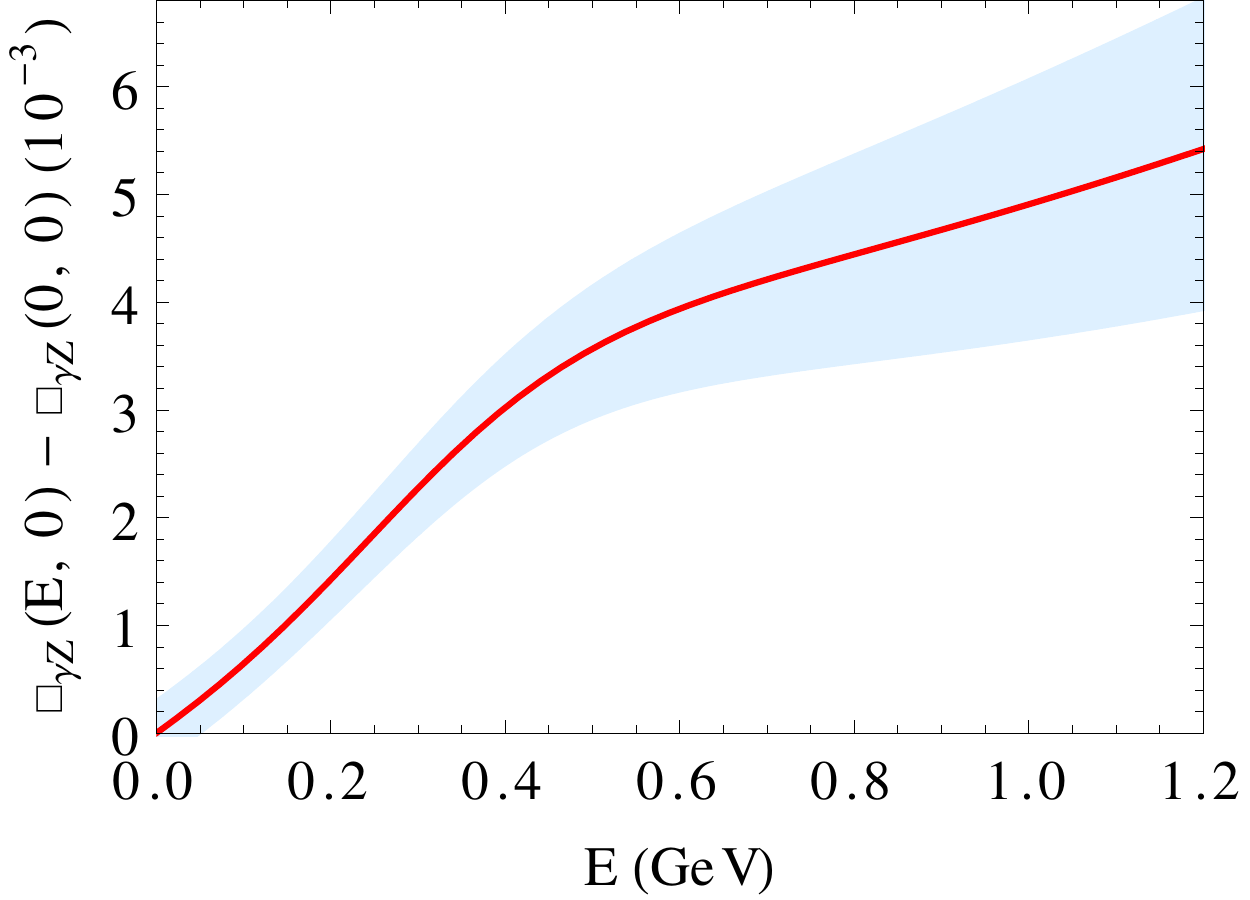}}
  \caption{
  Energy dependence of the $\gamma Z$ box graph, 
  Eq.~(\ref{eq:BoxgZ-Edep}) at $Q^2=0$, and its uncertainty band.  
  }
  \label{fig:BoxGammaZ_vs_Ebeam}
\end{figure}

Subsequently, the energy dependence of the $\gamma Z$ box 
was addressed in Ref.~\cite{Gorchtein:2008px} in the 
framework of forward dispersion relations. It was shown that 
the energy dependence of $\Box_{\gamma Z}$ is much more 
significant than anticipated. It has been the subject of 
active scrutiny in the theory community
\cite{Gorchtein:2011mz,Sibirtsev:2010zg,Rislow:2010vi,Blunden:2011rd,Gorchtein:2015naa,Hall:2015loa}. 
The dispersive method for calculating $\Box_{\gamma Z}$ is per 
se model-independent, relating the $\gamma Z$ box to an integral 
over measurable unpolarized interference structure functions 
$F_{1,2,3}^{\gamma Z}$. Nonetheless, due to the lack of reliable 
experimental data for these structure functions one is 
forced to introduce model assumptions to define the required 
input in unmeasured regions. While different groups agree on 
the central value of $\Box_{\gamma Z}(E_i)$ within errors, 
this model dependence leads to a discrepancy in the uncertainty 
estimate.

In Fig.~\ref{fig:BoxGammaZ_vs_Ebeam} the energy dependence 
of the $\gamma Z$ box is shown. It is obtained as a sum of 
its vector part $\Box_{\gamma Z}^V$ calculated in 
Ref.~\cite{Gorchtein:2015naa} and its axial-vector part, 
$\Box_{\gamma Z}^A$ obtained in 
Ref.~\cite{Zhou:2009nf,Blunden:2011rd,Gorchtein:2016qtl} at zero 
momentum transfer. The respective uncertainties are added in 
quadrature. The extrapolation from the actual value of $Q^2$ 
corresponding to the kinematics at P2 down to $Q^2=0$ is done 
according to Ref.~\cite{Gorchtein:2011mz}. Due to the tiny 
value of $Q^2\approx0.0045$ GeV$^2$ this extrapolation leads 
to a numerically negligible effect, both on the central value 
and its uncertainty. For the kinematics at P2, the 
energy-dependent contribution amounts to 
\begin{eqnarray}
&& 
\Box_{\gamma Z}(E_i = \SI{155}{MeV}, Q^2 = 0) 
 - \Box_{\gamma Z}(0, 0)
\nonumber \\ && 
~~~~~ = (1.06 \pm 0.32) \times 10^{-3}  
\end{eqnarray}
and the uncertainty is dominated by that due to the effective 
axial charge of the nucleon seen by charged leptons, also 
referred to as the anapole moment, 
\begin{eqnarray}
\delta \Box_{\gamma Z}^A &=& 0.27 \times 10^{-3},  
\\
\delta \Box_{\gamma Z}^V &=& 0.18 \times 10^{-3}. 
\end{eqnarray}
A measurement at backward angles as described in section 
\ref{sec:BackwardAngle} will allow to reduce the uncertainty 
due to the anapole moment considerably, $\delta \Box_{\gamma Z}^A 
\to 0.07 \times 10^{-3}$. Assuming that this precision goal is 
achieved, the energy-dependent correction from the $\gamma Z$ box 
will change to 
\begin{eqnarray}
&& 
\Box_{\gamma Z}(E_i = \SI{155}{MeV}, Q^2 = 0) 
 - \Box_{\gamma Z}(0, 0)
\nonumber \\ && 
~~~~~ = (1.06 \pm 0.19) \times 10^{-3}  
\end{eqnarray}
with a reduced uncertainty. This estimate was used in 
Sect.~\ref{sec:SinThetaW}, Tab.~\ref{tab:achievable_precision} 
in the summary of the uncertainty budget.

%% file: qedcorrections.tex

Electromagnetic corrections are parity conserving and do 
not affect the proton's weak charge. However, the relation 
to the measured helicity asymmetry $A^\text{PV}$ receives 
corrections since extra radiated photons lead to a shift of 
the observed momentum transfer relative to the true one. 
$Q^2$ can not be determined from the electron scattering 
angle alone, but the momentum of unobserved photons has to 
be taken into account.  

The tracking detectors described in 
Sect.~\ref{sec:TrackingDetectors} will allow one to determine 
the momentum of the scattered electron, i.e., its energy $E_f$ 
and the electron scattering angle $\theta_f$. From this 
information one can determine $Q^2 = - (k - k')^2$ 
(corresponding to Eq.~(\ref{eq:Q2_electron})), where 
$k$ and $k'$ are the momentum 4-vectors of the initial and 
final electron. In the presence of bremsstrahlung, a photon 
with 4-momentum $k_\gamma$ emitted from the electron will 
shift $Q^2$ to the true momentum transfer $Q^2_{\text{true}} 
= - (k - k' - k_\gamma)^2$. This true $Q^2$ value has to be 
used in the equation relating the measured asymmetry with 
the proton's weak charge, Eq.~(\ref{eq:Apv}).

\begin{figure}[tb]
  \centering
  \resizebox{0.4\textwidth}{!}{\includegraphics{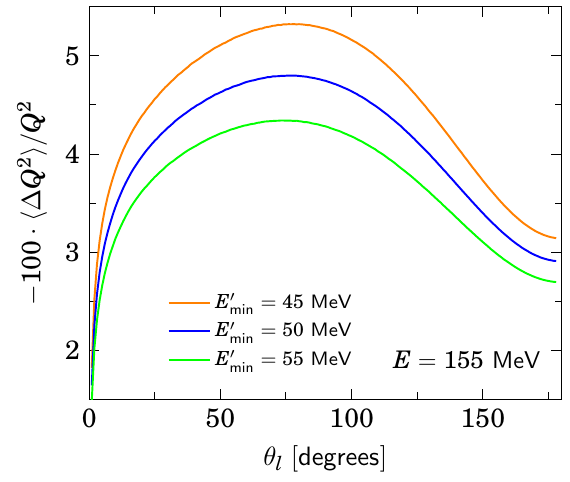}}
  \caption{
  The average relative shift of the momentum transfer due 
  to photon radiation.}
  \label{fig:qed-shiftQ2}
\end{figure}

In Fig.~\ref{fig:qed-shiftQ2} we show the average relative 
shift of $Q^2$ due to photon radiation including one-photon 
bremsstrahlung. The $Q^2$ shift depends strongly on the beam 
energy and the scattering angle, as well as on a possible 
cutoff of the energy of photons radiated into the final 
state. Preliminary results of a calculation including 
two-photon radiation show that order $O(\alpha^2)$ corrections 
are much less important. This can be understood since we deal 
with a kinematic effect: the relation between scattering angle 
and momentum transfer is not unique anymore in the presence of 
an additional photon. A second photon does not add considerably 
more freedom, but only adds corrections of order $O(\alpha^2)$ 
to the cross section. 
Eventually, bremsstrahlung effects will be included in the 
detector simulation.

%% file: theorysummary.tex

The current SM prediction for the parity-violating 
asymmetry in elastic ep scattering, summarized in this 
section, includes the complete set of NLO corrections. 
At this order, no further theoretical uncertainties will 
affect the interpretation of a high-precision measurement 
of $A^{\text{PV}}$ in terms of $\sin^2 \theta_\text{W}$ 
beyond the level of $10^{-4}$. In particular, uncertainties 
from the hadronic structure entering the $\gamma Z$-box 
graph at the low energy and small $Q^2$ values relevant 
for the P2 experiment are well under control. 

A conclusive test of the SM and analyses of the anticipated 
experimental result at P2 in terms of New Physics can be 
obtained by comparing with other high-precision determinations 
of the weak mixing angle from LEP, SLC, and future measurements 
at the LHC. This will require the inclusion of two-loop 
electroweak corrections and work on the corresponding NNLO 
calculations is underway.

%% file: carbon.tex

High precision measurements of the weak charges of different 
particles and nuclei offer complementary sensitivities to 
physics beyond the Standard Model in the form of new tree 
level and quantum loop correction parameters. For example, 
one may consider ratios of polarization asymmetries in which 
the polarization uncertainty mostly cancels. 

Here we summarize the results of first feasibility studies 
of a measurement of the weak charge of the $^{12}$C nucleus 
with the P2 setup. As a spin-zero nucleus, $^{12}$C 
can be described by a single form factor and is thus 
theoretically easy to handle \cite{Souder:1990ia}. 
Moreover, its QED cross section is 36 times larger than that 
of the proton, and its weak charge is 78 times as large, which 
significantly reduces beam time requirements. The SM prediction 
for the helicity asymmetry at leading order can be written as 
in Eq.~(\ref{eq:Apv}) with 
\begin{equation}
Q_\text{W}({}^{12}\text{C}) = - 24 \sin^2 \theta_W . 
\end{equation}

At low momentum transfer $Q^2\ll M^2_Z$ weak charges can be 
parametrized with respect to the so-called oblique parameters, 
such as the $S$, $T$ and $U$ parameters introduced by Peskin 
and Takeuchi~\cite{Peskin:1990zt,Peskin:1991sw}.
However, $S$, $T$ and $U$ are already very precisely determined 
from $Z$-pole observables, whereas weak charges are able to 
constrain some of the higher-order oblique parameters. For 
example, the $X$ parameter~\cite{Burgess:1993mg} describes 
the difference of new physics contributions to the $\gamma Z$ 
mixing at the $Z$ pole and at low energies, and cannot 
be determined by $Z$-pole physics alone. Likewise, in the 
absence of mass mixing, $Z$-pole observables are virtually 
blind to extra heavy gauge bosons $Z^\prime$ as new amplitudes 
are suppressed relative to the $Z$ resonance. By contrast, at 
low energies $Z^\prime$ amplitudes are merely suppressed by 
the square of the ratio of the $Z$ and $Z^\prime$ masses.
One has~\cite{Kumar:2013yoa} 
\begin{equation}
\begin{array}{l}
Q_\text{W}({}^{12}\text{C}) 
= -5.510[1 - 0.003 T + 0.016 S - 0.033 X - \chi ] , 
\\[1ex]
Q_\text{W}(\text{p}) 
= +0.0707[1 + 0.15 T - 0.21 S + 0.43 X + 4.3 \chi ] , 
\\[1ex]
Q_\text{W}(\text{e}) 
= -0.0435[1 + 0.25 T - 0.34 S + 0.7 X + 7 \chi ] , 
\\[1ex]
Q_\text{W}({}^{133}\text{Cs}) 
= -73.24[1 + 0.011 S - 0.023 X - 0.9 \chi ]
\end{array}\end{equation}
for the weak charges of ${}^{12}\text{C}$, the proton, the 
electron and ${}^{133}\text{Cs}$, and $\chi = M^2_Z /M^2_{Z_\chi}$ 
was used where $Z_\chi$ is the extra 
$Z$ boson predicted by $SO(10)$ Grand Unified Theories (in the 
absence of gauge kinetic mixing). The different pre-factors in 
this parametrization show the complementarity of the different 
weak charges to physics beyond the Standard Model. Low-$Q^2$ 
measurements also have unique sensitivity to certain beyond 
the Standard Model scenarios such as those involving so-called
dark $Z$ bosons~\cite{Davoudiasl:2012ag}, which are light (on 
the order of tens of MeV) and very weakly coupled extra neutral 
gauge bosons which may mix with the ordinary $Z$ boson, and 
which may be parametrized by taking $X$ as a function of $Q^2$. 

\begin{figure}
\centering
\includegraphics[width=0.48\textwidth]{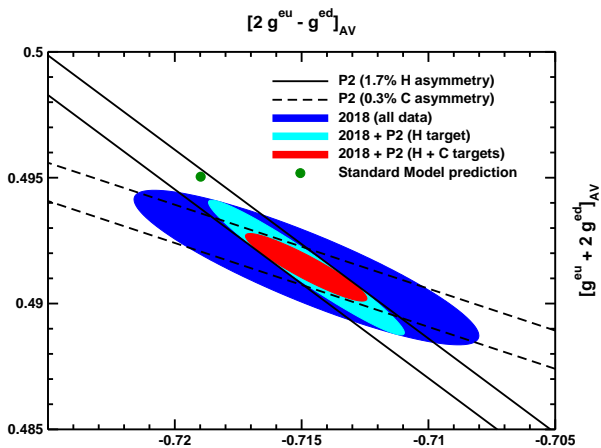}
\caption{
  1~$\sigma$ constraints on the 
  quark-vector and electron-axial-vector couplings, where the 
  ordinate corresponds to the valence quark combination of the 
  neutron, and the abscissa to the charge weighted sum of the 
  up and down quarks entering the photon-interference term in 
  polarized deep-inelastic scattering (DIS). The solid and 
  dashed lines indicate the constraints of anticipated P2 
  measurements on hydrogen and carbon, respectively. The blue 
  contour shows the present constraints from atomic parity 
  violation in Cs and Tl, DIS (SLAC and JLab) and the 
  anticipated result of the QWeak experiment~\cite{Carlini-PANIC}. 
  The cyan-colored contour adds the future 
  hydrogen measurement at P2 (assuming the central value remains
  unchanged), while the red contour includes 
  in addition the possible P2 carbon measurement. The Standard 
  Model prediction is also shown.}
\label{fig:JEcouplingsplot}
\end{figure}

One can also discuss the implications of weak charges in 
a model-independent way. In the effective field theory picture, 
the Standard Model may be defined by the most general Lagrangian 
consistent with gauge and Lorentz invariance built from the 
known particles up to dimension four, while the weak charges 
probe specific (combinations of) dimension six operators. In 
photon-interference experiments only vector and axial-vector 
Lorentz structures are important, and in the elastic regime 
the nucleus couples vector-like and parity-violation then 
forces the electron to enter axial-vector-like. Constraints 
on the quark-vector and electron-axial-vector couplings are 
illustrated in Fig.~\ref{fig:JEcouplingsplot}.

\subsubsection{Achievable precision}

The achievable precision of $\sin^2 \theta _W$ was determined 
numerically as described in section \ref{sec:MethodUncertainty}. 
The underlying Eqs.~\eqref{eq:Aexp} and \eqref{eq:Apv_averaged} 
have been modified 
appropriately for the case of scattering with a ${}^{12}\text{C}$ 
target. The beam energy was varied in the range from 
\SI{100}{~MeV} to \SI{300}{~MeV}, and we used a beam current of 
\SI{150}{\micro A}, a data taking time of \SI{2500}{h}, and 
for detector acceptance angles between \ang{2} and \ang{20}. 
We assumed a beam polarization of \SI{85}{\percent} with a 
relative error of \SI{0.3}{\percent}.

\begin{figure}
\centering
\includegraphics[width=0.48\textwidth]{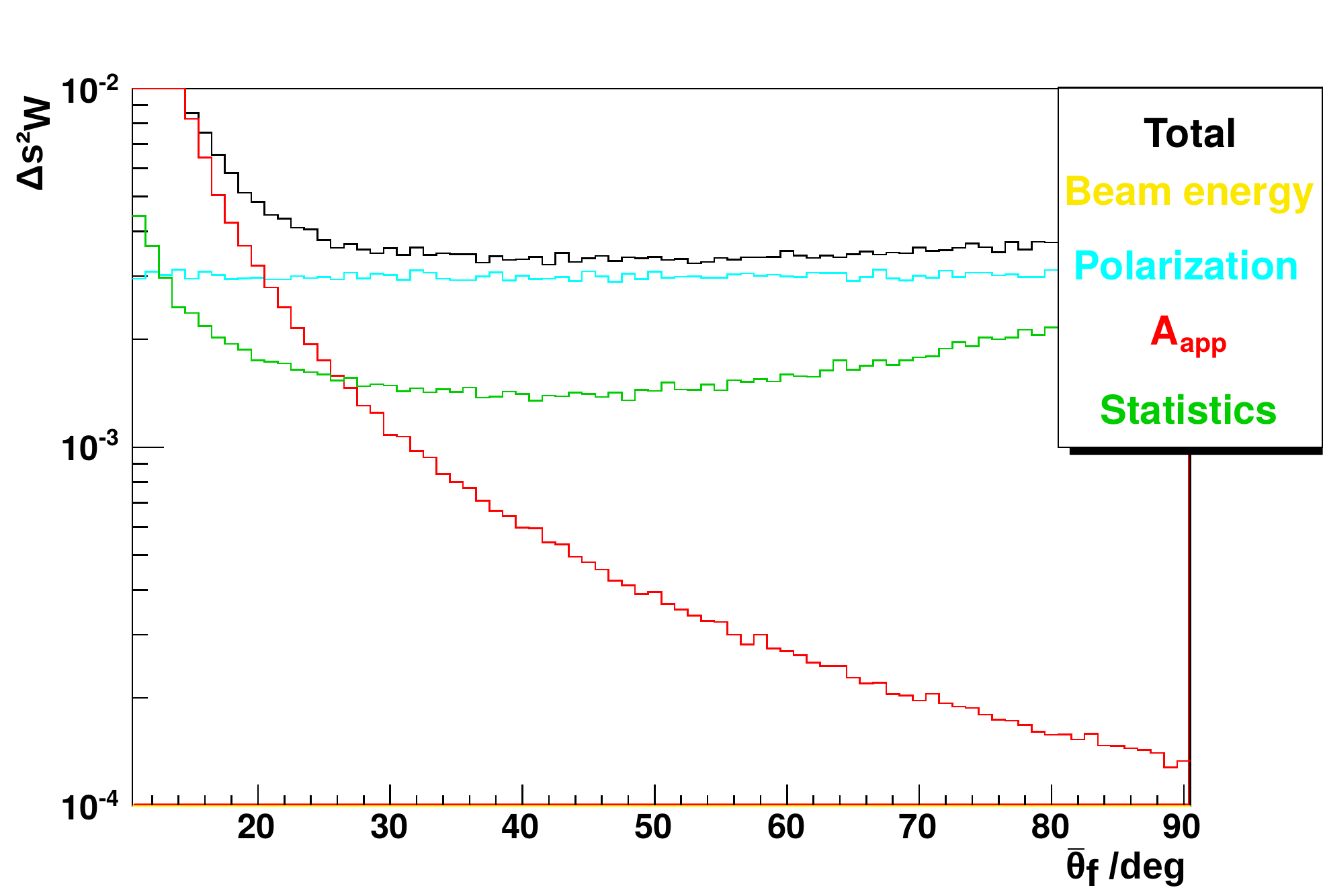}
\caption{
Monte Carlo determination of the achievable precision in 
$\sin^2 \theta_W$. Example graph for the simulated scattering 
angle dependence of the relative precision of $\sin^2 \theta_W$ 
at fixed beam energy (\SI{150}{MeV}) and detector acceptance 
angle (\ang{18}).
}
\label{fig:precisiontheta}
\end{figure}

Figure~\ref{fig:precisiontheta} shows the achievable precision 
for a fixed beam energy of \SI{150}{MeV} and an acceptance 
angle of $\delta\theta_\text{f} = \ang{18}$ as a function of 
the average scattering angle $\bar{\theta}_\text{f}$. It 
demonstrates that the total error in $\sin^2 \theta_W$ is 
dominated by the contribution of the beam polarization 
uncertainty (cyan) rather than by the statistical error (green).

\begin{figure}
\centering
\includegraphics[width=0.48\textwidth]{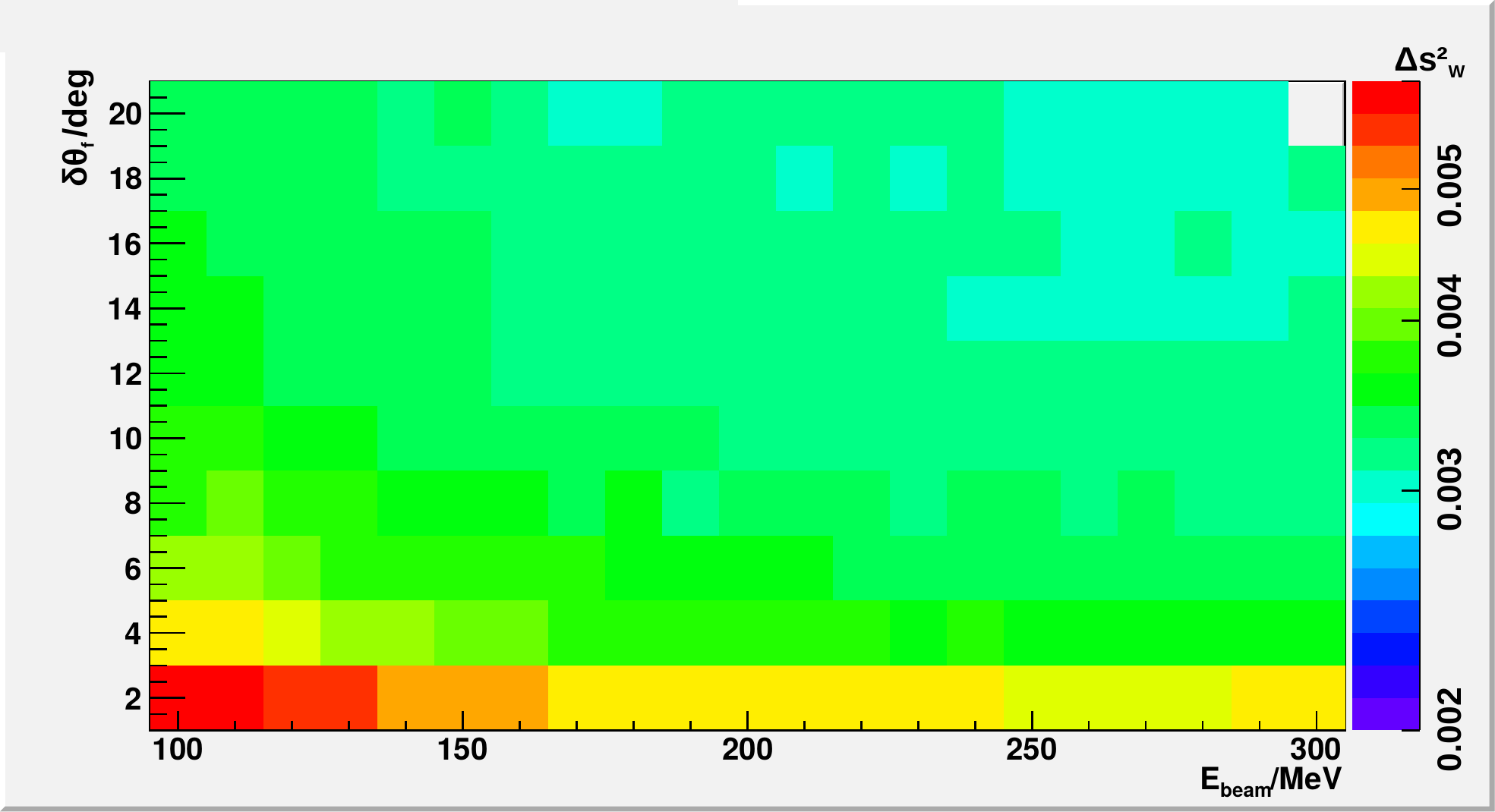}
\caption{
Monte Carlo determination of the achievable precision 
in $\sin^2 \theta_W$. Best values of $\Delta \sin^2 
\theta_W$ for a range of beam energies and acceptance 
angles are shown.
}
\label{fig:precisionsummary}
\end{figure}

Figure~\ref{fig:precisionsummary} summarizes the minimum values 
for the total error in $\sin^2 \theta_W$ for all simulated beam 
energies and scattering angles. It shows that we can obtain a 
relative error of 0.3\,\% with high detector acceptance angles 
and beam energies of \SI{150}{~MeV}. As the weak charge of the 
$^{12}$C nucleus is proportional to $\sin^2 \theta_W$, this 
corresponds to a relative error in the weak charge of $^{12}$C 
of 0.3\,\%.

\subsubsection{Experimental setup}

For the $^{12}$C experiment, the P2 hydrogen target in Fig.~\ref{fig:tgtloop}
can be replaced by a 5-finger graphite target to ensure high 
luminosities while suppressing double scattering inside the target.

\begin{figure}
\centering
\includegraphics[width=0.48\textwidth]{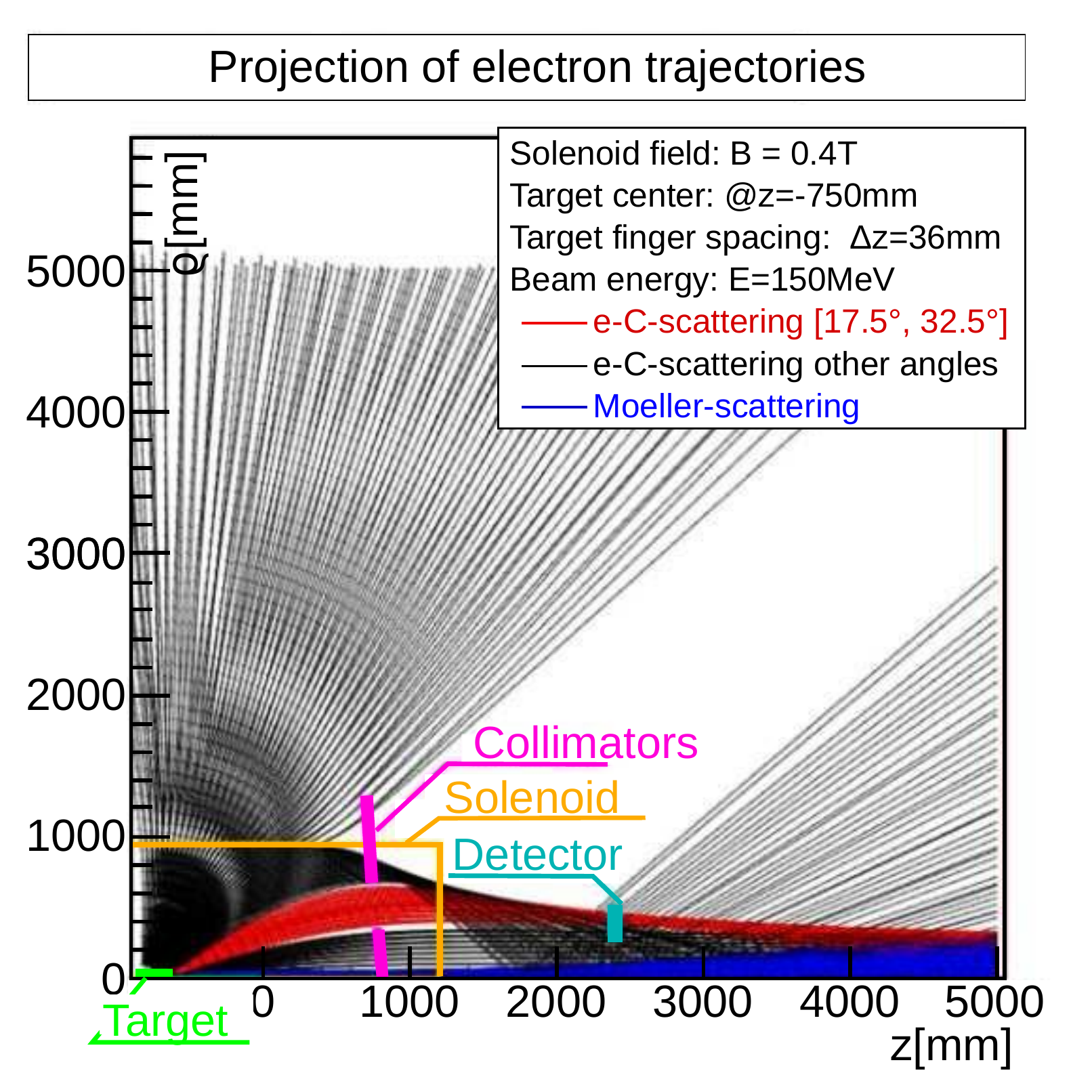}
\caption{
Geant4 ray tracing plot showing the distance of the electron 
trajectories from the $z$-axis. Separation of M\o ller background 
(blue) and electrons that were elastically scattered by other 
angles (black).
}
\label{fig:raytracer1}
\end{figure}

In order to find a suitable spectrometer, detector, and 
collimator setup, a Geant4 ray tracing study was carried out. 
Figure~\ref{fig:raytracer1} shows the distance of the electron 
trajectories from the beam axis $\rho$. Such plots were created 
for a variety of target positions and for spectrometer magnetic 
fields ranging from 0.1 T to 1 T. When studying the results of 
the simulation, a setting with which we can place the detector 
at a focal point of the electrons of interest and with which 
the undesired background can be collimated was sought. Possible 
detector and collimator positions are included in 
Fig.~\ref{fig:raytracer1}.

\begin{figure}
\centering
\includegraphics[width=0.48\textwidth]{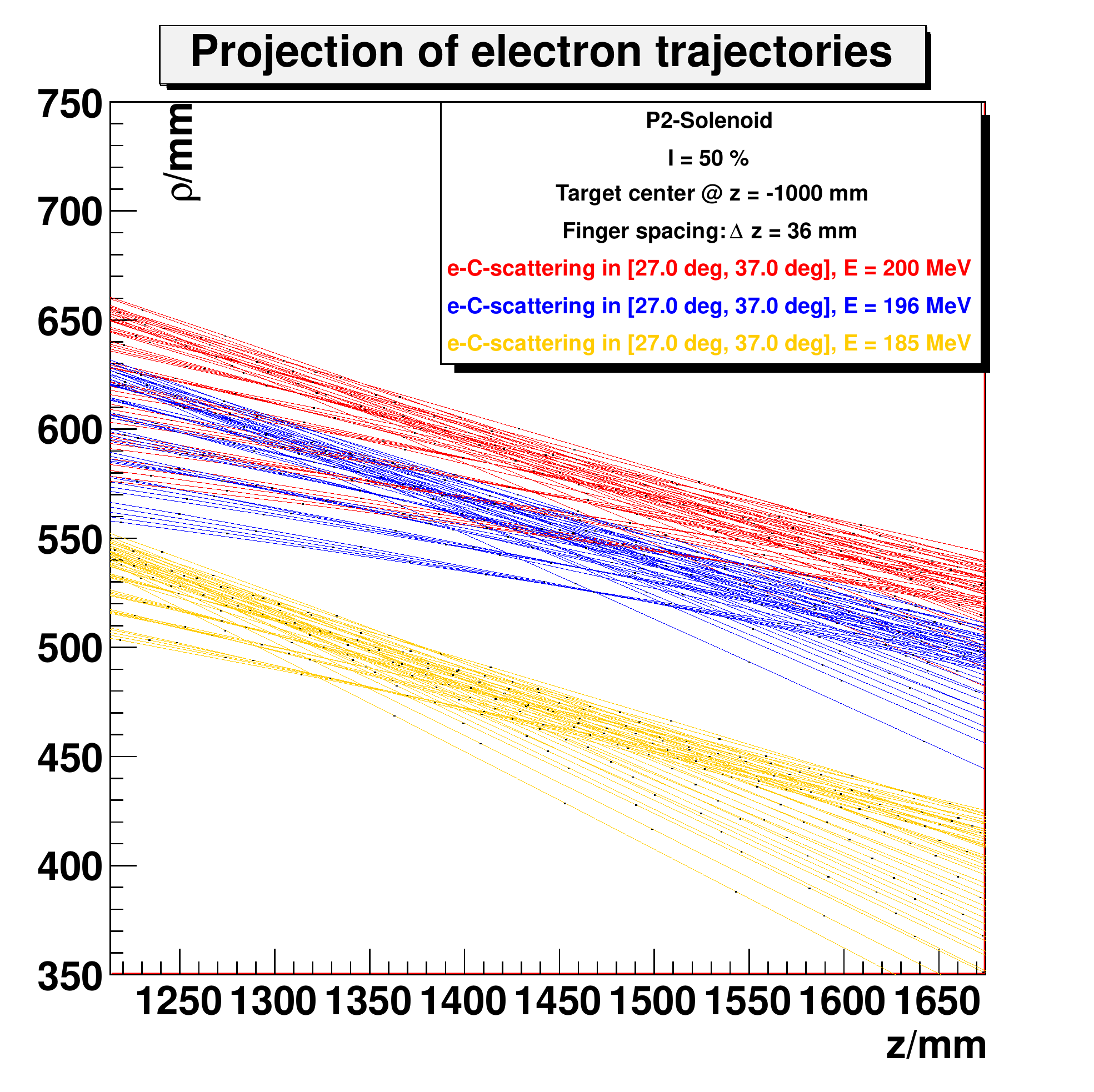}
\caption{
Magnification of the electron trajectories close to the 
position of the detector in Fig.~\ref{fig:raytracer1}. 
Trajectories of electrons scattered off the $^{12}$C ground 
state (red) are separated from those where $^{12}$C makes a 
transition into its first (blue) or second (yellow) excited 
state. 
}
\label{fig:raytracer2}
\end{figure}

Furthermore it is crucial to sort out inelastically scattered 
electrons. The plot in Fig.~\ref{fig:raytracer2} shows the
trajectories from elastically scattered electrons with a beam 
energy of \SI{200}{MeV} in red. Trajectories of the electrons 
that have excited the nucleus to the first and second excited 
states are shown in blue and yellow, respectively. In these 
cases, the energy of the electrons is effectively reduced to 
\SI{196}{MeV} (for the first excited state) or \SI{185}{MeV} 
(for the second excited state). The simulations show that the 
trajectories are well separated in all cases.

\subsubsection{Conclusion}

A measurement of the weak charge of the $^{12}$C nucleus at 
MESA can be made with a relative precision of 0.3\,\% assuming 
a measurement time of one fourth of that for P2 with a proton 
target. This corresponds to a precision in the weak mixing 
angle of 0.3\,\%. The $^{12}$C experiment will provide 
additional and complementary sensitivity to certain classes 
of new physics models. Furthermore, already individually 
the carbon and proton measurements by P2 will provide the 
strongest constraints (95\%~CL sensitivities of around 50~TeV 
in strong coupling scenarios \cite{Erler:2014fqa}) on any 
CP-allowed four-fermion operator built from first generation 
fermions. Combined, the operator corresponding to the coupling 
combination along the minor axis of the red ellipse in 
Fig.~\ref{fig:JEcouplingsplot} would be probed up to  
\SI{60}{TeV}. First conceptual studies regarding the 
feasibility of the project give promising results.

%% file: neutronskin.tex

High precision parity-violating electron scattering experiments 
on nuclei provide a portal to the properties of neutron-rich 
matter. Unfortunately, it is difficult to study neutron matter 
directly in the laboratory because the neutron and extremely 
neutron-rich isotopes are unstable. However, heavy nuclei are 
expected to develop a neutron-rich skin where many neutrons 
collect near the surface. Since the parity-violating asymmetry 
is particularly sensitive to the neutron density, it provides 
a clean and model independent measurement of the neutron skin 
of nuclei.

The neutron-skin thickness of a nucleus $\Delta R_\text{np}$, 
defined as the difference between the neutron and proton 
rms-radii is strongly related to the poorly-known symmetry 
energy at saturation density $\rho_{0}$. The symmetry energy 
is a key parameter of the nuclear Equation of State 
(EoS)~\cite{RocaMaza:2011pm}, since it quantifies the changes 
in nuclear energy associated with modifications of the 
neutron-proton asymmetry. It determines as well the properties 
of spectacular astrophysical phenomena such as supernovae 
explosions or neutron star mergers. 

After the first observations of two colliding neutron 
stars~\cite{TheLIGOScientific:2017qsa}, 
the emergent multi-messengers astronomy field will provide 
us with new ways to constrain the EoS and the properties of 
neutron-rich matter. Even stronger constraints can be imposed 
by combining this information with independent measurements 
like the precise determination of the neutron-skin thickness 
of heavy nuclei.

Though intensive experimental effort has been made to determine 
the neutron-skin thickness, a precise measurement of this 
quantity remains elusive. Several observables sensitive to 
the neutron skin have been proposed and recent experiments 
have been successful in measuring giant- and pygmy-resonance 
modes~\cite{Carbone:2010az} on a variety of nuclei as well as 
the electric-dipole polarizability~\cite{Tamii:2011pv} and 
coherent pion photoproduction~\cite{Tarbert:2013jze}. These 
data together with data from hadron scattering experiments 
(involving protons~\cite{Zenihiro:2010zz}, 
anti-protons~\cite{Klos:2007is} and pions~\cite{GarciaRecio:1991wk}) 
are valuable, but interpretations contain implicit model 
dependences.

The measurement of parity-violating asymmetries provides a 
clean and model-independent determination of neu\-tron-skin 
thicknesses. In Born approximation the parity-violating 
asymmetry $A^\text{PV}$ is proportional to the weak form 
factor $F_\text{W}(Q^{2})$, 
\begin{equation}
A^\text{PV} = 
\frac{\sigma^{+}-\sigma^{-}}{\sigma^{+}+\sigma^{-}} 
\approx 
\frac{G_\text{F}Q^{2}}{4\pi\alpha_\text{em}\sqrt{2}} 
\frac{F_\text{W}(Q^{2})}{F_\text{Ch}(Q^{2})}.
\end{equation} 
$F_\text{Ch}(Q^{2})$ is the Fourier transform of the charge 
density which is known. A measurement of the parity-violating 
asymmetry will therefore allow one to determine the weak-charge 
density $F_\text{W}(Q^2)$, from which the weak radius 
\begin{equation}
R_\text{W}^{2} = 
- \frac{6}{F_\text{W}(0)} 
\left. \frac{d F_\text{W}}{d Q^2} \right|_{Q^2 = 0}
\end{equation} 
and thus the neutron radius $R_\text{n}$ can be obtained. 
Taking the known charge radius into account, the neutron-skin 
thickness can be determined, for details see 
Refs.~\cite{Horowitz:1999fk,Horowitz:2013wha}.
The $^{208}$Pb Radius EXperiment (PREX) at JLab has provided the 
first proof-of-principle of the application of parity-violating 
electron scattering for the measurement of the neutron-skin 
thickness~\cite{Abrahamyan:2012gp}. The experiment achieved the 
systematic error goals of \SI{2}{\%} and was a major accomplishment 
as a first measurement of its kind. However, because of various 
problems, the experiment took only $\approx$\SI{15}{\%} of the 
planned statistics. With all corrections, the measured asymmetry 
is $A^\text{PV}=0.656\pm 0.060(\mathrm{stat.}) 
\pm 0.014(\mathrm{sys.})$ ppm at $Q^2$ = 0.00906 
GeV$^2$. This corresponds to a value for the neutron 
skin of $^{208}$Pb of $\Delta R_\text{np} = (0.33^{+0.16}_{-0.18})$ 
fm and confirmed the existence of a neutron-radius excess with a 
2$\sigma$ statistical significance.

\begin{figure}
	\centering
	\includegraphics[width=0.48\textwidth]{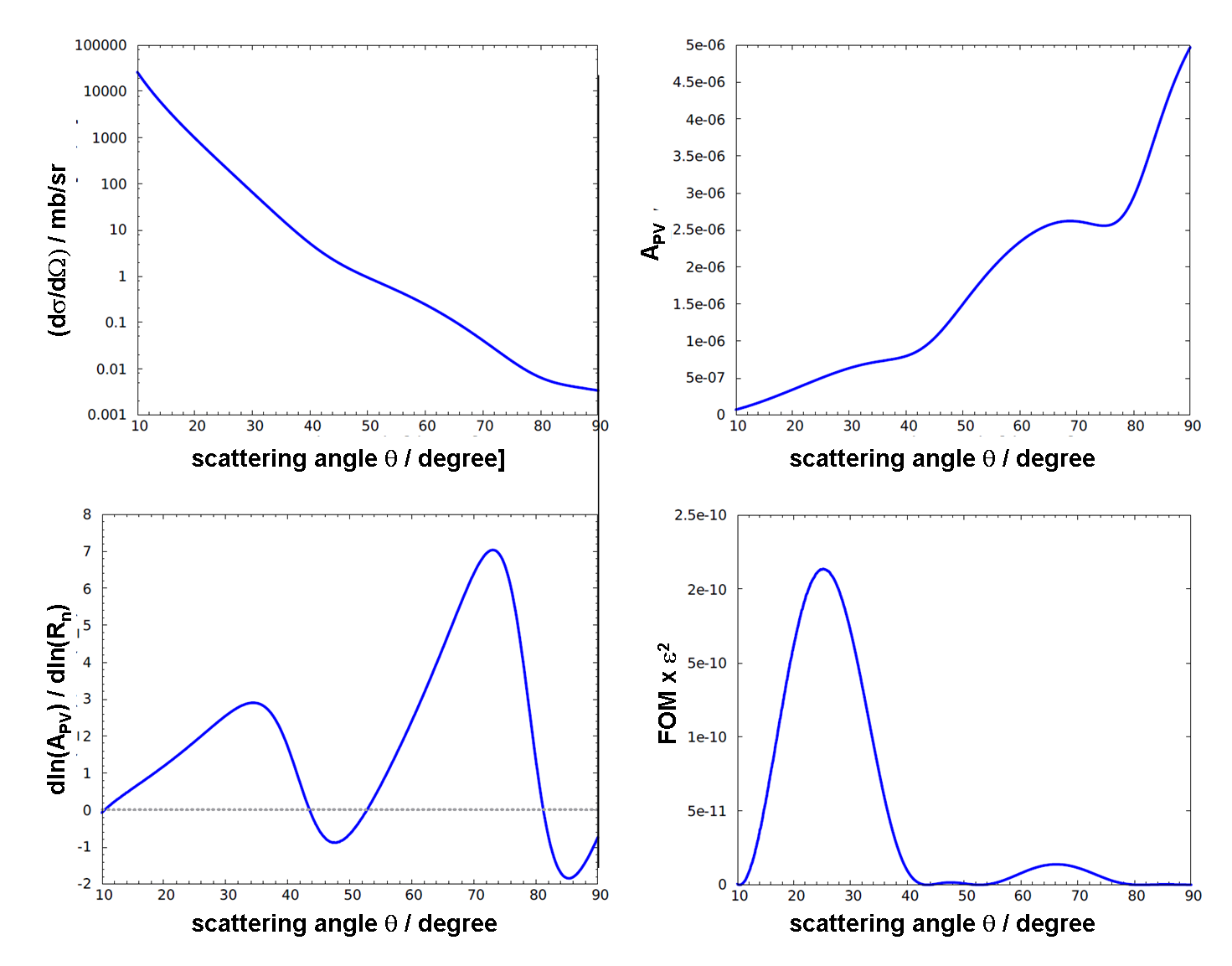}
	\caption{
	All plots are for the elastic scattering of electrons off 
	$^{208}$Pb at a beam energy of \SI{155}{MeV} and are based 
	on calculations described in Ref.~\cite{horowitz2015}. 
	Top left: Cross section as a function of the scattering 
	angle $\theta$. Top right: Parity-violating asymmetry 
	$A^\text{PV}$ as a function of scattering angle $\theta$. 
	Bottom left: Sensitivity $\varepsilon$ of the parity-violating 
	asymmetry $A^\text{PV}$ to changes in the neutron radius 
	$R_\text{n}$ as a function of the scattering angle $\theta$. 
	Bottom right: Resulting figure of merit times $\varepsilon^{2}$.
	}
	\label{fig:FoM}
\end{figure}

The P2 experiment will open the window for a new generation 
of high-precision parity-violating electron-scat\-tering experiments. 
Within the scope of the P2 experimental setup, the Mainz Radius 
EXperiment (MREX) will determine the neutron-skin thickness of 
$^{208}$Pb with ultimate precision. Figure~\ref{fig:FoM} shows 
the expected cross section, parity-violating asymmetry, and 
sensitivity as well as the resulting figure of merit (FOM) for 
a beam energy of \SI{155}{MeV}. 

\begin{figure}
	\centering
	\includegraphics[width=0.48\textwidth]{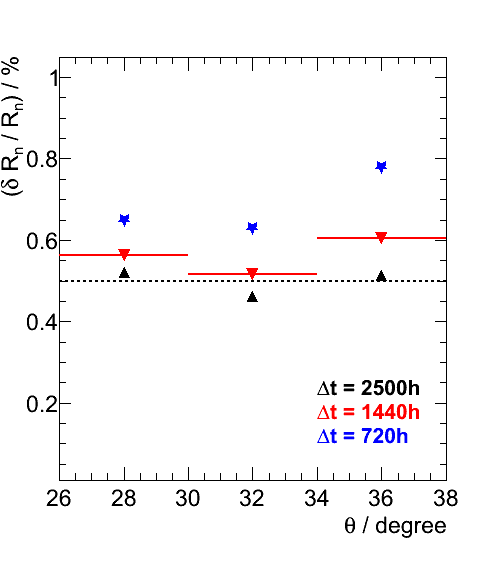}
	\caption{
	Running time estimate for a detector acceptance of 
	$\Delta\theta = \ang{4}$ indicated by the width of the 
	horizontal red lines. A systematic error of \SI{1}{\%} 
	was included in the calculation.}
	\label{fig:TimeEstimate}
\end{figure}

Here, the relevant figure of merit for a neutron skin measurement, 
for a given energy, is:
\begin{equation}
\text{FOM}\times \varepsilon^{2} = 
\frac{d\sigma}{d\Omega}\times \left(A^\text{PV}\right)^{2} 
\times \varepsilon^{2}.
\end{equation} 
In addition to the typical FOM for parity-violation experiments, 
the sensitivity $\varepsilon$ of $A^\text{PV}$ to changes in 
the neutron radius $R_\text{n}$ was taken into 
account~\cite{Horowitz:1999fk}:
\begin{equation}
\varepsilon = 
\frac{d\ln(A^\text{PV})}{d\ln(R_\text{n})} = 
\frac{R_\text{n}}{A^\text{PV}} 
\frac{\delta A^\text{PV}}{\delta R_\text{n}}.
\end{equation}
To estimate the best possible achievable precision with MREX, 
calculations within the peak sensitivity in the polar angular 
range have been performed using three different settings for 
the detector acceptance: $\Delta \theta=\ang{2},\ang{4},$ and 
$\ang{8}$. A beam polarization of at least \SI{85}{\%} and a 
systematic uncertainty of \SI{1}{\%} in the determination of 
$A^\text{PV}$ was taken into account for the calculation. 
Compared to the given error contribution in 
Tab.~\ref{tab:achievable_precision}, a rather conservative 
estimation for the size of the systematic uncertainty was used. 
The best sensitivity was obtained with a detector acceptance 
of $\Delta\theta=\ang{4}$ in the angular range between \ang{30} 
and \ang{34}. The running time for the latter one was restricted 
to 2500 hours to fit the overall beam-time planning of the MESA 
accelerator. Already after 1500 hours a determination of the 
neutron radius with a sensitivity of $\delta R_\text{n}/R_\text{n} 
= \SI{0.52}{\%}$ can be achieved (see Fig.~\ref{fig:TimeEstimate}). 
The main parameters used for the calculation are summarized in 
Tab.~\ref{tab:overview}.
\begin{table}
  \centering
  \begin{tabular}{cc}
	\toprule[1.5pt]
	Beam energy & $\SI{155}{MeV}$ \\
	Beam current & $\SI{150}{\micro A}$ \\
	Target density & $\SI{0.28}{g/cm^2}$ \\
	Polar angle step size & $\Delta\theta=\ang{4}$ \\
	Polar angular range & \ang{30} to \ang{34} \\
	Degree of polarization & \SI{85}{\%} \\
	Parity-violating asymmetry & \SI{0.66}{ppm}\\
	Running time & 1440 hours \\
	\hline
	Systematic uncertainty & \SI{1}{\%}\\
	$\delta A^\text{PV}/A^\text{PV}$ & \SI{1.39}{\%} \\
	$\delta R_\text{n}/R_\text{n}$ & \SI{0.52}{\%}\\
  \bottomrule[1.5pt]
  \end{tabular}
  \caption{
  Kinematical values and general parameters used for the run time 
  estimate.}
  \label{tab:overview}
\end{table}

\begin{figure}
	\centering
	\includegraphics[width=0.48\textwidth]{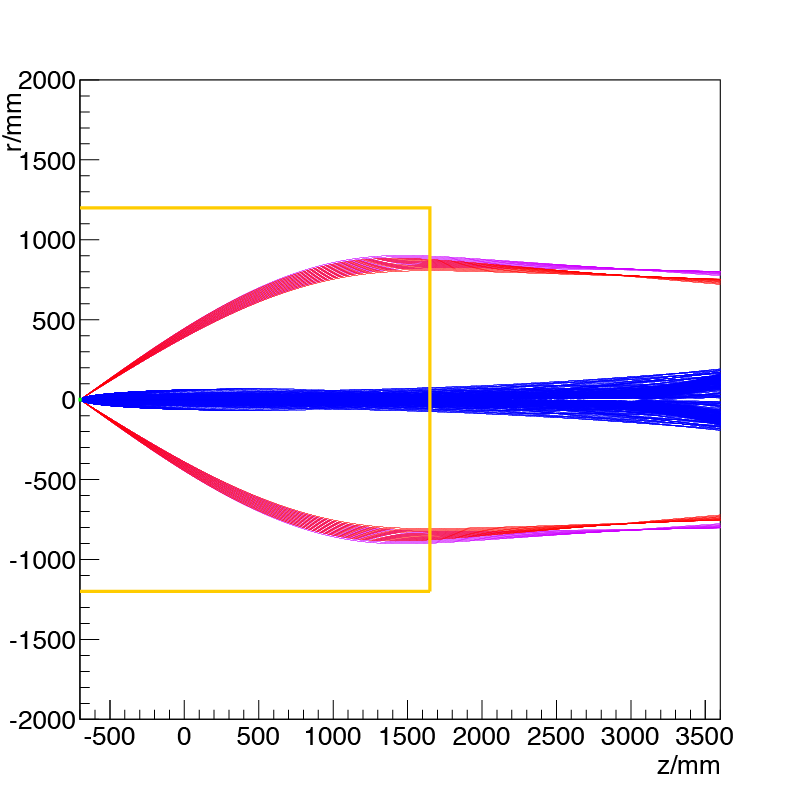}
	\caption{
	Detector configuration (option I) determined by the tracking 
	simulation. Electrons originating from the target at $z$ = 
	\SI{-700}{mm} with \SI{155}{MeV} (from elastic scattering on 
	the $^{208}$Pb target) are shown in magenta, electrons with 
	\SI{152}{MeV} (belonging to the first excited state) in red 
	and M\o{}ller electrons in blue. The yellow lines represent 
	the dimension of the solenoid.}
	\label{fig:option1}
\end{figure}

\begin{figure}
	\centering
	\includegraphics[width=0.48\textwidth]{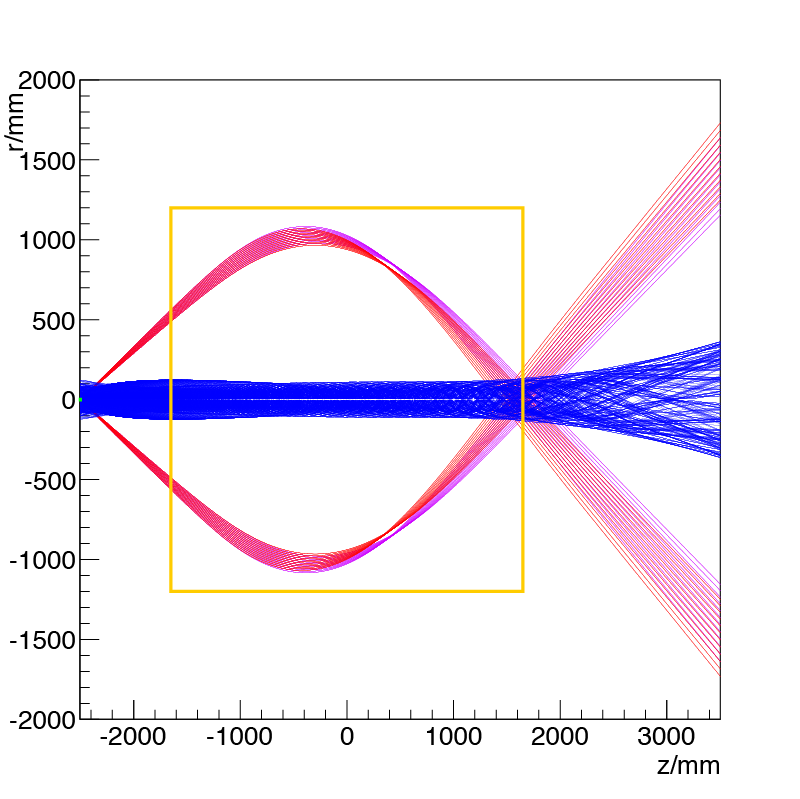}
	\caption{
	Detector configuration (option II) determined by the tracking 
	simulation. Electrons originating from the target at $z$ = 
	\SI{-2500}{mm} with \SI{155}{MeV} (from elastic scattering on 
	the $^{208}$Pb target) are shown in magenta, electrons with 
	\SI{152}{MeV} (belonging to the first excited state) in red 
	and M\o{}ller electrons in blue. The yellow lines represent 
	the dimension of the solenoid.}
	\label{fig:option2}
\end{figure}

A crucial requirement for the experiment to correctly measure 
the parity violating asymmetry is the separation of the ground 
state of $^{208}$Pb from its first excited state ($\Delta E 
\approx \SI{3}{MeV}$). Therefore tracking simulations of signal 
and background particles in the angular range of interest 
(\ang{30} to \ang{34}) have been performed for a solenoid with 
a magnetic field strength of $B=\SI{0.6}{T}$. For this study the 
existing trajectory simulation (see Sect.~\ref{sec:Simulation 
of trajectories in the magnetic field} for details) was adapted 
and the extended hydrogen target was replaced by a point-like 
lead target. Trajectories were calculated for the elastically 
scattered electrons with an energy of \SI{155}{MeV} as well as 
for electrons with an energy of \SI{152}{MeV}, mimicking events 
from the first excited state of $^{208}$Pb. The target position 
was varied between $z=\SI{-2500}{mm}$ and $z=\SI{-500}{mm}$ 
using a step size of \SI{100}{mm}. With two configurations it 
was possible to get a clean separation between the elastic and 
inelastic events. The two options with the target placed at 
$z=\SI{-700}{mm}$ and $z=\SI{-2500}{mm}$ are illustrated in 
Fig.~\ref{fig:option1} and Fig.~\ref{fig:option2}, respectively.

In both cases it is possible to block the inelastic events with a set of 
collimators. While both options need the same amount of space behind the 
solenoid to position the detectors, option~II requires additional space 
in front of the solenoid, adding a further experimental requirement to 
the changes in the P2 hall. Moreover, option~I is similar to the detector 
configuration intended 
for the measurement of the weak mixing angle and could allow the 
use of a combined scattering chamber. This would simplify the structural 
alteration works during the two experimental campaigns. Implementation of the 
lead target inside the planned scattering chamber as well as a more detailed 
Geant4 simulation, including radiative corrections, are currently being 
performed to finalize the design.

The preliminary results reported here show that a \SI{0.5}{\%} measurement 
($\delta R_{n}/R_{n}$) of the neutron skin thickness of $^{208}$Pb using the P2 
setup is feasible. 
\begin{figure}
	\centering
	\includegraphics[width=0.48\textwidth]{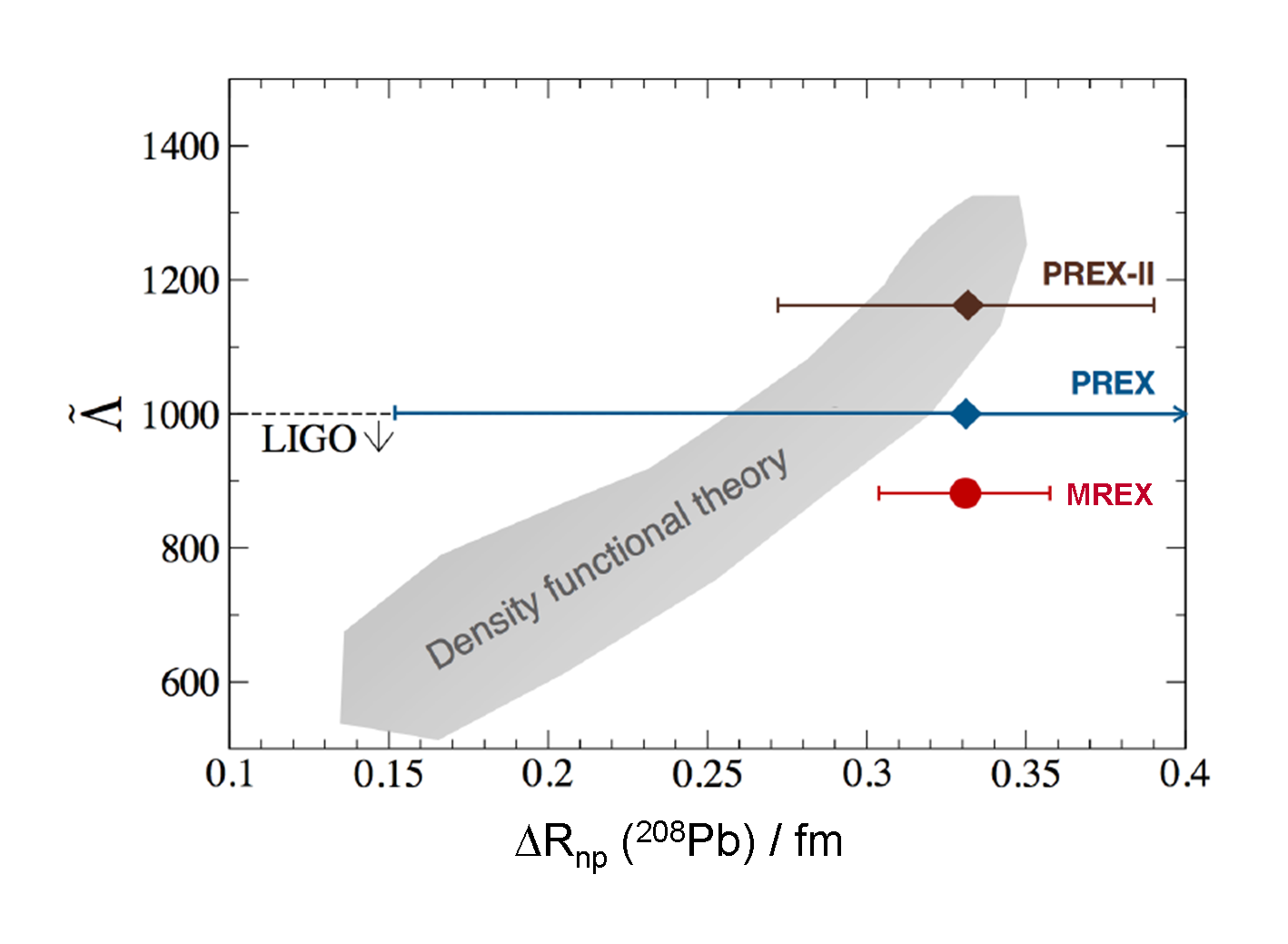}
	\caption{
	Constraints provided on density functional theory by combining 
	the tidal deformability parameter $\Tilde{\Lambda}$ of two 
	spiraling neutron stars observed at 
	LIGO~\cite{TheLIGOScientific:2017qsa} with present and future 
	$^{208}$Pb neutron skin measurements. Projected precisions of 
	PREX-II~\cite{prexIIproposal} and MREX are shown centered at 
	the measured PREX value~\cite{Abrahamyan:2012gp}. 
	Figure: Courtesy of C.~J.~Horowitz.}
	\label{fig:ligo-skin}
\end{figure}

Figure~\ref{fig:ligo-skin} shows the predicted MREX sensitivity 
together with the first calculation of the tidal deformability 
parameter $\Tilde{\Lambda}$ of two spiraling neutron stars observed 
by LIGO~\cite{TheLIGOScientific:2017qsa} as a function of the neutron 
skin thickness of $^{208}$Pb within different density functional 
models. Of course such correlations are only approximate since the 
neutron skin in $^{208}$Pb depends on the EoS at about $0.7\rho_{0}$, 
while the radius of a neutron star and its deformability depend on 
the EoS at about twice the nuclear density.  However, in general 
the higher the pressure at low density the larger the skin and the 
higher the pressure at higher densities the bigger the deformability.  
The upper \SI{90}{\%} bound observed by LIGO is consistent with 
rather thin neutron-skin thicknesses. The future PREX-II experiment 
at JLab~\cite{prexIIproposal} with its anticipated precision of 
\SI{1}{\%} together with additional measurement of neutron star 
mergers at LIGO will clarify the current picture.  

Moreover, the improvement in accuracy aimed for by the MREX will 
provide stringent constraints to such correlation and thus lead 
to a deeper understanding of neutron-rich matter.

%% file: backwardangle.tex

As seen in Eq.~(\ref{eq:Apv}), the parity-violating asymmetry $A^\text{PV}$
depends not only on the weak charge of the proton, but also on its hadronic structure.
Since $F(Q^2=0)=0$, a measurement at low momentum transfer $Q^2\ll 0.1$~GeV$^2$ 
is mainly sensitive to the weak charge $Q_\text{W}(\text{p})$. Nevertheless, 
for any $Q^2>0$ the hadronic contribution cannot be neglected. $F(Q^2)$ can 
be split up into the three terms $F^\text{EM}(Q^2)$, $F^\text{A}(Q^2)$ and 
$F^\text{S}(Q^2)$, see Eqs.~(\ref{eq:F(Q2)}-\ref{eq:Fstrange}). 
In order to extract the weak charge, the term $F^\text{EM}(Q^2)$ does not 
limit the achievable
precision because the electromagnetic form factors are known at sufficiently
high precision. On the other hand, the axial form factor $G_\text{A}^{\text{p},Z}$ 
and the strange magnetic form factor $G_\text{M}^s$ have relatively
large uncertainties such that the terms $F^\text{A}(Q^2)$ and $F^\text{S}(Q^2)$ make
non-negligible contributions to the uncertainty, depending on the scattering
angle or the momentum transfer respectively. Therefore we plan a dedicated 
measurement of these
form factors within the P2 experiment. A backward-angle measurement is much more
sensitive to $G_\text{M}^s$ and $G^{\text{p},Z}_\text{A}$ compared to a 
measurement at forward angles. Two different scenarios can be considered: 
either a backward-angle
measurement in parallel to the main forward-angle experiment or dedicated
measurements at backward angles alone. The first option depends on the available
space in the experimental hall as well as on the availability of additional
detectors, readout etc. The second option would require additional beam time
in the order of \SI{2.000}{hours}.

\begin{table}
\centering
\begin{tabular}{l r}
\toprule[1.4pt]
\multicolumn{2}{c}{P2 backward-angle experiment}\\\hline
Integrated luminosity&$8.7\cdot10^7\rm{fb}^{-1}$\\
Statistical uncertainty&$\Delta A_{\rm{stat}}=0.03$~ppm\\
False asymmetries&$\Delta A_{\rm{HC}}<0.01$~ppm\\
Polarimetry&$\Delta A_{\rm{pol}}=0.04$~ppm\\\hline
{\bf Total uncertainty}&$\Delta A_{\rm{tot}}=0.05$~ppm\\
\bottomrule[1.4pt]
\end{tabular}
\caption{Performance of a possible P2 backward-angle measurement parallel 
to the P2 forward
experiment. The beam energy used for this calculation is \SI{200}{MeV}, 
the Standard Model
expectation for the asymmetry is $A^\text{PV}\approx\SI{7.5}{ppm}$.}
\label{tab:Comparison}
\end{table}

A backward-angle measurement parallel to the main experiment could be done in 
principle for the whole experiment time,
i.e., \SI{10000}{hours}. The beam energy used for the following discussion is
\SI{200}{MeV}. The backward detector covers polar and azimuthal scattering
angles of $\ang{140}\leq\theta\leq\ang{150}$ and $0\leq\phi\leq2\pi$, the
momentum transfer is $Q^2=0.1$~GeV$^2$. Table~\ref{tab:Comparison} shows what could
be achieved with such a P2 backward-angle measurement.
One can see that the asymmetry could be measured to a precision at the 
sub-percent level with the beam polarization as the main source of
uncertainty. From this asymmetry, one can derive a value for the linear
combination: 
\begin{eqnarray}
  F^\text{S} + F^\text{A} &=& 
  0.398\cdot\left(G_\text{M}^s+0.442G^{\text{p},Z}_\text{A}\right)\,\pm\, 0.011 \, .
\end{eqnarray}
Here, the form factor input $F_{\rm{EM}}=0.558\pm0.010$ is the predominant
source of uncertainty. For the P2 forward measurement, one needs as input the
linear combination $F^\text{S}+F^\text{A}=0.0040\cdot(G_\text{M}^s + 
0.691 G^{\text{p},Z}_\text{A})$. If one
scales down the linear combination from the backward-angle measurement to the P2
forward conditions, one has to keep in mind that the linear combinations
are slightly different and the momentum transfers do not match
exactly. Therefore we add an additional error for this transformation of \SI{100}{\percent}
of the error of the measured linear combination. The benefit of the backward-angle
measurement can be clearly seen: The uncertainty which is used as an input to
the P2 main experiment analysis drops from $\Delta(F^\text{S}+F^\text{A})=0.00076$ 
if no backward-angle measurement is performed down to $\Delta(F^\text{S} + 
F^\text{A})=0.00016$ using the results of the backward-angle measurement. This would 
mean an improvement by a factor of 4. 

We also considered two dedicated backward-angle measurements with \SI{1000}{hours} 
of data taking each using a hydrogen and a deuterium target. The beam energy for this
calculation is $E=\SI{150}{MeV}$, which corresponds to a momentum transfer of
$Q^2=0.06$~GeV$^2$. Combining hydrogen and deuterium results, one could obtain 
$G_\text{M}^s$ and $G^{\text{p},Z}_\text{A}$ separately, which is a valuable 
physics result by itself. 
\begin{figure}[htb]
  \centering
  \resizebox{0.48\textwidth}{!}{\includegraphics{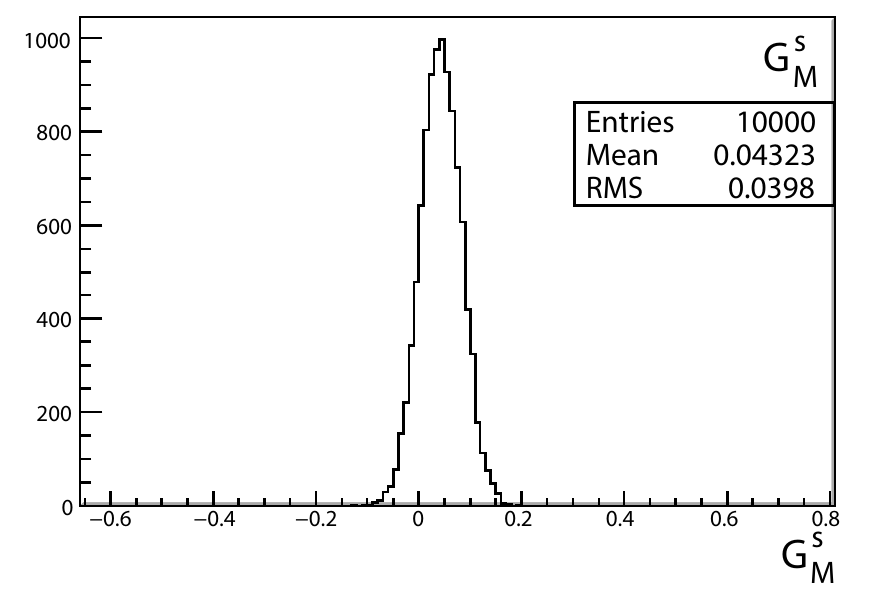}}
  \caption{Values for $G_\text{M}^s$ from dedicated backward-angle measurements 
  with a hydrogen and a deuterium target.}
  \label{fig:Backangle_GMs}
\end{figure}
\begin{figure}[htb]
  \centering
  \resizebox{0.48\textwidth}{!}{\includegraphics{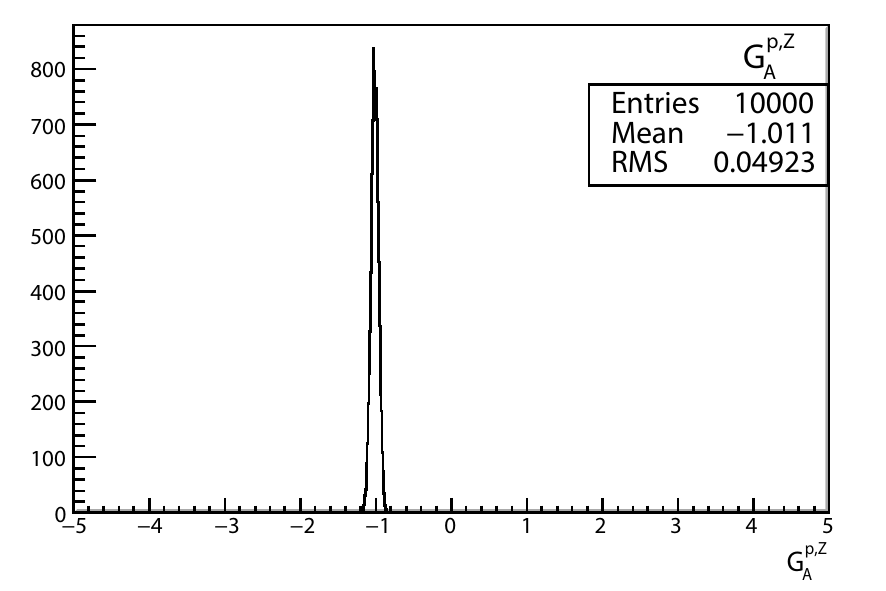}}
  \caption{Values for $G^{\text{p},Z}_\text{A}$ from dedicated backward-angle 
  measurements with a hydrogen and a deuterium target.}
  \label{fig:Backangle_GAp}
\end{figure}
In order to estimate the achievable precision in $G_\text{M}^s$ and 
$G^{\text{p},Z}_\text{A}$, all quantities that enter into their determination 
were varied according to their uncertainties. The width of the distributions 
are displayed in Fig.~\ref{fig:Backangle_GMs} and Fig.~\ref{fig:Backangle_GAp}. 
It turns out that the possible uncertainties would be $\Delta G_\text{M}^s = 
0.04$ and $\Delta G^{\text{p},Z}_\text{A} = 0.05$. The impact on the forward 
P2 experiment would be even better compared to a parallel backward-angle 
measurement, because the momentum transfer would match better and the required 
linear combination can be calculated directly from the separated form factors. 

To conclude, backward-angle measurements within the P2 experiment seem promising. 
Two options were discussed. A backward-angle measurement parallel to the 
forward P2 experiment doesn't require additional beam time, but depends on 
the available space in the experimental hall. The uncertainty contribution 
of axial and strange magnetic form factors, expressed by $F^\text{S} + 
F^\text{A}$ would drop by a factor of 4 compared to the assumptions without 
such a measurement. Separate measurements on hydrogen and deuterium targets 
seem even more promising and would yield the most precise determination of 
$G_\text{M}^s$ and $G^{\text{p},Z}_\text{A}$ at low momentum transfer.

%% file: summary.tex
This work summarizes the research and development work for the 
P2 experimental facility at the upcoming energy recovering 
recirculating accelerator MESA in Mainz. It is optimized for 
the measurement of an order 10$^{-8}$ parity-violating cross 
section asymmetry in electron scattering. This is the smallest 
asymmetry ever measured in electron scattering. Many new 
experimental techniques will be used for the first time in 
order to reach the high precision goal needed to obtain 
physics results with a high impact on the field of research. 
These are: 
\begin{itemize}
\renewcommand{\labelitemi}{$\bullet$}
\item 
a superconducting solenoid as a spectrometer for scattered 
electrons; 
\item 
HVMAPS as thin silicon tracking detectors for the $Q^2$ measurement; 
\item 
a so called Hydro-M{\o}ller polarimeter, a magnetic trap 
for M{\o}ller scattering off atomic hydrogen; 
\item 
a low energy recirculating accelerator with low number of 
recirculations and energy recovery capability. 
\end{itemize}

The P2 experimental facility has a rich program covering different 
fields like particle physics and nuclear physics. The building hosting 
the MESA accelerator will be finished according to the present construction 
plan in mid of 2020. The start of the accelerator and experiment 
commissioning is foreseen mid of the year 2021. A detailed beam-time 
plan for the experimental program has not yet been set up. This needs 
to be discussed in order to achieve a good compromise between 
commissioning of MESA and P2 and the high demands of the parity-violation 
program. Ideally, one would run experiments with large asymmetries 
and short measurement times first in order to meet the increasing 
demands of the experiments in parallel to the performance increase 
achieved by careful commissioning of accelerator, beam control systems, 
polarimetry and experimental setup. The present physics topics 
of P2 identified as topics with a high impact on the respective 
research fields, are:
\begin{itemize}
\renewcommand{\labelitemi}{$\bullet$}
\item 
Measurement of the weak charge of the proton for the determination 
of the weak mixing angle \sw with an accuracy of \SI{0.14}{\percent}. 
The expected asymmetry is $\langle A^\text{exp} \rangle_\text{Cherenkov} 
= \SI{-28.77}{ppb}$, averaged over cross section and acceptance. 
The necessary measurement time is \SI{10000}{h}. This is the most 
demanding measurement due to the small size of the asymmetry and 
the long measurement time. The target accuracy will go beyond the 
present state of the art by a factor three, corresponding to a 
factor of 10 higher statistics and a factor of three higher in all 
beam-related apparative asymmetries, and polarimetry on the level 
of 0.5\%. It has a sensitivity for new physics beyond the standard 
model with mass scales ranging from \SI{70}{MeV} up to \SI{50}{TeV} 
and complementary to the new physics searches at the LHC. 
\item 
Measurement of the weak charge of the $^{12}$C nucleus. The asymmetry 
is of order \SI{3}{ppm} due to the large weak charge of the neutrons 
in $^{12}$C and the measurement would need about \SI{2500}{h}. The 
precision goal of \SI{0.3}{\percent} is a challenge for polarimetry and 
requires additional research and development work. In combination 
with the measurement of the weak charge one can reach a still higher 
sensitivity for new physics of mass scales up to about \SI{60}{TeV}. 
\item 
Measurement of the neutron skin of the $^{208}$Pb nucleus. The asymmetry 
is of order \SI{3}{ppm} and the measurement would need about \SI{1500}{h}. 
The result is expected to yield an error on the neutron radius of 
$\delta R_\text{n}/R_\text{n}$ of \SI{0.52}{\%}. Such a measurement 
will be important for a better understanding of the physics of  
neutron stars. 
\item 
A backward-angle measurement of the parity-violating elastic scattering 
to measure the weak form factor of the nucleon. From measurements  
with hydrogen and deuterium targets, the error bars for $G_\text{M}^s$ 
and $G^{\text{p},Z}_\text{A}$ can be reduced substantially. 
\end{itemize}
The experimental program described in this manuscript comprises 
several years of pure data taking, calibration and the study of 
systematic effects like a parity-violating asymmetry from the 
aluminum windows. Together with the beam-time demand from the 
second MESA flagship facility MAGIX, one can safely say that we 
are facing 6 to 10 years of a research program at MESA. 

Technically, the beam energy of MESA could be increased up to 
\SI{200}{MeV}. The discussion of a possible research program 
for this enhanced MESA setup is ongoing. It would bring the 
pion-production threshold, both for scattering off protons and 
for heavier nuclei, into the reach of the experimental facilities 
at MESA. In particular the study of hadronic parity violation which 
has seen enormous progress in the theory sector over the last 
years will be interesting~\cite{Gardner:2017xyl,deVries:2015gea}. 

At energies above the pion production threshold, it is possible 
to directly address hadronic parity violation effects, in 
particular the parity-violating $\pi N$ coupling constant 
$h^1_{\pi NN}$. As pointed out in Refs.~\cite{Chen:2000km,Chen:2000hb} 
these effects become prominent in the near-threshold charged pion 
production with a polarized electron beam and with only the charged 
pion in the final state detected. The origin of this parity-violating 
coupling lies in the effective four-quark operators which are 
also responsible for generating the anapole moment. The anapole 
moment is the source of the largest uncertainty in the upcoming 
P2 experiment. Furthermore, measuring the parity-violating 
asymmetry in inelastic electron scattering between the 
pion production threshold and the $\Delta(1232)$ resonance, 
as argued in Ref.~\cite{Gorchtein:2015naa}, offers an enhanced 
sensitivity to strange form factors of the nucleon.  

Finally, the increased energy reach of MESA with a polarized beam 
would make it possible to perform a new measurement of the 
parity-conserving, beam normal spin asymmetry in elastic ep 
scattering at backward angles. This observable is a single-spin 
asymmetry with the electron polarization normal to the scattering 
plane. It is a purely electromagnetic effect which contributes an 
important systematic uncertainty to parity-violating electron 
scattering. The earlier measurement at \SI{200}{MeV} and \ang{150} 
at MIT-Bates~\cite{Wells:2000rx} are not fully understood: the 
observed large asymmetry is at variance with the much smaller 
theory expectation~\cite{Pasquini:2004pv} which is, at this 
energy, essentially free of uncertainties. 
\\

The upcoming new Mainz P2 experimental facility will bring 
very interesting times in the measurement of parity violating 
electron scattering.

%% file: Acknowledgements.tex
This work is supported by the Deutsche Forschungsgemeinschaft 
(DFG) in the framework of the collaborative research center 
SFB1044 ``The Low-Energy Frontier of the Standard Model: From 
Quarks and Gluons to Hadrons and Nuclei'', and in the 
framework of the PRISMA cluster of excellence ``Precision 
Physics, Fundamental Interactions and structure of Matter''. 
Silviu Covrig acknowledges support by an Early Career 
Award from the DOE. 
Jens Erler acknowledges support in part by the German-Mexican 
research collaboration grant 278017 (CONACyT) and SP 778/4-1 
(DFG), as well as by CONACyT project 252167-F. 
Michael Gericke is partially supported by funding received 
through the Natural Sciences and Engineering Research Council 
of Canada (NSERC) and the Canadian Foundation for Innovation 
(CFI). 
Mikhail Gorchtein acknowledges support by the DFG under the 
personal grant GO 2604/2-1. 
Boxing Gou would like to thank the Office of China Postdoctoral 
Council (OCPC) for financial support. 
The P2 tracker group would like to thank the Mu3e group in 
Heidelberg, in particular Frank Meier-Aeschbacher and Dirk 
Wiedner for many helpful discussions.
\\
This work comprises part of the Phd thesis work of Dominik 
Becker, Razvan Bucoveanu, Carsten Grzesik, Ruth Kempf, 
Kathrin Imai, Matthias Molitor, Alexey Tyukin, and Marco 
Zimmermann.

%% file: SinThetaW_appendix.tex


It has been assumed in Eq.~(\ref{eq:sw2_functional}) that the nucleon form
factors can be parametrized as functions $F(\{\kappa_l\}, ~Q^2)$, where
$\{\kappa_l\}$ is a set of independent, real parameters. The parametrizations
used are discribed in 
Sect.~\ref{sec:InputparameterstothecalculationsofDeltaSz2}. Here we present the
expected values $\langle\kappa_l\rangle$ and standard deviations
$\Delta\kappa_l$ of the parameters $\kappa_l$, which have been used in the
error propagation calculations.

\begin{table}[ht]
  \begin{center}
    \begin{tabular}{ccc}
      \toprule[1.5pt]
      i & $\kappa_i^\text{E}$/${(\mathrm{GeV}/c)}^{-2i}$ &
      $\Delta\kappa_i^\text{E}$/${(\mathrm{GeV}/c)}^{-2i}$ \\
      \midrule[1.5pt]
      1 & \SI{-4,701987e-01}{} & \SI{1,133586e-02}{} \\
      \midrule
      2 & \SI{4,342292e+00}{}  & \SI{6,849265e-02}{} \\
      \midrule
      3 & \SI{-2,068202e+01}{} & \SI{1,718847e-01}{} \\
      \midrule
      4 & \SI{4,406141e+01 }{} & \SI{3,152484e-01}{} \\
      \midrule
      5 & \SI{-2,474794e+01}{} & \SI{5,080538e-01}{} \\
      \midrule
      6 & \SI{-5,087120e+01}{} & \SI{7,708359e-01}{} \\
      \midrule
      7 & \SI{8,101379e+01 }{} & \SI{1,055087e+00}{} \\
      \midrule
      8 & \SI{-3,302248e+01}{} & \SI{1,047902e+00}{} \\
      \bottomrule[1.5pt]
    \end{tabular}
  \end{center}
  \caption{Parameter values used to parametrize
  $G^{\text{p},\gamma}_\text{E}$.}
  \label{tab:kappa_GpE}
\end{table}

\begin{table}[ht]
  \begin{center}
    \begin{tabular}{ccc}
      \toprule[1.5pt]
      $i$ & $\kappa_i^\text{M}$/${(\mathrm{GeV}/c)}^{-2i}$ &
      $\Delta\kappa_i^\text{M}$/${(\mathrm{GeV}/c)}^{-2i}$ \\
      \midrule[1.5pt]
      1  & \SI{2,445791e-01}{} & \SI{1,285954e-02}{} \\
      \midrule
      2  & \SI{-4,387620e+00}{} & \SI{4,832165e-02}{} \\
      \midrule
      3  & \SI{2,244408e+01}{} & \SI{8,019477e-02}{} \\
      \midrule
      4  & \SI{-4,477354e+01}{} & \SI{1,120105e-01}{} \\
      \midrule
      5  & \SI{2,507312e+01}{} & \SI{1,455939e-01}{} \\
      \midrule
      6  & \SI{3,475912e+01}{} & \SI{1,827526e-01}{} \\
      \midrule
      7  & \SI{-5,305466e+01}{} & \SI{2,105056e-01}{} \\
      \midrule
      8  & \SI{1,976824e+01}{} & \SI{1,874455e-01}{} \\
      \bottomrule[1.5pt]
    \end{tabular}
  \end{center}
  \caption{Parameter values used to parametrize $G^{\text{p},\gamma}_\text{M}$.}
  \label{tab:kappa_GpM}
\end{table}

\begin{table}[ht]
  \begin{center}
    \begin{tabular}{ccc}
      \toprule[1.5pt] i &
      $\kappa_i$ & $\Delta\kappa_i$ \\
      \midrule[1.5pt]
      1 & \SI{1,770221e+00}{} & \SI{1,454643e-02}{} \\
      \midrule
      2 & \SI{3,425350e+00}{} & \SI{2,075773e-01}{} \\
      \bottomrule[1.5pt]
    \end{tabular}
  \end{center}
  \caption{Parameter values used to parametrize $G^{\text{n},\gamma}_\text{E}$.}
  \label{tab:kappa_GnE}
\end{table}

\begin{table}
  \begin{center}
    \begin{tabular}{ccc}
      \toprule[1.5pt]
      i & $\kappa_i$/${(\mathrm{GeV}/c)}^{-2i}$ &
      $\Delta\kappa_i$/${(\mathrm{GeV}/c)}^{-2i}$ \\
      \midrule[1.5pt]
      0 & \SI{-1,916029e+00}{} & \SI{4,589687e-04}{} \\
      \midrule
      1 & \SI{7,092145e+00}{} & \SI{3,229584e-02}{} \\
      \midrule
      2 & \SI{-3,329785e+01}{} & \SI{1,602581e-01}{} \\
      \midrule
      3 & \SI{1,574668e+02}{} & \SI{4,007755e-01}{} \\
      \midrule
      4 & \SI{-4,144474e+02}{} & \SI{9,176047e-01}{} \\
      \midrule
      5 & \SI{1,627159e+02}{} & \SI{2,025616e+00}{} \\
      \midrule
      6 & \SI{1,152293e+03}{} & \SI{4,366665e+00}{} \\
      \midrule
      7 & \SI{-2,117386e+02}{} & \SI{9,120974e+00}{} \\
      \midrule
      8 & \SI{-4,908379e+03}{} & \SI{1,819254e+01}{} \\
      \midrule
      9 & \SI{5,114440e+03}{} & \SI{3,374769e+01}{} \\
      \bottomrule[1.5pt]
    \end{tabular}
  \end{center}
  \caption{Parameter values used to parametrize $G^{\text{n},\gamma}_\text{M}$.}
  \label{tab:kappa_GnM}
\end{table}

\begin{table}[ht]
  \begin{center}
    \begin{tabular}{ccc}
      \toprule[1.5pt]
      i & $\kappa_i$ &
      $\Delta\kappa_i$ \\
      \midrule[1.5pt]
      1 & \SI{3,231461e-01}{} & \SI{8,871228e-01}{} \\
      \midrule
      2 & \SI{4,704640e+00}{} & \SI{3,000726e+01}{} \\
      \bottomrule[1.5pt]
    \end{tabular}
  \end{center}
  \caption{Parameter values used to parametrize $G^\text{s}_\text{E}$.}
  \label{tab:kappa_GsE}
\end{table}

\begin{table}[ht]
  \begin{center}
    \begin{tabular}{ccc}
      \toprule[1.5pt]
      i & $\kappa_i$/${(\mathrm{GeV}/c)}^{-2i}$ &
      $\Delta\kappa_i$/${(\mathrm{GeV}/c)}^{-2i}$ \\
      \midrule[1.5pt]
      0 & \SI{4,411866e-02}{} & \SI{1,393027e-01}{} \\
      \midrule
      1 & \SI{9,312301e-01}{} & \SI{1,016812e+00}{} \\
      \bottomrule[1.5pt]
    \end{tabular}
  \end{center}
  \caption{Parameter values used to parametrize $G^\text{s}_\text{M}$.}
  \label{tab:kappa_GsM}
\end{table}

\begin{table}[ht]
  \begin{center}
    \begin{tabular}{ccc}
      \toprule[1.5pt]
      i & $\kappa_i$/${(\mathrm{GeV}/c)}^{i}$ &
      $\Delta\kappa_i$/${(\mathrm{GeV}/c)}^{i}$ \\
      \midrule[1.5pt]
      0 & \SI{-1,136}{} & \SI{0,411}{} \\
      \midrule
      1 & \SI{1,032}{} & \SI{0,036}{} \\
      \bottomrule[1.5pt]
    \end{tabular}
  \end{center}
  \caption{Parameter values used to parametrize $G^\text{p,Z}_\text{A}$.}
  \label{tab:kappa_GpA}
\end{table}

\begin{table}[ht]
  \begin{center}
    \begin{tabular}{ccc}
      \toprule[1.5pt]
      i & $\kappa_i^\text{E}$/${(\mathrm{GeV}/c)}^{-2i}$ &
      $\Delta\kappa_i^\text{E}$/${(\mathrm{GeV}/c)}^{-2i}$ \\
      \midrule[1.5pt]
      0 & \SI{1,344573e-13}{} & \SI{1,000000e-07}{}  \\
      \midrule
      1 & \SI{5,669833e-02}{} & \SI{2,772295e-02}{}  \\
      \midrule
      2 & \SI{-2,465694e-01}{} & \SI{6,866436e-01}{} \\
      \midrule
      3 & \SI{5,813392e-01}{} & \SI{4,856379e+00}{}  \\
      \midrule
      4 & \SI{-7,002228e-01}{} & \SI{1,023000e+01}{} \\
      \bottomrule[1.5pt]
    \end{tabular}
  \end{center}
  \caption{Parameter values used to parametrize $G^\text{ud}_\text{E}$.}
  \label{tab:kappa_GudE}
\end{table}

\begin{table}[ht]
  \begin{center}
    \begin{tabular}{ccc}
      \toprule[1.5pt]
      i & $\kappa_i^\text{M}$/${(\mathrm{GeV}/c)}^{-2i}$ &
      $\Delta\kappa_i^\text{M}$/${(\mathrm{GeV}/c)}^{-2i}$ \\
      \midrule[1.5pt]
      0 & \SI{2,474684e-02}{} & \SI{1,824655e-02}{}  \\
      \midrule
      1 & \SI{6,787448e-02}{} & \SI{7,769135e-01}{}  \\
      \midrule
      2 & \SI{-3,042028e-02}{} & \SI{1,003967e+01}{} \\
      \midrule
      3 & \SI{-4,367643e-01}{} & \SI{4,767653e+01}{} \\
      \midrule
      4 & \SI{8,468409e-01}{} & \SI{7,470339e+01}{}  \\
      \midrule
      \bottomrule[1.5pt]
    \end{tabular}
  \end{center}
  \caption{Parameter values used to parametrize $G^\text{ud}_\text{M}$.}
  \label{tab:kappa_GudM}
\end{table}